\documentclass[rmp,aps,twocolumn,amssymb,amsfonts,nofootinbib,showkeys,floatfix]{revtex4}
\tighten
\usepackage{float}
\usepackage{graphicx}
\usepackage{mathrsfs}
\usepackage{bm}
\usepackage{chapterbib}
\usepackage{amsmath}

\begin{document}
\title{Google matrix analysis of directed networks}
\author{Leonardo Ermann}
\affiliation{Departamento de F\'\i sica Te\'orica, GIyA, Comisi\'on Nacional 
de Energ\'ia At\'omica, Buenos Aires, Argentina}
\author{Klaus M. Frahm}
\affiliation{Laboratoire de Physique Th\'eorique du CNRS, IRSAMC, 
Universit\'e de Toulouse, UPS, 31062 Toulouse, France}
\author{Dima L. Shepelyansky}
\affiliation{Laboratoire de Physique Th\'eorique du CNRS, IRSAMC, 
Universit\'e de Toulouse, UPS, 31062 Toulouse, France}

\date{July 25, 2014; Revised: January 27, 2015}

\begin{abstract}
In the past decade modern societies have developed enormous
communication and social networks. Their classification and 
information retrieval processing has become a formidable task for the society. 
Due to the rapid growth of the World Wide Web, and social and communication
networks, new mathematical
methods have been invented to characterize the properties of these networks
in a more detailed and precise way.
Various search engines use extensively such methods.
It is highly important to develop new tools to classify and 
rank massive amount of network information in a way that is 
adapted to internal 
network structures and characteristics. 
This review describes
the Google matrix analysis of directed complex networks
demonstrating its efficiency using various examples
including World Wide Web,
Wikipedia, software architectures, world trade,
social and citation networks,
brain neural networks, DNA sequences
 and Ulam networks.
The analytical and numerical matrix methods
used in this analysis originate from the fields of Markov chains,
quantum chaos and Random Matrix theory.
\end{abstract}



\maketitle

``The Library  exists {\it ab aeterno}.''\\
Jorge Luis Borges
{\it The Library of Babel}

\tableofcontents

\section{Introduction}
\label{s1}

   In the past ten years, modern societies  have developed enormous 
communication and social networks. The World Wide Web (WWW) alone has about 
50 billion indexed web pages, so that their classification and 
information retrieval processing
becomes a formidable task.  
Various search engines 
have been developed by private companies such as Google, Yahoo! and others 
which are extensively used by Internet users.  
In addition, social networks (Facebook, LiveJournal, Twitter, etc) 
have gained huge popularity in the last few years.  
In addition, use of social networks has spread
beyond their initial purpose, making them important for
political or social events.
   
      To handle such massive databases, fundamental mathematical tools and 
algorithms related to centrality measures and network matrix properties 
are actively being developed.  Indeed, the PageRank algorithm, 
which was initially 
at the basis of the development of the Google search engine 
\cite{brin:1998,langville:2006}, 
is directly linked to 
the mathematical properties of Markov chains \cite{markov:1906} and
Perron-Frobenius operators \cite{brin:2002,langville:2006}.  
Due to its mathematical foundation, this algorithm determines a ranking order 
of nodes that can be applied to various types of directed networks. 
However, the recent rapid development of WWW and communication networks 
requires 
the creation of new tools and algorithms to characterize the properties of 
these networks on a more detailed and precise level. For example, such networks 
contain weakly coupled or secret communities which may correspond to 
very small values of the PageRank and are hard to detect.  It is therefore 
highly important to have new methods to classify and rank hige amounts of 
network information in a way adapted 
to internal network structures and characteristics.

     This review describes matrix tools and algorithms which 
facilitate classification and information retrieval from large networks 
recently created by human activity.    The Google matrix, formed 
by links of the network has, is typically huge 
(a few tens of billions of webpages).
 Thus, the analysis of its 
spectral properties including complex eigenvalues and
eigenvectors represents a challenge for analytical and numerical methods.
It is rather surprising, but the class of such matrices, which  belong
to the class of Markov chains and Perron-Frobenius operators,
has been essentially overlooked in physics.
Indeed, physical problems typically belong to the class of
Hermitian or unitary matrices. Their properties
have been actively studied in the frame of 
Random Matrix Theory (RMT) \cite{akemann:2011,guhr:1998,mehta:2004}
and quantum chaos \cite{haake:2010}.
The analytical and numerical tools developed
in these research fields have paved the way for understanding 
many universal and peculiar features
of  such matrices in the limit of large
matrix size corresponding to many-body quantum
systems \cite{guhr:1998}, quantum computers \cite{shepelyansky:2001}
and a semiclassical limit of large quantum numbers in the
regime of quantum chaos \cite{haake:2010}.
In contrast to the Hermitian problem, the Google matrices of 
directed networks have complex eigenvalues. 
The only physical systems where similar
matrices had been studied analytically
and numerically correspond to models of quantum chaotic
scattering whose spectrum is known to have such unusual properties 
as the fractal Weyl law 
\cite{gaspard:2014,sjostrand:1990,nonnenmacher:2007,shepelyansky:2008,zworski:1999}.

\begin{figure}[H]
\begin{center}
\includegraphics[width=0.48\textwidth]{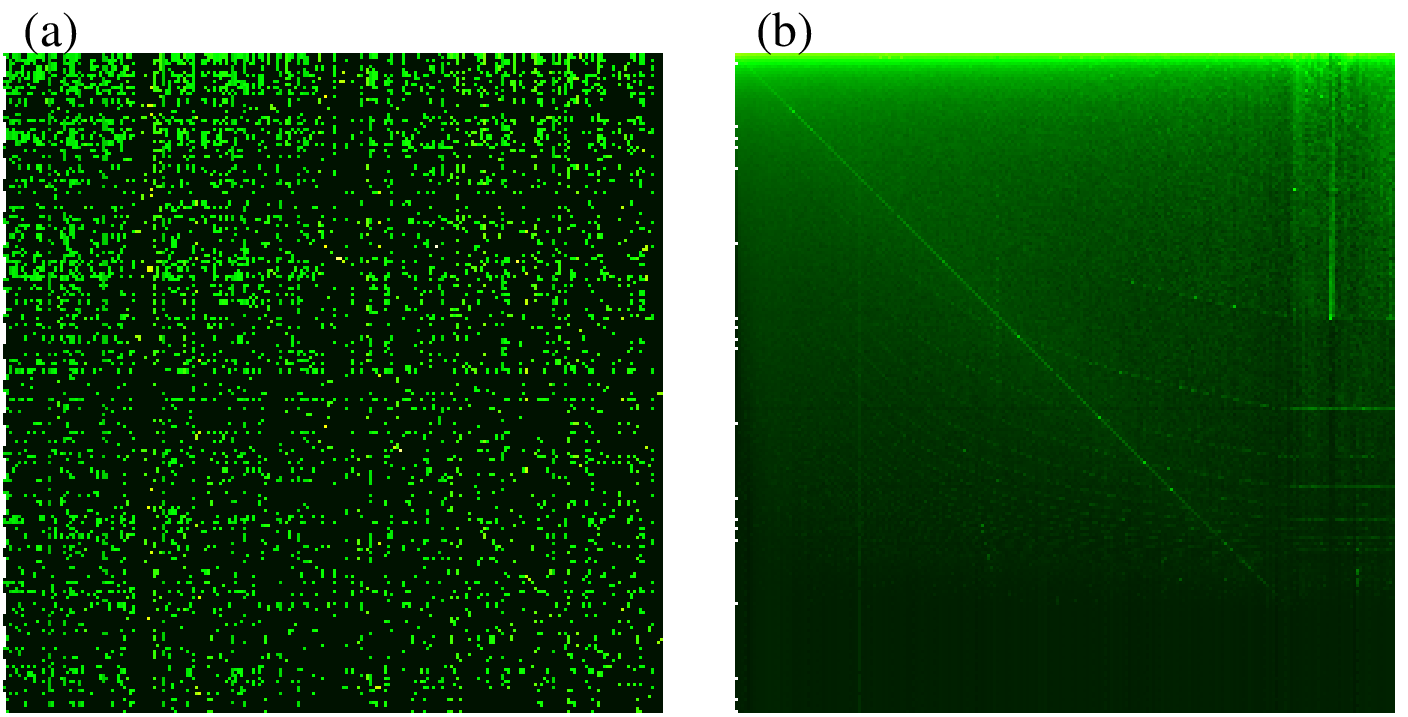}
\end{center}
\vglue -0.3cm
\caption{(Color online)
Google matrix of the network Wikipedia English articles for Aug 2009
in the basis of PageRank index $K$ 
(and $K^\prime$). Matrix $G_{KK^\prime}$ corresponds 
to $x$ (and $y$) axis with $1 \leq K,K^\prime \leq 200$
on panel (a), and with $1\leq K, K^\prime \leq N$ 
on  panel (b); all nodes are ordered by PageRank 
index $K$ of matrix $G$ and thus we have two matrix 
indexes $K, K^\prime$ for matrix elements in this basis. 
Panel (a) shows the first $200 \times 200$ matrix elements of $G$ 
matrix  (see Sec.~\ref{s3}).
Panel (b) shows density of all matrix elements
coarse-grained on $500\times 500$ cells where its elements, 
$G_{K,K^\prime}$, are written in the PageRank basis $K(i)$ 
with indexes $i \rightarrow K(i)$ (in $x$-axis) and 
$j \rightarrow K^\prime(j)$ (in a usual matrix representation 
with $K = K^\prime = 1$ on the top-left corner).
 Color shows the density
of matrix elements changing from black for minimum value 
($(1-\alpha)/N$) to white for maximum value via green (gray) and 
yellow (light gray);
here  the damping factor is $\alpha = 0.85$
After \cite{ermann:2012a}.
\label{fig1_1}}
\end{figure}

In this review we present an extensive analysis of a variety
of Google matrices emerging from real networks in various sciences
including WWW of UK universities,  
Wikipedia, Physical Review citation network,
Linux Kernel network, world trade network from the UN COMTRADE database, brain
neural  networks, networks of DNA sequences and many others. 
As an example,  the Google matrix of the Wikipedia network 
of English articles (2009) is shown in Fig.~\ref{fig1_1}.
We demonstrate that
the analysis of the spectrum and eigenstates
of a Google matrix of a given network provides a detailed
understanding  about the information flow and ranking.
We also show that such types of matrices naturally appear for 
Ulam networks of dynamical maps 
\cite{frahm:2012b,shepelyansky:2010a}
in the framework of the Ulam method
\cite{ulam:1960}.

Currently, Wikipedia, a free online encyclopaedia,
stores more and more information and has become the largest
database of human knowledge.
In this respect it is similar
to {\it the Library of Babel}, described by Jorge Luis Borges \cite{borges}.
The understanding of hidden relations between 
various areas of knowledge on the basis of Wikipedia
can be improved with the help of Google matrix analysis of
directed hyperlink networks of Wikipedia articles
as described in this review.

The specific tools of RMT and quantum chaos,
combined with  the efficient numerical methods 
for large matrix diagonalization like the Arnoldi method 
\cite{stewart:2001}, allow to analyze the spectral properties of
such large matrices as the entire Twitter network of 41 millions users
\cite{frahm:2012b}.
In 1998 Brin and Page pointed out that
{\it ``despite the importance of large-scale search engines on the web,
very little academic research has been done on them''} \cite{brin:1998}.
The Google matrix of a directed network,
like {\it the Library of Babel} of Borges \cite{borges},
contains all the information about a network.
The PageRank eigenvector of this matrix finds a broad range of
applications being at the mathematical foundations
of the Google search engine \cite{brin:1998,langville:2006}.
We show below that the spectrum of this matrix and its other eigenvectors
also provide interesting information
about network communities and the formation of PageRank vector.
We hope that this review yields a solid scientific basis
of matrix methods for efficient analysis of directed networks
emerging in various sciences.
The described methods will find broad interdisciplinary
applications in mathematics, 
physics and computer science with the cross-fertilization 
of different research fields. 
Our aim is to combine the analytic tools and numerical 
analysis of concrete directed networks
to gain a better understanding of the properties of these
complex systems. 

An interested  reader can find a  general introduction
about complex networks (see also Sec.~\ref{s2}) in well established papers, 
reviews and books 
\cite{watts:1998}, 
\cite{albert:2002,caldarelli:2003,newman:2003},
\cite{dorogovtsev:2008,castellano:2009},
\cite{dorogovtsev:2010,fortunato:2010,newman:2010}. Descriptions of 
Markov chains and Perron-Frobenius operators are given in
\cite{brin:1998,brin:2002,gantmacher:2000,langville:2006}, while 
the properties of Random Matrix Theory (RMT) and
quantum chaos are described in
\cite{akemann:2011,haake:2010,guhr:1998,mehta:2004}.

The data sets for the main part of the networks considered
here are available at \cite{fetnadine} from Quantware group.

\section{Scale-free properties of directed networks}
\label{s2}

\begin{figure}[H]
\begin{center}
\includegraphics[width=0.48\textwidth]{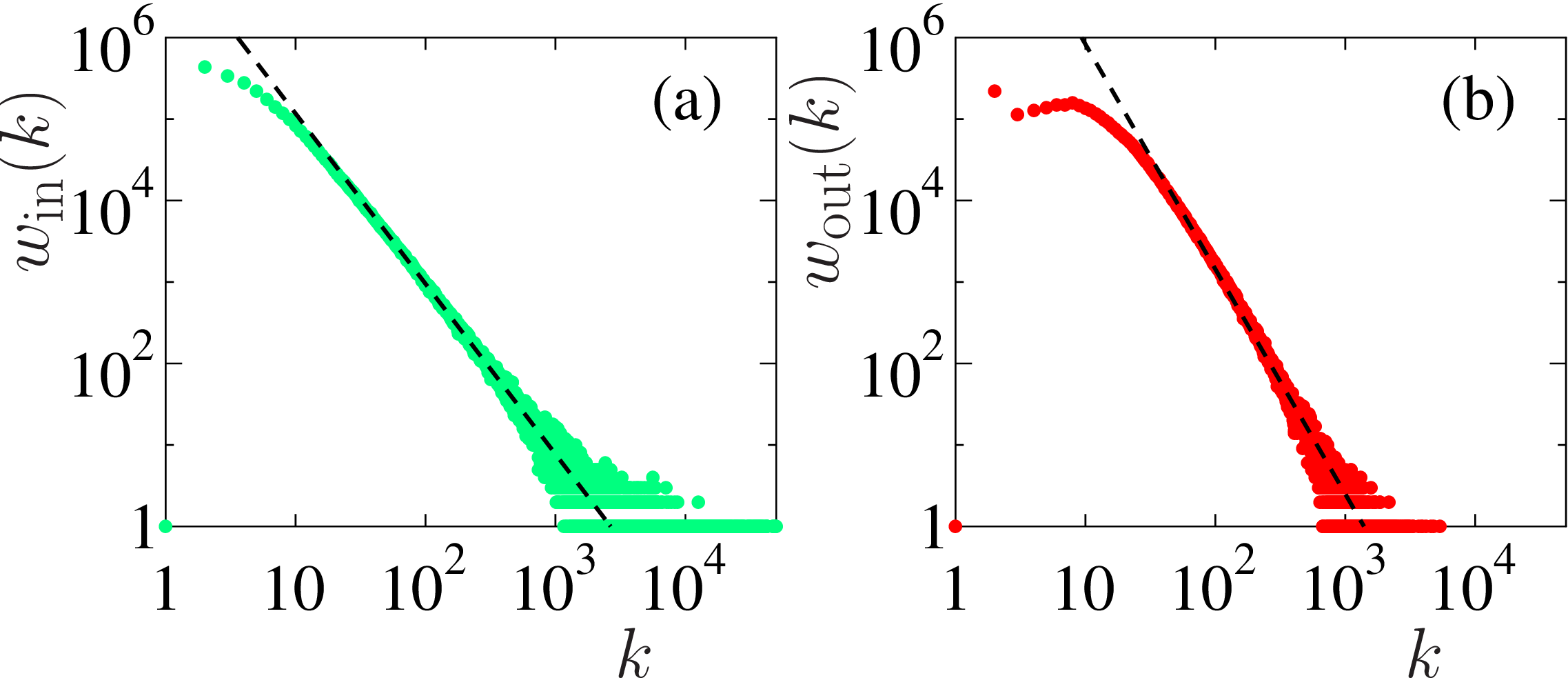}
\end{center}
\vglue -0.2cm
\caption{(Color online) 
Distribution $w_{\rm in,out}(k)$ of number  of ingoing (a)
and outgoing (b) links $k$ for $N=3282257$
Wikipedia English articles (Aug 2009) of Fig.~\ref{fig1_1}
with total number of links $N_\ell= 71 012 307$. 
The straight dashed fit line shows the slope
with $\mu_{\rm in}=2.09 \pm 0.04$ (a) and 
$\mu_{\rm out}=2.76 \pm 0.06$ (b).
After \cite{zhirov:2010}.
\label{fig2_1}}
\end{figure}

The distributions of the number of ingoing or outgoing links per node 
for directed networks with $N$ nodes and $N_\ell$ links are well known 
as indegree and outdegree distributions in the community of computer science 
\cite{caldarelli:2003,donato:2004,pandurangan:2005}.
A network is described by an adjacency matrix $A_{ij}$ of size $N\times N$
with $A_{ij}=1$ when there is a link 
from a node $j$ to a node $i$ in the network, i.~e. ``$j$ points to $i$'', 
and $A_{ij}=0$ otherwise. 
Real networks are often characterized by power law distributions
for the number of ingoing and outgoing links per node 
 $w_{\rm in,out}(k) \propto 1/k^{\mu_{\rm in,out}}$
with typical exponents $\mu_{\rm in} \approx 2.1 $
and $ \mu_{\rm out} \approx 2.7$ for the WWW. 
For example, for the Wikipedia network of Fig.~\ref{fig1_1}
one finds $\mu_{\rm in}=2.09 \pm 0.04$,
$\mu_{\rm out}=2.76 \pm 0.06$  as shown in Fig.~\ref{fig2_1} 
\cite{zhirov:2010}.

Statistical preferential attachment models
were initially developed for undirected networks
\cite{albert:2000}.
Their generalization to directed networks \cite{giraud:2009} 
generates a power law distribution for ingoing
links with $\mu_{\rm in} \approx 2$ but 
the distribution of outgoing links 
is closer to an exponential decay.
We will see below that these models are not
able to reproduce the spectral properties of $G$
in real networks.

The most recent studies of WWW, crawled
by  the Common Crawl Foundation in 2012 \cite{vigna:2014} for
$N \approx 3 .5 \times 10^9$ nodes and
$N_\ell \approx  1.29 \times 10^{11}$ links,
provide the exponents $\mu_{\rm in} \approx 2.24$, $\mu_{\rm out} \approx 2.77$,
even if the authors stress 
that these distributions describe 
probabilities at the tails which capture only about 
one percent of nodes.
Thus, at present the existing statistical
models of networks capture only
in an approximate manner the real situation
in large networks even if certain models are able to 
generate a power law decay of PageRank probability. 

\section{Construction of Google matrix and its properties}
\label{s3}

\subsection{Construction rules}
\label{s3.1}

The matrix $S_{ij}$ of Markov transitions \cite{markov:1906}
is constructed from the adjacency matrix $A_{ij} \rightarrow S_{ij}$
by normalizing elements of each column so that their sum is 
equal to unity ($\sum_i S_{ij}=1$) and replacing columns with only zero 
elements ({\em dangling nodes}) by $1/N$. 
Such matrices with columns sum normalized to unity and $S_{ij}\ge 0$ 
belong to the class of Perron-Frobenius operators with a possibly degenerate  
unit eigenvalue $\lambda=1$ and other eigenvalues obeying $|\lambda|\le 1$ 
(see Sec.~\ref{s3.2}). Then the Google matrix of the network is introduced as:
\cite{brin:1998}
\begin{equation}
   G_{ij} = \alpha  S_{ij} + (1-\alpha)/N \;\; .
\label{eq3_1} 
\end{equation} 
The damping factor $\alpha$ in the WWW context 
describes the probability 
$(1-\alpha)$ to jump to any node for a random surfer.
At a given node
a random surfer follows the available direction
of links making a random choice between them
with probability proportional to the weight of links. 
For WWW the Google search engine uses 
$\alpha \approx 0.85$ \cite{langville:2006}.
For $0\le\alpha\le 1$ the matrix $G$ also 
belongs to the class of Perron-Frobenius operators as $S$ and with its 
columns sum normalized. However, for $\alpha<1$ its largest eigenvalue 
$\lambda = 1$ is not degenerate and the other eigenvalues lie inside a 
smaller circle of radius $\alpha$, i.e. $|\lambda| \le \alpha$
\cite{brin:2002,langville:2006}.

\begin{figure}[H]
\begin{center}
\includegraphics[width=0.48\textwidth]{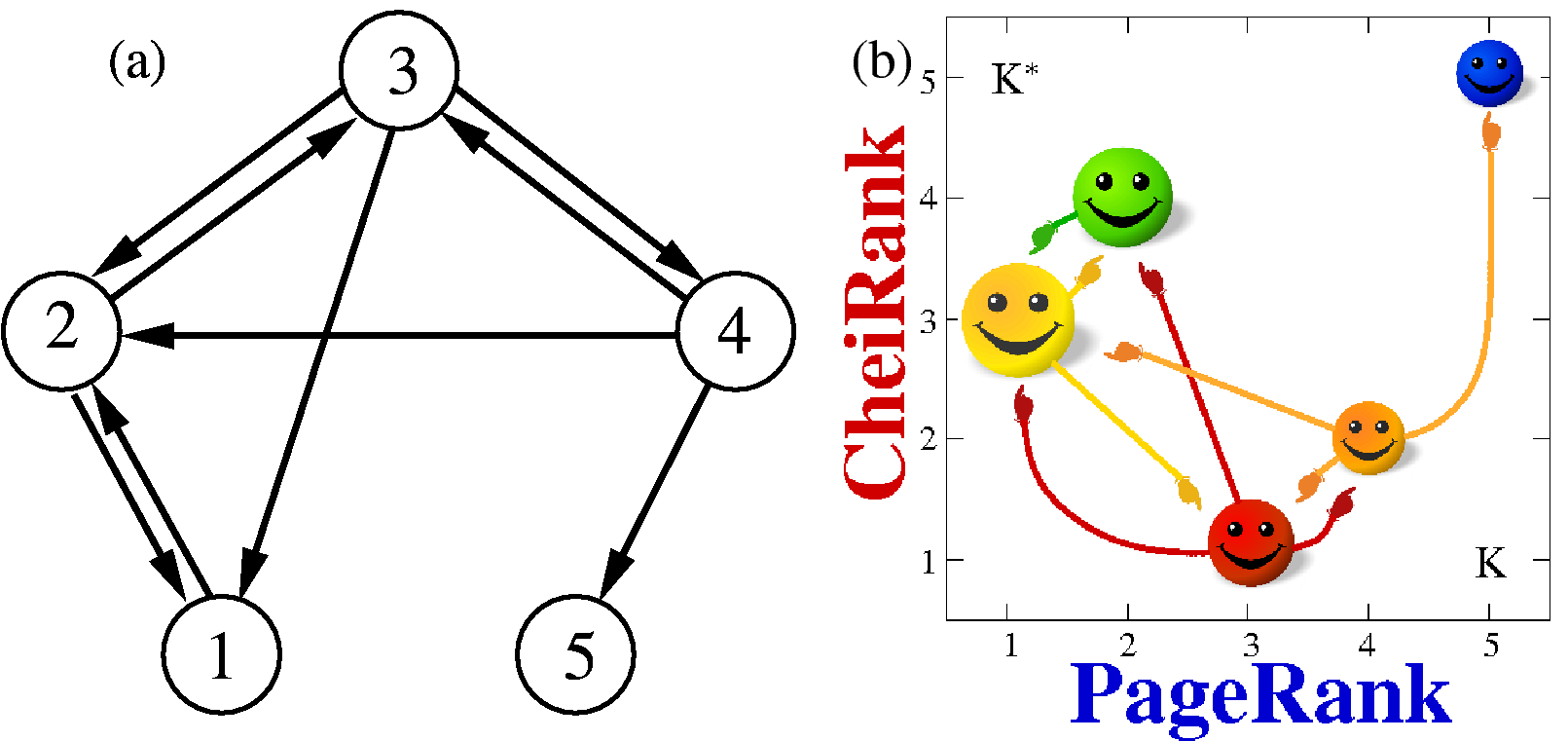}
\end{center}
\vglue -0.3cm
\caption{(Color online) 
(a) Example of simple network with 
directed links between 5 nodes.
(b) Distribution of 5 nodes from (a) 
on the PageRank-CheiRank plane
$(K,K^*)$, where the size of node is proportional
to PageRank probability $P(K)$ and color of node
is proportional to CheiRank probability $P^*(K^*)$,
with maximum at red/gray and minimum at blue/black;
the location of nodes of panel (a) on $(K_i,{K_i}^*)$ plane is:
$(2,4)$,  $(1,3)$, $(3,1)$, $(4,2)$, $(5,5)$
for original nodes $i=1,2,3,4,5$ respectively;
PageRank and CheiRank vectors are computed from
the Google matrices $G$ and $G^*$ shown in Fig.~\ref{fig3_2}
at a damping factor $\alpha=0.85$.
\label{fig3_1}}
\end{figure}

The right eigenvector at $\lambda = 1$, which is called the PageRank, 
has real nonnegative elements $P(i)$
and gives the stationary probability $P(i)$ to find a random surfer at site $i$. 
The PageRank can be efficiently determined by the power 
iteration method which consists of repeatedly multiplying $G$ to an 
iteration vector which is initially chosen as a given random or uniform 
initial vector. 
Developing the initial vector in a basis 
of eigenvectors of $G$ one finds that the other eigenvector coefficients 
decay as $\sim\lambda^n$ and only the 
PageRank component, with $\lambda=1$, survives in the limit 
$n\to\infty$. The finite gap 
$1-\alpha\approx 0.15$ between the largest eigenvalue and other eigenvalues 
ensures, after several tens of iterations, the fast exponential convergence 
of the method also called the ``PageRank algorithm''. 
A multiplication of $G$ to a vector requires
only $O(N_\ell)$ multiplications due to the links 
and the additional contributions due to dangling nodes and 
damping factor can be efficiently performed with $O(N)$ operations. 
Since often the average number of links per node 
is of the order of a few tens for WWW and many other networks one has 
effectively $N_\ell$ and $N$ of the same order of magnitude. 
At $\alpha=1$ the matrix $G$ coincides with the matrix $S$ and we will 
see below in Sec.~\ref{s8} that for this case the largest eigenvalue 
$\lambda=1$ is usually highly degenerate due to many invariant subspaces which 
define many independent Perron-Frobenius operators with at least 
one eigenvalue $\lambda=1$ for each of them.

Once the PageRank is found, e.g. at $\alpha=0.85$, 
all nodes can be sorted by decreasing probabilities $P(i)$. 
The node rank is then given by the index $K(i)$ which
reflects the  relevance of the node $i$. The top 
PageRank nodes, with largest probabilities, 
are located at small values of $K(i)=1,2,...$.

It is known that on average the PageRank probability 
is proportional to the number of ingoing links 
\cite{langville:2006,litvak:2008},
characterizing how popular or known a given node is.
Assuming that the Page\-Rank probability decays algebraically
as $P_i \sim 1/K_i^{\beta}$ we obtain that
the number of nodes $N_P$ with Page\-Rank probability $P$
scales as $N_P \sim 1/P^{\mu_{\rm in}}$ with
$\mu_{\rm in}=1+1/\beta$ so that $\beta \approx 0.9$ for 
$\mu_{\rm in} \approx 2.1$ being in a agreement with
the numerical data for WWW \cite{donato:2004,pandurangan:2005,vigna:2014}
and Wikipedia network \cite{zhirov:2010}.
More recent mathematical studies on the relation between
PageRank probability decay and ingoing links
are reported in \cite{jelenkovic:2013,chen:2014}.
At the same time the proportionality relation
between PageRank probability and ingoing links
assumes certain statistical properties of networks
and works only on average. We note that
there are examples of Ulam networks
generated  by dynamical maps where 
such proportionality is not 
working (see \cite{ermann:2010a} and Sec.~\ref{s6.5}).

In addition to a given directed network with adjacency matrix $A$
it is useful to analyze an inverse network where links are inverted 
and whose adjacency matrix $A^*$ is the transpose of $A$, i.e.
$A^*_{ij}=A_{ji}$. The matrices $S^*$ and the Google matrix $G^*$ 
of the inverse network are then constructed in the same way from $A^*$ 
as described above and according to the relation 
(\ref{eq3_1}) using the same value of $\alpha$ as for the $G$ matrix.
The right eigenvector of $G^*$ at eigenvalue $\lambda=1$
is called CheiRank giving a complementary rank index $K^*(i)$ of network nodes
\cite{chepelianskii:2010,zhirov:2010,ermann:2012a}.
The CheiRank probability $P^*(K^*)$ is proportional to the 
number of outgoing links highlighting  node communicativity
(see e.g. \cite{ermann:2012a,zhirov:2010}).
In analogy with the PageRank we obtain that
$P^* \sim 1/{{K^*}^{\beta}}$ with $\beta=1/(\mu_{\rm out}-1) \approx 0.6$
for typical $\mu_{\rm out} \approx 2.7$.
The statistical properties of distribution of nodes
on the PageRank-CheiRank plane are described 
in \cite{ermann:2012a} for various directed networks.
We will discuss them below.

\begin{figure}[H]
\begin{center}
\includegraphics[width=0.48\textwidth]{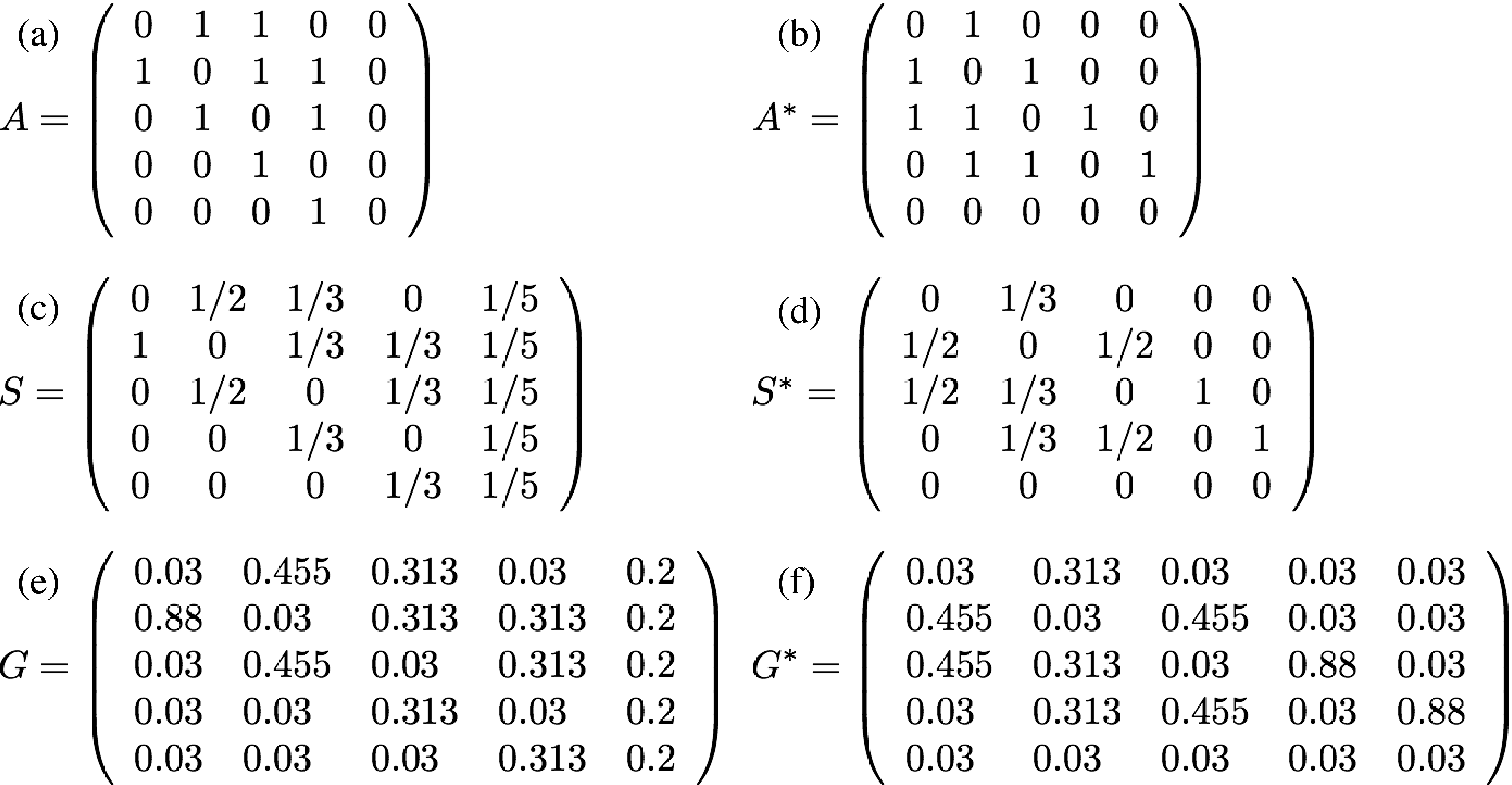}
\end{center}
\caption{(a) Adjacency matrix $A$ of network of Fig.~\ref{fig3_1}(a)
with indexes used there,
(b) adjacency matrix $A^*$ for the network with inverted links;
matrices $S$ (c) and $S^*$ (d) corresponding
to the matrices $A$, $A^*$;
the Google matrices $G$ (e) and $G^*$ (f)
corresponding to matrices  $S$ and $S^*$
for $\alpha=0.85$ (only 3 digits of matrix elements are shown).
\label{fig3_2}}
\end{figure}

For an illustration we consider an example of a simple
network of five nodes shown in Fig.~\ref{fig3_1}(a).
The corresponding adjacency matrices $A$, $A^*$
are shown in Fig.~\ref{fig3_2} for the indexes
given in Fig.~\ref{fig3_1}(a). The matrices of Markov transitions 
$S$, $S^*$ and Google matrices are computed as described above and 
from Eq.~(\ref{eq3_1}). 
The distribution of nodes on $(K,K^*)$ plane is shown in Fig.~\ref{fig3_1}(b).
After permutations the matrix $G$ can be rewritten in the
basis of PageRank index $K$ as it is done in Fig.~\ref{fig1_1}.

\subsection{Markov chains and Perron-Frobenius operators}
\label{s3.2}

Matrices with real non-negative elements and column sums normalized to 
unity belong to the class of Markov chains \cite{markov:1906}
and Perron-Frobenius operators \cite{brin:2002,gantmacher:2000,langville:2006},
which have been used in a mathematical 
analysis of dynamical systems and theory of matrices. 
A numerical analysis
of finite size approximants of such operators
is closely linked with the Ulam method \cite{ulam:1960}
which naturally generates such matrices
for dynamical maps \cite{ermann:2010a,ermann:2010b,shepelyansky:2010a}.
The Ulam method generates Ulam networks whose properties are 
discussed in Sec.\ref{s6}. 

Matrices $G$ of this type have at least 
(one) unit eigenvalue $\lambda=1$ since the vector $e^T=(1,\,\ldots,\,1)$ 
is obviously a left eigenvector for this eigenvalue. Furthermore one 
verifies easily that for any vector $v$ the inequality $\|G\,v\|_1\le \|v\|_1$ 
holds where the norm is the standard 1-norm. From this inequality 
one obtains immediately that all eigenvalues $\lambda$ 
of $G$ lie in a circle of radius unity: $|\lambda| \le 1$. 
For the Google matrix $G$ as given in (\ref{eq3_1}) one can furthermore show 
for $\alpha<1$ that the unity eigenvalue is not degenerate and the other 
eigenvalues obey even $|\lambda|\le \alpha$ \cite{langville:2006}. 
These and other mathematical results  about properties of 
matrices of such type can be found at
\cite{gantmacher:2000,langville:2006}.

It should be pointed out 
that due to the asymmetry of links on directed networks 
such matrices have in general a complex eigenvalue spectrum and sometimes 
they are not even diagonalizable, i.e. there may also be generalized 
eigenvectors associated to non-trivial Jordan blocks. 
Matrices of this type rarely appear in physical problems which 
are usually characterized by Hermitian or unitary matrices
with real eigenvalues or located on the unitary circle.
The universal spectral properties of such hermitian or unitary matrices
are well described by RMT \cite{akemann:2011,haake:2010,guhr:1998}.
In contrast to this non-trivial complex spectra appear in physical systems
only in problems of quantum chaotic scattering
and systems with absorption.  In such cases 
it may happen that the number of states 
$N_{\gamma}$, with 
finite values  $0<\lambda_{\rm min} \leq |\lambda| \leq 1$
($\gamma=-2 \ln|\lambda|$),
can grow algebraically $N_{\gamma} \propto N^{\nu}$
 with increasing matrix size $N$,
with an exponent $\nu <1$ corresponding to
a fractal Weyl law proposed first in mathematics 
\cite{sjostrand:1990}. 
Therefore most of eigenvalues drop to $\lambda=0$ with $N\to\infty$.
We discuss this unusual property in Sec.\ref{s5}.

\subsection{Invariant subspaces}
\label{s3.3}

For typical networks the set of nodes can be decomposed in 
invariant {\em subspace nodes} and fully connected {\em core space nodes} 
leading to a block structure of the matrix $S$ in (\ref{eq3_1}) 
which can be represented  as \cite{frahm:2011}: 
\begin{equation}
\label{eq3_2}
S=\left(\begin{array}{cc}
S_{ss} & S_{sc}  \\
0 & S_{cc}  \\
\end{array}\right) \; .
\end{equation}
The core space block $S_{cc}$ contains the links between core space 
nodes and the coupling block $S_{sc}$ may contain links from certain 
core space nodes to certain invariant subspace nodes. By construction 
there are no links from  nodes of invariant subspaces 
to the nodes of core space.
Thus the subspace-subspace block $S_{ss}$ is actually composed of 
many diagonal blocks for many invariant subspaces
whose number can generally be rather large. Each of these blocks 
corresponds to a column sum normalized matrix with positive elements 
of the same type as $G$ 
and has therefore at least one unit eigenvalue. This leads to a 
high degeneracy $N_1$ of the eigenvalue $\lambda=1$ of $S$, 
for example $N_1\sim 10^3$ as for the case of UK universities 
(see Sec.~\ref{s8}). 

In order to obtain the invariant subspaces, 
we determine iteratively for each node the set
of nodes that can be reached by a chain of non-zero matrix elements of $S$. 
If this set contains all nodes (or at least a macroscopic fraction) 
of the network, 
the initial node belongs to the {\em core space} $V_c$. 
Otherwise, the limit set defines a subspace which 
is invariant with respect to applications of the matrix $S$. 
At a second step all subspaces with common 
members are merged resulting in a sequence of disjoint subspaces $V_j$ of 
dimension $d_j$ and which are invariant by applications of $S$.
This scheme, which can be efficiently implemented in a 
computer program, provides a subdivision over $N_c$ 
core space nodes (70-80\% of $N$ for UK university networks) 
and $N_s=N-N_c$ subspace nodes 
belonging to at least one of the invariant subspaces $V_j$. This procedure
generates  the block triangular structure (\ref{eq3_2}). One may note that 
since a dangling node is connected by construction to all other nodes it 
belongs obviously to the core space as well as all nodes which are linked 
(directly or indirectly) to a dangling node. As a consequence the 
invariant subspaces do not contain dangling nodes nor nodes linked to 
dangling nodes. 

The detailed algorithm for an efficient computation of the 
invariant subspaces is described in \cite{frahm:2011}. 
As a result the total number of all subspace nodes $N_s$, 
the number of independent subspaces $N_d$, the maximal subspace dimension
$d_{\rm max}$ etc. can be determined. The statistical properties for the 
distribution of subspace dimensions are discussed in
Sec.~\ref{s8} for UK universities and Wikipedia networks.
Furthermore it is possible to determine numerically with a very low effort 
the eigenvalues of $S$ associated to each subspace by separate diagonalization 
of the corresponding diagonal blocks in the matrix $S_{ss}$. For this, 
either exact diagonalization or, in rare cases of quite large subspaces, 
the Arnoldi method (see the next subsection) can be used.

After the subspace eigenvalues are determined one can apply the 
Arnoldi method to the projected core space matrix block $S_{cc}$ 
to determine the leading core space eigenvalues. In this way one obtains 
accurate eigenvalues because the 
Arnoldi method does not need to compute the numerically 
very problematic highly degenerate unit eigenvalues of $S$ since the 
latter are already obtained from the separate and cheap subspace 
diagonalization. Actually the alternative and 
naive application of the Arnoldi method on the 
full matrix $S$, without 
computing the subspaces first, does not provide the correct number $N_1$ of 
degenerate unit eigenvalues and also the obtained 
clustered eigenvalues, close to 
unity, are not very accurate. Similar problems hold for the full matrix $G$ 
(with damping factor $\alpha<1$) since here only the first eigenvector, 
the PageRank, can be determined accurately but there are still many 
degenerate (or clustered) eigenvalues at (or close to) $\lambda=\alpha$. 

Since the columns sums of $S_{cc}$ are less than unity, 
due to non-zero matrix elements in the block
$S_{sc}$, the leading core space eigenvalue of $S_{cc}$ is also below unity 
$|\lambda_1^{(\rm core)}|<1$ even though in certain cases the gap to 
unity may be very small (see Sec.~\ref{s8}). 

We consider concrete examples of such decompositions in Sec.~\ref{s8} 
and show in this review spectra with subspace and core space eigenvalues 
of matrices $S$ for several network examples. 
The  mathematical results for properties of the 
matrix $S$ are discussed in \cite{serra:2005}.

\subsection{Arnoldi method for numerical diagonalization}
\label{s3.4}

The most adapted numerical method to determine the largest 
eigenvalues of large sparse matrices is the Arnoldi method
\cite{arnoldi:1951,frahm:2010,golub:2006,stewart:2001}.
Indeed, usually the matrix $S$ in Eq. (\ref{eq3_1}) 
is very sparse with only a few tens
of links per node $\zeta =N_\ell/N \sim 10$. 
Thus, a multiplication of a vector by $G$ or $S$ is numerically 
cheap. The Arnoldi method is similar in spirit to the 
Lanzcos method, but is adapted to non-Hermitian 
or non-symmetric matrices. Its main idea is to 
determine recursively an orthonormal set of 
vectors $\xi_0,\,\ldots\,\xi_{n_{\rm A}-1}$, which define 
a {\em Krylov space}, by orthogonalizing $S\xi_k$ on the previous vectors 
$\xi_0,\,\ldots\xi_k$ by the Gram-Schmidt procedure to 
obtain $\xi_{k+1}$ 
and where $\xi_0$ is some normalized initial vector. The 
dimension $n_{\rm A}$ of the Krylov space (in the following called the 
{\em Arnoldi-dimension}) should be ``modest'' but not too small. 
During the Gram-Schmidt procedure one obtains furthermore the 
explicit expression: $S\xi_k=\sum_{j=0}^{k+1} h_{jk}\,\xi_j$ 
with matrix elements $h_{jk}$, of the Arnoldi representation matrix of 
$S$ on the Krylov space, given by the scalar products or inverse normalization 
constants calculated during the orthogonalization. In order to obtain a 
closed representation matrix one needs to replace the last coupling element 
$h_{n_{\rm A},n_{\rm A}-1}\to 0$ which introduces a mathematical 
approximation. The eigenvalues of the $n_{\rm A}\times n_{\rm A}$ matrix 
$h$ are called the {\em Ritz eigenvalues} and represent often 
very accurate approximations of the exact eigenvalues of $S$, at least 
for a considerable fraction of the Ritz eigenvalues with largest modulus.

In certain particular cases, when $\xi_0$ belongs to an $S$ invariant subspace 
of small dimension $d$, the element $h_{d,d-1}$ vanishes automatically 
(if $d\le n_{\rm A}$ and assuming that numerical rounding errors are not 
important) and the Arnoldi iteration stops at $k=d$ and provides $d$ exact 
eigenvalues of $S$ for the invariant subspace. One can mention that there are 
more sophisticated variants of the Arnoldi method \cite{stewart:2001} 
where one applies (implicit) modifications on the initial vector 
$\xi_0$ in order to force this vector to be in some small dimensional 
invariant subspace which results in such a vanishing coupling matrix element. 
These 
variants known as (implicitly) restarted Arnoldi methods allow to 
concentrate on certain regions on the complex plane to determine a few but 
very accurate eigenvalues in these regions. However, for the cases of 
Google matrices, where one is typically interested in the largest 
eigenvalues close to the unit circle, only the basic variant described 
above was used but choosing larger values of $n_{\rm A}$ as would have 
been possible with the restarted variants. The initial vector was 
typically chosen to be random or as the vector with unit entries. 

Concerning the numerical resources the Arnoldi method requires $\zeta N$ 
double precision registers 
to store the non-zero matrix elements of $S$, 
$n_{\rm A} N$ registers to store the vectors 
$\xi_k$ and const.$\times n_{\rm A}^2$ registers to store 
$h$ (and various copies of $h$). The computational time 
scales as $\zeta \,n_{\rm A} \, N_d$ 
for the computation of $S\,\xi_k$, with 
$N_d\,n_{\rm A}^2$ for the Gram-Schmidt 
orthogonalization procedure (which is typically dominant) and with 
const.$\times n_{\rm A}^3$ for the diagonalization of $h$. 

The details of the Arnoldi method are described in Refs. given above.
This method has problems with degenerate or strongly clustered eigenvalues 
and therefore for typical examples of Google matrices it is applied 
to the core space block $S_{cc}$ where the effects of the invariant 
subspaces, being responsible for most of the degeneracies, are exactly 
taken out according to the discussion of the previous subsection. 
In typical examples it is possible to find 
about $n_{\rm A} \approx 640$ eigenvalues with largest $|\lambda|$
for the entire Twitter network with $N \approx 4.1 \times 10^7$
(see Sec.~\ref{s10}) and about $n_{\rm A} \approx 6000$ eigenvalues 
for Wikipedia networks with $N \approx 3.2 \times 10^6$ 
(see Sec.~\ref{s9}). For the two university networks of Cambridge and 
Oxford 2006 with $N\approx 2\times 10^5$ it is possible to compute 
$n_{\rm A} \approx 20000$ eigenvalues (see Sec.~\ref{s8}). 
For the case of the Citation network of Physical Review (see Sec.~\ref{s12}) 
with $N \approx 4.6 \times 10^5$ 
it is even possible and necessary to use high precision computations 
(with up to 768 binary digits) to determine accurately the Arnoldi matrix $h$ 
with $n_{\rm A} \approx 2000$ \cite{frahm:2014b}. 

\subsection{General properties of eigenvalues and eigenstates}
\label{s3.5}

According to the Perron-Frobenius theorem 
all eigenvalues $\lambda_i$ of $G$ are distributed inside the unitary
circle $|\lambda| \leq 1$. It can be shown that
at $\alpha <1$ there is only one 
eigenvalue $\lambda_0=1$ and all other $|\lambda_i| \leq \alpha$
having a simple dependence on $\alpha$:
$\lambda_i \rightarrow \alpha \lambda_i$
(see e.g. \cite{langville:2006}).
The right eigenvectors $\psi_i(j)$ are defined by the equation
\begin{equation}
 \sum_{j'}  G_{jj'} \psi_i(j') = \lambda_i \psi_i(j) \; .
\label{eq3_3}  
\end{equation} 
Only the PageRank vector is affected by $\alpha$
while other eigenstates are independent of $\alpha$
due to their orthogonality to the left unit eigenvector 
at $\lambda=1$. Left eigenvectors are orthonormal 
to right eigenvectors \cite{langville:2006}.

It is useful to characterize the eigenvectors by their 
Inverse Participation Ratio (IPR)
$\xi_i=(\sum_j |\psi_i(j)|^2)^2/\sum_j |\psi_i(j)|^4$
which gives an effective number of nodes
populated by an eigenvector $\psi_i$. This characteristics is
broadly used for description of 
localized or delocalized eigenstates of electrons 
in a disordered potential 
with  Anderson transition (see e.g. \cite{guhr:1998,evers:2008}).
We discuss the specific properties of eigenvectors in next Secs.

\begin{figure}[H]
\begin{center}
\includegraphics[width=0.48\textwidth]{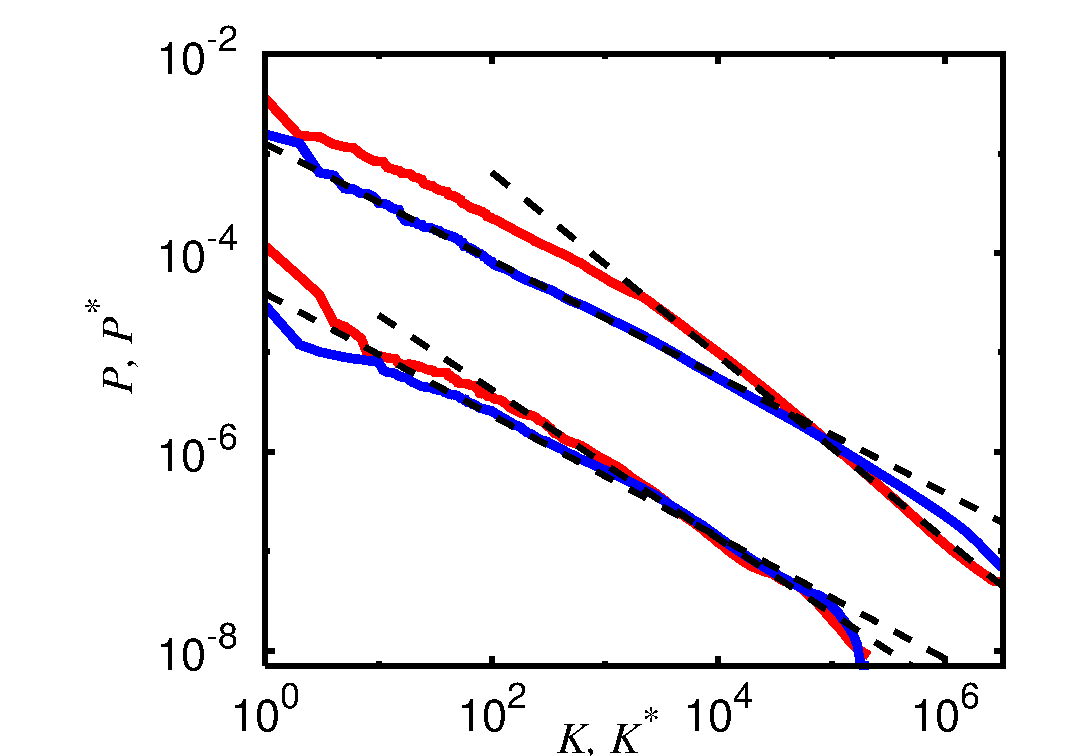}
\end{center}
\caption{(Color online)
Dependence of probabilities of PageRank $P$ (red/gray curve) and
CheiRank $P^*$ (blue/black curve) vectors
on the corresponding rank indexes $K$ and $K^*$
for networks of Wikipedia Aug 2009 (top curves)
and   University of Cambridge (bottom curves,
moved down by a factor $100$).
The straight dashed lines show
the power law fits for PageRank and CheiRank with the slopes
$\beta=0.92; 0.58$ respectively,  corresponding to 
$\beta=1/(\mu_{\rm in,out}-1)$ for Wikipedia (see Fig.~\ref{fig2_1}),
and $\beta = 0.75, 0.61$ for Cambridge.
After \cite{zhirov:2010} and \cite{frahm:2011}.
\label{fig4_1}}
\end{figure} 

\section{CheiRank versus PageRank}
\label{s4}

It is established that ranking of network nodes based on PageRank order
works reliably not only for WWW but also for other directed networks.
As an example it is possible to quote
 the citation network of Physical Review 
\cite{radicchi:2009,redner:1998,redner:2005}, Wikipedia network
\cite{aragon:2012,eom:2013a,zhirov:2010,skiena:2014} and 
even the network of world commercial trade \cite{ermann:2011b}.
Here we describe the main properties of PageRank and CheiRank
probabilities using a few real networks.
More detailed presentation for concrete networks follows in next Secs.

\subsection{Probability decay of PageRank and CheiRank}
\label{s4.1}

Wikipedia is a useful example of a scale-free network.
An article quotes other Wikipedia articles
that generates a network of directed links.
For Wikipedia of English articles dated by Aug 2009
we have $N=3 282 257$, $N_\ell= 71 012 307$
(\cite{zhirov:2010}). The dependencies of PageRank $P(K)$
and CheiRank $P^*(K^*)$ probabilities
 on  indexes $K$ and $K^*$ are shown in Fig.~\ref{fig4_1}.
In a large range the decay can be satisfactory described by an algebraic law
with an exponent $\beta$.
The obtained $\beta$ values are in a reasonable agreement
with the expected relation $\beta = 1/(\mu_{\rm in,out}-1)$
with the exponents of distribution of links given above.
However,  the decay is algebraic only on a tail,
showing certain nonlinear variations well visible for $P^*(K^*)$
at large values of $P^*$.

Similar data for network of University of Cambridge (2006)
with $N=212710$, $N_\ell=2015265$
\cite{frahm:2011} are shown in the same Fig.~\ref{fig4_1}.
Here, the exponents $\beta$ have different values
with approximately the same statistical accuracy of $\beta$.

Thus we come to the same conclusion as \cite{vigna:2014}:
the probability decay of PageRank and CheiRank
is only approximately algebraic, the relation between 
exponents $\beta$ and $\mu$ also works only approximately. 

\subsection{Correlator between PageRank and CheiRank}
\label{s4.2}

Each network node $i$ has both PageRank $K(i)$ and CheiRank $K(i)^*$
indexes so that it is interesting to know what is a correlation between
the corresponding vectors of PageRank and CheiRank. 
It is convenient to characterized this 
by a correlator introduced in \cite{chepelianskii:2010}
\begin{equation}
  \kappa =N \sum^N_{i=1} P(K(i)) P^*(K^*(i)) - 1 \;\; .
\label{eq4_1} 
\end{equation}

\begin{figure}[H]
\begin{center}
\includegraphics[width=0.48\textwidth]{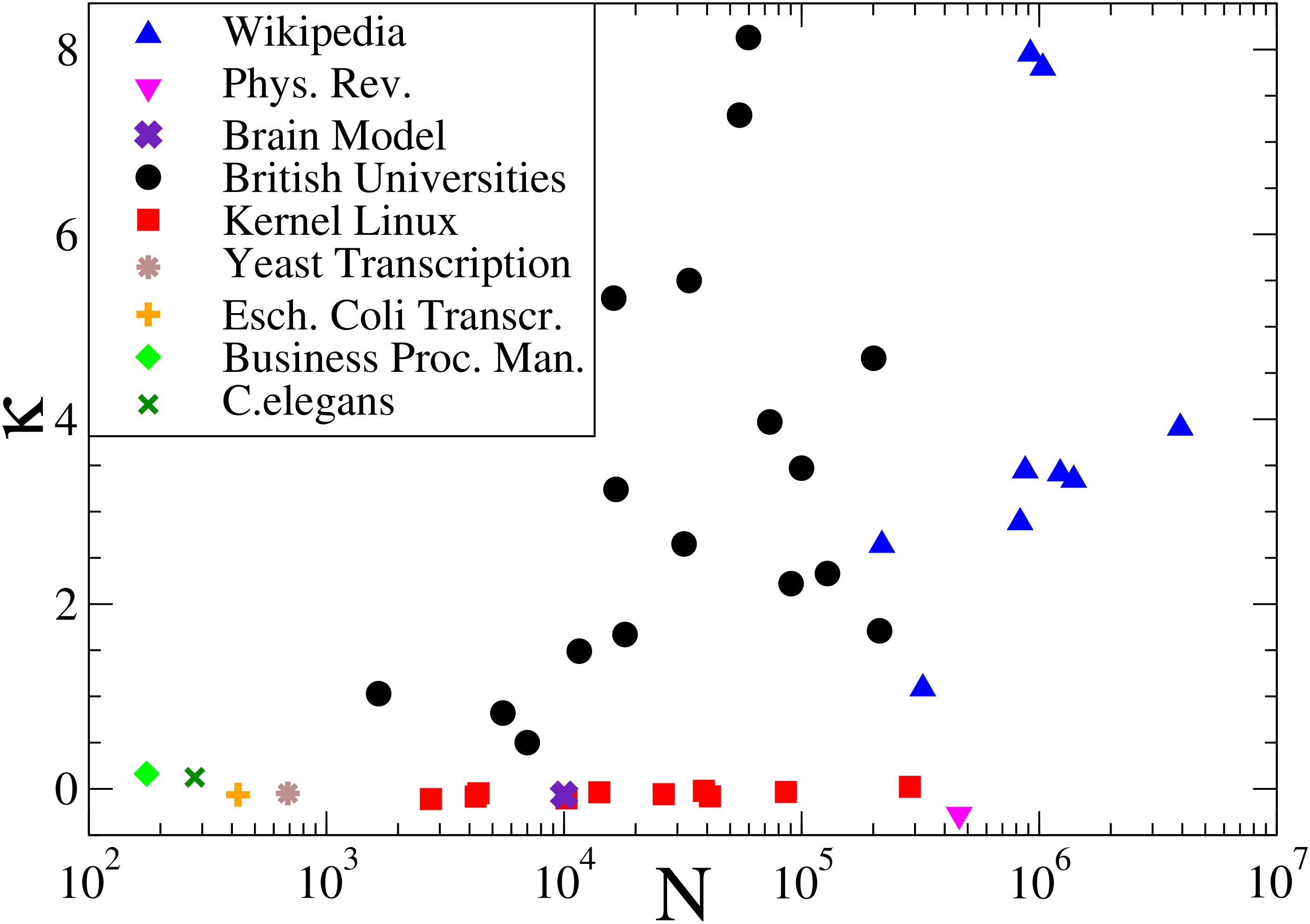}
\end{center}
\caption{(Color online)
Correlator $\kappa$ as a function of the number of nodes $N$
for different networks:  Wikipedia networks, Phys Rev network,
17 UK universities, 10 versions of Kernel 
Linux Kernel PCN, 
Escherichia Coli and Yeast Transcription Gene networks, 
Brain Model Network, C.elegans  neural network  and
Business Process Management Network.
After \cite{ermann:2012a} with additional data from
\cite{abel:2011}, \cite{eom:2013a}, 
\cite{kandiah:2014a}, \cite{frahm:2014b}.
\label{fig4_2}}
\end{figure}

\begin{figure}[H]
\begin{center}
\includegraphics[width=0.48\textwidth]{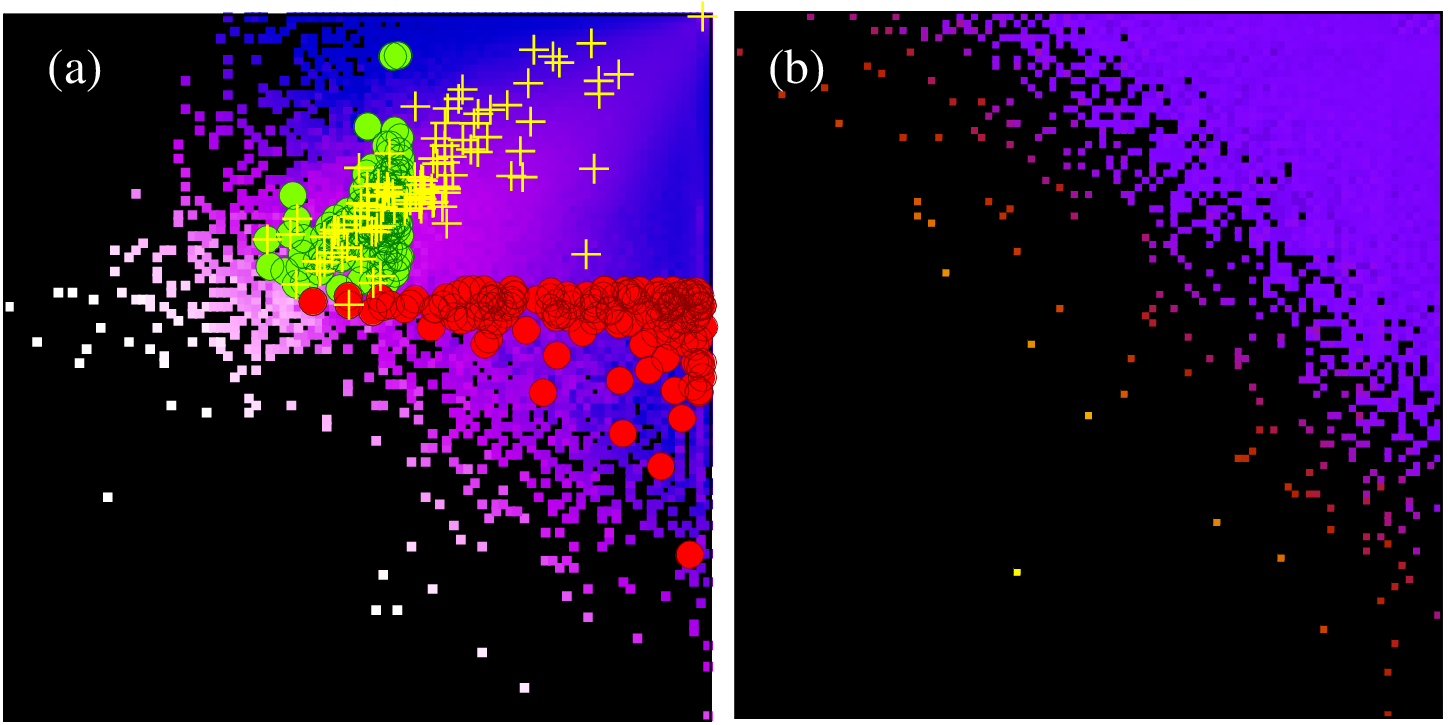}
\end{center}
\caption{(Color online)
Density distribution of network nodes
$W (K,K^*) = d N_i/dK dK^* $  shown on
the plane of PageRank 
and CheiRank indexes in logscale $(\log_N K,\log_N K^*)$
for all $1 \leq K,K^* \leq N$, density 
is computed over equidistant grid in plane $(\log_N K, \log_N K^*)$
with $100 \times 100$ cells; color shows average 
value of $W$ in each cell, the normalization condition is
$\sum_{K,K^*} W(K,K^*)=1$.
Density $W(K,K^*)$ is shown by color
with blue (dark gray) for minimum in (a),(b) and 
white (a) and yellow (white) (b) for maximum  (black for zero).
Panel (a): data for Wikipedia Aug (2009),
$N=3282257$, green/red (light gray/dark gray)
points show top 100 persons from
PageRank/CheiRank, yellow (white) pluses show top 100 persons
from \cite{hart:1992};
after \cite{zhirov:2010}.
Panel (b):
Density distribution $W(K,K^*)=dN_i/dKdK^*$ for 
Linux Kernel V2.4 network with $N = 85757$,
after \cite{ermann:2012a}.
\label{fig4_3}}
\end{figure}

Even if all the networks from Fig.~\ref{fig4_2}
have similar algebraic decay of PageRank probability with $K$
and similar $\beta \sim 1$ exponents
we see that the correlations between PageRank
and CheiRank vectors are drastically different in these 
networks. Thus the networks of UK universities and
9 different language editions of Wikipedia have the correlator
$\kappa \sim 1 - 8$ while all other networks have $\kappa \sim 0$.
This means that there are significant differences hidden in the network
architecture which are no visible from PageRank analysis.
We will discuss the possible origins of such a difference
for the above networks in next Secs. 

\subsection{PageRank-CheiRank plane}
\label{s4.3}

A more detailed characterization of correlations 
between PageRank and CheiRank
vectors can be obtained from a distribution of network nodes on the 
two-dimensional plane (2D) of indexes $(K,K^*)$.
Two examples for Wikipedia and Linux networks
are shown in Fig.~\ref{fig4_3}. A qualitative difference between two
networks is obvious. For Wikipedia we have
a maximum of density along the line
$\ln K^* \approx 5+ (\ln K)/3$ that results from
a strong correlation between PageRank and CheiRank
with $\kappa = 4.08$. In contrast to that 
for the Linux network V2.4 we have a homogeneous 
density distribution of nodes along lines
$\ln K^* = \ln K +const$ corresponding to 
uncorrelated probabilities $P(K)$ and $P^*(K^*)$
and even slightly negative value of $\kappa = -0.034$.
We note that if for Wikipedia we generate nodes with 
independent probabilities distributions $P$ and $P^*$,
obtained from this network 
at the corresponding value of $N$, 
then we obtain a homogeneous node distribution
in $(K,K^*)$ plane (in $(\log K, \log K^*)$
plane it takes a triangular form, see Fig.4 at \cite{zhirov:2010}).

In Fig.~\ref{fig4_3}(a) we also show the distribution
of top 100 persons from PageRank and CheiRank
compared with the top 100 persons from \cite{hart:1992}.
There is a significant overlap between PageRank and Hart ranking
of persons while CheiRank generates mainly another 
listing of people. We discuss the 
Wikipedia ranking of historical figures in Sec.~\ref{s9}.

\subsection{2DRank}
\label{s4.4}

PageRank and CheiRank indexes $K_i {K_i}^*$ order all network nodes
according to a monotonous decrease of corresponding
probabilities $P(K_i)$ and $P^*({K_i}^*)$. While
top $K$ nodes are most popular or known in the
network, top $K^*$ nodes are most communicative nodes with many 
outgoing links.  It is useful to consider an additional
ranking $K_2$, called 2DRank, which combines properties of
both ranks $K$ and $K^*$ \cite{zhirov:2010}.

The ranking list   $K_2(i)$ is constructed by  
increasing $K \rightarrow K+1$ 
and increasing 2DRank index 
$K_2(i)$ by one if a new entry is present in the list of
first $K^*<K$ entries of CheiRank, then the one unit step is done in
$K^*$ and $K_2$ is increased by one if the new entry is
present in the list of first $K<K^*$ entries of CheiRank.
More formally, 2DRank $K_2(i)$ gives the ordering 
of the sequence of sites, that $\;$ appear
inside  $\;$ the squares  $\;$
$\left[ 1, 1; \;K = k, K^{\ast} = k; \; \-... \right]$ when one runs
progressively from $k = 1$ to $N$. In fact, at each step
$k \rightarrow k + 1$ there are tree possibilities: 
(i) no new sites on two edges of square, 
(ii) only one site is on these two edges 
and it is added in the listing of $K_2(i)$
and (iii) two  sites are on the edges and both are added 
in the listing $K_2(i)$, first with $K > K^{\ast}$ 
and second with $K < K^{\ast}$. For (iii) the choice of order
of addition in the list $K_2(i)$ affects only some pairs of  
neighboring sites and does
not change the main structure of ordering.
An illustration example of 2DRank 
algorithm is given in Fig.7 
at \cite{zhirov:2010}. For Wikipedia 2DRanking of persons  
is discussed in Sec.~\ref{s9}. 

\subsection{Historical notes on spectral ranking}
\label{s4.5}

Starting from the work of Markov  \cite{markov:1906}
many scientists contributed to the development
of spectral ranking of Markov chains. Research of
Perron (1907) and Frobenius (1912) 
led to the Perron-Frobenius theorem
for square matrices with positive entries
(see e.g. \cite{brin:2002}).
 The detailed historical description
of spectral ranking research is reviewed by
\cite{franceschet:2011} and \cite{vigna:2013}.
 As described there, the important
steps have been done by researchers in
psychology, sociology and mathematics including
J.R.Seeley (1949), T.-H.Wei (1952), L.Katz (1953),
C.H.Hubbell (1965) (see Refs. in 
the above publications by \cite{franceschet:2011,vigna:2013}).
In the WWW context, the Google matrix in the form (\ref{eq3_1}),
with regularization of dangling nodes and
damping factor $\alpha$, was introduced 
by \cite{brin:1998}.

A PageRank vector of a Google matrix $G^*$ with inverted directions
of links has been considered by \cite{fogaras:2003}
and \cite{hrisitidis:2008},
but no systematic statistical analysis
of 2DRanking was presented there. An important step was done by
\cite{chepelianskii:2010} who analyzed $\lambda=1$ eigenvectors
of $G$ for directed network and of $G^*$ for
network with inverted links. The comparative analysis
of Linux Kernel network and WWW of University of Cambridge 
demonstrated a significant differences in correlator $\kappa$
values on these networks and different functions of top nodes
in $K$ and $K^*$. The term CheiRank 
was coined in \cite{zhirov:2010}
to have a clear distinction 
between eigenvectors of $G$ and $G^*$.  
We note that top PageRank and CheiRank nodes
have certain similarities with authorities and
hubs appearing in the HITS algorithm \cite{kleinberg:1999}.
However, the HITS is query dependent while the rank probabilities
$P(K_i)$ and $P^*({K_i}^*)$ classify all nodes of the network.

\section{Complex spectrum and fractal Weyl law}
\label{s5}

The Weyl law \cite{weyl:1912} gives a fundamental link between
the properties of quantum eigenvalues in closed Hamiltonian systems,
the Planck constant $\hbar$
and the classical phase space volume. 
The number of states in this case is determined by the 
phase volume of a
system with dimension
$d$. The case of Hermitian 
operators is now well understood both on mathematical and 
physical grounds \cite{dimassi:1999,landau:1989}. 
Surprisingly, only recently
it has been realized that the case of nonunitary
operators describing open systems in the semiclassical limit
has a number of new interesting properties
and the concept of the fractal Weyl law 
\cite{sjostrand:1990,zworski:1999}
has been introduced
to describe the dependence of number of 
resonant Gamow eigenvalues \cite{gamow:1928}
on $\hbar$.  

The Gamow eigenstates
find important applications for decay of radioactive nuclei, 
quantum chemistry reactions, chaotic scattering and 
microlasers with chaotic resonators, open quantum maps
(see \cite{gaspard:1998,gaspard:2014,shepelyansky:2008} and Refs. therein).
The spectrum of corresponding operators has a complex
spectrum $\lambda$. The spread width
$\gamma=-2\ln|\lambda|$ of eigenvalues $\lambda$
determines the life time of a corresponding eigenstate.
The understanding of the spectral  properties 
of related operators in the semiclassical limit represents
an important challenge.

According to the fractal Weyl law \cite{sjostrand:1990,lu:2003} the number
of Gamow eigenvalues $N_\gamma$, which have escape rates $\gamma$
in a finite band width  $0 \leq \gamma \leq \gamma_b$,
scales as
\begin{equation}
N_\gamma \propto \hbar^{-d/2} \propto N^{d/2}
\label{eq5_1} 
\end{equation}
where $d$ is a fractal dimension
of a classical strange repeller formed by classical orbits
nonescaping in future and past times. 
In the context of eigenvalues $\lambda$ of the Google matrix we have
$\gamma=-2 \ln|\lambda|$.
By numerical simulations 
it has been shown that the law (\ref{eq5_1})  works
for a scattering problem in
3-disk system \cite{lu:2003} and quantum chaos maps
with absorption when the fractal dimension $d$ is 
changed in a broad range $0<d<2$
\cite{shepelyansky:2008,ermann:2010b}.

The fractal Weyl law (\ref{eq5_1}) of open
systems with a fractal dimension $d<2$ 
leads to a striking consequence: only a relatively
small fraction of eigenvalues 
$\mu_W \sim N_\gamma/N \propto \hbar^{(2-d)/2}
\propto N^{(d-2)/2} \ll 1$ has finite values of $|\lambda|$
while almost all
eigenstates of the matrix operator of size $N \propto 1/\hbar$
have $\lambda \rightarrow 0$. The eigenstates with 
finite $|\lambda| >0$ are related to the classical fractal
sets of orbits non-escaping neither in the future neither in the past.
A fractal structure of these quantum fractal eigenstates
has been investigated in \cite{shepelyansky:2008}.
There it was conjectured that the eigenstates of a Google matrix
with finite $|\lambda| >0$  will select interesting specific communities
of a network. We will see below that the fractal Weyl law 
can indeed be observed in certain directed networks and in particular we 
show in the next section that 
it naturally appears for Perron-Frobenius operators of dynamical systems
and Ulam networks.

It is interesting to note that nontrivial 
complex spectra also naturally appear 
in systems of quantum chaos in presence of a contact with a 
measurement device \cite{bruzda:2010}. The properties of 
complex spectra of small size orthostochastic 
(unistochastic) matrices are analyzed in \cite{zyczkowski:2003}.
In such matrices the elements can be presented in a form
$S_{ij}=O^2_{ij}$ ($S_{ij}=|U_{ij}|^2$) where 
$O$ is an orthogonal matrix ( $U$ is a unitary matrix).
We will see  certain similarities of
their spectra with the spectra of 
directed networks discussed in Sec. VIII.

Recent mathematical results for the fractal Weyl law are presented in
\cite{nonnenmacher:2007,nonnenmacher:2014}. 

\begin{figure}[H]
\begin{center}
\includegraphics[width=0.48\textwidth]{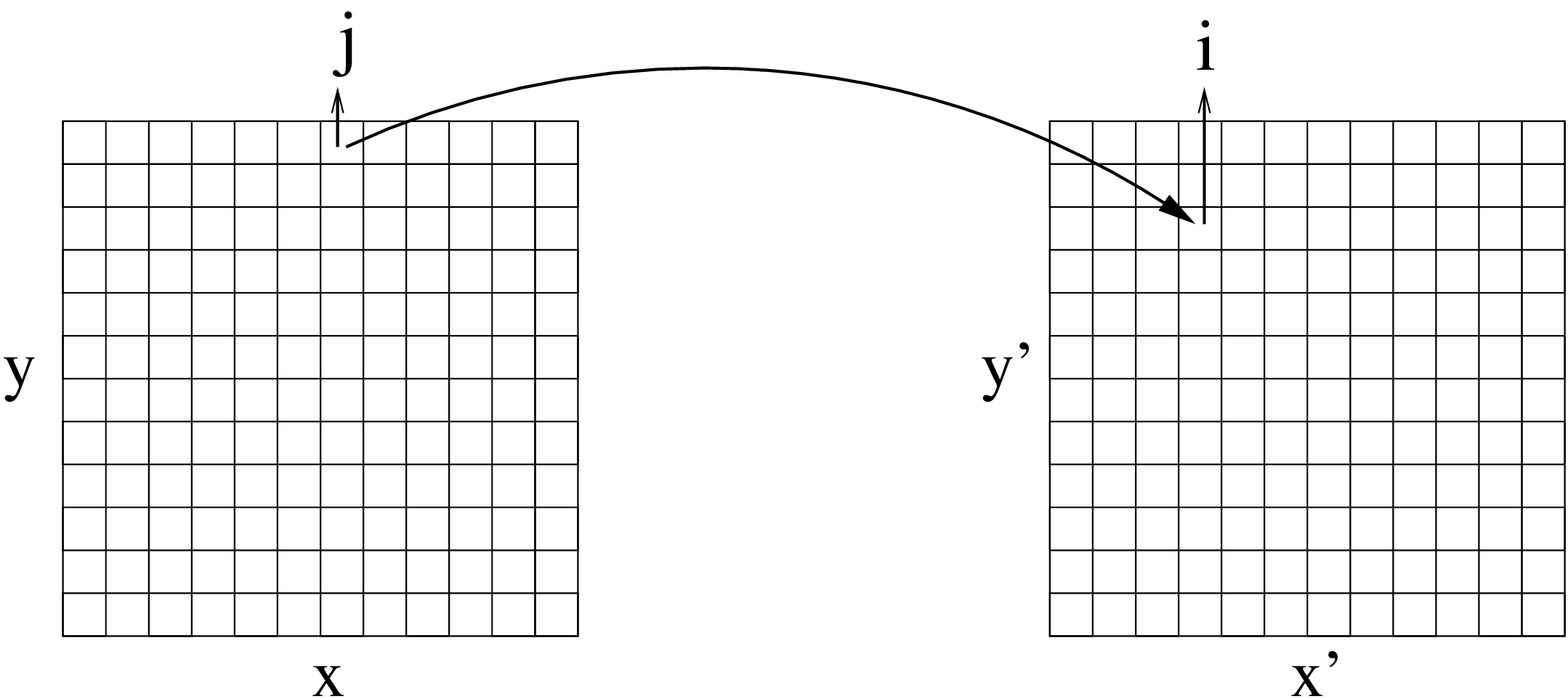}
\end{center}
\caption{Illustration of operation of the Ulam method:
the phase space $(x,y)$ is divided in 
$N=N_x \times N_y$ cells, $N_c$ trajectories start from cell
$j$ and the number of trajectories $N_{ij}$ 
arrived to a cell $i$ from a cell $j$ 
is collected after a map iteration. Then the 
matrix of Markov transitions is defined as
$S_{ij}=N_{ij}/N_c$, by construction $\sum_{i=1}^{N} S_{ij}=1$.
\label{fig6_1}}
\end{figure} 

\section{Ulam networks}
\label{s6}

By construction the Google matrix belongs
to the class of Perron-Frobenius operators
which naturally appear in ergodic theory
\cite{cornfeld:1982}
and  dynamical systems
with Hamiltonian or dissipative dynamics \cite{brin:2002}.
In 1960 Ulam \cite{ulam:1960} 
proposed a method, now known as the Ulam method,
for a construction of finite size approximants
for the Perron-Frobenius operators of 
dynamical maps. 
The method is based on discretization 
of the phase space and construction of
a Markov chain based on probability transitions 
between such discrete cells
given by the dynamics. Using as an example a simple chaotic map
Ulam made a conjecture that the 
finite size approximation
converges to the continuous limit
when the cell size goes to zero. 
Indeed, it has been proven that for hyperbolic maps 
in one and higher dimensions
the Ulam method converges to the spectrum of 
continuous system \cite{li:1976,blank:2002}. 
The probability flows in dynamical systems 
have rich and nontrivial features
of general importance, like simple and strange attractors
with localized and delocalized dynamics 
governed by simple dynamical rules \cite{lichtenberg:1992}.
Such objects are generic for nonlinear dissipative dynamics and
hence can have relevance for actual WWW structure.
The analysis of  Ulam networks,
generated by the Ulam method, allows to
obtain a better intuition about 
the spectral properties of Google matrix.
The term Ulam networks was introduced in 
\cite{shepelyansky:2010a}.

\subsection{Ulam method for dynamical maps}
\label{s6.1} 

In Fig.~\ref{fig6_1} we show how the Ulam method works.
The phase space of a dynamical map is divided in equal cells
and a number of trajectories $N_c$ is propagated by a map iteration.
Thus a number of trajectories $N_{ij}$ 
arrived from cell $j$ to
cell $i$ is determined. Then the matrix of Markov transition is
defined as $S_{ij}=N_{ij}/N_c$. By construction
this matrix belongs to the class of Perron-Frobenius
operators which includes the Google matrix.

The physical meaning of the coarse grain
description by a finite number of cells
is that it introduces in the system a noise 
of cell size amplitude.
Due to that an exact time reversibility of
dynamical equations of chaotic maps is destroyed due 
to exponential instability of chaotic dynamics.
This time reversibility breaking
is illustrated by an example of the Arnold cat map
by \cite{ermann:2012b}. For the Arnold cat map on a long torus it
is shown that the spectrum of the Ulam approximate of the Perron-Frobenius
(UPFO) is composed of a large group
of complex eigenvalues with $\gamma \sim 2 h \approx 2$,
and real eigenvalues with $|1-\lambda| \ll 1$
corresponding to a statistical relaxation to the ergodic state
at $\lambda=1$ described by the Fokker-Planck equation
(here $h$ is the Kolmogorov-Sinai entropy of the map
being here equal to the Lyapunov exponent, 
see e.g. \cite{chirikov:1979}). 

For fully chaotic maps the finite cell size, corresponding to
added noise, does not significantly affect the 
dynamics and the discrete UPFO converges
to the limiting case of 
continuous Perron-Frobenius operator
\cite{li:1976,blank:2002}. The Ulam method finds 
useful applications in studies of dynamics of molecular systems
and coherent structures in dynamical flows
\cite{froyland:2009}. Additional Refs. can be found 
in \cite{frahm:2010}.

\subsection{Chirikov standard map}
\label{s6.2} 

However, for  symplectic maps with a divided phase 
space, a noise present in the Ulam method
significantly affects the original dynamics
leading to a destruction of islands of stable motion
and Kolmogorov-Arnold-Moser (KAM) curves.
A famous example of such a map is the Chirikov standard
map which describes the dynamics of many physical systems
\cite{chirikov:1979,chirikov:2008}:
\begin{equation}
\label{eq6_1}
{\bar y} = \eta y + \frac{K_s}{2\pi} \sin (2\pi x) \; , \;\; 
{\bar x} = x + {\bar y} \;\; ({\rm mod} \; 1) \;.
\end{equation}
Here bars mark the variables after one map iteration
and we consider the dynamics to be periodic on  a torus so that
$0 \leq x \leq 1$, $-1/2 \leq y \leq 1/2$;
$K_s$ is a dimensionless parameter of chaos.
At $\eta=1$ we have area-preserving  symplectic map,
considered in this SubSec., for $0< \eta <1$
we have a dissipative dynamics analyzed in next
SubSec.

\begin{figure}[H]
\begin{center}
\includegraphics[width=0.48\textwidth]{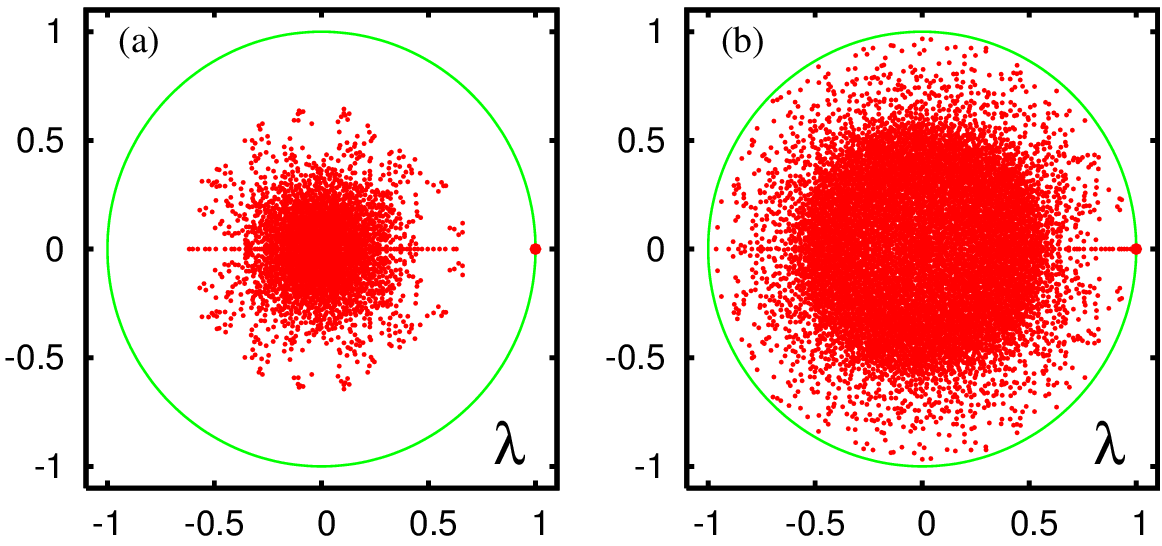}
\end{center}
\caption{(Color online)
Complex spectrum of eigenvalues $\lambda_j$,
shown by red/gray dots, for the UPFO of two variants 
of the Chirikov standard map (\ref{eq6_1});
the unit circle $|\lambda|=1$ is shown by a green (light gray) curve,
the unit eigenvalue at 
$\lambda=1$ is shown as larger red/gray dot. Panel
(a) corresponds to 
the Chirikov standard map at dissipation $\eta=0.3$ and
$K_s=7$;
the phase space is covered by $110\times 110$ cells 
and the UPFO is constructed by many trajectories with random initial 
conditions generating transitions
from one cell into another
(after \cite{ermann:2010b}). 
Panel (b) corresponds to the Chirikov standard map 
without dissipation at $K_s=0.971635406$ with an UPFO 
constructed from a single trajectory of length $10^{12}$ in the 
chaotic domain and $280\times 280/2$ 
cells to cover the phase space (after \cite{frahm:2010}).
\label{fig6_2}}
\end{figure}

\begin{figure}[H]
\begin{center}
\includegraphics[width=0.48\textwidth]{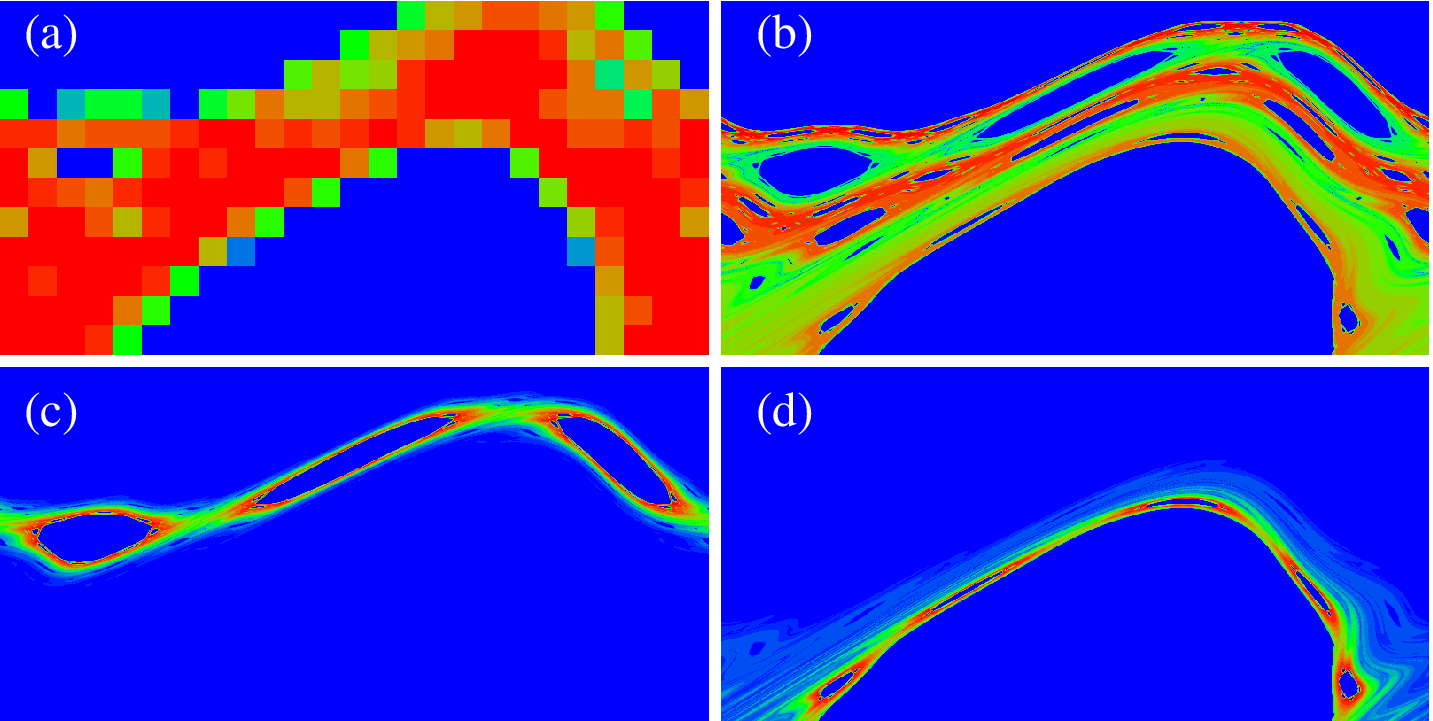}
\end{center}
\caption{(Color online)
Density plots of absolute values
of the eigenvectors of the UPFO obtained by the 
generalized Ulam method with a single trajectory of 
$10^{12}$ iterations of the Chirikov standard map 
at $K_s=0.971635406$. 
The phase space is shown in the area
$0\leq x \leq 1$, $0 \leq y \leq 1/2$; 
the UPFO is obtained from $M \times M/2$ cells
placed in this area.
Panels represent:
(a) eigenvector $\psi_0$ with eigenvalue $\lambda_0=1$; 
(b) eigenvector $\psi_2$ with real 
eigenvalue $\lambda_2=0.99878108$; 
(c) eigenvector $\psi_6$ with complex eigenvalue 
$\lambda_6=-0.49699831+i\,0.86089756\approx |\lambda_6|\,e^{i\,2\pi/3}$;
(d) eigenvector $\psi_{13}$ with complex eigenvalue 
$\lambda_{13}=0.30580631+i\,0.94120900\approx |\lambda_{13}|\,e^{i\,2\pi/5}$.
Panel (a) corresponds to $M=25$ while (b), (c) and (d) have $M=800$. 
Color is proportional to amplitude with blue (black) for zero
and red (gray) for maximal value.
After \cite{frahm:2010}.
\label{fig6_3}}
\end{figure} 

Since the finite cell size generates noise and destroys the KAM curves
in the map (\ref{eq6_1}) at $\eta=1$, 
one should use the generalized 
Ulam method \cite{frahm:2010},
where the transition probabilities $N_{ij}/N_c$
are collected along one chaotic trajectory.
In this construction a  trajectory
visits only those cells which belong to one connected
chaotic component. Therefore the noise induced
by the discretization of the phase space
does not lead to a destruction of invariant curves,
in contrast to the original Ulam method \cite{ulam:1960},
which uses all cells in the available phase space.
Since a trajectory is generated by a continuous map it cannot
penetrate inside the stability islands and on a physical 
level of rigor one can expect that, due to ergodicity
of dynamics on one connected chaotic component,
the UPFO constructed in such a way should converge
to the Perron-Frobenius operator of the continuous map
on a given subspace of chaotic component.
The numerical confirmations of this convergence are
presented in \cite{frahm:2010}.

We consider the map (\ref{eq6_1}) at $K_s=0.971 635 406$
when the golden KAM curve is critical. Due to the symmetry 
of the map with respect to $x\to 1-x$ and $y\to -y$ 
we can use only the upper part of
the phase space with $y \geq 0$ dividing it
in $M \times M/2$ cells.
At that $K_s$ we   find that the number of
cells visited by the trajectory in this half square scales as
$N_d \approx C_d M^2/2$ with $C_d \approx 0.42$.
This means that the chaotic component contains
about $40\%$ of the total area which is in good agreement 
with the known result of \cite{chirikov:1979}.

The spectrum of the UPFO matrix $S$ for the 
phase space division by $280 \times 208/2$ cells
is shown in Fig.~\ref{fig6_2}(b).
In a first approximation the spectrum $\lambda$ of $S$
is more or less homogeneously distributed
in the polar angle $\varphi$ defined as
$\lambda_j = |\lambda_j| \exp(i\varphi_j)$.
 With the increase of matrix size $N_d$ the two-dimensional 
density of states $\rho(\lambda)$ 
converges to a limiting distribution 
\cite{frahm:2010}. With the help of the Arnoldi method
it is possible to compute a few thousands of eigenvalues
with largest absolute values $|\lambda|$ for
maximal $M=1600$ with the total matrix size
$N = N_d \approx  5.3 \times 10^5$.

The eigenstate at $\lambda=1$ is homogeneously 
distributed over the chaotic component at $M=25$
(Fig.~\ref{fig6_3})
and higher $M$ values \cite{frahm:2010}.
This results from the ergodicity of motion and the fact that
for symplectic maps the measure
is proportional to the phase space area 
\cite{chirikov:1979,cornfeld:1982}.
Examples of other right eigenvalues of $S$ 
at real and complex eigenvalues $\lambda$ 
with $|\lambda| <1$ are also
shown in Fig.~\ref{fig6_3}. For $\lambda_2$
the eigenstate corresponds to some diffusive mode
with two nodal lines, while other two eigenstates are localized 
around certain resonant structures in phase space.
This shows that eigenstates of the matrix $G$ (and $S$)
are related to specific communities of a network.

With the increase of number of cells $M^2/2$
there are eigenvalues which become more and more
close to the unit eigenvalue. This is shown to be related to 
an algebraic statistics of Poincar\'e recurrences 
and long time sticking of trajectories
in a vicinity of critical KAM curves.
At the same time for symplectic maps the measure is proportional
to area so that we have dimension $d=2$ and hence
we have a usual Weyl law with $N_\gamma \propto N$.
More details can be found at \cite{frahm:2010,frahm:2013a}.

\begin{figure}[H]
\begin{center}
\includegraphics[width=0.48\textwidth]{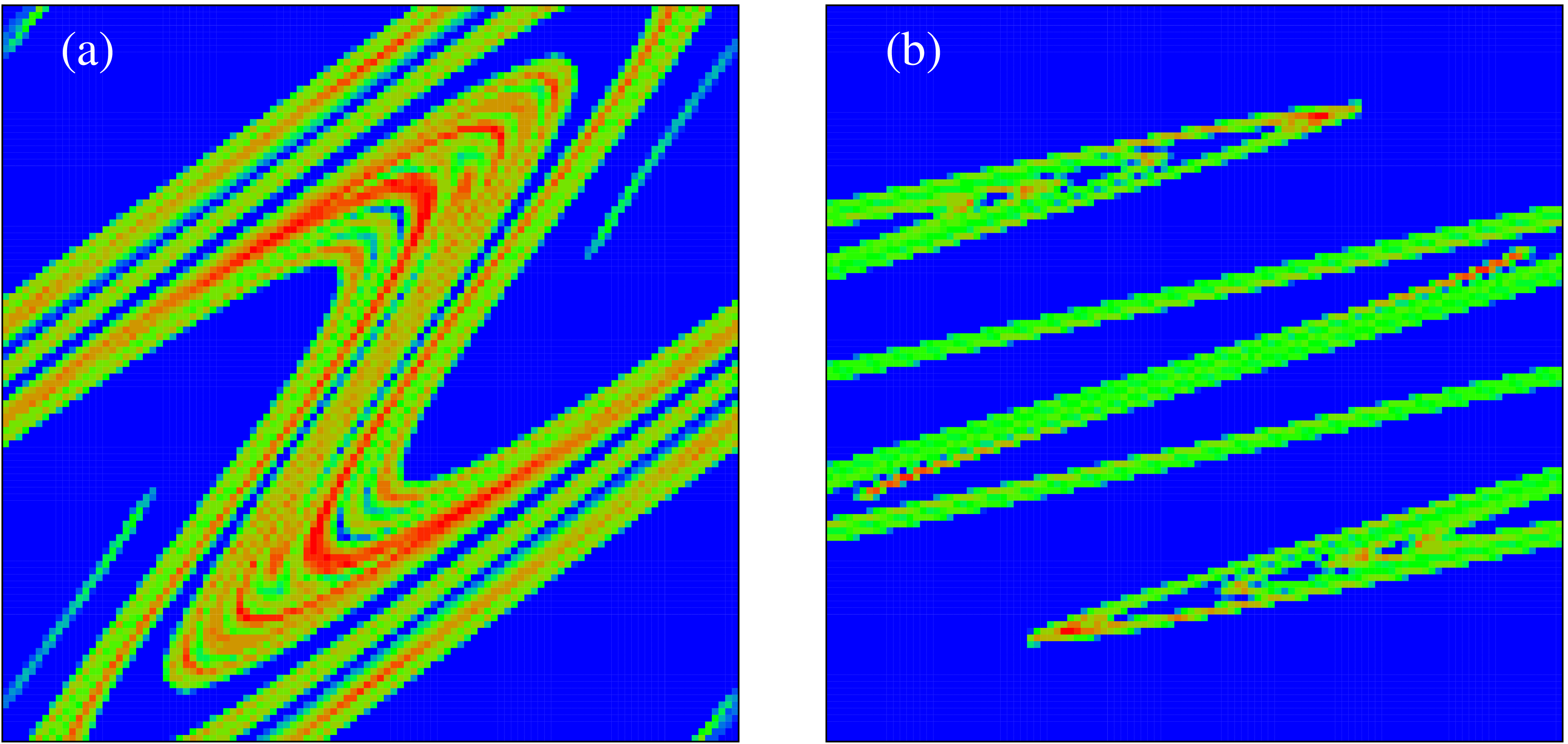}
\end{center}
\caption{
(Color online) Phase space representation of eigenstates 
of the UFPO $S$ 
for $N = 110 \times 110$ cells (color is proportional to absolute 
value $|\psi_i|$ with red/gray for maximum and blue/black for zero).
Panel (a) shows an eigenstate 
with maximum eigenvalue $\lambda_1 = 0.756$
of the UFPO of map (\ref{eq6_1}) with absorption
at $K_s = 7$, $a=2$, $\eta=1$, 
the space region is ($-aK_s/4\pi \leq y \leq aK_s/4\pi$, $0 \leq x \leq 1$),
the fractal dimension of 
the strange repeller set nonescaping in future 
is $d_e=1+d/2 = 1.769$.
Panel (b) shows an eigenstate at $\lambda = 1$
of the UFPO of map (\ref{eq6_1}) without absorption
at $K_s = 7$,  $\eta = 0.3$, 
the shown space region is 
($-1/\pi \leq y \leq 1/\pi$, $0 \leq x \leq 1$) and the fractal
dimension of the strange attractor is $d = 1.532$.  
After \cite{ermann:2010b}.
\label{fig6_4}}
\end{figure}

\subsection{Dynamical maps with strange attractors}
\label{s6.3}

The fractal Weyl law (\ref{eq5_1}) has initially been proposed for
quantum systems with chaotic scattering. However, it is natural
to assume that it should also work for
Perron-Frobenius operators of dynamical systems.
Indeed, the mathematical results for the Selberg zeta function 
indicated that the law  (\ref{eq5_1}) should remain valid for the UFPO
(see Refs. at \cite{nonnenmacher:2014}).
A detailed test of this conjecture \cite{ermann:2010b} 
has been performed 
for the map (\ref{eq6_1}) with dissipation 
at $0<\eta<1$, when at large $K_s$ the dynamics converges to
a strange attractor in the range $-2 <y<2$, 
and for the nondissipative case
$\eta=1$ with absorption where all orbits
leaving the interval $-aK_s/4\pi \leq y \leq aK_s/4\pi$ are 
absorbed after one iteration (in both cases
there is no modulus in $y$).

An example of the spectrum of UPFO for 
the model with dissipation is shown in 
Fig.~\ref{fig6_2}(a). We see that now, in contrast to 
the symplectic case of Fig.~\ref{fig6_2}(b), the spectrum has a
significant gap which separates the eigenvalue $\lambda=1$
from the other eigenvalues with $|\lambda|<0.7$.
For the case with absorption the spectrum has a similar structure
but now with $|\lambda|<1$ for the leading eigenvalue $\lambda$ since the 
total number of initial trajectories decreases with the
number of map iterations due to absorption implying that for this case 
$\sum_i S_{ij}<1$ with $S$ being the UPFO.

It is established that 
the distribution of density of states $d W/d \gamma$ 
(or $d W/d|\lambda|$) converges to a fixed
distribution in the limit of large $N$ 
or cell size going to zero \cite{ermann:2010b}
(see Fig.4 there).
This demonstrates the validity of the Ulam conjecture
for considered systems.

Examples of two eigenstates of the UFPO for these two models
are shown in Fig.~\ref{fig6_4}. The fractal structure of eigenstates 
is well visible. 
For the dissipative case without absorption we have eigenstates
localized on the strange attractor.
For the case with absorption 
eigenstates are located on a strange repeller 
corresponding to an invariant set of nonescaping orbits.
The fractal dimension $d$ of these classical invariant sets
can be computed by the usual box-counting method 
for dynamical systems. It is important to note that
for the case with absorption it is more natural to measure
the dimension $d_e$ of the set of orbits nonescaping in future.
Due to the time reversal symmetry of the continuous map
the dimension of the set of orbits nonescaping in the past is also
$d_e$. Thus the phase space dimension $2$ 
is composed of $2=d_e+d_e-d$ and $d_e=1+d/2$
where $d$ is the dimension of the invariant set of orbits
nonescaping neither in the future neither in the past.
For the case with dissipation without absorption all orbits
drop on a strange attractor 
and we have the dimension of invariant set
$d_e=d$.

\subsection{Fractal Weyl law for Perron-Frobenius operators}
\label{s6.4}

\begin{figure}[H]
\begin{center}
\includegraphics[width=0.48\textwidth]{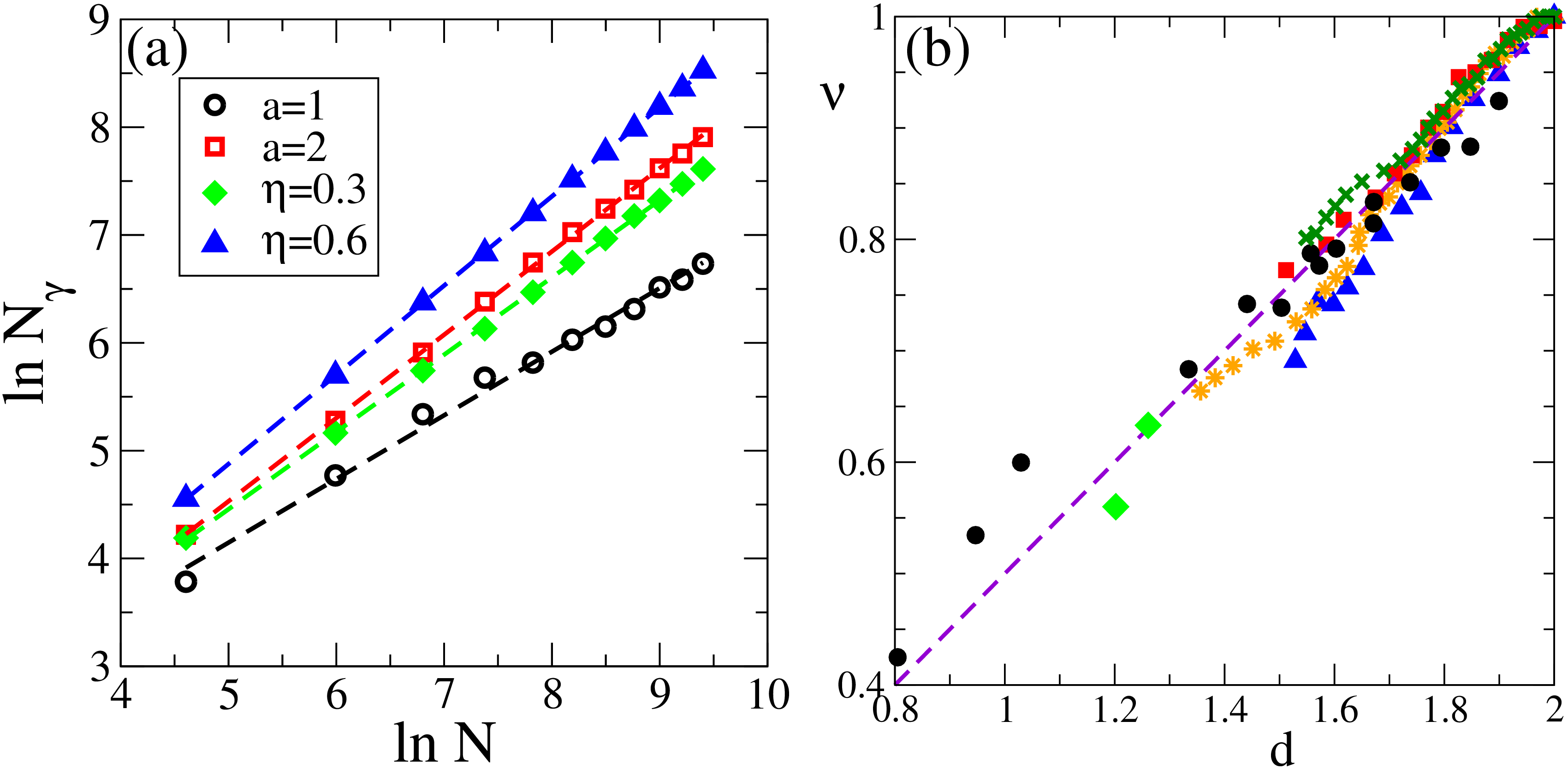}
\end{center}
\caption{
(Color online) 
Panel (a) shows the dependence of the integrated 
number of states $N_{\gamma}$ with decay rates 
$0 \leq \gamma\leq\gamma_b=16$ on the size $N$ of the UFPO matrix $ S$
for the map (\ref{eq6_1}) at $K_s=7$. 
The fits of numerical data, 
shown by dashed straight lines, give
$\nu=0.590, d_e=1+d/2= 1.643$ (at $a=1$);
$\nu=0.772, d_e=1+d/2=1.769$ (at $a=2$);
$\nu=0.716, d=1.532$ (at $\eta=0.3$);
$\nu=0.827, d=1.723$ (at $\eta=0.6$).
Panel (b) shows
the fractal Weyl exponent $\nu$ 
as a function of fractal dimension $d$ of the 
invariant fractal set for the map  (\ref{eq6_1})
with a strange attractor $(\eta<1)$
at $K_s=15$ (green/gray crosses),  $K_s=12$ (red/gray squares),
$K_s=10$ (orange/gray stars), $K_s=7$ blue/black triangles;
for a  strange repeller  $(\eta=1)$
at $K_s=7$ (black points) and 
for a strange attractor for the H\'enon map 
at standard parameters  $a=1.2; 1.4$, $b=0.3$ 
(green diamonds).
The straight dashed line shows the 
fractal Weyl law dependence $\nu=d/2$.
After \cite{ermann:2010b}.
\label{fig6_5}}
\end{figure}

The direct verification of the validity of the fractal Weyl law 
(\ref{eq5_1}) is presented in Fig.~\ref{fig6_5}. 
The number of eigenvalues
$N_\gamma$ in a range with 
$0 \leq \gamma \leq \gamma_b$ ($\gamma=-2\ln |\lambda|$)
is numerically computed as a function of matrix size $N$.
The fit of the dependence $N_\gamma(N)$,
as shown in Fig.~\ref{fig6_5}(a),
allows to determine the exponent $\nu$
in the relation $N_\gamma \propto N^\nu$. The dependence of $\nu$
on the fractal dimension $d$, computed from the 
invariant fractal set by the box-counted method,
is shown in  Fig.~\ref{fig6_5}(b). 
The numerical data are in good agreement with the theoretical fractal Weyl law
dependence $\nu=d/2$. This law works for a variety of parameters
for the system (\ref{eq6_1}) with absorption and dissipation,
and also for a strange attractor in the H\'enon map
(${\bar x}=y+1 -a x^2, {\bar y}=b x$). We attribute
certain deviations, visible in Fig.~\ref{fig6_5}
especially for $K_s=7$, to 
the fact that at $K_s=7$ there is a small island of stability
at $\eta=1$, which  can produce 
certain influence on the dynamics.

The physical origin of the law (\ref{eq5_1})
can be understood in a simple way: the number of states $N_\gamma$
with finite values of $\gamma$ is proportional to the number of
cells $N_f \propto N^{d/2}$ on the fractal set of strange attractor.
Indeed, the results for the overlap measure 
show that the eigenstates $N_\gamma$ have a strong overlap with the
steady state while the states with $\lambda \rightarrow 0$
have very small overlap. Thus almost all $N$
states have eigenvalues  $\lambda \rightarrow 0$
and only a small fraction of states on a strange attractor/repeller
$N_\gamma \propto N_f \propto N^{d/2} \ll N$
has finite values of $\lambda$. We also checked that the participation
ratio $\xi$ of the eigenstate  at $\lambda=1$,
grows as $\xi \sim N_f \propto N^{d/2}$
in agreement with the fractal Weyl law \cite{ermann:2010b}.

\subsection{Intermittency maps}
\label{s6.5}

The properties of the Google matrix
generated by one-dimensional intermittency maps
are analyzed in \cite{ermann:2010a}. 
It is found that for such Ulam networks
there are many eigenstates 
 with eigenvalues $|\lambda|$ being very close to
unity.  The PageRank of such networks at $\alpha=1$ is 
characterized by a power law decay with an exponent
determined by the parameters of the map.
It is interesting to note that usually for WWW 
the PageRank probability is proportional 
to a number of ingoing links distribution 
(see e.g. \cite{litvak:2008}).
For the case of intermittency maps
 the decay of PageRank is independent of 
number of ingoing links. In addition,  
 for $\alpha$ close to unity
a decay of the PageRank has an exponent $\beta \approx 1$
but at smaller values $\alpha \leq 0.9$
the PageRank becomes completely delocalized.
It is shown that the delocalization depends on
the intermittency exponent of the map.
This indicates that a rather dangerous phenomenon
of PageRank delocalization can appear for certain
directed networks. At the same time the one-dimensional
intermittency map still generates a relatively simple
structure of links with a typical
number of links per node being close to unity.
Such a case is  probably not very
typical for real networks. Therefore it is
useful to analyze richer Ulam networks with a
larger number of links per node.

\subsection{Chirikov typical map}
\label{s6.6}

\begin{figure}[H]
\begin{center}
\includegraphics[width=0.48\textwidth]{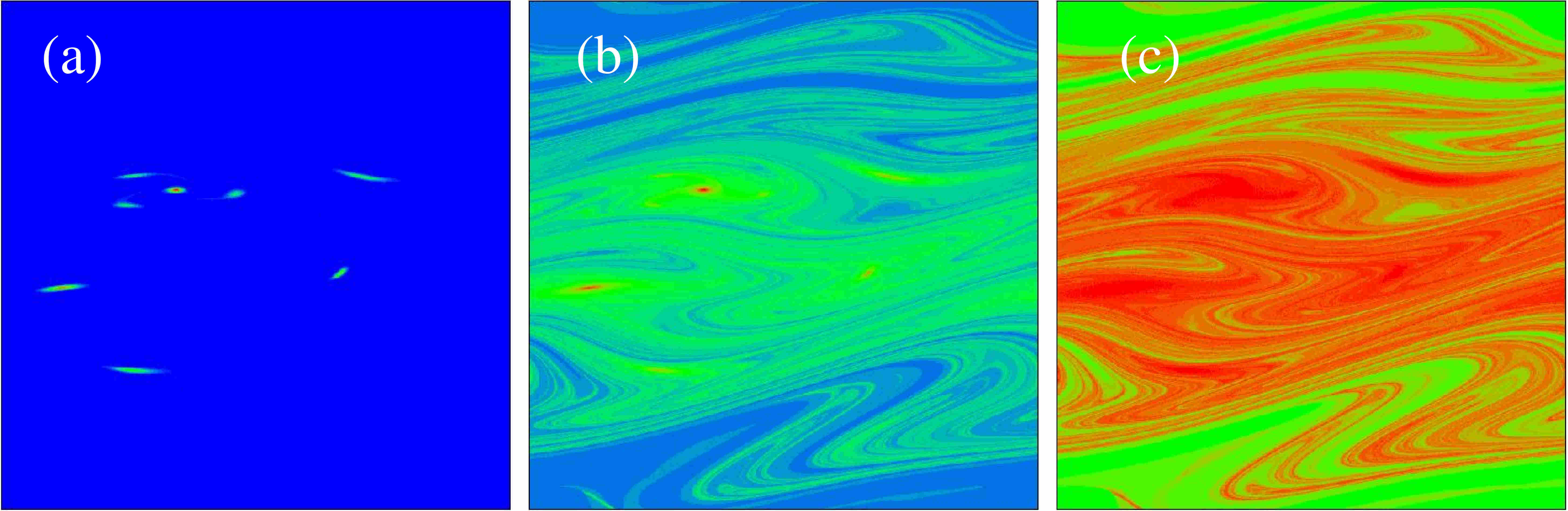}
\end{center}
\caption{(Color online) 
 PageRank probability $P_j$ 
for the Google matrix generated by
the Chirikov typical map  at
$T=10$, $k_s=0.22$, $\eta=0.99$ 
with  $\alpha=1$ (a), $\alpha=0.95$ (b), and $\alpha=0.85$ (c).
The probability $P_j$ is shown in 
the phase space region  
$0 \leq x < 2\pi; -\pi \leq y < \pi $
which is divided in
$N=3.6 \cdot 10^5$ cells; $P_j$
is zero for blue/black and maximal for red/gray.
After \cite{shepelyansky:2010a}.
\label{fig6_6}}
\end{figure}

With this aim we consider 
the Ulam networks generated by the Chirikov typical map
with dissipation studied by \cite{shepelyansky:2010a}. 
The map introduced,
by Chirikov in 1969 for description of continuous chaotic flows,
has the form:
\begin{equation}
y_{t+1} =\eta y_{t}+k_s \sin (x_t+\theta_t) \;, 
\;\; x_{t+1} = x_t+y_{t+1} \; .
\label{eq6_2}
\end{equation}
Here the dynamical variables $x,y$ are taken at 
integer moments of time $t$. Also
$x$ has a meaning of phase variable 
and $y$ is a conjugated
momentum or action. The  phases
$\theta_t=\theta_{t+T}$ are $T$ random phases periodically repeated
along  time $t$. We stress that their $T$ values 
are chosen and fixed once and they are not
changed during the dynamical evolution of $x,y$.
We consider the map in the region of Fig.~\ref{fig6_6} 
($0 \leq x < 2\pi, -\pi \leq y <\pi$)
with the $2\pi$-periodic boundary conditions. 
The parameter $0< \eta < 1$ gives a global dissipation. 
The properties of the symplectic map at $\eta=1$
have been studied  in detail in \cite{frahm:2009}.
The dynamics is globally chaotic for 
$k_s > k_c \approx 2.5/T^{3/2}$ and
the Kolmogorov-Sinai entropy is $h \approx 0.29 {k_s}^{2/3}$
(more details about 
the Kolmogorov-Sinai entropy can be found in
\cite{brin:2002,chirikov:1979,cornfeld:1982}).
A bifurcation diagram at $\eta <1$
shows a series of transitions between
fixed points, simple and strange attractors. 
Here we present results for $T=10$, $k_s=0.22$, $\eta=0.99$
and a specific random set of  $\theta_t$ 
given in \cite{shepelyansky:2010a}.

Due to exponential instability of motion 
one cell in the Ulam method gives transitions
approximately to $k_{\rm cl} \approx \exp(hT)$
other cells. According to this relation a large
number of cells $k_{\rm cl}$
can be coupled at large $T$ and $h$.
For parameters of Fig.~\ref{fig6_6} 
one finds an approximate power law distribution
of ingoing and outgoing links in the corresponding
Ulam network with 
the exponents $\mu_{\rm in} \approx \mu_{\rm out} \approx 1.9$.
The variation of the PageRank vector with the damping factor $\alpha$
is shown in Fig.~\ref{fig6_6} on the phase plane
$(x,y)$. For $\alpha=1$ the PageRank is concentrated in a vicinity
of a simple attractor composed of  several fixed points on
the phase plane. 
Thus  the dynamical attractors are the most popular nodes
from the network view point.
With a decrease of $\alpha$ down to $0.95, 0.85$
values we find a stronger and stronger
delocalization of PageRank
over the whole phase space.

\begin{figure}[H]
\begin{center}
\includegraphics[width=0.48\textwidth]{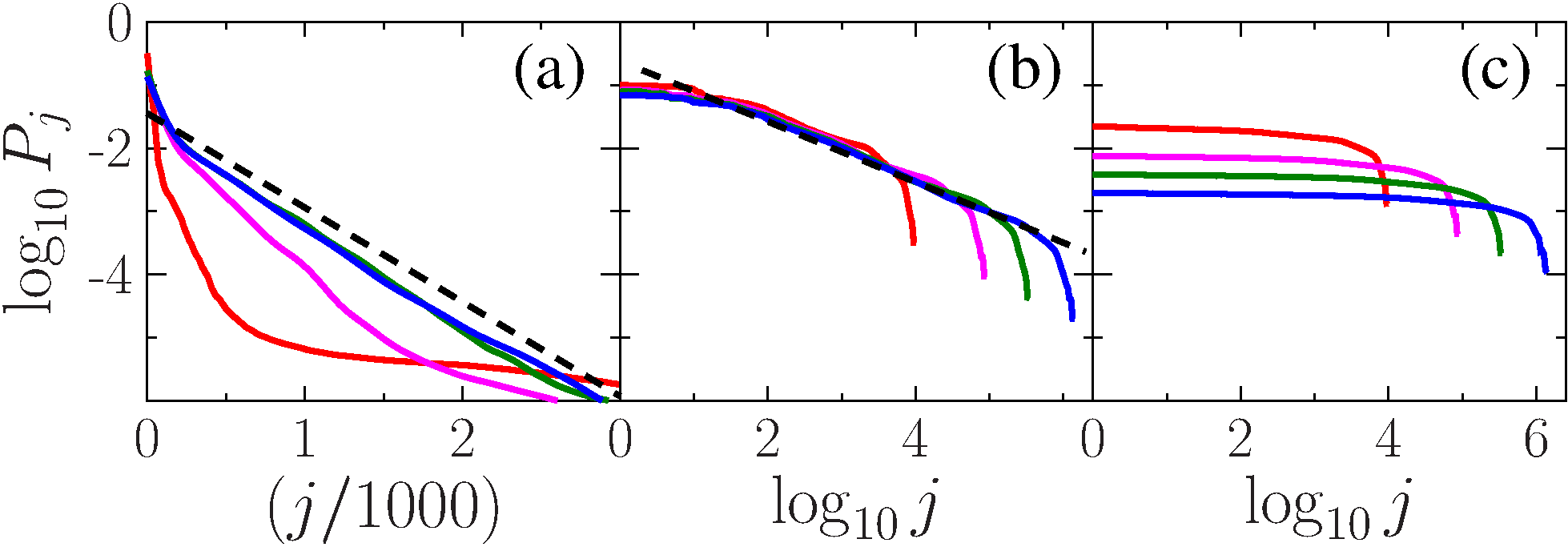}
\end{center}
\caption{(Color online) 
 Dependence of PageRank probability $P_j$  on PageRank index $j$ 
for number of cells in the UFPO being
$N=10^4$, $9 \times 10^4$, $3.6 \times 10^5$ and $1.44 \times 10^6$ 
(larger $N$ have more dark and more long curves in (b), (c);
in (a) this order of $N$ is for curves from bottom to top 
(curves for $N=3.6 \times 10^5$ and $1.44 \times 10^6$
practically coincide in this panel; for online version we note that
the above order of $N$ values corresponds to red, magenta, green, blue 
curves respectively).
Dashed line in (a)
shows an exponential Boltzmann decay (see text, line is shifted
in $j$ for clarity).
The dashed straight line in (b) shows the fit 
$P_j \sim 1/j^\beta$ with $\beta=0.48$. 
Other parameters, including the values of $\alpha$, and panel order 
are as in Fig.~\ref{fig6_6}. 
After \cite{shepelyansky:2010a}.
\label{fig6_7}}
\end{figure}

The delocalization with a decrease of $\alpha$
is also well seen in Fig.~\ref{fig6_7}
where we show $P_j$ dependence on PageRank index
$j$ with a monotonic decreasing probability $P_j$.
At $\alpha=1$ we have an exponential decay 
of $P_j$ with $j$ that corresponds to a Boltzmann type distribution
where a noise produced by a finite cell size in the Ulam
method is compensated by dissipation.
For $\alpha=0.95$ the random jumps of a network surfer,
induced by the term $(1-\alpha)/N$ in (\ref{eq3_1}),
produce an approximate
power law decay of $P_j \propto 1/j^\beta$ with
$\beta \approx 0.48$. For $\alpha=0.85$ the 
PageRank probability is flat
and completely delocalized over the whole phase space.  

The analysis of the spectrum of $S$ for the map (\ref{eq6_2})
for the parameters of Fig.~\ref{fig6_7} shows the existence
of eigenvalues being very close to $\lambda=1$,
however, there is no exact degeneracy 
as it is the case for UK universities which we will discuss below.
The spectrum is characterized by the fractal Weyl law with the 
exponent $\nu \approx 0.85$.
For eigenstates with $|\lambda| <1$
the values of IPR $\xi$ are less than 
$300$ for a matrix size $N \approx 1.4 \times 10^4$
showing that eigenstates are localized.
However, for the PageRank the computations can be done 
with larger matrix sizes reaching a maximal value of 
$N=  6.4 \times 10^5$. The dependence of $\xi$ on $\alpha$
shows that a delocalization transition of PageRank vector
takes place for $\alpha < \alpha_c \approx 0.95$.
Indeed, at $\alpha =0.98$ we have $\xi \approx 30$
while at
$\alpha \approx 0.8$ the IPR value of PageRank
becomes comparable with the whole system size
$\xi \approx 5 \times 10^5 \sim N =6.4 \times 10^5$
(see Fig.9 at \cite{shepelyansky:2010a}).

The example of Ulam networks
considered here shows that a dangerous phenomenon
of PageRank delocalization can take place under certain conditions.
This delocalization may represent a serious danger
for efficiency of search engines since for a 
delocalized flat PageRank the ranking of nodes becomes
very sensitive to small perturbations and fluctuations.

\section{Linux Kernel networks}
\label{s7}

Modern software codes represent now complex
large scale structures and analysis and optimization
of  their architecture become a challenge.
An interesting approach to this problem,
based on a directed network construction,
has been proposed by \cite{chepelianskii:2010}.
Here we present results obtained for such networks.

\subsection{Ranking of software architecture}
\label{s7.1}

Following \cite{chepelianskii:2010} we consider the
Procedure Call Networks (PCN) for open source programs
with emphasis on the code of Linux Kernel \cite{linux:2010}
written in the C programming language \cite{kernighan:1978}.
In this language the code is structured as a sequence 
of procedures calling each other. Due to that feature the
organization of a code can be naturally represented as
a PCN, where each node represents 
a procedure and each directed link corresponds
to a procedure call. For the Linux source code
such a directed network is built by its lexical scanning
with the identification of all the
defined procedures. For each of them a list keeps track of
the procedures calls inside their definition. 

\begin{figure}[H]
\begin{center}
\includegraphics[width=0.48\textwidth]{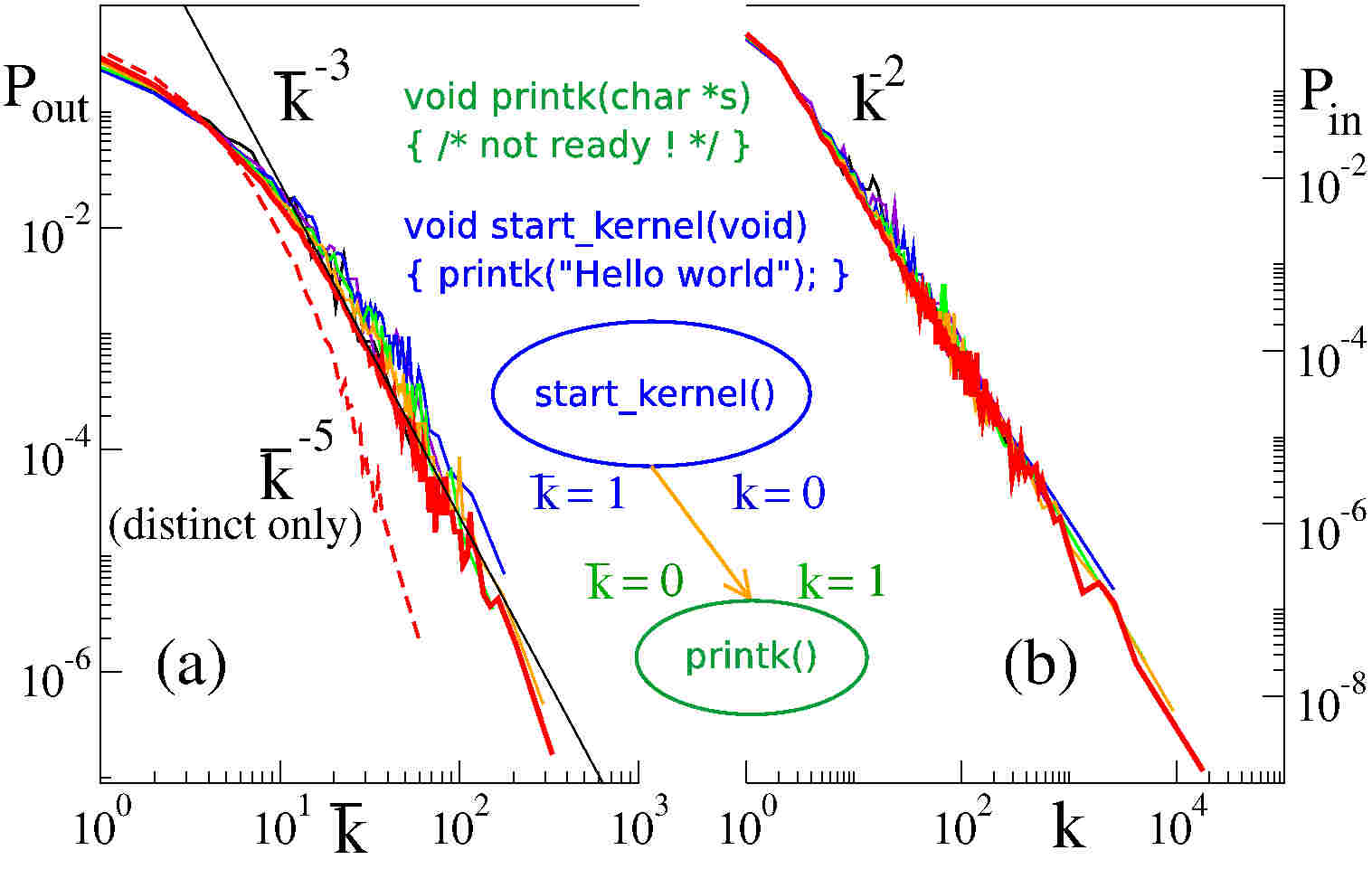}
\end{center}
\vglue -0.3cm
\caption{(Color online)
The diagram in the center represents the PCN of 
a toy kernel with two procedures 
written in  C-programming language. 
The data on panels (a) and (b)
show  outdegree and indegree probability 
distributions $P_{\rm \,out}({\bar k})$ and $P_{\rm \,in}(k)$ respectively. 
The colors correspond to different Kernel releases.
The most recent version 2.6.32, with $N = 285509$ and 
an average $3.18$ calls per 
procedure, is represented in red/gray.
Older versions (2.4.37.6, 2.2.26, 2.0.40, 1.2.12, 1.0) 
with $N$ respectively equal to (85756, 38766, 14079, 4358, 2751) 
follow the same behavior. 
The dashed curve in (a) shows the outdegree probability 
distribution if only calls to distinct 
destination procedures are kept.
After \cite{chepelianskii:2010}.
\label{fig7_1}}
\end{figure}

An example
of the obtained network for a toy code with 
two procedures {\it start\_kernel}
and {\it printk } is shown in Fig.~\ref{fig7_1}.
The in/out-degrees of this model, noted as
$k$ and ${\bar k}$, are shown in Fig.~\ref{fig7_1}.
These numbers correspond to the number of out/in-going 
calls for each procedure.
The obtained in/out-degree probability distributions
$P_{\rm \,in}(k)$, $P_{\rm \,out}({\bar k})$ 
are shown Fig.~\ref{fig7_1} for 
different Linux Kernel releases. 
These distributions are well described by power law 
dependencies $P_{\rm \,in}(k) \propto 1/k^{\mu_{\rm in}}$ and
$P_{\rm \,out}({\bar k}) \propto 1/{\bar k}^{\mu_{\rm out}}$ 
with $\mu_{\rm in}=2.0 \pm 0.02$,
and $\mu_{\rm out} = 3.0 \pm 0.1$. 
These values of exponents are close to those found for
the WWW \cite{donato:2004,pandurangan:2005}.
If only calls to distinct functions are counted in the outdegree
distribution then the exponent drops to $\mu_{\rm out} \approx 5$ 
whereas $\mu_{\rm in}$ remains unchanged. 
It is important that the distributions 
for the different kernel releases remain stable
even if the network size increases from $N = 2751$ for version
V1.0 to $N = 285509$ for the latest version V2.6.32 taken into account 
in this study.
This confirms the free-scale structure of software architecture
of Linux Kernel network.

The probability distributions of PageRank and CheiRank vectors
are also well described by power laws with exponents
$\beta_{\rm in} \approx 1$ and $\beta_{\rm out} \approx 0.5$
being in good agreement with the usual relation
$\beta=1/(\mu-1)$ (see Fig.2 in \cite{chepelianskii:2010}). 
For V2.6.32 the top three procedures of PageRank 
at $\alpha=0.85$ are
{\it printk, memset, kfree} with probabilities
$0.024, 0.012, 0.011$ respectively, while at the 
top of CheiRank we have 
{\it start\_kernel, btrfs\_ioctl, menu\_finalize}
with respectively $0.000280, 0.000255, 0.000250$.
These procedures perform rather different tasks
with {\it printk} reporting messages and {\it start\_kernel}
initializing the Kernel and managing the repartition of tasks.
This gives an idea that both Page\-Rank and CheiRank order
can be useful to highlight en different aspects of directed and inverted
flows on our network. Of course, in the context of WWW
ingoing links related to PageRank
are less vulnerable as compared to 
outgoing links related to CheiRank,
which can be modified by a user rather easily.
However, in other type of networks
both directions of links appear in a natural manner
and thus both vectors of PageRank and CheiRank
play an important and useful role.

For the Linux Kernel network the correlator $\kappa$
(\ref{eq4_1}) between PageRank and CheiRank
vectors is close to zero (see Fig.~\ref{fig4_2}).
This confirms the independence of two vectors.
The density distribution of nodes of the Linux Kernel network, shown 
in Fig.~\ref{fig4_3}(b), has a homogeneous distribution
along $\ln K + \ln K^*=const$ lines demonstrating once more
absence of correlations between $P(K_i)$ and 
$P^*({K_i}^*)$.
Indeed, such homogeneous distributions appear if nodes are generated
randomly with factorized probabilities
$P_i {P_i}^*$ \cite{chepelianskii:2010,zhirov:2010}.
Such a situation seems to be rather
generic for software architecture.
Indeed, other open software codes also have a small values
of correlator, e.g. OpenSource software including Gimp 2.6.8 
has $\kappa=-0.068$ at $N = 17540$ and 
X Windows server R7.1-1.1.0 has
$\kappa = -0.027$ at $N = 14887$.
In contrast to these software codes the Wikipedia networks
have large values of $\kappa$ and inhomogeneous distributions
in $(K,K^*)$ plane (see Figs.~\ref{fig4_2},\ref{fig4_3}).

The physical reasons for absence of correlations
between $P(K)$ and $P^*(K^*)$ 
have been explained in \cite{chepelianskii:2010} 
on the basis of the concept of ``separation of concerns''
in software architecture \cite{dijkstra}. 
It is argued that a good
code should decrease the number of procedures that have high values 
of both PageRank
and CheiRank since such procedures will play a critical role 
in error propagation since
they are both popular and highly communicative at the same time. 
For example in the Linux Kernel, {\it do\_fork}, 
that creates new processes, belongs to this class.
Such critical procedures may introduce subtle errors 
because they entangle otherwise
independent segments of code. 
The above observations suggest that the independence
between popular procedures, which have high $P(K_i)$ and 
fulfill important but well 
defined tasks, and communicative procedures, which have high
$P^*({K_i}^*)$ and organize
and assign tasks in the code, is an important ingredient 
of well structured software.

\subsection{Fractal dimension of Linux Kernel Networks}
\label{s7.2}

\begin{figure}[H]
\begin{center}
\includegraphics[width=0.48\textwidth]{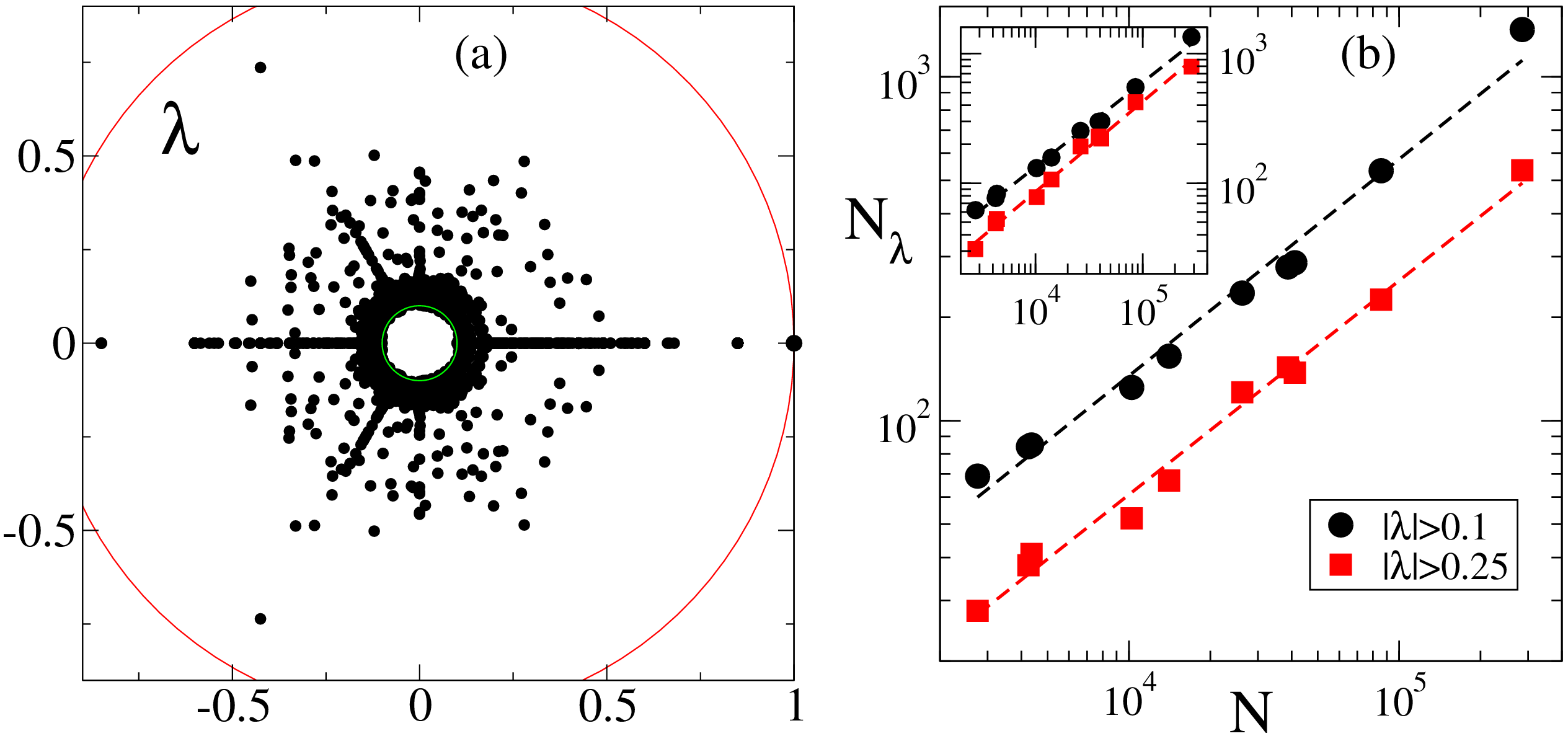}
\end{center}
\caption{(Color online)
Panel (a) shows distribution of eigenvalues 
$\lambda$ in the complex plane 
for the Google matrix $G$ of the 
Linux Kernel version $2.6.32$ with $N=285509$
and $\alpha=0.85$;
the solid curves represent the unit circle and 
the lowest limit of computed eigenvalues. 
Panel (b) shows dependence of the integrated number 
of eigenvalues $N_\lambda$ with 
$\vert\lambda\vert>0.25$ (red/gray squares) 
and $\vert\lambda\vert>0.1$ (black circles)
as a function of the total number of processes 
$N$ for versions of Linux Kernels. 
The values of $N$ correspond (in increasing order) 
to Linux Kernel versions 
$1.0$, $1.1$, $1.2$, $1.3$, $2.0$, $2.1$, $2.2$, $2.3$, $2.4$ and $2.6$.
The power law $N_\lambda\propto N^{\nu}$ has fitted values 
$\nu_{\vert\lambda\vert>0.25}=0.622 \pm 0.010$ and
$\nu_{\vert\lambda\vert>0.1}=0.630 \pm 0.015$.
Inset shows data for the Google matrix $G^*$ 
with inverse link directions,
the corresponding  exponents are 
$\nu^*_{\vert\lambda\vert>0.25}=0.696 \pm 0.010$ and
$\nu^*_{\vert\lambda\vert>0.1}=0.652 \pm 0.007$. 
After \cite{ermann:2011a}.
\label{fig7_2}}
\end{figure}

The spectral properties the Linux Kernel network
are analyzed in \cite{ermann:2011a}. At large $N$
the spectrum  is obtained with the help of Arnoldi method from
ARPACK library. This allows to find eigenvalues with
$|\lambda| > 0.1$ for the maximal $N$ at V2.6.32.
An example of complex spectrum $\lambda$ of $G$ is shown in
Fig.~\ref{fig7_2}(a). There are clearly visible 
lines at real axis and 
polar angles $\varphi = \pi/2, 2\pi/3, 4\pi/3, 3\pi/2$.
The later are related to certain cycles in procedure calls,
e.g. an eigenstate at $\lambda_i =0.85 \exp(i 2\pi/3)$ is 
located only on 6 nodes. The spectrum of $G^*$ has a similar structure.

The network size $N$ grows with the version number of Linux Kernel
corresponding to its evolution in time.
We determine the total number of states
$N_\lambda$ with $0.1 < |\lambda| \leq 1$
and $0.25 < |\lambda| \leq 1$.
The dependence of $N_\lambda$ on $N$, shown in 
Fig.~\ref{fig7_2}(b),
clearly demonstrates the validity of the fractal Weyl law
with the exponent $\nu \approx  0.63$ for $G$
(we find $\nu^* \approx 0.65$ for $G^*$). 
We take the values of $\nu$ for $\lambda=0.1$ where the number 
of eigenvalues $N_\lambda$ gives a better statistics. 
Within statistical errors the value of $\nu$ is not sensitive
to the cutoff value at small $\lambda$. The matrix $ G^*$
has slightly higher values of $\nu$. These results show that the PCN
of Linux Kernel has a fractal dimension $d=2\nu \approx 1.26$
for $G$ and $d=2\nu \approx 1.3$ for $G^*$.

To check that the fractal dimension of the PCN indeed has this value
the dimension of the network is computed by another direct method
known as the cluster growing method (see e.g. \cite{song:2005}). 
In this method the
average mass or number of nodes $\langle M_c \rangle$ 
is computed as a function of
the {\it network distance $l$} counted from an initial
seed node with further averaging over all seed nodes.
For a dimension $d$ the mass $\langle M_c \rangle$
should grow as $\langle M_c \rangle \propto l^d$
that allows to determine the value of $d$ for a given network.
It should be noted that the above method should be generalized
for the case of directed networks. For that
 the network distance $l$ is computed
following only outgoing links. The average 
of $\langle M_c(l) \rangle$ is done over all nodes.
Due to global averaging the method gives
the same result for the matrix with inverted link direction
(indeed, the total number of outgoing links is equal to the number
of ingoing links). However, as established in \cite{ermann:2011a},
the fractal  dimension obtained by this generalized method
is very different from the case of converted undirected network,
when each directed link is replaced by an undirected one.
The average dimension obtained with this method for PCN
is $d =1.4$ even if a certain 20\%  increase of $d$ 
appears for the latest Linux versions V2.6.
We attribute this deviation for the version V2.6
to the well known fact that significant 
rearrangements in the Linux Kernel have been done
after version V2.4 \cite{linux:2010}.

Thus in view of the above restrictions we consider that there is
a rather good agreement of the fractal dimension
obtained from the fractal Weyl law with $d \approx 1.3$
and the value obtained with the cluster growing method
which gives an average $d \approx 1.4$.
The fact that $d$ is approximately the same for
all versions up to V2.4  means that the  Linux Kernel
is characterized by a self-similar fractal
growth in time. The closeness of $d$  to unity
signifies that procedure calls are almost linearly ordered 
that corresponds to a good code organization.
Of course, the fractal Weyl law gives the dimension $d$
obtained during time evolution of the network.
This dimension is not necessary the same as 
for a given version of the network of fixed size.
However, one can expect that the growth 
goes in a self-similar way \cite{dorogovtsev:2008} 
and that the static dimension is close to the 
dimension value emerging 
during the time evolution. This  can be viewed 
as a some kind of ergodicity 
conjecture. Our data show that this conjecture
works with a good accuracy up to the Linux Kernel V.2.6.

Thus the results obtained in \cite{ermann:2011a}
and described here confirm the validity of the fractal Weyl law
for the Linux Kernel network with the exponent $\nu \approx 0.65$
and the fractal dimension $ d \approx 1.3$. 
It is important to note that the fractal Weyl exponent $\nu$
is not sensitive to the exponent 
$\beta$ characterizing the decay of the PageRank.
Indeed, the exponent $\beta$  
remains practically the same for the WWW \cite{donato:2004}
and the PCN of Linux Kernel \cite{chepelianskii:2010}
while the values of fractal dimension are different
with $d \approx 4$ for WWW and $d \approx 1.3$ for PCN
(see \cite{ermann:2011a} and Refs. therein).

The analysis of the eigenstates of $G$ and $G^*$
shows that their IPR values  remain small
($\xi < 70$)
compared to the matrix size $N \approx 2.8 \times 10^5$
showing that they are well localized on certain
selected nodes.

\section{WWW networks of UK universities}
\label{s8}

The WWW networks of certain UK universities for years between 2002 
and 2006 are publicly available at \cite{ukuniv:2011}.
Due to their modest size, these networks are well
suitable for a detail study of 
PageRank, CheiRank, complex eigenvalue spectra and eigenvectors 
\cite{frahm:2011}.

\subsection{Cambridge and Oxford University networks}
\label{s8.1}

We start our analysis of WWW university networks from 
those of Cambridge and Oxford 2006.
For example, in Fig.~\ref{fig4_1} we show the dependence of
PageRank (CheiRank) probabilities $P (P^*)$ on rank index $K$ ($K^*$) 
for the WWW of Cambridge 2006 
at $\alpha=0.85$. The decay is satisfactory described by
a power law with the exponent $\beta=0.75$ ($\beta=0.61$). 

The complex eigenvalue spectrum and the invariant 
subspace structure (see section \ref{s3.3}) have been studied in great 
detail for the cases of Cambridge 2006 and Oxford 2006.
For Cambridge 2006 (Oxford 2006) the network size is 
$N=212710$ ($200823$) and the number of links is 
$N_\ell=2015265$ ($1831542$). There are $n_{\rm inv}=1543 (1889)$
invariant subspaces, with maximal dimension $d_{\rm max}=4656 (1545)$, 
together they contain $N_s=48239$ ($30579$) subspace nodes 
leading to 3508 (3275) eigenvalues 
(of the matrix $S$) with $|\lambda_j|=1$ 
of which $n_1=1832 (2360)$ are at $\lambda_j=1$ 
(about 1\% of $N$). The last number $n_1$ is larger 
than the number of invariant subspaces $n_{\rm inv}$ since
each of the subspaces has at least one unit eigenvalue because 
 each subspace is described by a 
full representation matrix of the Perron-Frobenius type. 
To determine the complex eigenvalue spectrum one can apply exact 
diagonalization on each subspace and the Arnoldi method 
on the remaining core space. 

\begin{figure}[H]
\begin{center}
\includegraphics[width=0.9\columnwidth]{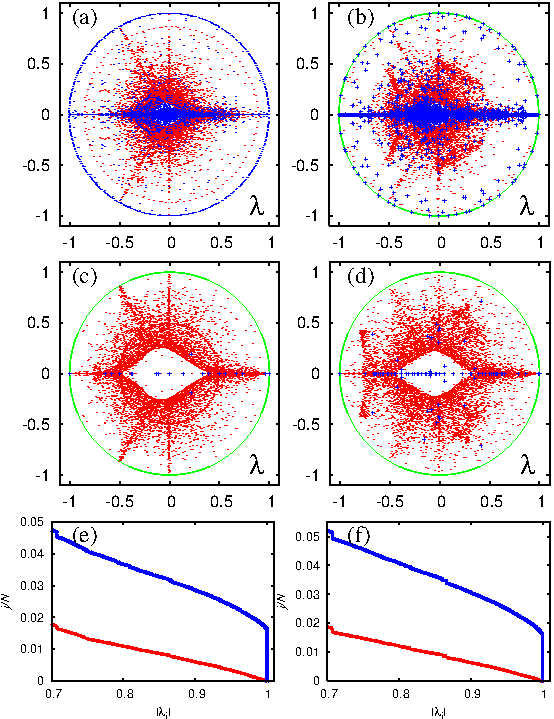}
\end{center}
\vglue -0.2cm
\caption{(Color online) 
Panels (a) and (b) show the complex eigenvalue 
spectrum $\lambda$ of matrix $S$   
for the University of Cambridge 2006 and Oxford 2006 respectively.
The spectrum $\lambda$ of matrix $S^*$ for Cambridge 2006 and Oxford 2006  
are shown in panels  (c) and (d).
Eigenvalues $\lambda$ of the core space are shown by red/gray points,
eigenvalues of isolated subspaces are shown by blue/black points and 
the green/gray curve (when shown) is the unit circle. Panels
(e) and (f) show the fraction $j/N$ of eigenvalues with 
$|\lambda| >  |\lambda_j|$ for the 
core space eigenvalues 
(red/gray bottom curve) and all eigenvalues (blue/black top curve)
from top row data for Cambridge 2006 and Oxford 2006. 
After \cite{frahm:2011}.
}
\label{fig8_1}
\end{figure}

The spectra of all subspace eigenvalues and $n_A=20000$ core 
space eigenvalues of the matrices $S$ and $S^*$ 
are shown in Fig.~\ref{fig8_1}. Even if the decay of 
PageRank and CheiRank probabilities with rank index is rather similar 
for both universities (see Fig.1 in \cite{frahm:2011}) the spectra
of two networks are very different. Thus the spectrum contains
much more detailed information about the network features
compared to the rank vectors.

\begin{figure}[H]
\begin{center}
\includegraphics[width=0.9\columnwidth]{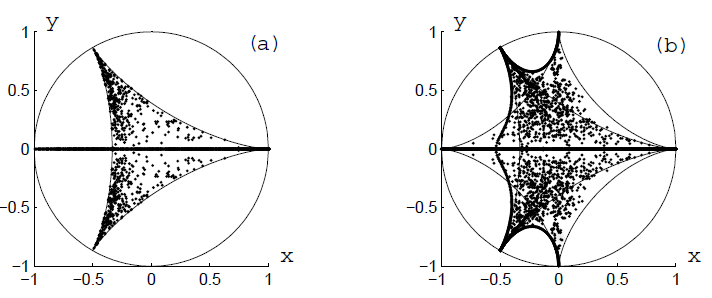}
\end{center}
\vglue -0.2cm
\caption{Spectra $\lambda$ of  800
random orthostochastic matrices of size
$N=3$ (a) and $N=4$ (b) ($Re \lambda=x, Im \lambda=y$). 
Thin lines denote 3- and 4-hypocycloids, while the thick
lines represent the 3-4 interpolation arc.
After \cite{zyczkowski:2003}.
}
\label{fig8bis}
\end{figure}

At the same time the spectra of two universities have
certain similar features. Indeed, one can identify 
cross and triple-star structures.
These structures are very similar to those seen in 
the spectra of random 
orthostochastic matrices of small size $N=3,4$
shown in Fig.~\ref{fig8bis} from \cite{zyczkowski:2003} 
(spectra of unistochastic
matrices have a similar structure). 
The spectrum borders, determined analytically
in \cite{zyczkowski:2003}
for these $N$ values, are also shown.
The similarity is more visible for  the spectrum of $S^*$ case
((c) and (d) of Fig.~\ref{fig8_1}).
We attribute this to a larger randomness
in outgoing links which have more fluctuations
compared to ingoing links,
as discussed in \cite{eom:2013b}.
The similarity of spectra of 
Fig.~\ref{fig8_1} with those
of random matrices in Fig.~\ref{fig8bis}
indicates that there are dominant
triple and quadruple structures 
of nodes present in the University networks
which are relatively weakly connected to other nodes.

The core space submatrix $S_{cc}$ of Eq.~(\ref{eq3_2}) 
does not obey to the column sum normalization due to 
non-vanishing elements 
in the block $S_{sc}$ which allow for a small but finite 
escape probability from core space to 
subspace nodes. Therefore the maximum eigenvalue of the core space (of 
the matrix $S_{cc}$) is below unity. For Cambridge 2006 (Oxford 2006) 
it is given by  $\lambda^{\rm (core)}_1=0.999874353718$ (0.999982435081) 
with a quite clear gap 
$1-\lambda^{\rm (core)}_1 \sim 10^{-4}$ ($\sim 10^{-5}$).

\subsection{Universal emergence of PageRank}
\label{s8.2}

\begin{figure}[H]
\begin{center}
\includegraphics[width=0.48\textwidth]{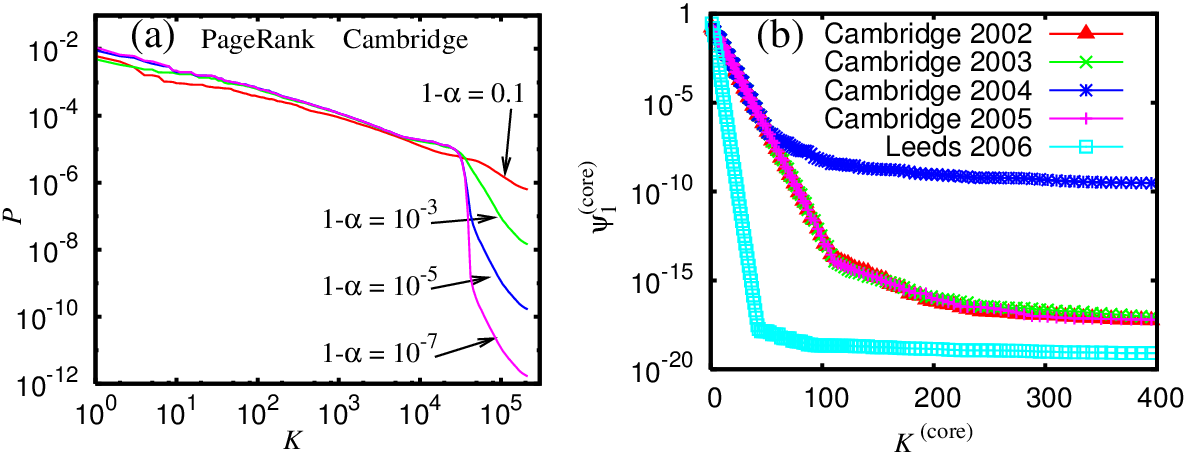}
\end{center}
\caption{(Color online) 
(a) PageRank $P(K)$ of Cambridge 2006 for 
$1-\alpha=0.1,\,10^{-3},\,10^{-5},\,10^{-7}$. 
(b) First core space eigenvector $\psi_1^{\rm (core)}$ 
versus its rank index $K^{\rm (core)}$ for the UK university 
networks with a small core space gap $1-\lambda_1^{\rm (core)}<10^{-8}$.
After \cite{frahm:2011}.
\label{fig8_2}}
\end{figure}

For $\alpha=1$ the leading eigenvalue $\lambda=1$ is highly degenerate 
due to the subspace structure. This degeneracy is lifted for $\alpha<1$ 
with a unique eigenvector, the Page\-Rank, 
for the leading eigenvalue. The 
question arises how the PageRank emerges if $1-\alpha\ll 1$. 
Following \cite{frahm:2011}, an answer is obtained 
from a formal matrix expression:
\begin{equation}
\label{PageRank1}
P=(1-\alpha)\,(I-\alpha S)^{-1}\,e/N ,
\end{equation}
where the vector $e$ has 
unit entries on each node and $I$ is the  unit matrix.
Then, assuming that $S$ is diagonalizable 
(with no nontrivial Jordan 
blocks) we can use the expansion:
\begin{equation}
\label{PageRank2}
P=\sum_{\lambda_j=1} c_j\,\psi_j + \sum_{\lambda_j\neq 1} 
\frac{1-\alpha}{(1-\alpha)+\alpha(1-\lambda_j)}\, c_j\,\psi_j \ .
\end{equation}
where $\psi_j$ are the eigenvectors of $S$ and $c_j$ coefficients 
determined by the expansion $e/N=\sum_j c_j\psi_j$. 
Thus Eq.~(\ref{PageRank2}) indicates that 
in the limit $\alpha\to 1$ the Page\-Rank
converges to a particular linear combination of the eigenvectors 
with $\lambda_j=1$, which 
are all localized in one of the subspaces. For a finite but very small 
value of $1-\alpha \ll 1-\lambda_1^{\rm (core)}$ 
the corrections for the contributions of the core space nodes are 
$\sim (1-\alpha)/(1-\lambda_1^{\rm (core)})$. This behavior is 
indeed confirmed by Fig.~\ref{fig8_2} (a) showing 
the evolution of the PageRank for different values of $1-\alpha$ for the 
case of Cambridge 2006 and using a particular method, 
based on an alternate combination of the power iteration method and the 
Arnoldi method \cite{frahm:2011}, to determine numerically the PageRank 
for very small values of $1-\alpha\sim 10^{-8}$. 

However, for certain of the university 
networks the core space gap $1-\lambda_1^{\rm (core)}$ 
is particularly small, for example 
$1-\lambda_1^{\rm (core)}\sim 10^{-17}$, such that 
in standard double precision arithmetic the Arnoldi method, applied 
on the matrix $S_{cc}$, does not 
allow to determine this small gap. For these particular cases it is 
possible to determine rather accurately the core space gap and the 
corresponding eigenvector by another numerical approach called 
``projected power method'' \cite{frahm:2011}. 
These eigenvectors, shown in Fig.~\ref{fig8_2} (b), 
are strongly localized on a modest number of 
nodes $\sim 10^2$ and with very small 
but non-vanishing values on the other nodes. 
Technically these vectors 
extend to the whole core space but practically 
they define small quasi-subspaces (in 
the core space domain) where the escape probability is extremely small
\cite{frahm:2011} and in the range $1-\alpha\sim 10^{-8}$ they still 
contribute to the PageRank according to Eq.~(\ref{PageRank2}). 

\begin{figure}[H]
\begin{center}
\includegraphics[width=0.48\textwidth]{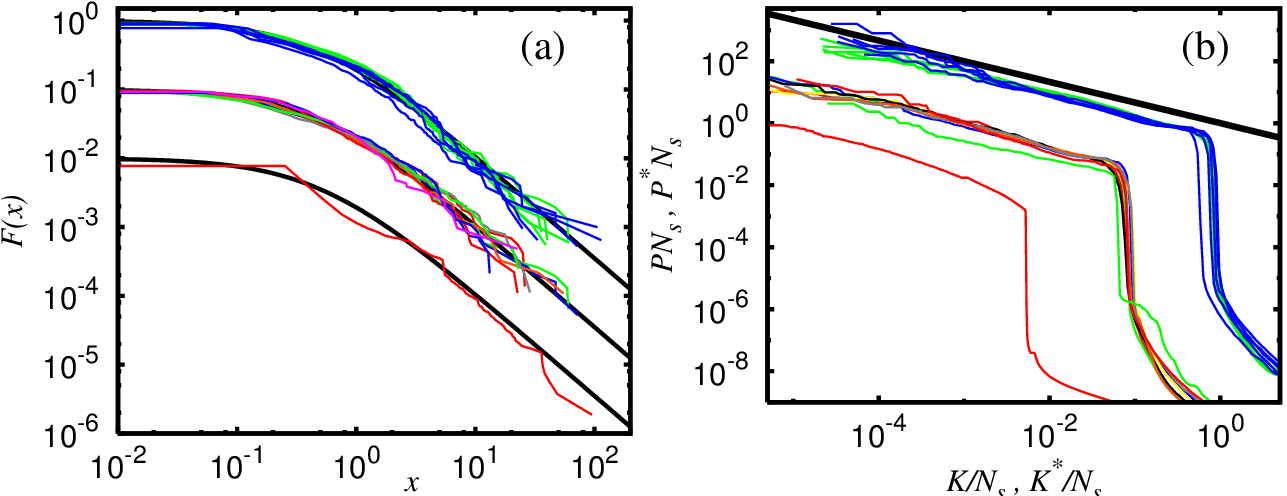}
\end{center}
\caption{(Color online)
(a) Fraction of invariant subspaces $F$ with dimensions larger than $d$ 
as a function of the rescaled variable $x=d/\langle d\rangle$. Upper
curves correspond to Cambridge (green/gray) and Oxford (blue/black) 
for years 2002 to 2006 and middle curves (shifted down by a factor of 10) 
correspond
to the university networks of Glasgow, Cambridge, Oxford, Edinburgh, 
UCL, Manchester, Leeds, Bristol and Birkbeck for year 2006 
with $\langle d\rangle$ between 14 and 31.
Lower curve (shifted down by a factor of 100) corresponds to the 
matrix $S^*$ of Wikipedia with $\langle d\rangle=4$. 
The thick black line is 
$F(x)=(1+2x)^{-1.5}$. 
(b) Rescaled PageRank $P\,N_s$ versus rescaled 
rank index $K/N_s$ for $1-\alpha=10^{-8}$ and 
$3974 \leq N_s \leq 48239$ 
for the same university networks as in (a) 
(upper and middle curves, the latter shifted down 
and left by a factor of 10). 
The lower curve (shifted down and left by a factor of 100) 
shows the rescaled 
CheiRank of Wikipedia $P^*\,N_s$ versus $K^*/N_s$ with $N_s=21198$. 
The thick black line 
corresponds to a power law with exponent $-2/3$.
After \cite{frahm:2011}.
\label{fig8_3}}
\end{figure} 

In Fig.~\ref{fig8_3}(b) we show that for several of 
the university networks the PageRank at $1-\alpha=10^{-8}$ 
has actually  a universal form when 
using the rescaled variables $P\,N_s$ versus $K/N_s$ with a power law 
behavior close to $P \propto K^{-2/3}$ for $K/N_s<1$. The rescaled data
of Fig.~\ref{fig8_3} (a) show that 
the fraction of subspaces with dimensions 
larger than $d$ is well described by the power law  
$F(x)\approx (1+2x)^{-1.5}$
with the dimensionless variable
$x=d/\langle d\rangle$ where $\langle d\rangle$ is an average 
subspace dimension computed for WWW of a given university.
The tables of all considered UK universities
with the parameters of their WWW are given in \cite{frahm:2011}.
We note that the CheiRank of $S^*$ of Wikipedia 2009 also
approximately follows the above universal distributions.
However, for $S$ matrix of Wikipedia the number of 
subspaces is small and statistical analysis cannot be performed
for  this case.

The origin of the universal distribution $F(x)$
still remains a puzzle. Possible links with a percolation 
on directed networks (see e.g. \cite{dorogovtsev:2008}) 
are still to be elucidated. It also remains unclear
how stable this distribution really is. It works well
for UK university networks 2002-2006.
However, for the Twitter network \cite{frahm:2012b}
such a distribution becomes rather approximate.
Also for the network of Cambridge in 2011,
analyzed in \cite{ermann:2012a,ermann:2013b}
with $N \approx 8.9 \times 10^5$, $N_{\ell} \approx 1.5 \times 10^7$,
the number of subspaces is significantly reduced
and a statistical analysis of their size distribution
becomes not relevant.
It is possible that an increase of 
number of links per node 
$N_{\ell}/N$ from a typical value of $10$ for
UK universities to $35$ for Twitter 
affects this distribution. For Cambridge 2011
the network entered in a regime when many
links are generated by robots
that apparently leads to a change of its
statistical properties.

\subsection{Two-dimensional ranking for University networks}
\label{s8.3}

\begin{figure}[H]
\begin{center}
\includegraphics[width=0.48\textwidth]{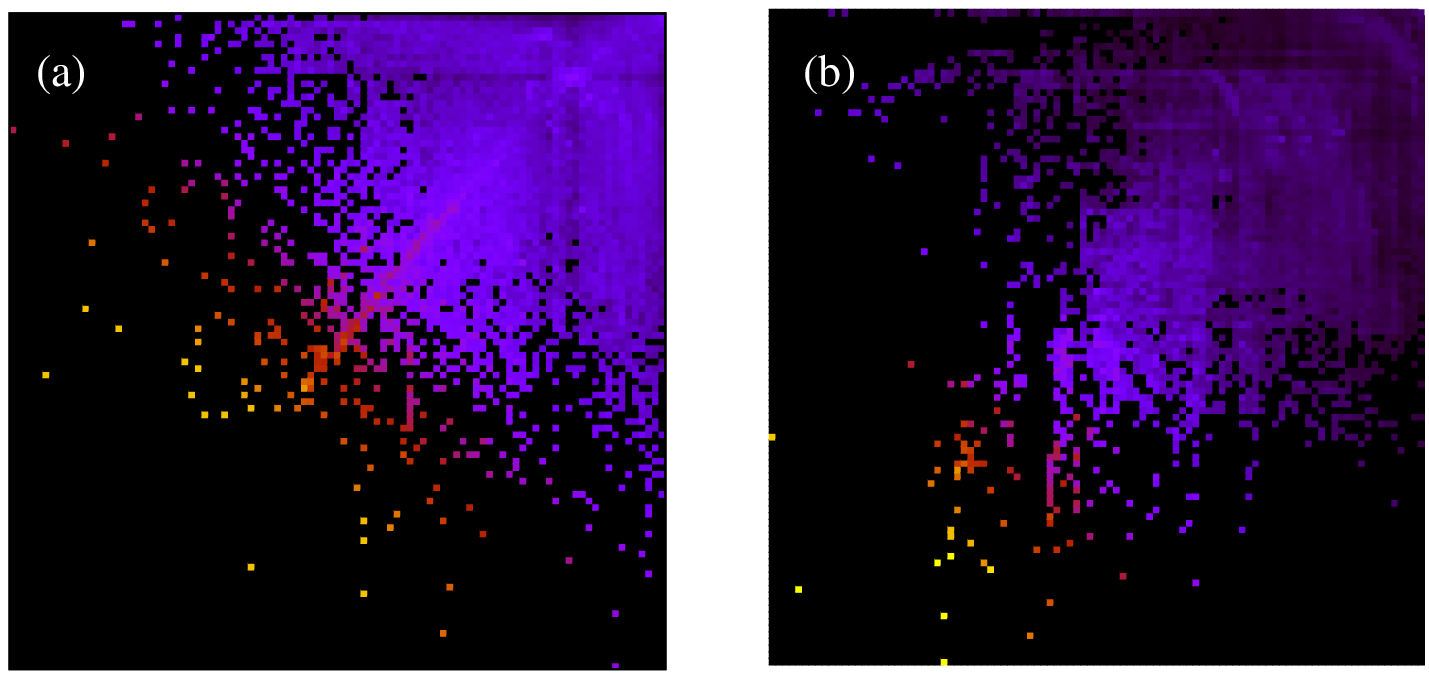}
\end{center}
\caption{(Color online)
Density distribution $W(K,K^*)=dN_i/dKdK^*$ 
for networks of Universities
in the plane of PageRank $K$ and CheiRank $K^*$ 
indexes in log-scale $(\log_N K,\log_N K^*)$.
The density is shown for 
$100\times100$ equidistant grid in $\log_N K,\log_N K^*\in[0,1]$,
the density is averaged over all nodes inside each cell of the grid,
the normalization condition is $\sum_{K,K^*}W(K,K^*)=1$.
Color varies from black for zero to yellow/gray 
for maximum density value $W_M$ with a saturation value of 
$W_s^{1/4}=0.5W_M^{1/4}$ so that the same color
is fixed for $0.5W_M^{1/4} \leq  W^{1/4} \leq W_M^{1/4}$
to show in a better way low densities.
The panels show networks of University of Cambridge 2006
with $N=212710$ (a) and ENS Paris 2011 for crawling 
level 7 with $N=1 820 015$ (b).
After \cite{ermann:2012a}.
\label{fig8_4}}
\end{figure}

Two-dimensional ranking of network nodes
provides a new characterization of directed networks.
Here we consider a density distribution of nodes (see Sec.~\ref{s4.3}) 
in the PageRank-CheiRank plane for examples of 
two WWW networks of Cambridge 2006 and ENS Paris 2011
shown in Fig.~\ref{fig8_4} from  \cite{ermann:2012a}. 

The density distribution for Cambridge 2006 clearly shows 
that nodes with high PageRank have low CheiRank
that corresponds to zero density at low $K$, $K^*$
values. At large $K$, $K^*$ values there is a maximum line
of density which is located not very far from 
the diagonal $K \approx K^*$. 
The presence of correlations between $P(K_i)$ and  $P^*({K_i}^*)$
leads to a probability distribution with one main
maximum along a diagonal at $\ln K + \ln K^*=const$.
This is similar to the properties of the density distribution
for the Wikipedia network shown in Fig.~\ref{fig4_3}(a).

The 2DRanking might give new possibilities for 
information retrieval from large databases which 
are growing rapidly with time. Indeed, for example 
the size of the Cambridge network increased
by a factor 4 from 2006 to 2011. At present, 
web robots start automatically to generate 
new web pages. These features can be responsible for the 
appearance of gaps in the density distribution in
$(K,K^*)$ plane at large $K, K^* \sim N$ values
visible for large scale university
networks such as ENS Paris in 2011 
(see Fig.~\ref{fig8_4}). Such an automatic generation of links
can change the scale-free properties of networks.
Indeed, for  ENS Paris a large step in the PageRank distribution 
appears \cite{ermann:2012a} possibly indicating 
a delocalization transition tendency
of the PageRank that can destroy the efficiency 
of information retrieval from the WWW. 

\section{Wikipedia networks}
\label{s9}

The free online encyclopedia Wikipedia is a huge
repository of human knowledge. 
Its size is growing permanently
accumulating huge amount of information
and becoming a modern version
of {\it Library of Babel},
described by Jorge Luis Borges \cite{borges}. 
The hyperlink citations between
Wikipedia articles provides an important example of
directed networks evolving in time  for many different languages. 
In particular, the English edition of August 2009 has been 
studied in detail \cite{zhirov:2010,ermann:2012a,ermann:2013b}. 
The effects of time evolution \cite{eom:2013b} and 
entanglement of cultures in multilingual Wikipedia editions
have been investigated in 
\cite{aragon:2012,eom:2013a,eom:2014}.

\subsection{Two-dimensional ranking of Wikipedia articles}
\label{s9.1}

The statistical distribution of links in Wikipedia
networks has been found to follow a power law
with the exponents $\mu_{\rm in}, \mu_{\rm out}$
(see e.g. \cite{zlatic:2006,capocci:2006,muchnik:2007,zhirov:2010}).
The probabilities of
PageRank and CheiRank are shown in Fig.~\ref{fig4_1}.
They are satisfactory described by a power law decay
with exponents $\beta_{PR,CR} =1/(\mu_{\rm in,out}-1)$ \cite{zhirov:2010}.

The density distribution of articles over PageRank-CheiRank plane
$(\log_N K, \log_N K^*)$ is shown in Fig.~\ref{fig4_3}(a) 
for English Wikipedia Aug 2009. We stress that the density
is very different from those generated by the product of 
independent probabilities of $P$ and $P^*$
given in Fig.~\ref{fig4_1}. 
In the latter case
we obtain a density  homogeneous along lines 
$\ln K^* =  - \ln K + const$ being rather similar
to the distribution for Linux network also shown in 
Fig.~\ref{fig4_3}. This result is in good agreement
with a fact that the correlator $\kappa$ 
between PageRank and CheiRank vectors 
is rather large for Wikipedia $\kappa =4.08$
while it is close to zero for Linux network
$\kappa \approx -0.05$.

The difference between PageRank and CheiRank is clearly seen
from the names of articles with highest ranks
(ranks of all articles are given in \cite{zhirov:2010}).
 At the top of PageRank we have 
1. {\it United States}, 2. {\it United Kingdom}, 3. {\it France} 
while for CheiRank we find
1. {\it Portal:Contents/Outline of knowledge/Geography and places},
2. {\it List of state leaders by year}, 
3. {\it Portal:Contents/Index/Geography and places}. Clearly PageRank
selects first articles on a broadly known subject 
with a large number of ingoing links
while CheiRank selects first highly communicative articles with
many outgoing links. The 2DRank combines these two characteristics
of information flow on directed network.
At the top of 2DRank $K_2$ we find
 1. {\it India},
2. {\it Singapore}, 3. {\it Pakistan}. Thus, these articles
are most known/popular and most communicative
at the same time.

The top 100 articles in $K, K_2, K^*$ are determined
for several categories including countries, universities,
people, physicists. It is shown in \cite{zhirov:2010}
that PageRank
recovers about 80\% of top 100 countries
from SJR data base \cite{sjr:2007},
about 75\% of top 100 universities of 
Shanghai university ranking \cite{shanghai},
and, among physicists, about 50\% of top 100 Nobel winners
in physics. This overlap is lower for 2DRank
and even lower for CheiRank. However, 
as we will see below in more detail,
2DRank and CheiRank
highlight other properties being complementary to
PageRank.

Let us give an example of top three physicists among
those of 754 registered in Wikipedia in 2010:
 1. {\it Aristotle}, 2. {\it Albert Einstein},
3. {\it Isaac Newton} from PageRank;
1. {\it Albert Einstein}, 2. {\it Nikola Tesla},
3. {\it Benjamin Franklin} from 2DRank;
1. {\it Hubert Reeves}, 2. {\it Shen Kuo},
3. {\it Stephen Hawking} from CheiRank.
It is clear that PageRank gives most known,
2DRank gives most known and active in other areas,
CheiRank gives those who are known and contribute to popularization
of science. Indeed, e.g. {\it Hubert Reeves} and
{\it Stephen Hawking} are very well known for their popularization
of physics that increases their communicative power and 
place them at the top of CheiRank. {\it Shen Kuo} obtained recognized
results in an enormous variety of fields of science that 
leads to the second top position in CheiRank even if his activity
was about thousand years ago. 

According to Wikipedia ranking
the top universities are 
1.~{\it Harvard University}, 2.~{\it University of Oxford},
3.~{\it University of Cambridge} in PageRank;
1.~{\it Columbia University}, 2.~{\it University of Florida},
3.~{\it Florida State University} in 2DRank and CheiRank.
CheiRank and 2DRank  highlight connectivity degree of universities
that leads to appearance of significant
number of  arts, religious and military specialized colleges 
(12\% and 13\% respectively for CheiRank and 2DRank)
while PageRank has only 1\% of them. 
CheiRank and 2DRank introduce also
a larger number of relatively small universities
who are keeping links to their 
alumni in a significantly better way that gives an increase
of their ranks. It is established \cite{eom:2013b} that
top $10$ PageRank universities from English Wikipedia
in years  
$2003, 2005, 2007, 2009, 2011$
recover correspondingly
$9, 9, 8, 7, 7 $ from top 10 of 
\cite{shanghai}.

The time evolution of probability distributions
of PageRank, CheiRank and two-dimensional ranking
is analyzed in \cite{eom:2013b}
showing that they become
stabilized for the period 2007-2011.

On the basis of these results we can conclude that
the above algorithms provide
correct and important ranking 
of huge information and knowledge 
accumulated at Wikipedia.
It  is interesting that even Dow-Jones companies
are ranked  via Wikipedia networks in a
good manner \cite{zhirov:2010}.
We discuss ranking of top people of Wikipedia
a bit later.

\subsection{Spectral properties of Wikipedia network}
\label{s9.2}

\begin{figure}[H]
\begin{center}
\includegraphics[width=0.48\textwidth]{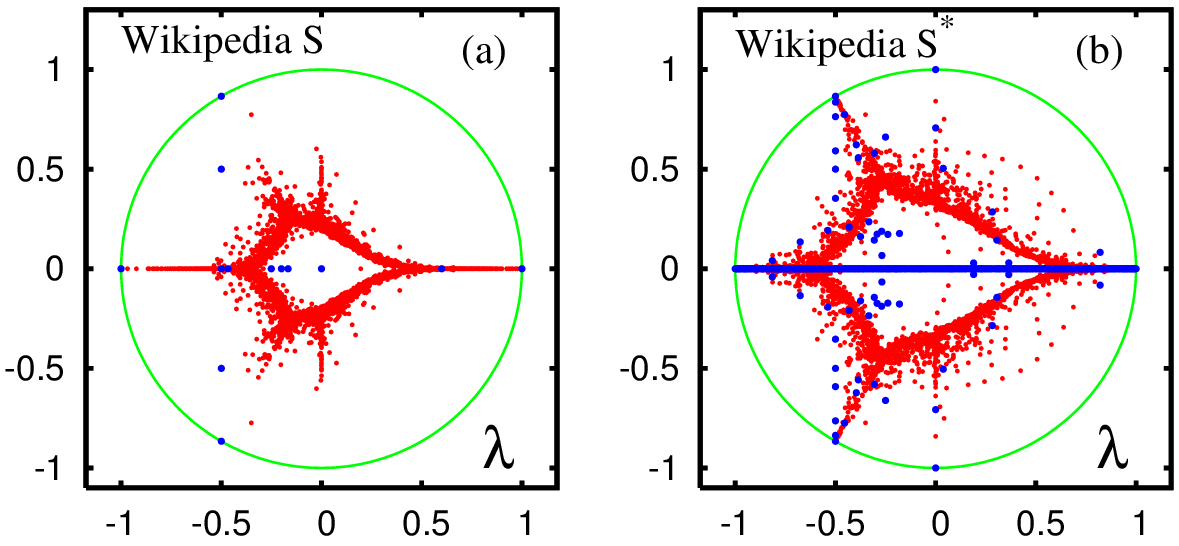}
\end{center}
\caption{(Color online)
Complex eigenvalue spectra $\lambda$ of $S$ (a) and 
$S^*$ (b) for English Wikipedia of Aug 2009
with $N=3 282 257$ articles and $N_\ell= 71 012 307$ links. 
Red/gray dots are core space eigenvalues, blue/black dots are subspace 
eigenvalues and the full green/gray curve shows the unit circle. 
The core space eigenvalues are computed 
by the projected Arnoldi method with 
Arnoldi dimension $n_A=6000$. 
After \cite{eom:2013b}.
\label{fig9_1}}
\end{figure}

The complex spectrum  
of eigenvalues of $G$ for English Wikipedia network of Aug 2009
is shown in Fig.~\ref{fig9_1}.
As for university networks,  the spectrum also has
some invariant subspaces resulting in degeneracies 
of the leading eigenvalue $\lambda=1$ of $S$ (or $S^*$). However, 
due to the stronger connectivity of the Wikipedia network these 
subspaces are significantly smaller compared to university networks 
\cite{ermann:2013b,eom:2013b}. For example of Aug 2009 
edition in  Fig.~\ref{fig9_1}
there are $255$ invariant subspaces (of the 
matrix $S$) covering $515$ nodes with $255$ unit 
eigenvalues $\lambda_j=1$ and 
$381$ eigenvalues on the complex unit circle with $|\lambda_j|=1$. 
For the matrix $S^*$ of Wikipedia there 
are $5355$ invariant subspaces with $21198$ nodes, 
$5365$ unit eigenvalues and 8968 eigenvalues on the unit circle 
\cite{ermann:2013b}. 
The complex spectra of all subspace eigenvalues and the first 
$n_A=6000$ core space eigenvalues of 
$S$ and $S^*$ are shown in Fig.~\ref{fig9_1}. 
As in the university cases, in the spectrum 
we can identify cross 
and triple-star structures similar to those of orthostochastic
matrices shown in Fig.~\ref{fig8bis}.
However, for Wikipedia (especially for $S$) the largest complex eigenvalues 
outside the real axis are more far away from the unit circle. 
For $S$ of Wikipedia the two largest core space eigenvalues are 
$\lambda^{\rm (core)}_{1}=0.999987$ and $\lambda^{\rm (core)}_{2}=0.977237$ 
indicating that the core space gap 
$|1-\lambda^{\rm (core)}_{1}|\sim 10^{-5}$ is much smaller 
than the secondary gap 
$|\lambda^{\rm (core)}_{1}-\lambda^{\rm (core)}_{2}|\sim 10^{-2}$. 
As a consequence the PageRank of Wikipedia (at $\alpha=0.85$) 
is strongly influenced by the leading core space eigenvector 
and actually both vectors select the same 5 top nodes. 

The time evolution of spectra of $G$ and $G^*$
for English Wikipedia is studied in \cite{eom:2013b}.
It is shown that the spectral structure remains stable
for years 2007 - 2011. 

\begin{figure}[H]
\begin{center}
\includegraphics[width=0.48\textwidth]{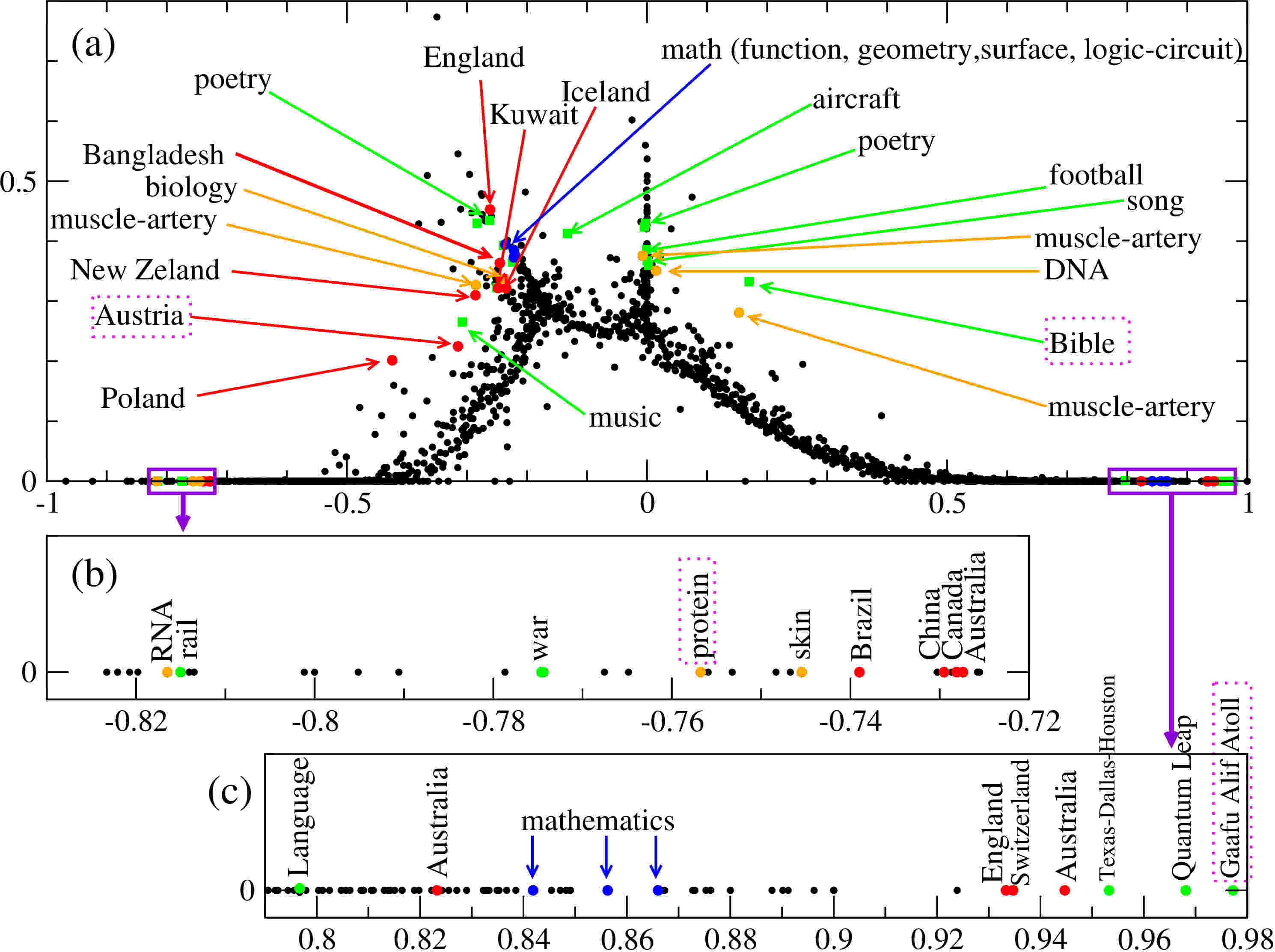}
\end{center}
\caption{(Color online)
Complex eigenvalue spectrum of the matrices $S$ 
for English Wikipedia Aug 2009.
Highlighted eigenvalues represent different communities 
of Wikipedia and are labeled by the most 
repeated and important words following word counting of first 1000 nodes.
Panel (a) shows complex plane for positive imaginary part
of eigenvalues, 
while panels (b) and (c) zoom in 
the negative and positive real parts. 
After \cite{ermann:2013b}.
\label{fig9_2}}
\end{figure}

\subsection{Communities and eigenstates of Google matrix}
\label{s9.3}

The properties of eigenstates of Gogle matrix of
Wikipedia Aug 2009 are analyzed in \cite{ermann:2013b}.
The global idea is that the eigenstates with large values
of $|\lambda|$ select certain specific communities.
If $|\lambda|$ is close to unity then a 
relaxation of probability from such nodes is rather
slow and we can expect that such eigenstates
highlight some new interesting information
even if these nodes are located on a tail of PageRank. 
The important advantage of the Wikipedia network
is that its nodes are Wikipedia articles with 
a relatively clear meaning allowing to
understand the origins of appearance of certain
nodes in  one community.

The localization properties of eigenvectors $\psi_i$ of the Google matrix 
can be analyzed with the help of IPR $\xi$ (see Sec. \ref{s3.5}). 
Another possibility is to fit a decay of an eigenstate amplitude 
by a power law  
$|\psi_i(K_i)|\sim \,K_i^b$ where $K_i$ is the 
index ordering $|\psi_i(j)|$ by monotonically 
decreasing amplitude (similar to $P(K)$ for PageRank).
 The  exponents $b$ on the tails of   $|\psi_i(j)|$
are found to be typically in the range $-2<b<-1$ \cite{ermann:2013b}.
At the same time the
eigenvectors with large complex eigenvalues or real eigenvalues 
close to $\pm 1$ are quite well localized on 
$\xi_i\approx 10^{2}-10^{3}$ nodes that is much smaller than the whole
network size $N \approx 3 \times 10^6$.

To understand the meaning of other eigenstates in the core space
we order selected eigenstates by their decreasing
value $|\psi_i(j)|$ and apply word frequency analysis 
for the first $1000$ articles with $K_i \leq 1000$.
The mostly frequent word of a given eigenvector  is
used to label the eigenvector name. These labels with 
corresponding eigenvalues are shown in Fig.~\ref{fig9_2}.
There are four main categories for the selected eigenvectors
belonging to countries (red/gray), biology and medicine 
(orange/very light gray), mathematics (blue/black) 
and others (green/light gray).
The category of others
contains rather diverse articles about poetry, Bible, football, music,
American TV series (e.g. Quantum Leap), small geographical
places (e.g. Gaafru Alif Atoll). Clearly these eigenstates
select certain specific communities which are relatively weakly
coupled with the main bulk part of  Wikipedia that
generates relatively large modulus of $|\lambda_i|$.

For example, for the article  {\it Gaafu Alif Atoll}
the eigenvector is mainly localized on names of small atolls 
forming {\it Gaafu Alif Atoll}. 
Clearly this case represents well localized community of 
articles mainly linked between themselves that gives
slow relaxation rate of this eigenmode with $\lambda=0.9772$ 
being rather close to unity. 
Another eigenvector has 
a complex eigenvalue with 
 $|\lambda|=0.3733$ and the top article
{\it Portal:Bible}. Another two
articles are 
{\it Portal:Bible/Featured chapter/archives},
{\it Portal:Bible/Featured article}.
These top $3$ articles 
have very close values of $|\psi_i(j)|$
that seems to be the reason why
we have $\varphi=\arg(\lambda_i) = 0.3496 \pi$
being very close to $\pi/3$.
Examples of other eigenvectors 
are discussed in \cite{ermann:2013b} in detail.

The analysis performed in \cite{ermann:2013b} for Wikipedia Aug 2009
shows that  the eigenvectors
of the Google matrix of Wikipedia 
clearly identify certain communities
which are relatively weakly connected with the
Wikipedia core when the modulus of corresponding
eigenvalue is close to unity. 
For moderate
values of $|\lambda|$ we still have 
well defined communities which are however
have stronger links with some popular articles
(e.g. countries) that leads to a more rapid decay
of such eigenmodes. Thus the 
 eigenvectors highlight
interesting features of communities and
network structure. However, a priori, it is not evident
what is a correspondence between
the numerically obtained eigenvectors  
and the specific community features 
in which someone has a specific interest.
In fact, practically each eigenvector
with a moderate value $|\lambda| \sim 0.5$
selects a certain community and there are many of them.
So it remains difficult to target 
and select from eigenvalues $\lambda$ a 
specific community
one is interested.

The spectra and eigenstates of other networks
like WWW of Cambridge 2011, Le Monde, BBC and
PCN of Python are discussed in \cite{ermann:2013b}.
It is found that IPR values
of eigenstates with large $|\lambda|$
are well localized with $\xi \ll N$. The spectra of each network have
significant differences from one another. 

\subsection{Top people of Wikipedia}
\label{s9.4}

There is always a significant public interest to know
who are most significant historical figures,
or persons, of humanity. 
The Hart list of the top 100 people who, according to him,
most influenced human history, is available at
\cite{hart:1992}. 
Hart ``ranked these 100 persons
in order of importance: that is, according to the total
amount of influence that each of them had on human history and on the
everyday lives of other human beings'' \cite{hart:1992}.
Of course, a human ranking can be always
objected arguing that an investigator has its own
preferences. Also investigators from different cultures can have
different view points on a same historical figure.
Thus it is important to perform
ranking of historical figures
on purely mathematical and statistical grounds
which exclude any cultural and personal 
preferences of investigators. 

A detailed two-dimensional ranking of
persons of English Wikipedia Aug 2009
has been done in \cite{zhirov:2010}.
Earlier studies had been done in a non-systematic way
without any comparison with established top $100$ lists
(see these Refs. in \cite{zhirov:2010,wiki100}).
Also at those times Wikipedia did not yet
entered in its stabilized phase of development.

The top people of Wikipedia Aug 2009
are found to be 
1. {\it Napoleon I of France}, 2. {\it George W. Bush},
3. {\it Elizabeth II of the United Kingdom} for PageRank;
1.{\it Michael Jackson}, 2. {\it Frank Lloyd Wright},
3. {\it David Bowie} for 2DRank;
1. {\it Kasey S. Pipes}, 2. {\it Roger Calmel},
3. {\it Yury G. Chernavsky} for CheiRank \cite{zhirov:2010}.
For the PageRank list of $100$ the overlap with 
the Hart list is at 35\% (PageRank),
10\% (2DRank) and almost zero for CheiRank.
This is attributed to a 
very broad distribution of historical figures on 2D plane, 
as shown in Fig.~\ref{fig4_3},
and a large variety of human activities. These activities 
are classified by $5$ main categories:
politics, religion, arts, science, sport. 
For the top $100$ PageRank persons we have
the following distribution 
over these categories: $58$, $10$, $17$, $15$, $0$ respectively.
Clearly PageRank overestimates the significance of politicians
which list is dominated by USA presidents not always much known to a broad
public.
For 2DRank we find respectively $24$, $5$,  $62$, $7$, $2$. 
Thus this rank 
highlights artistic sides of human activity.
For CheiRank we have $15$, $1$, $52$, $16$, $16$ so that the dominant
contribution comes from arts, science and sport. 
The interesting property 
of this rank is that it selects many composers, singers, writers, actors.
As an interesting feature of CheiRank we note
that among scientists it selects those who are not so much known to
a broad public but who  discovered new objects, e.g. 
George Lyell who discovered many Australian butterflies  
or  Nikolai Chernykh who discovered many asteroids. 
CheiRank also selects persons active in several 
categories of human activity.

For English Wikipedia Aug 2009 the distribution
of top 100 PageRank, CheiRank and Hart's persons
on PageRank-CheiRank plane is shown in Fig.~\ref{fig4_3} (a).

The distribution of Hart's top $100$
persons on $(K,K^*)$ plane
for English Wikipedia in years 
2003, 2005, 2007, Aug 2009, Dec 2009, 2011
is found to be stable
for the period 2007-2011
even if certain persons
change their ranks \cite{eom:2013b}.
The distribution of top $100$ persons
of Wikipedia Aug 2009
remains stable and compact for PageRank and 2DRank
for the period 2007-2011 while
for CheiRank the fluctuations
of positions are large. This is due to the fact 
that outgoing links are easily modified
and fluctuating.

The time evolution of distribution of top persons
over fields of human activity is established in \cite{eom:2013b}.
PageRank persons are
dominated by politicians whose percentage increases with
time, while the percent of arts decreases. For 2DRank 
the arts are dominant but  their percentage
decreases with time. We also see the appearance of sport
which is absent in PageRank. The mechanism of the qualitative 
ranking differences between two ranks is related to
the fact that 2DRank takes into account via CheiRank
a contribution of outgoing links. Due to that singers, 
actors, sportsmen improve their CheiRank and 2DRrank 
positions since articles about them contain various 
music albums, movies and sport competitions with many 
outgoing links. Due to that the component of arts gets
higher positions in 2DRank in contrast to 
dominance of politics in PageRank. 

The interest to ranking of people via 
Wikipedia network is growing, as shows the recent study
of English edition \cite{skiena:2014}.

\subsection{Multilingual Wikipedia editions}
\label{s9.5}

The English edition allows to obtain
ranking of historical people but as we saw 
the PageRank list is dominated by USA presidents
that probably does not correspond to the global world
view point. Hence, it is important to
study multilingual Wikipedia editions
which have now $287$ languages
and represent broader cultural views 
of the world.

One of the first cross-cultural study 
was done for $15$ largest language editions
constructing a network of links between set of
articles of people biographies for each edition. 
However, the number of nodes and links in such 
a biographical network
is significantly smaller compared to the whole 
network of Wikipedia articles and thus 
the fluctuations become rather large.
For example, from the biographical network
of the Russian edition one finds as the top person {\it Napoleon III}
(and even not {\it Napoleon I})~\cite{aragon:2012}, 
who has a rather low importance for Russia. 

Another approach was used in \cite{eom:2013a}
ranking top 30 persons by PageRank, 2DRank and CheiRank
algorithms 
for all articles of each of 9 editions 
and attributing each person to her/his native language.
The selected editions are
English (EN), French
(FR), German (DE), Italian (IT), Spanish (ES), Dutch (NL), Russian
(RU), Hungarian (HU) and Korean (KO). 
The aim here is to understand how different cultures evaluate a person? 
Is an important person in one culture is 
also important in the other culture? 
It is found that local heroes are dominant 
but also global heroes exist and
create an effective network representing entanglement of cultures. 

The top article of PageRank is usually
{\it USA} or the name of country of a given language
(FR, RU, KO). For NL we have at the top
 {\it beetle, species, France}.
The top articles of CheiRank are various listings.

\begin{figure}[H]
\begin{center}
\includegraphics[width=0.48\textwidth]{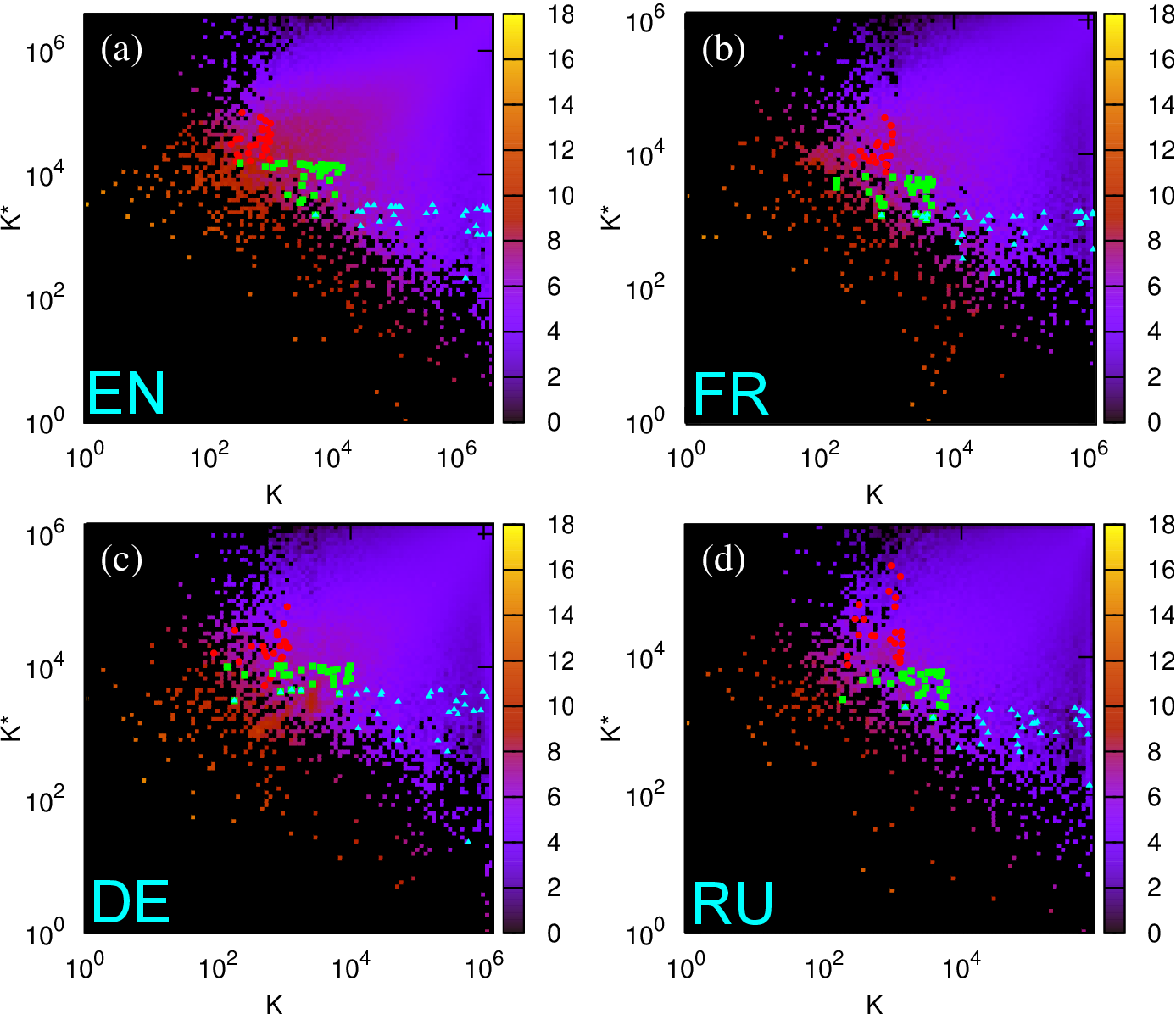}
\end{center}
\caption{(Color online)
Density of Wikipedia articles in the
PageRank-CheiRank plane $(K,K^*)$ for 
four different language Wikipedia editions. 
The red (gray) points are top PageRank articles of
persons, the green (light gray) squares are top 
2DRank articles of persons and
the cyan (dark gray) triangles are top CheiRank articles of persons.
Wikipedia language editions are English EN 
(a), French FR (b), German DE (c), and Russian RU (d).
Color bars show natural logarithm of density, changing
from minimal nonzero density (dark) to maximal one (white),
zero density is shown by black.
After \cite{eom:2013a}.
\label{fig9_3}}
\end{figure}

The distributions of articles density and top 30 persons for
each rank algorithm are shown in Fig.~\ref{fig9_3}
for four editions EN,  FR, DE, RU. We see that in global
the distributions have a similar shape that can be attributed
to a fact that all editions describe the same world.
However, local features of distributions are different
corresponding to different cultural views on the same world
(other 5 editions are shown in Fig.2 in \cite{eom:2013a}).
The top 30 persons for each edition are selected manually
that represents a weak point of this study.

\begin{figure}[H]
\begin{center}
\includegraphics[width=0.48\textwidth]{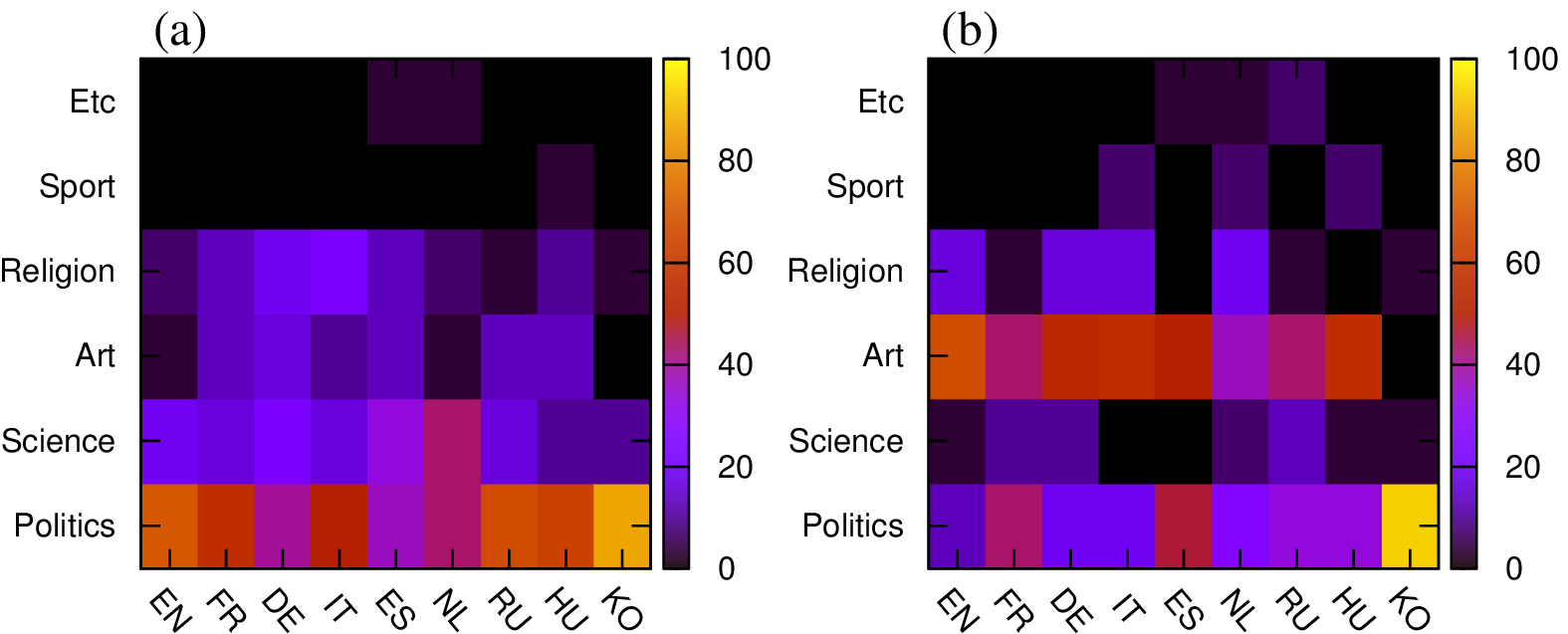}
\end{center}
\caption{(Color online)
Distribution of top 30 persons over activity fields for PageRank $(a)$ 
and 2DRank $(b)$ for each of 9 Wikipedia editions.
The color bar shows the values in percent.
After \cite{eom:2013a}.
\label{fig9_4}}
\end{figure}

From the lists of top persons,  the "fields" of activity
are identified 
for each top 30 rank persons in which he/she is active on.
The six activity fields are: politics,
art, science, religion, sport and etc
(here ``etc'' includes all other activities). 
As shown in Fig.~\ref{fig9_4}, for
PageRank, politics is dominant and science is secondarily
dominant. The only exception is Dutch where science
is the almost dominant activity field 
(politics has the same number of points).
In case of 2DRank in Fig.~\ref{fig9_4}, 
art becomes dominant and politics is
secondarily dominant. In case of CheiRank, art and sport are
dominant fields (see Fig.3 in \cite{eom:2013a}). 
Thus for example, in CheiRank top 30 list we find
astronomers who discovered a lot of asteroids,
e.g. Karl Wilhelm Reinmuth (4th position in RU
and 7th in DE),
who was a prolific discoverer of about 400 of them.
As a result, his article contains
a long listing of asteroids discovered by him and giving him
a high CheiRank.
The distributions of persons over
activity fields are shown in Fig.~\ref{fig9_4}
for 9 languages editions 
(marked by standard two letters used by Wikipedia).

The change of activity priority for different ranks is due to the
different balance between incoming and outgoing links there.
Usually the politicians are well known for a broad public,
hence, the articles about politicians
are pointed by many articles. However, 
the articles about politicians are not very communicative
since they rarely point to other articles.
In contrast, articles about
persons in other fields like science, art  and sport
are more communicative because of
listings of insects, planets, asteroids they discovered, or
listings of song albums or sport competitions they gain.

On the basis of this approach
one obtains local ranks of each of 30 persons
$1 \leq K_{P,E,A} \leq 30$
for each edition $E$ and algorithm $A$. Then an average ranking 
score of a person $P$ is determined as
$\Theta_{P,A} = \sum_{E} (31-K_{P,E,A})$
for each algorithm. This method determines
the global historical figures.
The top global persons are
 1.{\it Napoleon}, 2.{\it Jesus},
3.{\it Carl Linnaeus } for PageRank;
1.{\it Micheal Jackson }, 2.{\it Adolf Hitler},
3.{\it Julius Caesar} for 2DRank. For CheiRank the
lists of different editions have rather low overlap 
and such an averaging is not efficient.
The first positions reproduce top persons
from English edition discussed in Sec.~\ref{s9.4},
however, the next ones are different.

\begin{figure}[H]
\begin{center}
\includegraphics[width=0.48\textwidth]{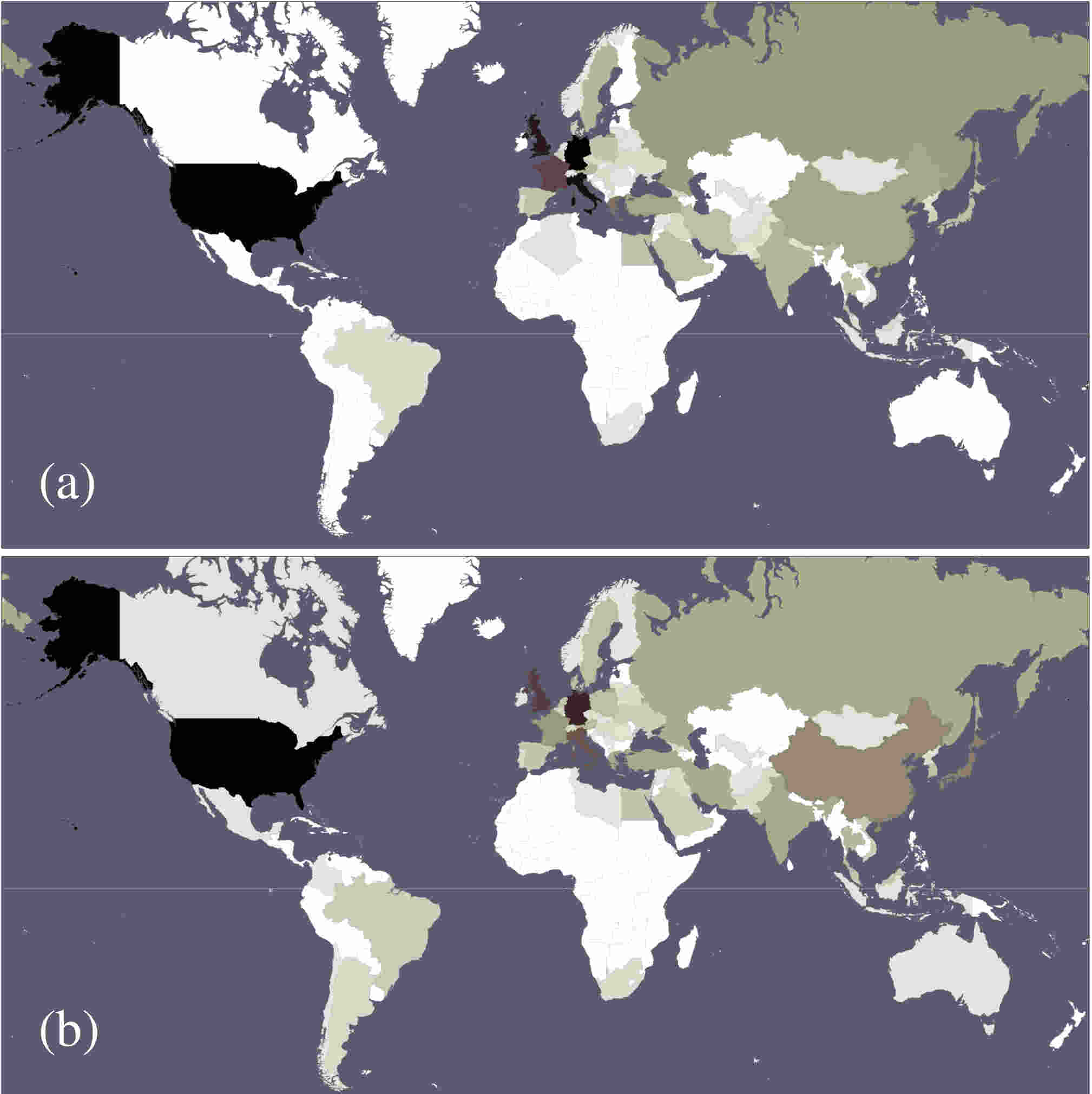}
\end{center}
\caption{
Number of appearances
of historical figures  of a given country, obtained
from 24 lists of top 100 persons of PageRank (a)
and 2DRank (b), shown
on the world map. Color changes from zero (white)
to maximum (black), it corresponds to average number of
person appearances per country.
After \cite{eom:2014}.
\label{fig9_5}}
\end{figure} 

Since each person is attributed to her/his
native language it is also possible for each edition
to obtain top local heroes who have native language of the edition.
For example, we find 
for PageRank for EN {\it George W. Bush},
{\it Barack Obama}, {\it Elizabeth II};
for FR
{\it Napoleon}, {\it  Louis XIV of France},
{\it Charles de Gaulle}; for DE
{\it Adolf Hitler}, {\it Martin Luther},
{\it Immanuel Kant};
for RU {\it Peter the Great},
{\it Joseph Stalin},  {\it Alexander Pushkin}.
For 2DRank we have
for EN {\it Frank Sinatra},
{\it Paul McCartney}, {\it Michael Jackson};
for FR {\it Francois Mitterrand},
{\it Jacques Chirac}, {\it Honore de Balzac};
for DE {\it Adolf Hitler}, 
{\it Otto von Bismarck}, {\it Ludwig van Beethoven};
for RU {\it Dmitri Mendeleev}, 
{\it Peter the Great}, 
{\it Yaroslav the Wise}. These ranking results
are rather reasonable for each language.
Results for other editions and CheiRank are given in
\cite{eom:2013a}.

A weak point of above study is a manual selection
of persons and a not very large number of editions.
A significant improvement has been reached in a recent study
\cite{eom:2014} where 24 editions have been analyzed.
These 24 languages cover 59 percent of world population,
and these 24 editions covers 68 percent
of the total number of Wikipedia articles in all 287
available languages. Also the selection of people from the rank
list of each edition is now done in an automatic computerized
way. For that a list of about 1.1 million biographical articles
about people with their English names is generated.
From this list of persons, with their biographical
article title in the English Wikipedia,
the corresponding titles in other language
editions are determined using the inter-language 
links provided by Wikipedia.

Using the corresponding articles,
identified by the inter-languages links in different 
language editions,  the top 100
persons are obtained from the rankings of all 
Wikipedia articles of each edition.
A birth place, birth date, and gender of each top 100 ranked
person are identified, based on DBpedia  
or a manual inspection of the corresponding 
Wikipedia biographical article, 
when for the considered person 
no DBpedia data were available.
In this way 24 lists of top 100 persons 
for each edition are obtained in PageRank 
with 1045 unique names and in 2DRank with 1616 unique names.
Each of the 100 historical figures 
is attributed to a birth place at the country level, 
to a birth date in year, to a gender, and 
to a cultural language group. The birth place is assigned 
according to the
current country borders. The cultural group of historical figures 
is assigned by the  most spoken
language of their birth place at the current country level. 
The considered editions are: English EN,
Dutch NL, German DE, French FR, Spanish, ES,
Italian IT, Potuguese PT, Greek, EL,
Danish DA, Swedish SV, Polish PL, 
Hungarian HU, Russian RU,
Hebrew HE, Turkish TR,
Arabic AR, Persian FA, Hindi HI,
Malaysian MS, Thai TH,
Vietnamese VI,
Chinese ZH,
Korean KO, Japanese JA
(dated by February 2013).
The size of network changes from 
maximal value $N=4212493$ for EN
to minimal one $N=78953$ for TH.

All persons are ranked by their average rank score
$\Theta_{P,A} = \sum_{E} (101-K_{P,E,A})$
with $1 \leq K_{P,E,A} \leq 100$
similar to the study of 9 editions described above.
For PageRank the top global historical figures are
{\it Carl Linnaeus}, {\it Jesus},
{\it Aristotle} and for 2DRank
we  obtain {\it Adolf Hitler},
{\it Michael Jackson}, {\it Madonna (entertainer)}.
Thus the averaging over 24 editions modifies the
top ranking. The list of top 100 PageRank global 
persons has overlap of 43 persons with the Hart list
\cite{hart:1992}. Thus the averaging over
24 editions gives a significant improvement
compared to 35 persons overlap
for the case of English edition only \cite{zhirov:2010}.
For comparison we note that
the top 100 list of historical figures
has been also determined recently by \cite{pantheon:2014}
having overlap of 42 persons with the Hart list.
This Pantheon MIT list is established on the basis of number of editions
and number of clicks on an article of a given person
without using rank algorithms discussed here.
The overlap between top 100 PageRank list and top 100 Pantheon list
is 44 percent.
More data are available in \cite{eom:2014}.

The fact that {\it Carl Linnaeus} is
the top historical figure of Wikipedia
PageRank list came out as a surprise for 
media and  broad public (see \cite{wiki100}).
This ranking is due to the fact that
{\it Carl Linnaeus} created a classification of
world species including, animals, 
insects, herbs, trees etc. Thus all articles 
of these species point to the article
 {\it Carl Linnaeus} in various languages.
As a result {\it Carl Linnaeus} appears
on almost top positions in all 24 languages.
Hence, even if a politician, like
{\it Barak Obama}, takes the second 
position in his
country language EN
({\it Napoleon} is at the first position in EN)
he is usually placed 
at low ranking in other language editions.
As a result {\it Carl Linnaeus} 
takes the first global PageRank position.

The number of appearances of historical persons
in 24 lists of top 100 for each edition
can be distributed over present world countries
according to the birth place of each person.
This geographical distribution is shown in Fig.~\ref{fig9_5}
for PageRank and 2DRank. In PageRank the top
countries are {\it DE, USA, IT} and in 2DRank
{\it US, DE, UK}. The appearance of many UK and US
singers improves the positions of English speaking countries
in 2DRank.

\begin{figure}[H]
\begin{center}
\includegraphics[width=0.48\textwidth]{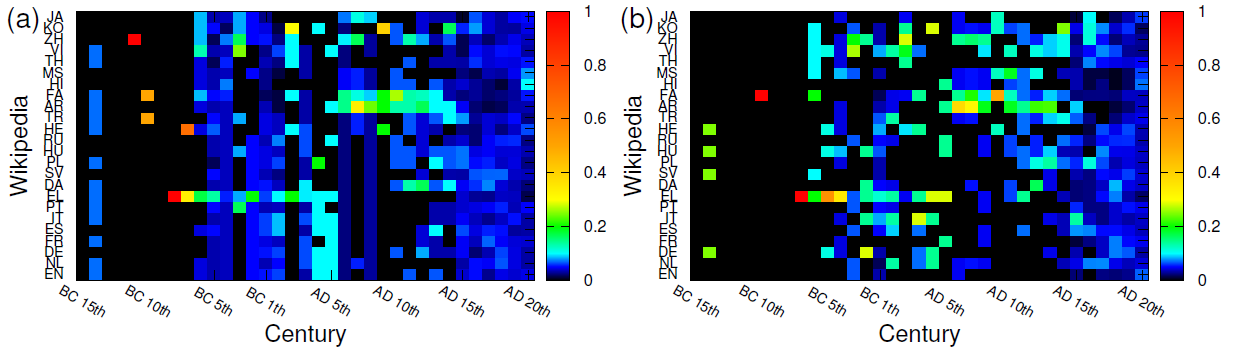}
\end{center}
\caption{(Color online)
Birth date distributions
over $35$ centuries of top historical figures
from each Wikipedia edition marked by 
two letters standard notation of Wikipedia. Panels:
(a) column normalized birth
date distributions of PageRank historical figures; 
(b) same as (a) for 2DRank historical figures.
After \cite{eom:2014}.
\label{fig9_6}}
\end{figure} 

The distributions of the top PageRank 
and 2DRank historical figures over 24 Wikipedia editions
for each century are shown in Fig.~\ref{fig9_6}.
Each person is attributed to 
a century according to the birth date
covering the range of $35$ centuries
from BC 15th to AD 20th centuries. For each 
century the number of persons for each century
is normalized to unity
to see more clearly
relative contribution of each 
language for each century.

The Greek edition has more historical figures 
in BC 5th century because of Greek philosophers. 
Also most of western-southern European language editions, 
including English,
Dutch, German, French, Spanish, Italian, Portuguese, and Greek, 
have more top historical
figures because they have Augustine the Hippo and 
Justinian I in common. 
The Persian (FA) and the Arabic (AR)
Wikipedia have more historical figures comparing to other 
language editions (in particular European
language editions) from the 6th to the 12th century
that is  
due to Islamic leaders and scholars. 
The data of Fig.~\ref{fig9_6} 
clearly show well pronounced patterns, corresponding to strong
interactions between cultures: from BC 5th century 
to AD 15th century for JA, KO, ZH,
VI; from AD 6th century to AD 12th century for FA, AR; 
and a common birth pattern in
EN,EL,PT,IT,ES,DE,NL (Western European languages) 
from BC 5th century to AD 6th century.
A detailed analysis shows that even in BC 20th century
each edition has a significant fraction of
persons of its own language so that even with 
on going globalization there is a significant dominance 
of local historical
figures for certain cultures.
More data on the above points and gender
distributions are available in \cite{eom:2014}.

\subsection{Networks and entanglement of cultures}
\label{s9.6}

We now know how a person of a given language
is ranked by editions of other languages.
Therefore, if a top person from
a language edition $A$ appears in another edition
$B$, we can consider this as a 'cultural' 
influence from culture $A$ to $B$. 
This generates entanglement in a network of cultures.
Here we associate a language edition with
its corresponding culture
considering that a language is a first element of culture,
even if a culture is not reduced only to a language.
In \cite{eom:2013a} a person is attributed to
a given language, or culture,
according to her/his native language
fixed via  corresponding Wikipedia article.
In \cite{eom:2014} the attribution to a culture is done
via a birth place of a person,
each language is considered as  a proxy for a cultural
group and a person 
is assigned to one of these cultural
groups based on the most spoken 
language of her/his birth place at the country level.
 If a person does
not belong to any of studied editions then
he/she is attributed to an additional
cultural group world WR.

After such an attributions of all persons
the two networks of cultures are
constructed  based on the top PageRank
historical figures and
 top 2DRank historical figures respectively.
Each culture (i.e. language) is represented 
as a node of the network, and the weight
of a directed link from culture $A$ to culture $B$ is given by the number of
historical figures belonging to culture $B$ (e.g. French)
appearing in the list of top 100 historical figures for a given
culture $A$ (e.g. English).

\begin{figure}[H]
\begin{center}
\includegraphics[width=0.48\textwidth]{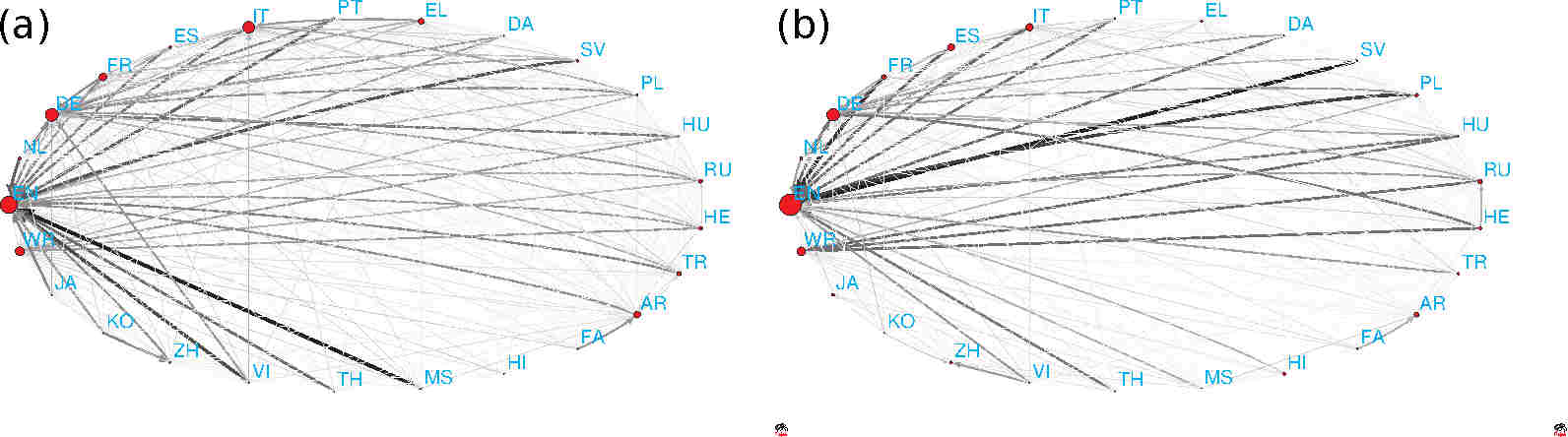}
\end{center}
\caption{(Color online)
 Network of cultures, obtained from 24
Wikipedia languages and the remaining world (WR), considering (a) top
100 PageRank historical figures and (b) top 100
2DRank historical figures. The
link width and darkness are proportional to a number of foreign
historical figures quoted in top 100 of a given culture, the link
direction goes from a given culture to cultures of quoted foreign
historical figures, quotations inside cultures are not considered.
The size of nodes is proportional to their PageRank.
After \cite{eom:2014}.
\label{fig9_7}}
\end{figure}

For example, according to \cite{eom:2014},
there are 5 French historical figures among the top 100 PageRank
historical figures of the English Wikipedia,
so we can assign weight 5 to the link from English to
French. Thus, Fig.~\ref{fig9_7}(a) and Fig.~\ref{fig9_7}(b)
represent the constructed networks of cultures 
defined by appearances of the top PageRank historical figures 
and top 2DRank historical figures, respectively. 

\begin{figure}[H]
\begin{center}
\includegraphics[width=0.48\textwidth]{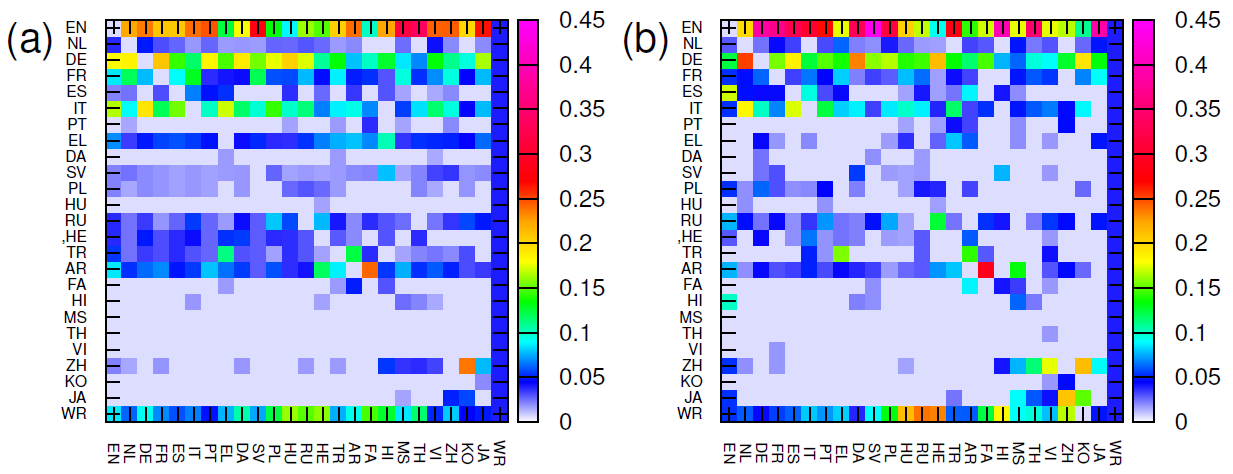}
\end{center}
\caption{(Color online)
Google matrix of network of cultures
shown in Fig~\ref{fig9_7} (a) and (b) respectively. 
The matrix elements $G_{ij}$
are shown by color with damping factor $\alpha=0.85$.
After \cite{eom:2014}.
\label{fig9_8}}
\end{figure}

In total we have two networks with 25 nodes which include our 24 editions
and an additional node WR for all other world cultures.
Persons of a given culture are not taken into
account in the rank list of language edition of this culture.
Then following the standard rules (\ref{eq3_1}) the Google matrix
of network of cultures is constructed by normalization of
sum of all elements in each column to unity.
The matrix $G_{KK'}$, written in the PageRank indexes $K, K'$
is shown in Fig.~\ref{fig9_8} for 
persons from PageRank $(a)$ and 2DRank $(b)$ lists.
The matrix $G^*$ is constructed in the same way as $G$ for the network
with inverted directions of links.

\begin{figure}[H]
\begin{center}
\includegraphics[width=0.48\textwidth]{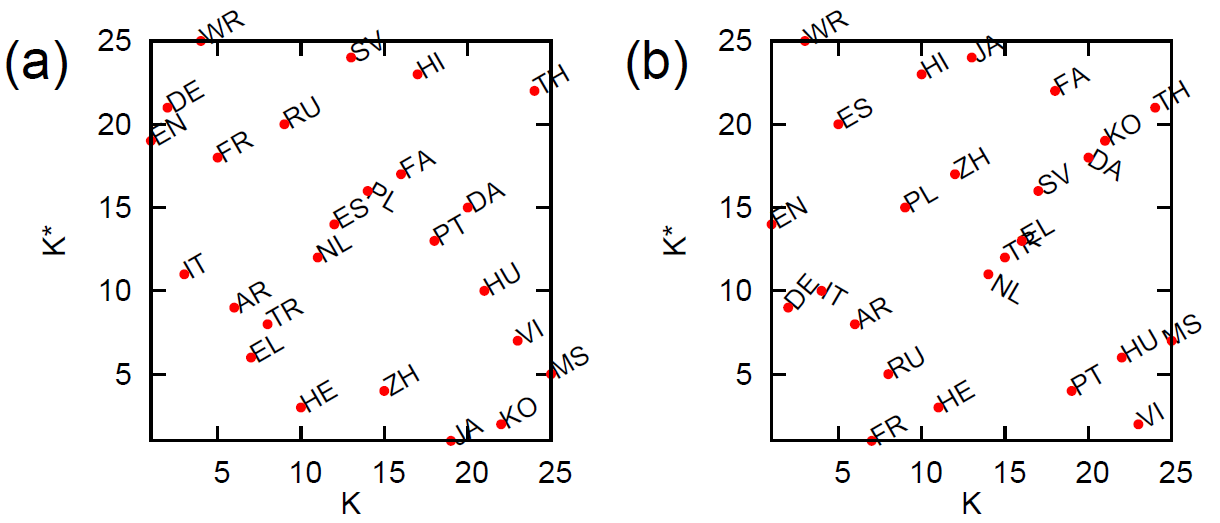}
\end{center}
\caption{(Color online)
 PageRank-CheiRank
 plane of cultures with corresponding indexes $K$ and $K^*$
obtained from the network of cultures based on (a) top 100 PageRank
historical figures, (b) top 100 2DRank historical figures.
After \cite{eom:2014}.
\label{fig9_9}}
\end{figure}

From the obtained matrix $G$ and $G^*$  
we determine PageRank and CheiRank vectors
and then the PageRank-CheiRank plane $(K,K^*)$,
shown in Fig.~\ref{fig9_9},
for networks of cultures from Fig.~\ref{fig9_7}.
Here $K$ indicates the ranking of a given culture ordered by 
how many of its own top historical
figures appear in other Wikipedia editions, and 
$K^*$  indicates the ranking of 
a given culture according to how many of 
the top historical figures in the considered culture are from other
cultures.  It is important to note that for 24 editions
the world node WR appears on positions
$K=3$ or $K=4$, for panels $(a), (b)$ in Fig.~\ref{fig9_9},
signifying that the 24 editions capture the 
main part of historical figures born
in these cultures. We note that for 9 editions
in \cite{eom:2013a} the node WR was
at the top position for PageRank
so that a significant fraction of historical figures was
attributed to other cultures.

From the data of Fig.~\ref{fig9_9}
we obtain at the top positions of $K$ cultures
EN, DE, IT showing that other cultures strongly point to them.
However, we can argue that for cultures it is also important to
have strong communicative property 
and hence it is important to have 2DRank
of cultures at top positions.
On the top 2DRank position  we have
Greek, Turkish and Arabic (for PageRank
persons) in Fig.~\ref{fig9_9}(a) 
and French, Russian and Arabic 
(for 2DRank persons) in Fig.~\ref{fig9_9}(b).
This demonstrates the important historical
influence of these cultures 
both via importance (incoming links) 
and communicative (outgoing links)
properties present in a balanced manner.

Thus  the described  research across
Wikipedia language editions suggests
a rigorous mathematical way, based on Markov chains and Google matrix,
for recognition of important historical figures
and analysis of interactions of cultures
at different historical periods and in different
world regions. 
Such an approach recovers 43 percent of persons
from the well established Hart historical study \cite{hart:1992},
that demonstrates the  reliability of this method.
We think that  a further extension of this approach
to a larger number of Wikipedia editions will provide
a more detailed and balanced analysis of interactions of
world cultures.

\section{Google matrix of social networks}
\label{s10}

Social networks like Facebook, LiveJournal, Twitter, Vkontakte
start to play a more and more important
role in modern society. The Twitter network is a directed one
and here we consider its spectral properties
following mainly the analysis reported in \cite{frahm:2012b}.

\subsection{Twitter network}
\label{s10.1}

\begin{figure}[H]
\begin{center}
\includegraphics[width=0.48\textwidth]{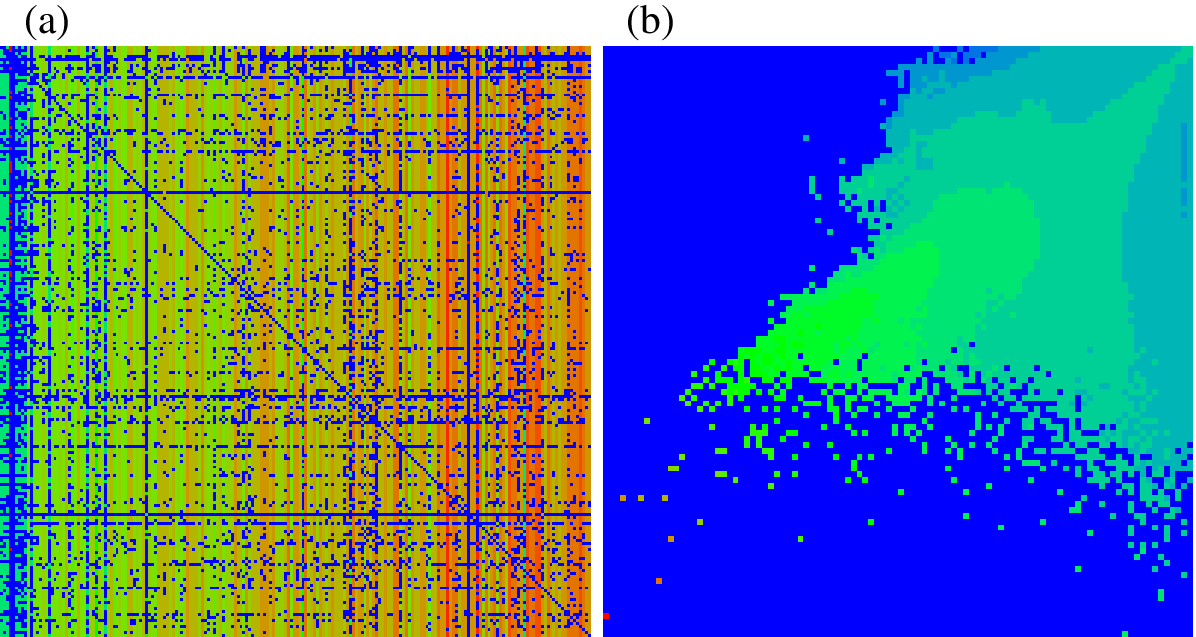}
\end{center}
\caption{(Color online) 
Panel (a): Google matrix of Twitter,
matrix elements of $G$ are shown in the basis of
PageRank index $K$ of matrix $G_{KK'}$. Here, $x$ (and $y$) axis 
show $K$ (and $K^{'}$) with the range $1 \leq K,K' \leq 200$. 
Panel (b)  shows the density of nodes $W(K,K^*)$  
of Twitter on PageRank-CheiRank plane $(K,K^*)$, 
averaged over $100\times100$ 
logarithmically equidistant grids for $0 \leq \ln K, \ln K^* \leq \ln N$ 
with the normalization condition $\sum_{K,K^*}W(K,K^*)=1$.
The $x$-axis corresponds to $\ln K$ and the $y$-axis to $\ln K^*$. 
In both panels color varies from blue/black at minimal value to red/gray 
at maximal value; here $\alpha=0.85$. 
After \cite{frahm:2012b}.
\label{fig10_1}}
\end{figure}

\begin{figure}[H]
\begin{center}
\includegraphics[width=0.48\textwidth]{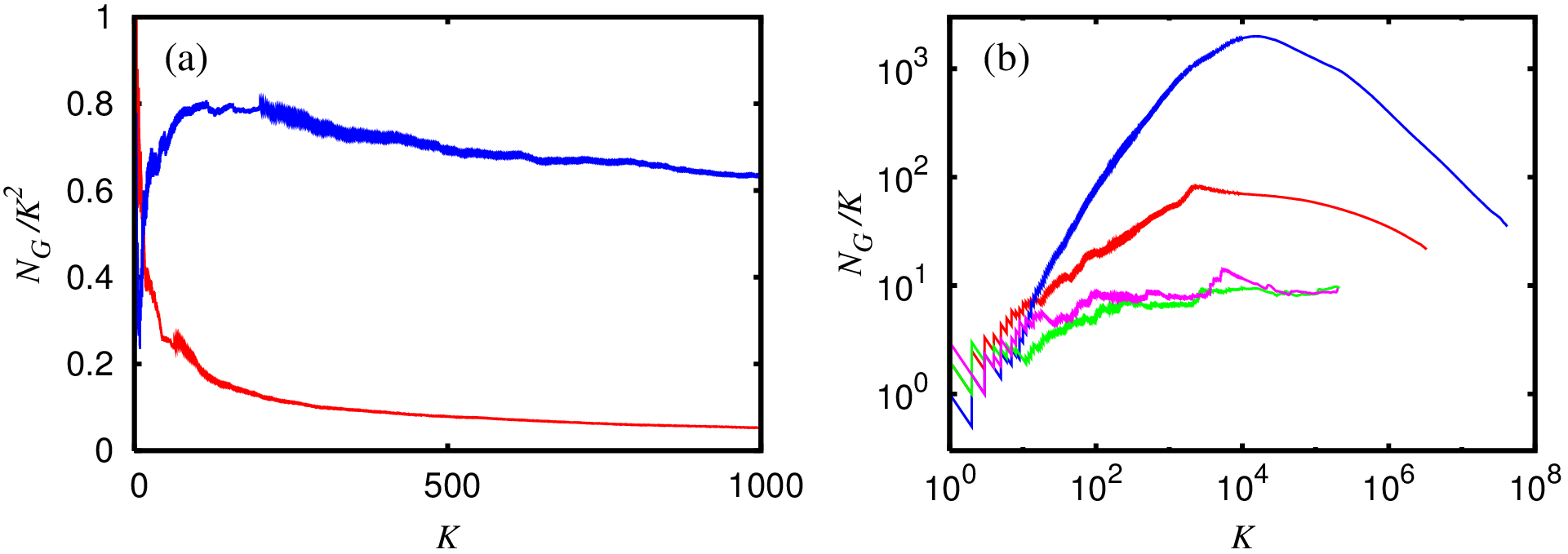}
\end{center}
\caption{(Color online)
(a) Dependence of the 
area density $g_K=N_G/K^2$  of nonzero elements of the adjacency
matrix among top PageRank nodes on the PageRank index $K$
for Twitter (blue/black curve) and Wikipedia (red/gray curve) networks,
data are shown in linear scale.
(b) Linear density $N_G/K$ 
of the same matrix elements shown for the whole range of $K$ in log-log scale
for Twitter (blue curve), Wikipedia (red curve), Oxford University 2006 
(magenta curve)
and Cambridge University 2006 (green curve) 
(curves from top to bottom at $K=100$).
After \cite{frahm:2012b}.
\label{fig10_2}}
\end{figure}

Twitter is a rapidly growing online directed social network. 
For July 2009 a data set of this entire network is available with 
$N=41652230$ nodes and $N_\ell=1468365182$ links
(for data sets see Refs. in \cite{frahm:2012b}). For this case 
the spectrum and eigenstate properties of the corresponding Google matrix 
have been analyzed in detail using the Arnoldi method and 
standard PageRank and CheiRank computations \cite{frahm:2012b}. 
For the Twitter network the average number of links per node 
$\zeta =N_\ell/N \approx 35$ and the general inter-connectivity between 
top PageRank nodes are considerably larger than for other networks such 
as Wikipedia 
(Sec.~\ref{s9}) or UK universities (Sec.~\ref{s8}) as can be seen in 
Figs.~\ref{fig10_1} and \ref{fig10_2}. 

The decay of PageRank probability can be approximately
described by an algebraic decay with the exponent
$\beta \approx 0.54$ while for CheiRank we have
a larger value $\beta \approx 0.86$ \cite{frahm:2012b}
that is opposite to the usual situation. The image of top
matrix elements of $G_{KK'}$ 
with $1 \leq K,K; \leq 200$ is shown in Fig.~\ref{fig10_1}.
The density distribution of nodes on $(K,K^*)$ plane
is also shown there. It is somewhat similar to 
those of Wikipedia case in Fig.~\ref{fig9_3},
may be with a larger density concentration
along the line $K \approx K^*$.

However, the most striking feature of $G$ matrix elements
is a very strong inteconnectivity between top PageRank nodes.
Thus for Twitter the
top $K \leq 1000$ elements fill about 70\% of the matrix and
about 20\% for size $K \leq 10^4$ . For Wikipedia the filling
factor is smaller by a factor $10 - 20$.
In particular the number $N_G$ of links between $K$ top PageRank nodes 
behaves for $K\le 10^3$ as $N_G\sim K^{1.993}$ while for Wikipedia 
$N_G\sim K^{1.469}$. The exponent for $N_G$, being close to 2 for Twitter, 
indicates that for the top PageRank nodes 
the Google matrix is macroscopically 
filled with a fraction $0.6-0.8$ of non-vanishing matrix elements (see 
also Figs.~\ref{fig10_1} and \ref{fig10_2}) and the very well connected 
top PageRank nodes can be considered as the Twitter elite 
\cite{kandiah:2012}. For Wikipedia the 
interconnectivity among top Page\-Rank nodes has an exponent $1.5$ 
being somewhat reduced 
but still stronger as compared to certain university networks where
typical exponents are close to unity (for the range $10^2 \le K \le 10^4$). 
The strong interconnectivity of Twitter is also visible in its global 
logarithmic density distribution of nodes in the 
PageRank-CheiRank plane $(K,K^*)$ (Fig.~\ref{fig10_1} (b))
which shows a maximal density along a certain ridge
along a line $\ln K^* =\ln K +\,$const. with a significant large number 
of nodes at small values $K,K^*<1000$.

The decay exponent of the PageRank is for Twitter $\beta=0.540$ (for 
$1\le K\le 10^6$), which indicates a precursor of a delocalization transition
as compared to Wikipedia ($\beta=0.767$) or WWW ($\beta\approx 0.9$),
caused by the strong interconnectivity \cite{frahm:2012b}. 
The Twitter network is also characterized by a large value of 
PageRank-CheiRank correlator
$\kappa =112.6$ that is by a factor $30 - 60 $ larger than
this value for Wikipedia and University networks.
Such a larger value of $\kappa$ results from certain
individual large values $\kappa_i = N P(K(i)) P^*(K^*(i)) \sim 1$.
It is argued that this is related to a
very strong inter-connectivity between top K PageRank
users of the Twitter network \cite{frahm:2012b}.

\begin{figure}[H]
\begin{center}
\includegraphics[width=0.48\textwidth]{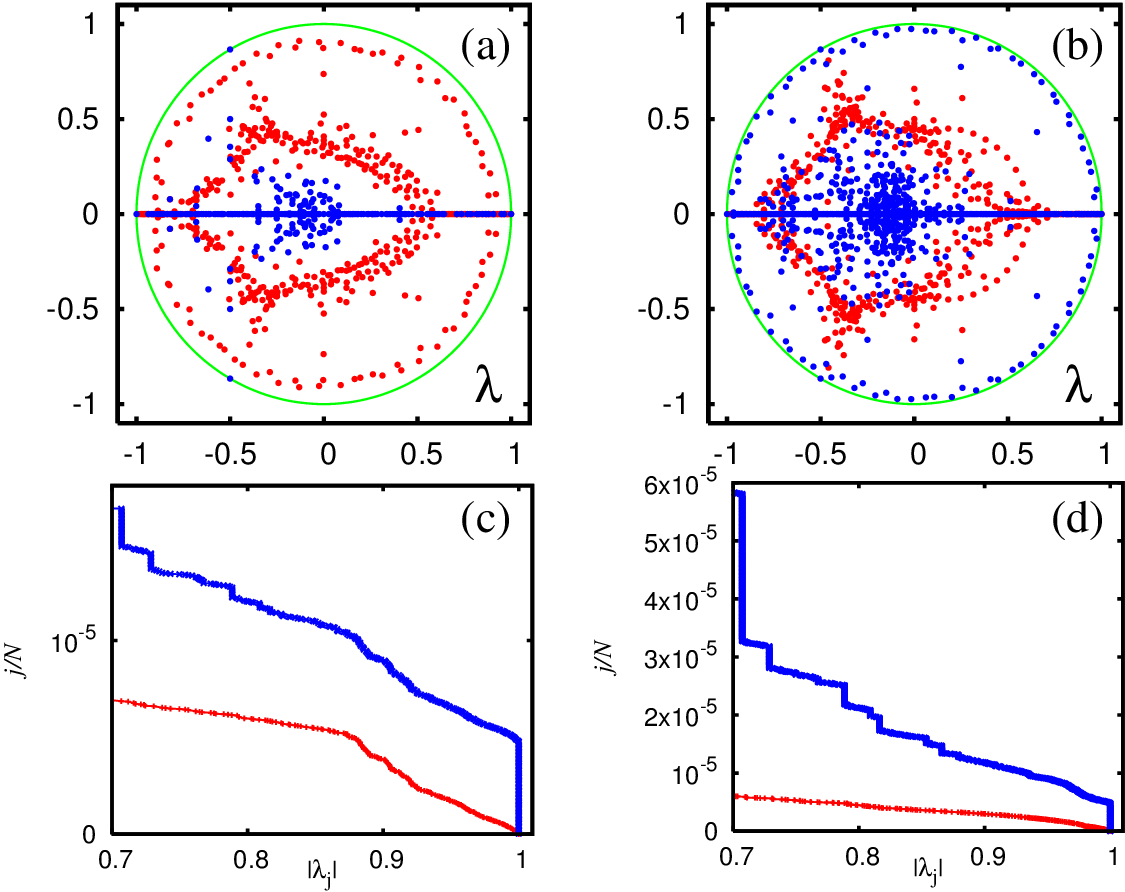}
\end{center}
\caption{(Color online)
Spectrum of the Twitter matrix $S$ (a) and (c), and $S^*$ (b) and (d). 
Panels (a) and (b) show subspace eigenvalues (blue/black dots)  and 
core space eigenvalues (red/gray dots) in $\lambda$-plane 
(green/gray curve shows unit circle);
there are 17504 (66316) invariant subspaces, with maximal dimension 44 
(2959) and the sum of all subspace dimensions is $N_s=40307$ (180414).
The core space eigenvalues are obtained from the Arnoldi method applied to 
the core space subblock $S_{cc}$ of $S$ with Arnoldi dimension $n_A=640$. 
Panels (c) and (d) show the  
fraction $j/N$ of eigenvalues with $|\lambda| >  |\lambda_j|$ for the 
core space eigenvalues (red/gray bottom curve) 
and all eigenvalues (blue/black top curve)
from raw data ((a) and (b) respectively). The number of 
eigenvalues with $|\lambda_j|=1$ is 34135 (129185) of which 
17505 (66357) are at $\lambda_j=1$; this number is (slightly) larger than the 
number of invariant subspaces which have each at least one unit eigenvalue.
Note that in panels (c) and (d) the number 
of eigenvalues with $|\lambda_j|=1$ is 
artificially reduced to 200 in order to have a better scale on the 
vertical axis. The correct numbers of those eigenvalues correspond to 
$j/N=8.195\times 10^{-4}$ (c) and $3.102\times 10^{-3}$ (d) 
which are strongly outside 
the vertical panel scale. 
After \cite{frahm:2012b}.
\label{fig10_3}}
\end{figure} 

The spectra of matrices $S$ and $S^*$ are obtained with the
help of the Arnoldi method for a relatively modest Arnoldi dimension
due to a very large matrix size. The largest 
$n_A$ modulus eigenvalues $|\lambda|$
are shown in Fig.~\ref{fig10_3}.
The invariant subspaces (see Sec.~\ref{s3.3}) for the Twitter network 
cover about $N_s=4\times 10^4$ ($1.8\times 10^5$) nodes for $S$ ($S^*$) 
leading to $1.7\times 10^4$ ($6.6\times 10^4$) eigenvalues 
with $\lambda_j=1$ or even $3.4\times 10^4$ ($1.3\times 10^5$) eigenvalues 
with $|\lambda_j|=1$. However, for Twitter the fraction of subspace nodes 
$g_1=N_s/N\approx 10^{-3}$ is smaller than 
the fraction $g_1\approx 0.2$ for the university networks of Cambridge 
or Oxford (with $N\approx 2\times 10^5$) since the size of the whole 
Twitter network is significantly larger.  
The complex spectra of $S$ and $S^*$ also show the cross and triple-star 
structures, as in the cases of Cambridge and Oxford 
2006 (see Fig.~\ref{fig8_1}), even though for the Twitter network 
they are significantly less pronounced. 

\subsection{Poisson statistics of PageRank probabilities}
\label{s10.2}

From a physical viewpoint one can conjecture 
that the PageRank probabilities
are described by a steady-state quantum Gibbs distribution
over certain quantum levels with energies $E_i$ by 
the identification $P(i) = \exp(-E_i/T)/Z$ with $Z=\sum_i \exp(-E_i/T)$
 \cite{frahm:2014a}. 
In some sense this conjecture assumes that the operator matrix $G$
can be represented as a sum of two operators $G_H$ and $G_{NH}$
where $G_H$ describes a Hermitian system while $G_{NH}$ represents
a non-Hermitian operator which creates a system thermalization at a 
certain effective temperature $T$ with the quantum Gibbs distribution 
over energy levels $E_i$ of the operator $G_H$. 

\begin{figure}[H]
\begin{center}
\includegraphics[width=0.48\textwidth]{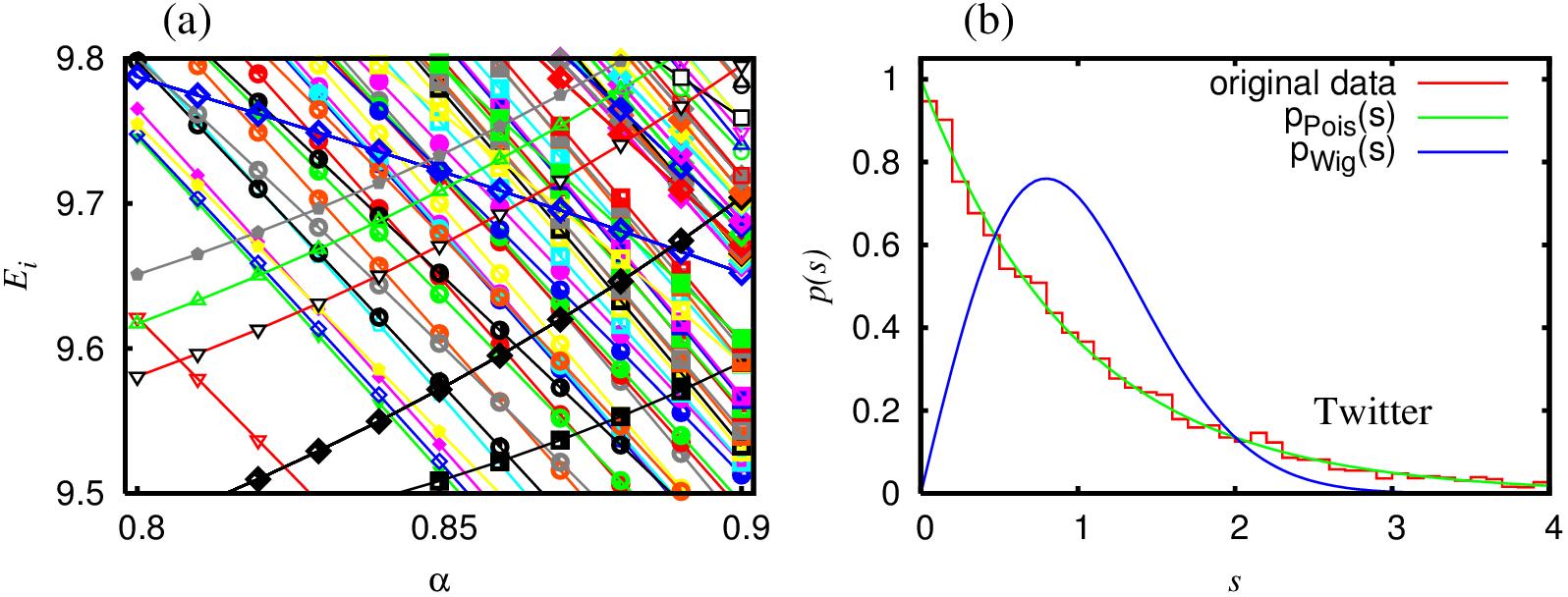}
\end{center}
\caption{(Color online)
Panel (a) shows the dependence of certain top 
PageRank levels $E_i=-\ln(P_i)$ on 
the damping factor $\alpha$ for Twitter network. 
Data points on curves with one 
color corresponds to the same node $i$; 
about 150  levels are shown close to the minimal energy
$E \approx 7.5$. Panel (b) represents the  
histogram of unfolded level spacing statistics for Twitter
at $10 < K \leq 10^4$. 
The Poisson distribution 
$p_{\rm Pois}(s)=\exp(-s)$ 
and the Wigner surmise 
$p_{\rm Wig}(s)=\frac{\pi}{2}\,s\,\exp(-\frac{\pi}{4}\,s^2)$ 
are also shown for comparison. 
After \cite{frahm:2014a}.
\label{fig10_4}}
\end{figure} 

The identification of PageRank with an energy spectrum allows to study 
the corresponding level statistics which represents a
well known concept in the framework of 
Random Matrix Theory \cite{mehta:2004,guhr:1998}. The most direct 
characteristic is the probability distribution $p(s)$ of 
unfolded level spacings $s$. Here $s=(E_{i+1}-E_i)/\Delta E$ 
is a spacing between nearest levels
measured in the units of average local energy spacing
$\Delta E$. The unfolding procedure 
\cite{mehta:2004,guhr:1998} requires the smoothed 
dependence of $E_i$ on the index $K$ which is obtained from a polynomial 
fit of $E_i\sim \ln(P_i)$ with $\ln(K)$ as argument \cite{frahm:2014a}.

The statistical properties of fluctuations of levels
have been extensively studied in the fields of RMT
\cite{mehta:2004}, quantum chaos \cite{haake:2010}
and disordered solid state systems \cite{evers:2008}.
It is known that integrable quantum systems 
have  $p(s)$ well described by the Poisson distribution 
$p_{\rm Pois}(s)=\exp(-s)$. In contrast the quantum systems, which
are chaotic in the classical limit (e.g. Sinai billiard),
have $p(s)$ given by the RMT
being close to the Wigner surmise
$p_{\rm Wig}(s)=\frac{\pi}{2}\,s\,\exp(-\frac{\pi}{4}\,s^2)$ 
\cite{bohigas}. Also the Anderson localized
phase is characterized by $p_{\rm Pois}(s)$
while in the delocalized regime one has
$p_{\rm Wig}(s)$ \cite{evers:2008}.

The results for the Twitter PageRank level statistics \cite{frahm:2014a} 
are shown in Fig.~\ref{fig10_4}. 
We find that $p(s)$ is  well described by the Poisson distribution. Furthermore, 
the evolution of energy levels $E_i$ with the variation of 
the damping factor $\alpha$ shows many level crossings which are 
typical for Poisson statistics. We  may note that here each level
has its own index so that it is rather easy to see 
if there is a real or avoided level crossing.

The validity of the Poisson statistics for PageRank probabilities 
is confirmed also for the networks of Wikipedia editions
 in English, French and German from Fig.~\ref{fig9_3} \cite{frahm:2014a}.
We argue that due to absence of level repulsion the PageRank
order of nearby nodes can be easily interchanged. 
The obtained Poisson law implies that 
the nearby PageRank probabilities fluctuate 
as random independent variables.

\section{Google matrix analysis of world trade}
\label{s11}

During the last decades the trade between countries 
has been developed in an extraordinary way.  
Usually countries are ranked in the world trade 
network (WTN) taking into account their exports 
and imports measured in {\it USD} \cite{cia:2010}. 
However, the use of these quantities, which are local
in the sense that countries know their total imports
and exports, could hide the information of the
centrality role that a country plays in this complex
network.
In this section we present the two-dimensional Google 
matrix analysis of the WTN introduced in \cite{ermann:2011b}.
Some previous studies of global network characteristics
were considered in \cite{garlaschelli,vespignani},
degree centrality measures were analyzed in \cite{benedictis}
and a time evolution of network global characteristics 
was studied in \cite{hedeem}. Topological and clustering
properties of multiplex network of
various commodities were discussed in \cite{garlaschelli:2010},
and an ecological ranking based on the nestedness of 
countries and products was presented in \cite{ermann:2013a}.

The money exchange between countries defines 
a directed network. 
Therefore Google matrix analysis 
can be introduced in a natural way. 
PageRank and CheiRank algorithms can be easily 
applied to this network with a straightforward 
correspondence with imports and exports. 
Two-dimensional ranking, introduced in 
Sec.~\ref{s4}, gives an illustrative representation 
of global importance of countries in the WTN.
The important element of Google ranking of WTN is
its democratic treatment of all world countries,
independently of their richness,
that follows the main principle of the United Nations
(UN).

\subsection{Democratic ranking of countries}
\label{s11.1}

The WTN is a directed network that can be constructed 
considering countries as nodes and money exchange 
as links. We follow the definition of the WTN 
of \cite{ermann:2011b} where trade information comes from 
\cite{uncomtrade:2011}.  
These data include all trades between countries 
for different products (using Standard International 
Trade Classification of goods, SITC1) from 1962 to 2009. 

All useful information of the WTN is expressed via  the 
\emph{money matrix} $M$, which definition, in terms of 
its matrix elements $M_{ij}$, is defined as the money 
transfer (in {\it USD}) from country $j$ to country 
$i$ in a given year. This definition can be applied to 
a given specific product or to \emph{all commodities},
which represent the sum over all products.

In contrast to the binary adjacency matrix 
$A_{ij}$ of WWW (as the ones analyzed in S\ref{s8} 
and S\ref{s10} for example) $M$ has weighted elements. 
This corresponds to a case when there are
in principle multiple number of links from $j$ to $i$
and this number is proportional to {\it USD} amount transfer. 
Such a situation appears in Sec.~\ref{s6} for Ulam networks and
Sec.~\ref{s7} for Linux PCN   with a main difference
that for the WTN case there is a very large variation 
of mass matrix elements
$M_{ij}$, related to the fact that there is a very strong 
variation of richness of various countries.

Google matrices $G$ and $G^*$ are constructed according 
to the usual rules and relation
(\ref{eq3_1}) with $M_{ij}$ and its transposed: 
$S_{ij}=M_{ij}/m_j$ and $S_{ij}=M_{ji}/m^*_j$
where $S_{ij}=1/N$ and $S^*_{ij}=1/N$, if for a given $j$ 
all elements $M_{ij}=0$ and $M_{ji}=0$ respectively.
Here $m_j= \sum_i M_{ij}$ and $m^*_j= \sum_i M_{ji}$ are 
the total export and import mass for country $j$. 
Thus the sum in each column of $G$ or $G^*$
is equal to unity.
In this way Google matrices $G$ and $G^*$
of WTN allow to treat all countries on equal grounds 
independently of the fact if a given country is rich or poor.
This kind of analysis treats in a democratic way all world 
countries in consonance with the standards of the UN.

The probability distributions of ordered PageRank $P(K)$
and CheiRank $P^*(K^*)$ depend on their indexes 
in a rather similar way with a power law decay given 
by $\beta$. For the fit of top 100 countries and 
\emph{all commodities} the average exponent value is 
close to $\beta=1$ 
corresponding to the Zipf law \cite{zipf}.

\begin{figure}[h]
\begin{center}
\includegraphics[width=0.48\textwidth]{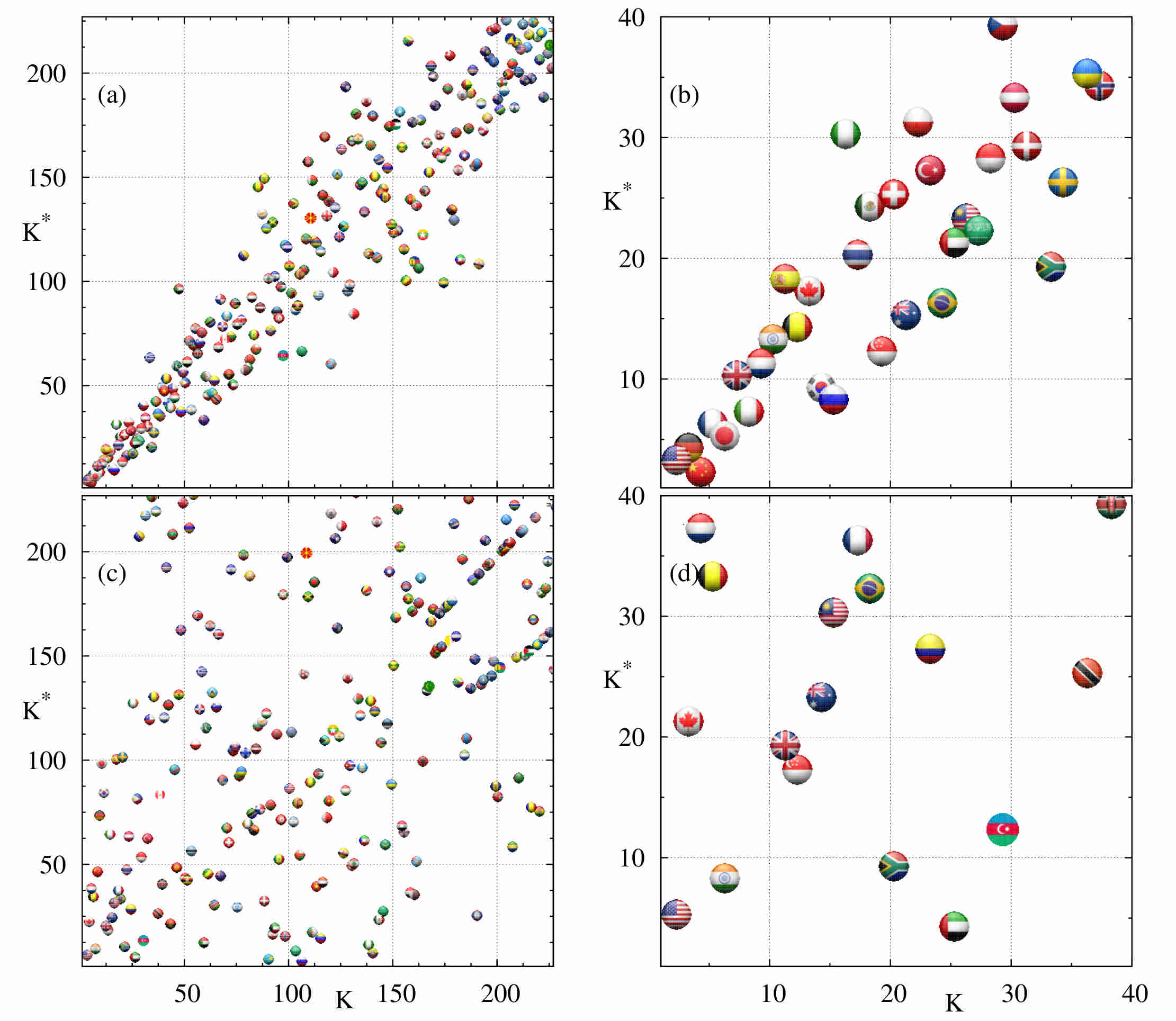}
\end{center}
\caption{(Color online)
Country positions in PageRank-CheiRank plane $(K,K^*)$ 
for world trade in various commodities in 2008.
Each country is shown by circle with its own flag
(for a better visibility
the circle center is slightly displaced from its
integer position $(K,K^*)$ along direction angle $\pi/4$).
The panels show the ranking for trade in the
following commodities: \emph{all commodities} (a) and (b);
and \emph{crude petroleum} (c) and (d). Panels
(a) and (c) show a global scale with all 227 countries, 
while (b) and (d) give a zoom in the region of $40\times40$ top ranks.
After \cite{ermann:2011b}.
\label{fig11_1}}
\end{figure} 

The distribution of countries
on PageRank-CheiRank plane for trade in \emph{all commodities}  
in year 2008  is shown in panels (a) and (b) of 
Fig.~\ref{fig11_1} at $\alpha=0.5$.
Even if the Google matrix approach
is based on a democratic ranking of
international trade, being independent of
total amount of export-import and PIB for a given 
country, the top ranks $K$ and $K^*$ belong to
the group of industrially developed countries.
This means that these countries have
efficient trade networks with optimally
distributed trade flows. Another striking feature 
of global distribution is that it is
concentrated along the main diagonal $K=K^*$.
This feature is not present in 
other networks studied before. 
The origin of this density concentration is related to 
a simple economy reason: for each country the 
total import is approximately equal to export 
since each country should keep
in average an economic balance.
This balance does not imply a symmetric money matrix,
used in gravity model of trade 
(see e.g. \cite{krugman2011,benedictis}), as can be seen
in the significant broadening of distribution of Fig.~\ref{fig11_1}
(especially at middle values of $K \sim 100$).

For a given country its trade is doing well
if its $K^*< K$ so that the country
exports more than it imports. The opposite
relation $K^*> K$ corresponds to a bad trade situation
(e.g.  Greece being significantly above the diagonal).
We also can say that local minima
in the curve of $(K^*-K) \; vs. \; K$ correspond to 
a successful trade while maxima mark
bad traders. In 2008 most successful were
China, R of Korea, Russia, Singapore,
Brazil, South Africa, Venezuela (in order of
$K$ for $K \leq 50$) while among bad traders we note
UK, Spain, Nigeria, Poland, Czech Rep, Greece, Sudan
with especially strong export drop for two last cases.

A comparison between local and global rankings of countries
for both imports and exports gives a new tool to analyze 
countries economy.
For example, in 2008 the most significant differences between
CheiRank and the rank given by total exports are for {\it Canada}
and {\it Mexico} with corresponding money export ranks 
$\tilde{K}^* = 11$ and $13$ and 
with $K^*=16$ and $K^*=23$ respectively. These variations
can be explained in the context that the export of these 
two countries is too strongly oriented on {\it USA}.
In contrast {\it Singapore}
moves up from $\tilde{K}^* = 15$ export position to $K^*=11$
that shows the stability and broadness of its 
export trade, a similar situation
appears for {\it India} moving up
from $\tilde{K}^*=19$ to $K^*=12$
(see \cite{ermann:2011b} for more detailed analysis).

\subsection{Ranking of countries by trade in products}
\label{s11.2}

If we focus on the two-dimensional distribution of countries
in a specific product we obtain a very different information.
The symmetry approximately visible for {\it all commodities}
is absolutely absent: the points are scattered 
practically over the whole square $N \times N$ (see Fig.~\ref{fig11_1}).
The reason of such a strong scattering is clear:
e.g. for  {\it crude petroleum} some countries 
export this product while other countries 
import it. Even if there is some flow from exporters
to exporters it remains relatively low.
This makes the Google matrix to be very asymmetric.
Indeed, the asymmetry of trade flow is well visible in panels
(c) and (d) of Fig.~\ref{fig11_1}.

\begin{figure}[h]
\begin{center}
\includegraphics[width=0.48\textwidth]{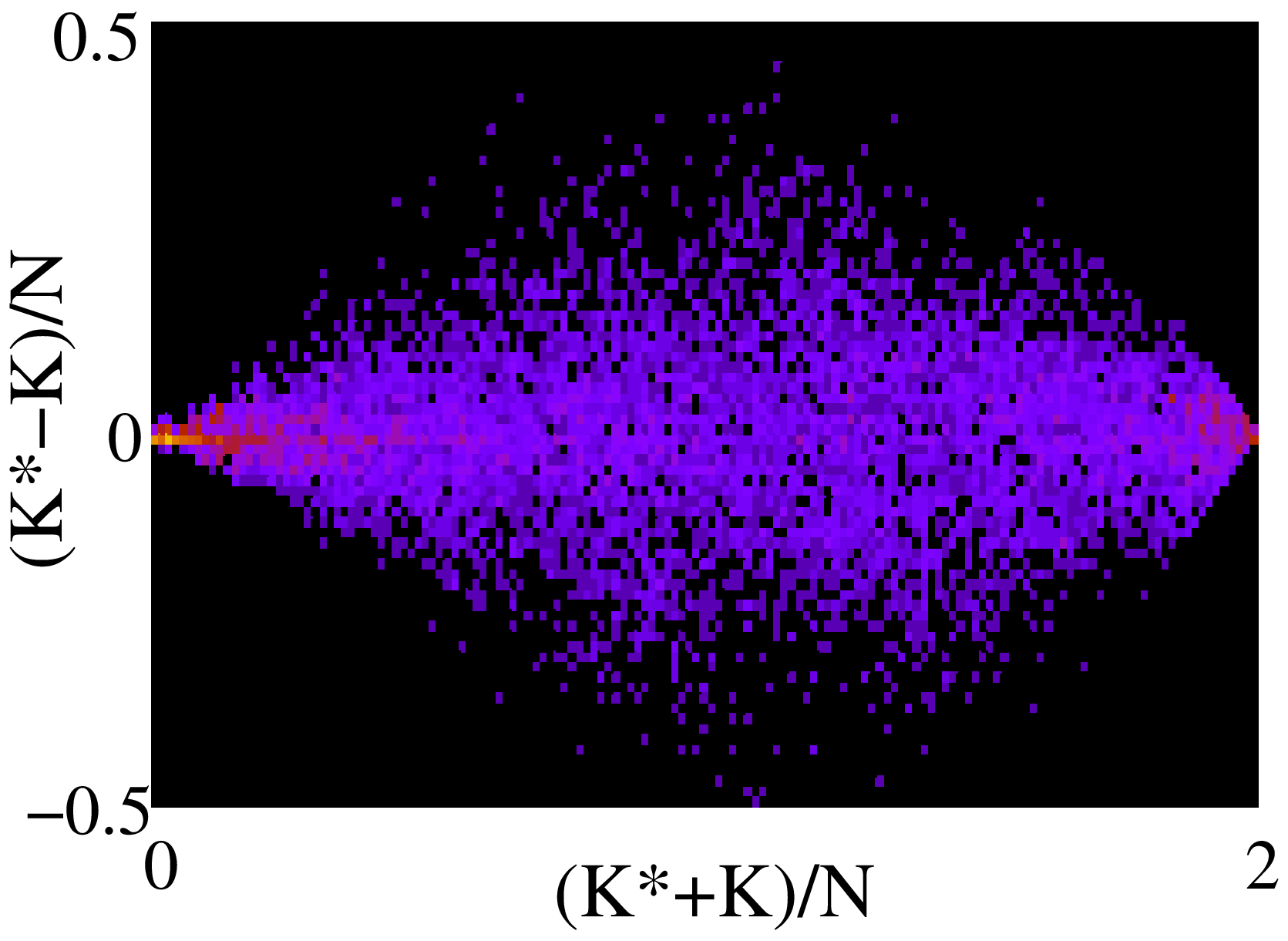}
\end{center}
\caption{(Color online) 
Spindle distribution for WTN
of \emph{all commodities} for all countries
in the period 1962 - 2009 shown in the plane of 
 $((K^*-K)/N, (K^*+K)/N)$
(coarse-graining inside
each of $76 \times 152$ cells); 
data from the UN COMTRADE database.
After \cite{ermann:2011b}.
\label{fig11_2}}
\end{figure} 

The same comparison of global and local rankings done before
for \emph{all commodities} can be applied to specific 
products obtaining even more strong differences.
For example for {\it crude petroleum}
Russia moves up from $\tilde{K}^*=2$ export position to $K^*=1$
showing that its trade network in this product is 
better and broader than the one of Saudi Arabia
which is at the first export position $\tilde{K}^*=1$ in money volume.
Iran moves in opposite direction
from $\tilde{K}^*=5$ money position down to $K^*=14$
showing that its trade network is 
restricted to a small number of nearby countries.
A significant improvement of ranking 
takes place for Kazakhstan moving up
from $\tilde{K}^*=12$ to $K^*=2$.
The direct analysis shows that this happens
due to an unusual fact that Kazakhstan
is practically the only country
which sells {\it crude petroleum}
to the CheiRank leader in this product Russia.
This puts Kazakhstan on the second position.
It is clear that such direction of trade
is more of political or geographical origin
and is not based on economic reasons.

The same detailed analysis can be applied 
to all specific products given by SITC1.
For example for trade of {\it cars}  
France goes up from  $\tilde{K}^*=7$ position 
in exports to $K^*=3$ due to its broad export 
network. 

\subsection{Ranking time evolution and crises}
\label{s11.3}

The WTN has evolved during the 
period 1962 - 2009.
The number of countries is increased by 38\%,
while the number of links per country
for {\it all commodities}
is increased in total by  140\% with a significant increase
from 50\% to 140\%  
during the period 1993 - 2009
corresponding to economy globalization.
At the same time for a specific commodity
the average number of links per country
remains on a level of 3-5 links
being by a factor 30 smaller 
compared to {\it all commodities} trade.
During the whole period the total amount
$M_T$ of trade in {\it USD}
shows an average
exponential growth by 2 orders of magnitude.

A statistical density
distribution of countries in the plane $(K^*-K, K^*+K)$
in the period 1962 - 2009 for {\it all commodities}
is shown in Fig.~\ref{fig11_2}.
The distribution has a form of {\it spindle}
with maximum density
at the vertical axis $K^*-K=0$.
We remind that good exporters are
on the lower side of this axis at
$K^*-K < 0$, while the good importers (bad exporters)
are on the upper side at $K^*-K > 0$.

\begin{figure}[h]
\begin{center}
\includegraphics[width=0.48\textwidth]{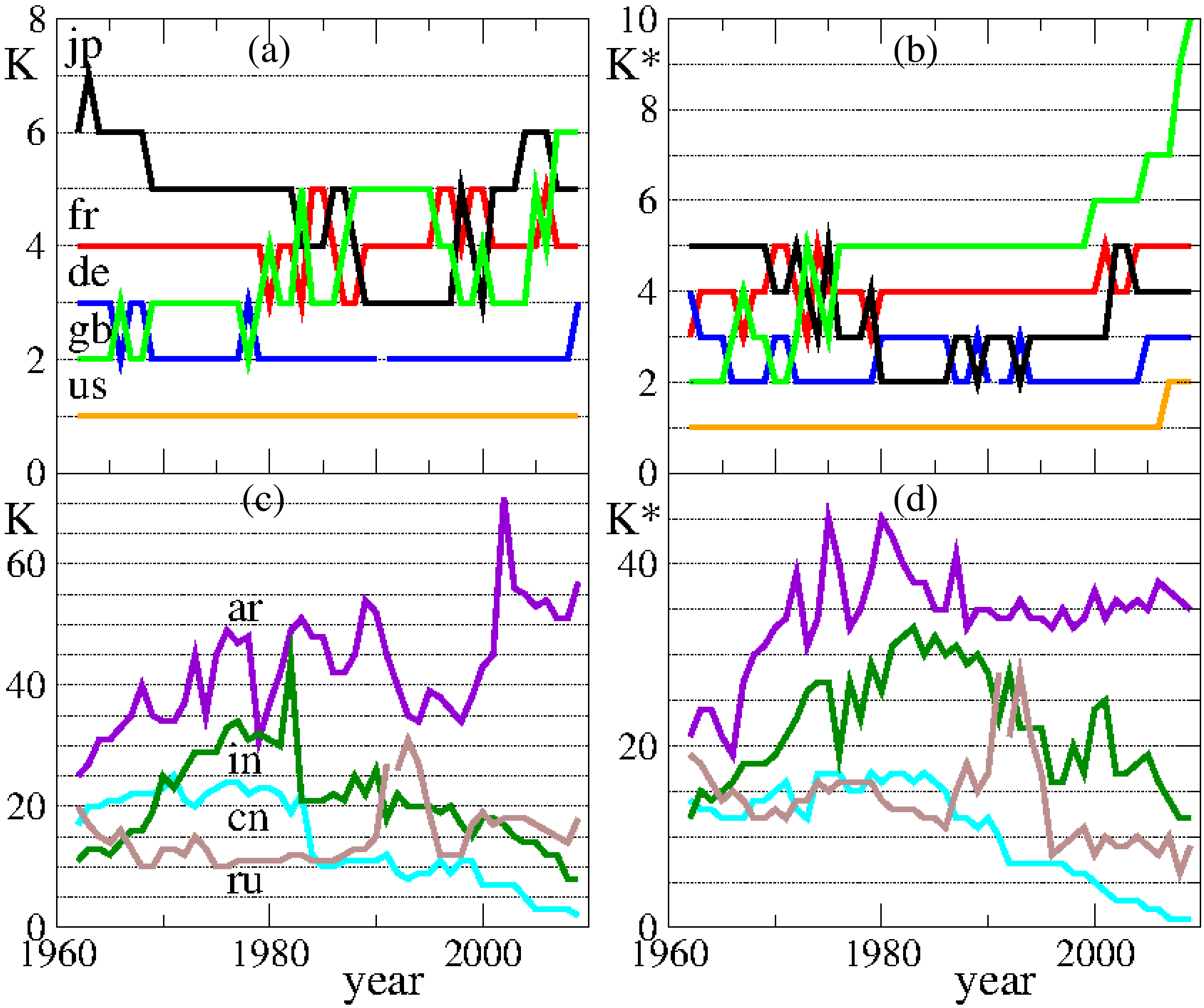}
\end{center}
\caption{(Color online)
Time evolution of 
CheiRank and PageRank indexes $K$, $K^*$
for some selected countries for \emph{all commodities}. 
The countries shown panels (a) and (b) are: 
Japan (jp-black), France (fr-red), Fed R of Germany 
and Germany (de - both in blue), Great Britain (gb - green), 
USA (us - orange) [curves
from top to bottom in 1962 in (a)].
The countries shown panels (c) and (d) are: 
Argentina (ar - violet), India (in - dark green), China (cn - cyan), 
USSR and Russian Fed (ru - both in gray)
[curves from top to bottom in 1975 in (c)].
After \cite{ermann:2011b}.
\label{fig11_3}}
\end{figure} 

The evolution of the ranking of countries for 
\emph{all commodities} reflects their economical changes.
The countries that occupy top positions tend to
move very little in their ranks and can be associated to 
a \emph{solid phase}. On the other hand,
the countries in the middle region of $K^*+K$ have a 
gas like phase with
strong rank fluctuations.

Examples of ranking evolution $K$ and $K^*$ 
for Japan, France, Fed R of Germany and Germany, 
Great Britain, USA, and for 
Argentina, India, China, 
USSR and Russian Fed are shown in Fig.~\ref{fig11_3}.
It is interesting to note that  sharp increases
in $K$ mark crises in 1991, 1998 for Russia 
and in 2001 for Argentina 
(import is reduced in period of crises).
It is also visible that in recent years the solid phase
is perturbed by entrance of new countries like China and India.
Other regional or global crisis could be highlighted 
due to the big fluctuations in the evolution of ranks. 
For example, in the range $81 \leq K+K^* \leq 120$,
during the period of 1992 - 1998 some financial crises as 
Black Wednesday, Mexico crisis,
Asian crisis and Russian crisis are appreciated
with this ranking evolution.

\subsection{Ecological ranking of world trade}
\label{s11.4}

Interesting parallels between multiproduct world trade 
and interactions between species in ecological systems has 
been traced in \cite{ermann:2013a}. This approach is
based on analysis of strength of transitions
forming the Google matrix for the
multiproduct world trade network.

Ecological systems are characterized by  high
complexity and biodiversity \cite{may:2001}
linked to nonlinear dynamics and chaos 
emerging in the process of their evolution \cite{lichtenberg:1992}.
The interactions between species
form a complex network whose properties can be analyzed 
by the modern methods of scale-free networks.
The analysis          
of their properties uses a concept of mutualistic networks and       
provides a detailed understanding of their features being            
linked to a high nestedness of these networks
\cite{burgos:2007,bastolla:2009,saverda:2011,burgos:2008}.                       
Using the UN COMTRADE database                           
we show that a similar ecological analysis                           
gives a valuable description of the world trade:                     
countries and trade products are analogous to plants                 
and  pollinators, and the whole trade network is characterized       
by a high nestedness  typical for ecological networks.

An important feature of ecological networks
is that they are highly structured, being very different from
randomly interacting species \cite{bascompte:2003}.
Recently is has been shown that the mutualistic
networks between plants and their pollinators 
\cite{bascompte:2003,vazquez:2004,memmott:2004,olesen:2007,rezende:2007}
are characterized by high nestedness  
which minimizes competition and increases biodiversity
\cite{burgos:2007,bastolla:2009,saverda:2011,burgos:2008}.

\begin{figure}[h]
\begin{center}
\includegraphics[width=0.48\textwidth]{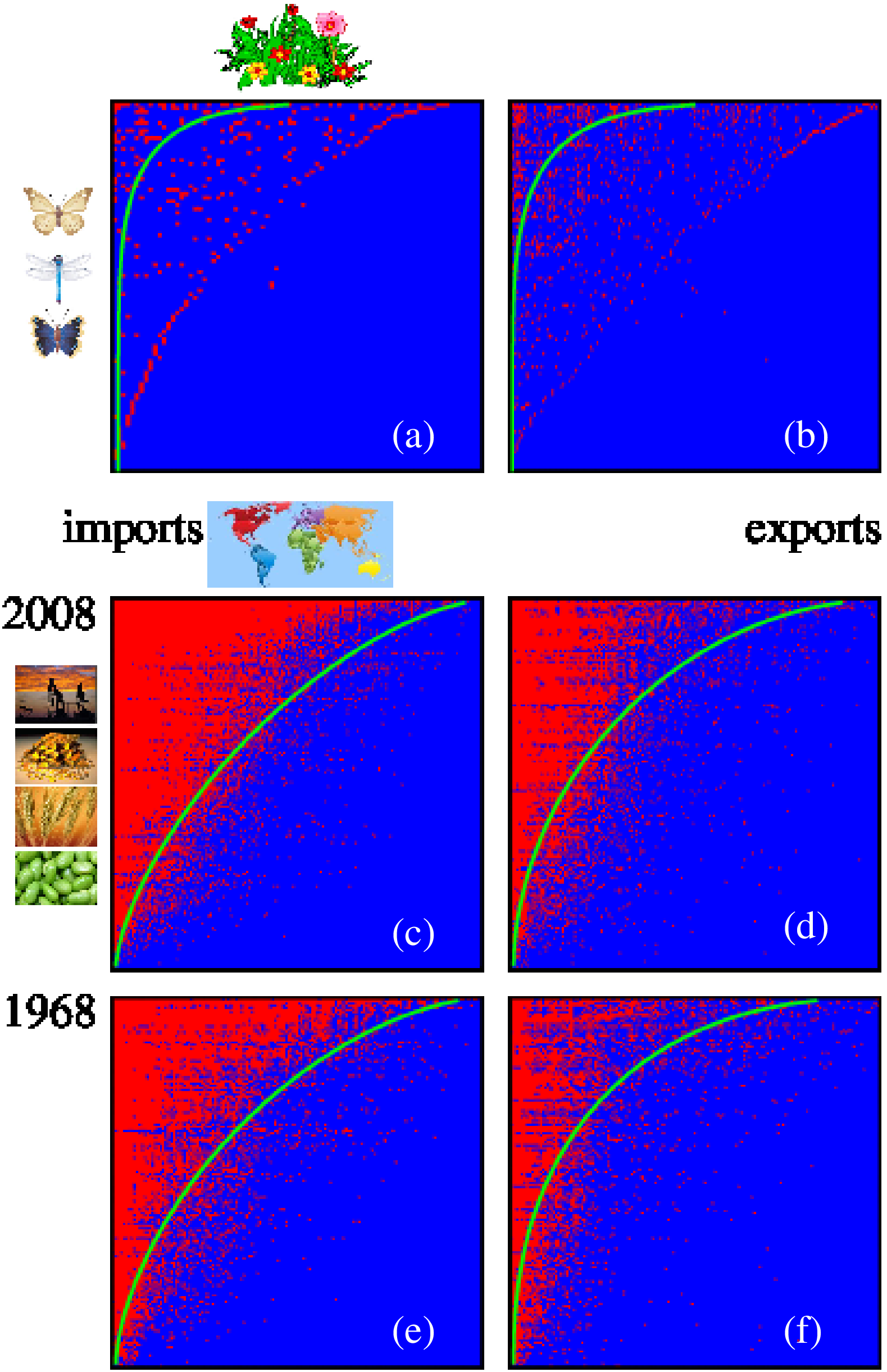}
\end{center}
\caption{(Color online)
Nestedness matrices for the plant-animal mutualistic 
networks on top panels,  and for the WTN of countries-products 
on middle and bottom panels. Panels (a) and (b) 
represent data of \emph{ARR1} and \emph{WES} networks from
\cite{rezende:2007}.
The WTN matrices are computed with the threshold $\mu=10^{-3}$
and corresponding $\varphi \approx 0.2$
for years 2008 (c,d) and 1968 (e,f) and 2008   
for import (c,e) and export (d,f) panels.
Red/gray and blue/black represent unit and zero elements respectively;
only lines and columns with nonzero elements are shown.
The order of plants-animals, countries-products is given by 
the nestedness algorithm \cite{rodriguez:2006}, 
the perfect nestedness is shown 
by green/gray curves for the corresponding values of $\varphi$.
After \cite{ermann:2013a}.
\label{fig11_4}}
\end{figure} 

The mutualistic WTN  is constructed on the
basis of the UN COMTRADE database  from the 
matrix of trade transactions $M^p_{c^\prime,c}$ expressed in USD
for a given product
(commodity) $p$ from country $c$ to country $c^\prime$
in a given year (from 1962 to 2009). For product
classification we use  3--digits SITC Rev.1 discussed above 
with the number of products $N_p=182$.
All these products are described in \cite{uncomtrade:2011}
in the commodity code document SITC Rev1.
The number of countries varies between $N_c=164$ in 1962
and $N_c=227$ in 2009. The import and export trade matrices
are defined as $M^{(i)}_{p,c}=\sum_{c^\prime=1}^{N_c} M^p_{c,c^\prime}$
and $M^{(e)}_{p,c}=\sum_{c^\prime=1}^{N_c} M^p_{c^\prime,c}$ respectively.
We use the dimensionless matrix elements
$m^{(i)}=M^{(i)}/M_{max}$ and $m^{(e)}=M^{(e)}/M_{max}$ where for a given year
$M_{max}=max\{max[M^{(i)}_{p,c}],max[M^{(e)}_{p,c}]\}$.
The distribution of matrix elements $m^{(i)}$, $m^{(e)}$
in the  plane of indexes $p$ and $c$,
ordered by the total amount of import/export
in a decreasing order, are shown and discussed in \cite{ermann:2013a}.
In global, the distributions of $m^{(i)}$, $m^{(e)}$
remain stable in time especially in a view of 
100 times growth of the total trade volume 
during the period 1962-2009. The fluctuations of $m^{(e)}$
are  larger compared to $m^{(i)}$ case since
certain products, e.g. petroleum, are exported by only a few countries 
while it is imported by almost all countries.

To use the methods of ecological analysis we 
construct the mutualistic network matrix
for import  $Q^{(i)}$ and export $Q^{(e)}$
whose matrix elements take binary value 
$1$ or $0$ if  corresponding elements
 $m^{(i)}$ and $m^{(e)}$ are  respectively larger or smaller
than a certain trade threshold value $\mu$. 
The fraction $\varphi$
of nonzero matrix elements
varies smoothly 
in the range $10^{-6} \leq \mu \leq 10^{-2}$ 
and the further analysis is not really sensitive
to the actual $\mu$ value inside this broad range.

\begin{figure}[h]
\begin{center}
\includegraphics[width=0.48\textwidth]{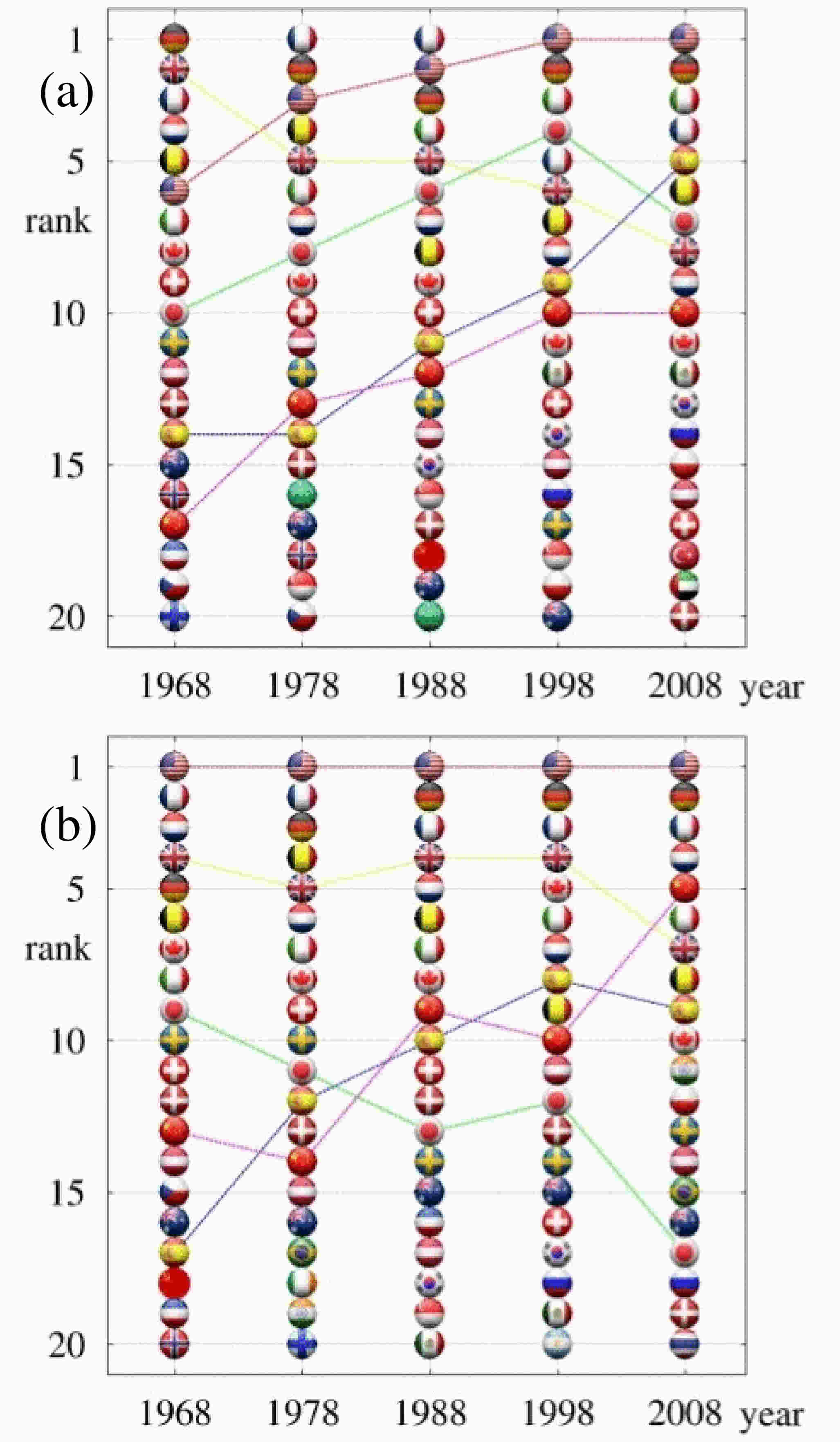}
\end{center}
\caption{(Color online)
Top 20 EcoloRank countries as a function of years 
 for the WTN
 import (a) and export (b)  panels.   
The ranking is given by the nestedness algorithm 
for the trade threshold $\mu=10^{-3}$; 
each country is represented by its corresponding flag.
As an example, dashed lines show time evolution of  
the following countries:
USA, UK, Japan, China, Spain.
After \cite{ermann:2013a}.
\label{fig11_5}}
\end{figure} 

In contrast to  ecological
systems \cite{bastolla:2009} the world trade is
described by a directed network and hence we 
characterize the system by two 
mutualistic matrices  $Q^{(i)}$ and $Q^{(e)}$ corresponding
to import and export. Using the standard nestedness 
BINMATNEST algorithm \cite{rodriguez:2006} 
we determine the nestedness parameter  $\eta$ of the 
WTN and the related nestedness temperature $T=100 (1-\eta)$.
The algorithm reorders lines and columns of
a mutualistic matrix   
concentrating nonzero elements as 
much as possible in the top left corner
and thus providing information about the role of
immigration and extinction in an ecological system.
A high level of nestedness and ordering
can be reached only for systems with low $T$.
It is argued that the nested architecture of real mutualistic networks
increases their biodiversity.

\begin{figure}[h]
\begin{center}
\includegraphics[width=0.48\textwidth]{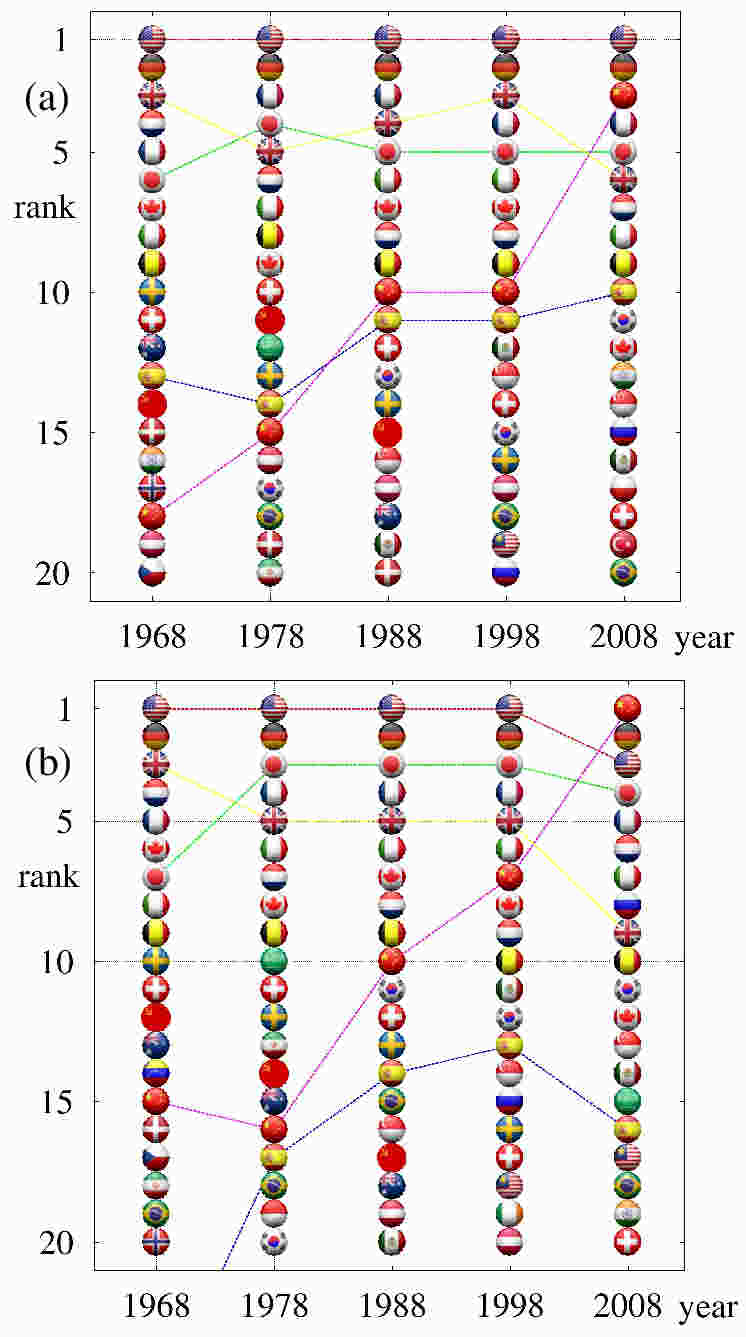}
\end{center}
\vglue +0.2cm
\caption{(Color online)
Top 20 countries as a function of years
ranked by the total monetary trade volume of the WTN
in  import (a) and export  (b) 
 panels respectively; 
each country is represented by its corresponding flag.
Dashed lines show time evolution of  the same countries
as in Fig.~\ref{fig11_5}.
After \cite{ermann:2013a}.
\label{fig11_6}}
\end{figure}

The nestedness matrices generated by 
the BINMATNEST algorithm \cite{rodriguez:2006} 
are shown in Fig.~\ref{fig11_4} for ecology networks
ARR1 ($N_{pl}=84$, $N_{anim}=101$,
$\varphi=0.043$, $T=2.4$) and 
WES ($N_{pl}=207$, $N_{anim}=110$,
$\varphi=0.049$, $T=3.2$) from \cite{rezende:2007}.
Using the same algorithm we generate the nestedness matrices
of WTN using the mutualistic matrices for import 
$Q^{(i)}$ and export $Q^{(i)}$
for the WTN in years 1968 and 2008 using 
a fixed typical threshold  $\mu=10^{-3}$
(see Fig.~\ref{fig11_4}). 
As for ecological systems, for the WTN data we
also obtain rather small nestedness temperature
($T \approx 6/8$ for import/export in 1968 
and $T\approx 4/8$ in 2008 respectively). These values are by
a factor 9/4 of times smaller than the corresponding $T$
values for import/export
from random generated networks with the corresponding 
values of $\varphi$. 

\begin{figure}[h]
\begin{center}
\includegraphics[width=0.48\textwidth]{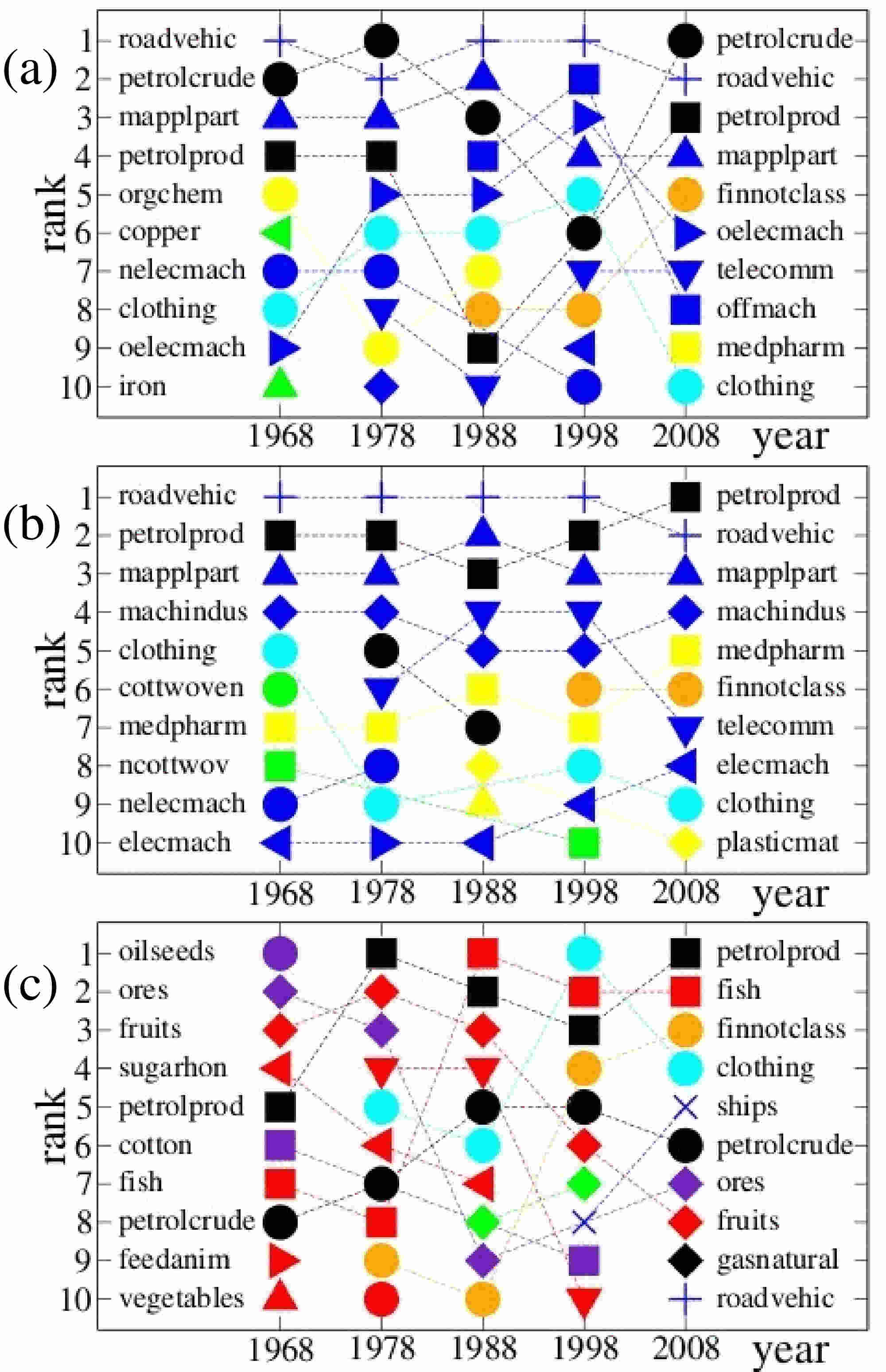}
\end{center}
\caption{(Color online)
Top 10 ranks of trade products as a function of years 
for the WTN. Panel (a): ranking of products by
monetary trade volume.
Panels (b), (c):  
ranking is given by the nestedness algorithm 
for import (b) and export (c)
with the trade threshold $\mu=10^{-3}$.
Each product is shown by its own symbol with
short name written at years 1968, 2008;
symbol color marks  1st SITC digit;
SITC codes of products and their names 
are given in \cite{uncomtrade:2011}
and Table 2 in \cite{ermann:2013a}.
After \cite{ermann:2013a}.
\label{fig11_7}}
\end{figure}

The small value of nestedness temperature obtained for the WTN
confirms the validity of the ecological analysis
of WTN structure: trade products play the role of pollinators
which produce exchange between world countries, which play the role 
of plants. Like in ecology the WTN
evolves to the state with very low
nestedness temperature that satisfies the
ecological concept
of system stability appearing as a result of high
network nestedness \cite{bastolla:2009}.

The nestedness algorithm creates effective ecological ranking  
(EcoloRanking) of all UN countries.
The evolution of 20 top ranks throughout the years is shown 
in Fig.~\ref{fig11_5} for import and export.
This ranking is quite different from the more commonly applied
ranking of countries by their total import/export monetary trade 
volume  \cite{cia:2010} (see corresponding data in Fig.~\ref{fig11_6}) 
or the  democratic ranking of WTN based on 
the Google matrix analysis discussed above.
Indeed, in 2008 China is at the top rank for total export volume
but it is only at 5th position in EcoloRank
(see Fig.~\ref{fig11_5}, Fig.~\ref{fig11_6}).
In a similar way Japan moves down from 4th to 17th position
while USA raises up from 3rd to 1st rank.

The same nestedness
algorithm generates not only the ranking of countries
but also the ranking of trade products for 
import and export which is presented in Fig.~\ref{fig11_7}.
For comparison we also show there the standard ranking of 
products by their trade volume. In  Fig.~\ref{fig11_7}
the color of symbol marks the 1st SITC digit
described in figure, \cite{uncomtrade:2011}
and Table 2 in \cite{ermann:2013a}.

The origin of such a difference between EcoloRanking 
and trade volume ranking of countries
is related to the main
idea of mutualistic ranking in ecological systems:
the nestedness ordering stresses the importance of mutualistic 
pollinators (products for WTN)
which generate links and exchange between plants
(countries for WTN). In this way generic products, 
which participate in the trade between many countries,
become of primary importance even if
their trade volume is not at the top lines
of import or export. In fact such mutualistic products
glue the skeleton of the world trade
while the nestedness concept allows to rank them
in order of their importance. The time evolution
of this EcoloRanking of products of WTN
is shown in Fig.~\ref{fig11_7} for import/export
in comparison with the product ranking by the monetary
trade volume
(since the trade matrix is
diagonal in product index the ranking of 
products in the latter case is the same for
import/export). The top and middle panels
have dominate colors corresponding to
machinery (SITC Rev. 1 code 7; blue) and mineral fuels (3; black)
with a moderate contribution of chemicals (5; yellow)
and manufactured articles (8; cyan) and a small 
fraction of goods classified by material (6; green).
Even if the global structure of product 
ranking by trade  volume has certain similarities
with import EcoloRanking there are also important 
new elements. Indeed, in 2008 the mutualistic significance of
petroleum products (code 332),
{\it machindus} (machines for special industries  code 718) 
and {\it medpharm}  (medical-pharmaceutic products  code 541) 
is much higher
compared to their volume ranking, while 
petroleum crude (code 331) and 
office machines (code 714) have smaller mutualistic significance
compared to their volume ranking.

The new element of EcoloRanking is that it differentiates 
between import and export products while for trade volume
they are ranked in the same way. Indeed, the dominant colors
for export (Fig.~\ref{fig11_7}  bottom panel)
correspond to food  (SITC Rev. 1 code 0; red) with contribution 
of black (present in import) and crude materials (code 2; violet);
followed by cyan (present in import) and more pronounced presence of
{\it finnotclass} (commodities/transactions not classified code 9; brown).
EcoloRanking of export shows a clear decrease tendency
of dominance of SITC codes 0 and 2 with time and increase
of importance of codes 3,7. It is interesting to note that
the  code 332 of petroleum products is vary vulnerable
in volume ranking
due to significant variations of petroleum prices  
but in EcoloRanking this product keeps the stable
top positions in all years showing its 
mutualistic structural importance for the world trade.
EcoloRanking of export shows also importance of
fish (code 031), clothing (code 841) and fruits (code 051)
which are placed on higher positions compared to their volume ranking.
At the same time {\it roadvehic} (code 732),
which are at top volume ranking, have 
relatively low ranking in export since only a few countries
dominate the production of road vehicles.

It is interesting to note that in Fig.~\ref{fig11_7}
petroleum crude is at the top of trade volume ranking e.g. in 2008 (top panel)
but it is absent in import EcoloRanking (middle panel)
and it is only on 6th position in export EcoloRanking (bottom panel).
A similar feature is visible for years 1968, 1978.
On a first glance this looks surprising but in fact for 
mutualistic EcoloRanking it is important that 
a given product is imported from top EcoloRank countries:
this is definitely not the case for petroleum crude
which practically is not produced inside
top 10 import EcoloRank countries (the only exception is USA,
which however also does not export much). Due to that reason
this product has low mutualistic significance.

The mutualistic concept of product importance 
is at the origin of significant difference
of EcoloRanking of countries compared to the usual trade
volume ranking 
(see Fig.~\ref{fig11_5}, Fig.~\ref{fig11_6}).
Indeed, in the latter case China and Japan are at the dominant
positions but their trade is concentrated in
specific products which mutualistic role is 
relatively low. In contrast USA, Germany and France
keep top three EcoloRank positions 
during almost 40 years clearly demonstrating their mutualistic
power and importance for the world trade.

Thus our results show the universal features of ecologic
ranking of complex networks with promising future applications
to trade, finance and other areas.

\subsection{Remarks on world trade and banking networks}
\label{s11.5}

The new approach to the world trade, based on the Google matrix analysis,
gives a democratic type of ranking
being independent of the trade amount of a given country.
In this way rich and poor countries are treated on equal democratic
grounds.   In a certain sense PageRank probability
for a given country  is proportional to its
rescaled import flows while CheiRank is proportional
to its rescaled export flows inside of the WTN.

The global characteristics of the world trade 
are analyzed on the basis of this new type of ranking.
Even if all countries are treated now on equal democratic grounds 
still we find at the top rank the group of 
industrially developed countries
approximately corresponding to {\it G-20} and
recover 74\% of countries listed in {\it G-20}.
The Google matrix analysis demonstrates
an existence of two solid state
domains of rich and poor countries which remain
stable during the years of consideration. 
Other countries correspond to a gas phase with 
ranking strongly fluctuating in time. We propose a simple
random matrix model which well describes the statistical properties
of rank distribution for the WTN \cite{ermann:2011b}.

The comparison between usual ImportRank--Export\-Rank 
(see e.g. \cite{cia:2010})
and our PageRank--CheiRank approach shows that the later
highlights the trade flows in a new useful manner which
is complementary to the usual analysis. 
The important difference between these two
approaches is due to the fact that  ImportRank--ExportRank
method takes into account only global amount of money
exchange between a country and the rest of the world while
Page\-Rank--CheiRank approach takes into account all links and money
flows between all countries.

The future developments should consider a
matrix with all countries and all products
which size becomes significantly larger
($N \sim 220 \times 10^4 \sim 2 \times 10^6$)
comparing to a modest size $N \approx 227$
considered here. However, some new problems
of this multiplex network analysis should be
resolved combining a democracy in countries with
volume importance of products which
role is not democratic. It is quite possible
that such an improved analysis will 
generate an asymmetric ranking of products
in contrast to their symmetric ranking by volume
in export and import. The ecological ranking 
of the WTN discussed in the previous SubSec.
indicates preferences and asymmetry
of trade in multiple products \cite{ermann:2013a}.
The first steps in the Google matrix 
analysis of multiproduct world trade network,
with $61$ products and up to $227$ countries,
have been done recently by  \cite{ermann:2015}
confirming this asymmetry. It is established there that
such multifunctional networks 
can be analyzed by the Google matrix 
approach, using 
a certain personalized vector, so that the world countries 
are treated on  democratic equal grounds
while the contribution of products 
remains proportional to their trade volume.
Such a multiproduct world trade network 
allows to investigate the sensitivity of trade to price
variation of various products. This approach can be also
applied to the world network of economic activities
obtained from the OECD-WTO database \cite{kandiah:2015}.
It allows to determine the sensitives of
economic balance of world countries in respect to
labor cost variations in certain selected countries.
In difference from the multiproduct WTN of UN COMTRADE,
where there are no direct transitions between products,
the   OECD-WTO database contains interactions between
various activity sectors of various countries
that opens new possibilities for a more advanced
analysis.

It is also important to note that 
usually in economy researchers analyze
time evolution of various indexes studying their correlations.
The results presented above for the WTN show
that in addition to time evolution there is also
evolution in space of the network.
Like for waves in an ocean time and space
are both important and we think that
time and space study of trade
captures important geographical factors
which will play a dominant role
for analysis of contamination propagation
over the WTN in case of crisis.
We think that the WTN data capture 
many essential elements
which will play a rather similar role
for financial flows in the interbank payment networks.
We expect that the analysis of financial flows
between bank units would prevent important financial
crisis shaking the world in last years.
Unfortunately, in contrast to
WWW and UN COMTRADE,
the banks keep hidden their financial
flows. Due to this 
secrecy of banks the society
is still suffering from financial crises.
And all this for a network of very small size
estimated on a level of 50 thousands bank units for the whole world
being by a factor million smaller than the 
present size of WWW
(e.g. Fedwire interbank payment network
of USA contains only 6600 nodes \cite{soramaki:2007}).
In a drastic contrast with bank networks the WWW
provided a public access to its nodes
changing the world on a scale of 20 years. 
A creation of the World Bank Web (WBW)
with information accessible for authorized investigators
would allow to understand and control
financial flows in an efficient manner
preventing the society from bank crises. 
We note that the methods of network
analysis and ranking start to attract interest
of researchers in various banks
(see e.g. \cite{craig:2010,garratt:2011}).

\section{Networks with nilpotent adjacency  matrix}
\label{s12}

\subsection{General properties}
\label{s12.0}

In certain networks \cite{frahm:2012a,frahm:2014b} it is possible 
to identify an ordering scheme for the nodes such that the adjacency 
matrix has non-vanishing elements $A_{mn}$ only for nodes $m<n$ providing 
a triangular matrix structure. In these cases it is possible to provide 
a semi-analytical theory \cite{frahm:2012a,frahm:2014b} 
which allows to simplify the numerical calculation 
of the non-vanishing eigenvalues of the matrix $S$ introduced in Sec. 
\ref{s3.1}. It is useful to write this matrix in the form
\begin{equation}
\label{eq_matrixS}
S=S_0+(1/N)\,e\,d^{\,T}
\end{equation}
where the vector $e$ has unit entries for all nodes 
and the {\em dangling vector} $d$ has unit entries for dangling nodes 
and zero entries for the other nodes. The extra contribution 
$e\,d^{\,T}/N$ just replaces the empty columns (of $S_0$) with 
$1/N$ entries at each element. For a triangular network structure the 
matrix $S_0$ is nilpotent, i.e. $S_0^l=0$ for some integer $l>0$ and 
$S_0^{l-1}\neq 0$. Furthermore for the network examples  studied previously 
\cite{frahm:2012a,frahm:2014b} we have $l\ll N$ which has important 
consequences for the eigenvalue spectrum of $S$. 

There are two groups of (right) 
eigenvectors $\psi$ of $S$ with eigenvalue $\lambda$. For the first group 
the quantity $C=d^{\,T}\,\psi$ vanishes and $\psi$ is also an eigenvector of 
$S_0$ and if $S_0$ is nilpotent we have $\lambda=0$ (there are also 
many higher order generalized eigenvectors associated to $\lambda=0$). 
For the second 
group we have $C\neq 0$, $\lambda\neq 0$ 
and the eigenvector is given by 
$\psi=(\lambda\openone-S_0)^{-1}\,C\,e/N$. Expanding the matrix inverse 
in a finite geometric series (for nilpotent $S_0$) and applying the 
condition $C=d^{\,T}\,\psi$ on this expression one finds that the 
eigenvalue must be a zero of the {\em reduced polynomial} of degree $l$:
\begin{equation}
\label{eq_polyred}
{\cal P}_r(\lambda)=\lambda^l-\sum_{j=0}^{l-1}\lambda^{l-1-j}\,c_j=0
\; ,\quad c_j=d^{\,T}\,S_0^j\,e/N \;.
\end{equation}
This shows that there are at most $l$ non-vanishing eigenvalues of 
$S$ with eigenvectors $\psi\propto \sum_{j=0}^{l-1} \lambda^{-j-1}\,v^{(j)}$ 
where $v^{(j)}=S_0^{j}\,e/N$ for $j=0,\,\ldots,\,l-1$. 
Actually, 
the vectors $v^{(j)}$ generate an $S$-invariant $l$-dimensional subspace and 
from $S\,v^{(j)}=c_j\,v^{(0)}+v^{(j+1)}$ (using the 
identification $v^{(l)}=0$) 
one obtains directly the $l\times l$ representation matrix $\bar S$ of 
$S$ with respect to $v^{(j)}$ \cite{frahm:2012a}. Furthermore, the 
characteristic polynomial of $\bar S$ is indeed given by the reduced 
polynomial (\ref{eq_polyred}) and 
the sum rule $\sum_{j=0}^{l-1} c_j=1$ ensures that $\lambda=1$ 
is indeed a zero of ${\cal P}_r(\lambda)$ \cite{frahm:2012a}.
The corresponding eigenvector (PageRank $P$ at $\alpha=1$) is given 
by $P\propto \sum_{j=0}^{l-1} v^{(j)}$. 
The remaining $N-l$ (generalized) eigenvectors of $S$ are associated 
to many different Jordan blocks of $S_0$ for the eigenvalue $\lambda=0$. 

These $l$ non-vanishing complex 
eigenvalues can be numerically computed 
as the zeros of the reduced polynomial 
by the Newton-Maehly method, by a numerical diagonalization of the ``small'' 
representation matrix $\bar S$ (or better a more stable transformed matrix 
with identical eigenvalues) or by the Arnoldi method using the uniform vector 
$e$ as initial vector. In the latter case the Arnoldi method should 
theoretically (in absence of rounding errors) 
exactly explore the $l$-dimensional subspace of the vectors 
$v^{(j)}$ and break off after $l$ iterations with $l$ exact eigenvalues. 

However, numerical rounding errors may have a strong effect due 
to the Jordan blocks for the zero eigenvalue \cite{frahm:2012a}.
Indeed, an error $\epsilon$ appearing in a left bottom corner
of a Jordan matrix of size $D$ with zero eigenvalue
leads to numerically induced  eigenvalues on a complex circle
of radius 
\begin{equation}
\label{eqjordan}
|\lambda_\epsilon| = \epsilon^{1/D} \;.
\end{equation}
Such an error can become
significant with $|\lambda| > 0.1$ even
for $\epsilon \sim 10^{-15}$ as soon as $D > 15$.
We call this phenomenon the Jordan error enhancement.
Furthermore, also the numerical determination of 
the zeros of ${\cal P}_r(\lambda)$ 
for large values of $l\sim 10^2$ can be numerically rather difficult.
Thus, it may be necessary to use a high precision 
library such as the GNU GMP 
library either for the determination of the zeros of 
${\cal P}_r(\lambda)$ 
or for the Arnoldi method \cite{frahm:2014b}.

\subsection{PageRank of integers}
\label{s12.1}

A network for integer numbers \cite{frahm:2012a} can be constructed 
by linking an integer number $n\in\{1,\,\ldots,\,N\}$ 
to its divisors $m$ different from $1$ and
$n$ itself by an adjacency matrix $A_{mn}=M(n,m)$ where 
the multiplicity $M(n,m)$ is the number of times 
we can divide $n$ by $m$, i.e. the largest integer such that 
$m^{M(n,m)}$ is a divisor of $n$, and $A_{mn}=0$ for all other cases. 
The number $1$ and the prime numbers are 
not linked to any other number and correspond to dangling nodes. 
The total size $N$ of the matrix is fixed by the maximal considered integer.
According to numerical data 
the number of links $N_\ell=\sum_{mn} A_{mn}$ is approximately given by 
$N_\ell=N\,(a_\ell+b_\ell\ln N$) with $a_\ell= -0.901 \pm 0.018$,
$b_\ell= 1.003 \pm 0.001$.

\begin{figure}[H]
\begin{center}
\includegraphics[width=0.48\textwidth]{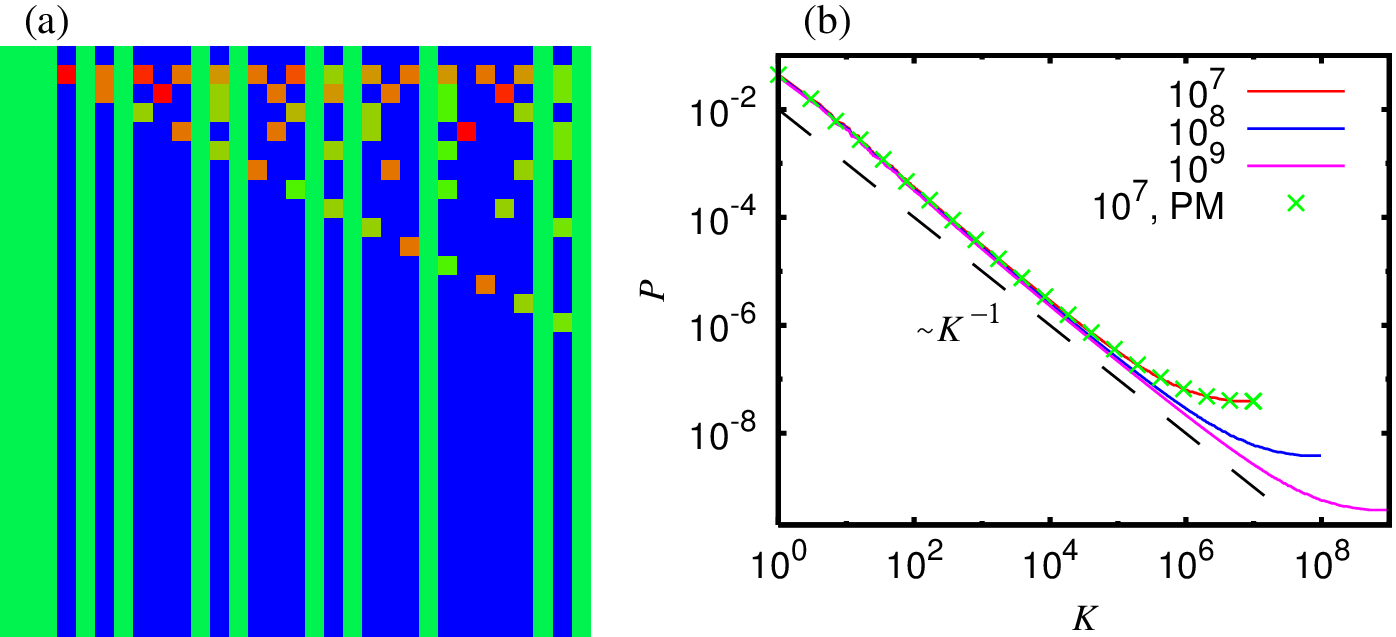}
\end{center}
\caption{(Color online)
Panel (a): the Google matrix of integers, 
the amplitudes of  matrix elements
$S_{mn}$ are shown by color with blue/black for minimal zero elements
and red/gray for maximal unity elements, with
$1 \leq n \leq 31$ corresponding to $x-$axis (with $n=1$ corresponding 
to the left column) and $1 \leq m \leq 31$ for $y-$axis 
(with $m=1$ corresponding to the upper row). 
Panel (b): the full lines correspond to the 
dependence of PageRank probability $P(K)$ 
on index $K$ for the matrix sizes 
$N=10^7$, $10^8$, $10^9$ with the PageRank evaluated by the exact 
expression $P\propto \sum_{j=0}^{l-1} v^{(j)}$. 
The green/gray crosses correspond to the PageRank obtained 
by the power method for $N=10^7$;
the dashed straight line shows the Zipf law dependence $P \sim 1/K$. 
After \cite{frahm:2012a}.
\label{fig12_1}}
\end{figure}

The matrix elements $A_{mn}$ are different from zero only 
for $n\ge 2m$ and the associated matrix $S_0$ is therefore 
nilpotent with $S_0^l=0$ and 
$l=[\log_2(N)]\ll N$. This triangular matrix structure can be seen in 
Fig.~\ref{fig12_1}(a) which shows the amplitudes of $S$. The vertical 
green/gray lines correspond to the extra contribution due to the dangling nodes. 
These $l$ non-vanishing eigenvalues of $S$ can be efficiently 
calculated as the zeros of the reduced polynomial (\ref{eq_polyred})
up to $N=10^9$ with $l=29$.
For $N=10^9$ the largest eigenvalues are $\lambda_1=1$, 
$\lambda_{2,3}\approx -0.27178\pm i\,0.42736$, $\lambda_4\approx -0.17734$ 
and $|\lambda_j|<0.1$ for $j\ge 5$. The dependence of the eigenvalues on 
$N$ seems to scale with the parameter $1/\ln(N)$ for $N\to\infty$ and 
in particular $\gamma_2(N)=- 2 \ln |\lambda_2(N)| \approx 1.020 + 7.14/\ln N$ 
\cite{frahm:2012a}. 
Therefore the first eigenvalue 
is clearly separated from the second eigenvalue and one can chose the 
damping factor $\alpha=1$ without any problems to define a unique PageRank.

\begin{figure}[H]
\begin{center}
\includegraphics[width=0.48\textwidth]{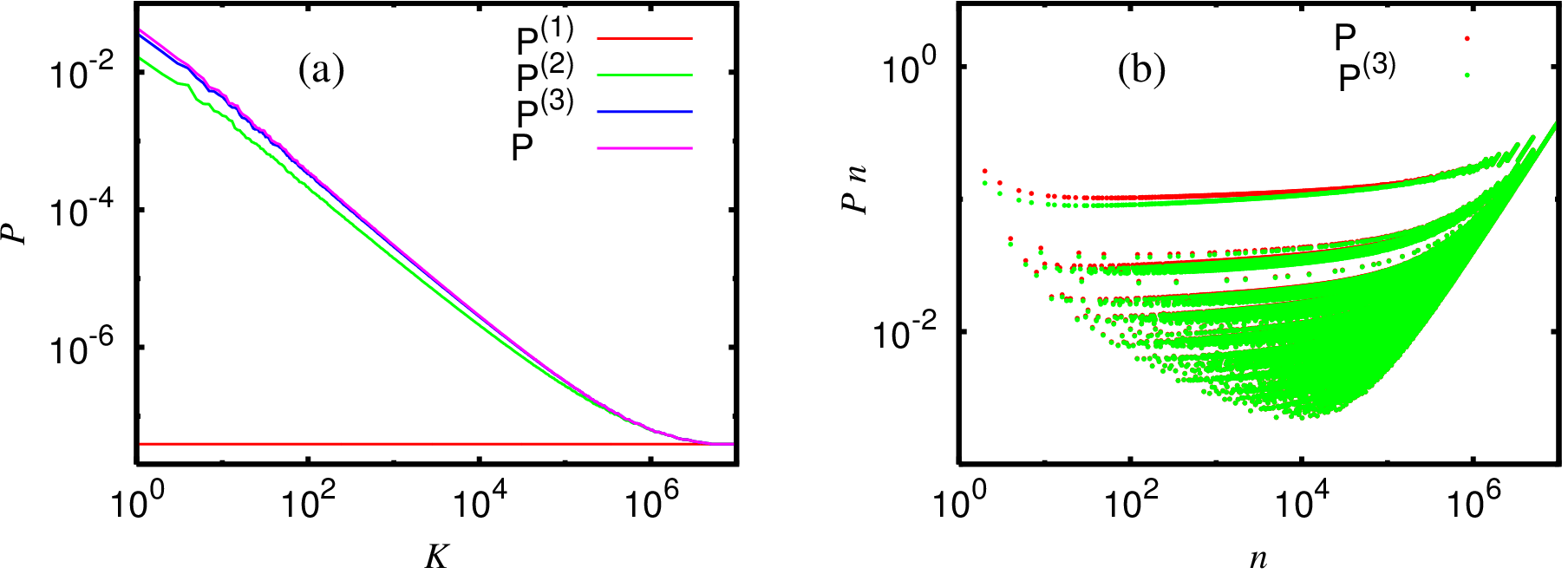}
\end{center}
\caption{(Color online)
Panel (a): comparison of the first three PageRank approximations 
$P^{(i)}\propto \sum_{j=0}^{i-1} v^{(j)}$ for $i=1,\,2,\,3$
and the exact PageRank dependence $P(K)$. 
Panel (b): comparison of the dependence of the rescaled probabilities 
$nP$ and $nP^{(3)}$ on $n$.
Both panels correspond to the case $N=10^7$. 
After \cite{frahm:2012a}.
\label{fig12_2}}
\end{figure} 

The large values of $N$ are possible because the vector iteration 
$v^{(j+1)}=S_0\,v^{(j)}$ can actually be computed without storing the 
$N_\ell \sim N \ln N$ non-vanishing elements of $S_0$ by using the relation:
\begin{equation}
\label{eq_efficient_iter}
v^{(j+1)}_n=\sum_{m=2}^{[N/n]}\,\frac{M(mn,m)}{Q(mn)}\,v^{(j)}_{mn}
\; , \quad{\rm if}\ n\ge 2
\end{equation}
and $v^{(j+1)}_1=0$ \cite{frahm:2012a}. The initial vector 
is given by $v^{(0)}=e/N$ and $Q(n)=\sum_{m=2}^{n-1}M(n,m)$ 
is the number of divisors of $n$ (taking into account the multiplicity). 
The multiplicity $M(mn,n)$ can be 
recalculated during each iteration and one needs only to store 
$N (\ll N_\ell)$ integer numbers $Q(n)$. It is also possible to 
reformulate (\ref{eq_efficient_iter}) in a different way without 
using $M(mn,n)$ \cite{frahm:2012a}. The vectors $v^{(j)}$ allow to 
compute the coefficients $c_j=d^{\,T} v^{(j)}$ in the reduced polynomial 
and the Page\-Rank $P\propto \sum_{j=0}^{l-1} v^{(j)}$.
Fig.~\ref{fig12_1}(b) shows the PageRank for $N\in\{10^7,\,10^8,\,10^9\}$ 
obtained in this way 
and for comparison also the result of the power method for $N=10^7$. 

Actually Fig.~\ref{fig12_2} shows that in the sum 
$P\propto \sum_{j=0}^{l-1} v^{(j)}$ already the first three terms 
give a quite satisfactory approximation to the PageRank allowing 
a further analytical simplified evaluation \cite{frahm:2012a} with 
the result $P(n)\approx C_N/(b_n\,n)$ for $n\ll N$, where $C_N$ is the 
normalization constant and $b_n=2$ for prime numbers $n$ and 
$b_n=6-\delta_{p_1,p_2}$ for numbers $n=p_1\,p_2$ being a product of two 
prime numbers $p_1$ and $p_2$. The behavior 
$P(n)\,n\approx C_N/b_n$, which  takes approximately 
constant values on several branches, 
is also visible in Fig.~\ref{fig12_2} with 
$C_N/b_n$ decreasing 
if $n$ is a product of many prime numbers. 
The numerical results up to $N=10^9$ show that the numbers $n$, 
corresponding to the leading PageRank values for $K=1,\,2,\,\ldots,\,32$, are
$n=2$, $3$, $5$, $7$, $4$, $11$, $13$, $17$, $6$, $19$, $9$, $23$, $29$, $8$, 
$31$, $10$, $37$, $41$, $43$, $14$, $47$, $15$, $53$, $59$, 
$61$, $25$, $67$, $12$, $71$, $73$, $22$, $21$ with about 30\%
of non-primes among these values \cite{frahm:2012a}. 

A simplified model for the network for integer numbers with $M(n,m)=1$ 
if m is divisor of n and $1<m<n$ has also been studied with similar 
results  \cite{frahm:2012a}.

\subsection{Citation network of Physical Review}
\label{s12.2}

Citation networks for Physical Review and other scientific journals 
can be defined by taking published articles as nodes and linking an article 
A to another article B if A cites B. PageRank and similar analysis 
of such networks are efficient to determine influential articles 
\cite{redner:1998,newman:2001,redner:2005,radicchi:2009}.

\begin{figure}[H]
\begin{center}
\includegraphics[width=0.48\textwidth]{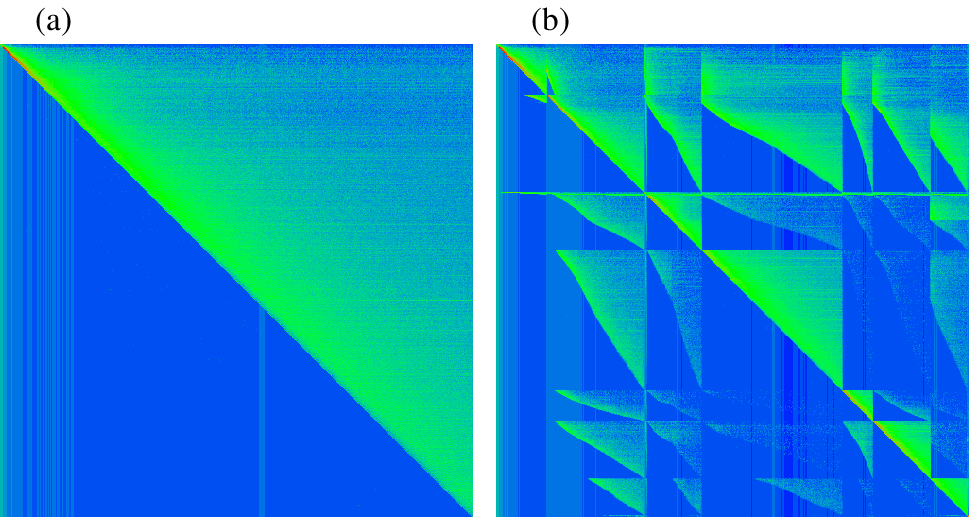}
\end{center}
\caption{(Color online)
Different representations of the Google matrix structure 
for the Physical Review network until 2009. 
(a) Density of matrix elements 
$G_{tt'}$ in the basis of the publication time index $t$ (and $t'$). 
(b) Density of matrix elements in the basis 
of journal ordering according to: Phys.~Rev.~Series~I, Phys.~Rev., 
Phys.~Rev.~Lett., Rev.~Mod.~Phys., Phys.~Rev.~A, B, C, D, E, Phys.~Rev.~STAB 
and Phys.~Rev.~STPER. and with time index ordering inside each journal.
Note that the journals Phys.~Rev.~Series~I, Phys.~Rev.~STAB and 
Phys.~Rev.~STPER are not clearly visible due to a small number of published 
papers. Also Rev.~Mod.~Phys. appears only as a thick line with 2-3 pixels 
(out of 500) due to a limited number of published papers. The different 
blocks with triangular structure correspond to 
clearly visible seven journals with considerable numbers of published papers. 
Both panels show the coarse-grained density of matrix elements on 
$500 \times 500$ square cells for the entire network.
Color shows the density of matrix elements (of $G$ at $\alpha=1$)
changing from blue/black for minimum zero value 
to red/gray at maximum value. 
After \cite{frahm:2014b}.
\label{fig12_3}}
\end{figure}

In citation network links go mostly 
from newer to older articles and therefore 
such networks have, apart from the dangling node contributions, typically 
also a (nearly) triangular structure as can be seen in Fig.~\ref{fig12_3} 
which shows a coarse-grained density of the corresponding Google matrix 
for the citation network of Physical Review 
from the very beginning until 2009 \cite{frahm:2014b}. 
However, due to the delay of the publication process 
in certain rare instances a published paper may cite another paper 
that is actually published a little later and sometimes two papers may even 
cite mutually each other. Therefore the matrix structure is not exactly 
triangular but in the coarse-grained density in Fig.~\ref{fig12_3} the 
rare ``future citations'' are not well visible. 

The nearly triangular matrix structure implies 
large dimensional Jordan blocks 
associated to the eigenvalue $\lambda=0$. This creates
the Jordan error enhancement  (\ref{eqjordan}) 
with severe numerical 
problems for accurate computation of eigenvalues in the range 
$|\lambda|<0.3-0.4$ when using the Arnoldi method with standard 
double-precision arithmetic \cite{frahm:2014b}. 

One can eliminate the small number of future citations ($12126$ which 
is $0.26$ \% of the total number of links $N_\ell=4691015$) and determine 
the complex eigenvalue spectrum of a triangular reduced citation network using 
the semi-analytical theory presented in previous subsection. 
It turns out that in this 
case the matrix $S_0$ is nilpotent $S_0^l=0$ with $l=352$ which is 
much smaller than the total network size $N=463348$. The 352 non-vanishing 
eigenvalues can be determined numerically as the zeros of the 
polynomial (\ref{eq_polyred}) but due to an alternate sign problem with a 
strong loss of significance it is necessary to use the 
high precision library GMP with 256 binary digits \cite{frahm:2014b}. 

The semi-analytical theory can also be generalized to the case of 
{\em nearly} triangular networks, i.e. the full citation network
including the future citations. 
In this case the matrix $S_0$ is no longer nilpotent 
but one can still generalize the arguments of previous subsection 
and discuss the two cases where the quantity $C=d^{\,T}\,\psi$ either vanishes 
(eigenvectors of first group) or is different from zero (eigenvectors 
of second group). The eigenvalues $\lambda$ for the first group, which 
may now be different from zero, can be determined by a quite complicated 
but numerically very efficient procedure using the subspace eigenvalues of 
$S$ and degenerate subspace eigenvalues of $S_0$ (due to absence of dangling 
node contributions the matrix $S_0$ produces much larger invariant subspaces 
than $S$) \cite{frahm:2014b}. The eigenvalues of the second group are 
given as the complex zeros of the rational function:
\begin{equation}
\label{eq_rationalfunction}
{\cal R}(\lambda)=1-d^{\,T}
\frac{\openone}{\lambda\openone-S_0}e/N=
1-\sum_{j=0}^\infty c_j \lambda^{-1-j}
\end{equation}
with $c_j$ given as in (\ref{eq_polyred}) and now the series is not finite 
since $S_0$ is not nilpotent. For the citation network of Physical Review 
the coefficients $c_j$ behave as $c_j\propto \rho_1^j$ where 
$\rho_1\approx 0.902$ is the largest eigenvalue of the 
matrix $S_0$ with an eigenvector non-orthogonal to $d$. Therefore the 
series in (\ref{eq_rationalfunction}) converges well for $|\lambda|>\rho_1$ 
but in order to determine the spectrum the rational function 
${\cal R}(\lambda)$ needs to be evaluated for smaller values of $|\lambda|$. 
This problem can be solved by interpolating ${\cal R}(\lambda)$ 
with (another) rational function using a certain number of support points 
on the complex unit circle, where (\ref{eq_rationalfunction}) converges very 
well, and determining the complex zeros, well inside the unit circle, 
of the numerator polynomial 
using again the high precision library GMP \cite{frahm:2014b}. In this 
way using 16384 binary digits one may obtain 2500 reliable eigenvalues of
the second group.

\begin{figure}[H]
\begin{center}
\includegraphics[width=0.48\textwidth]{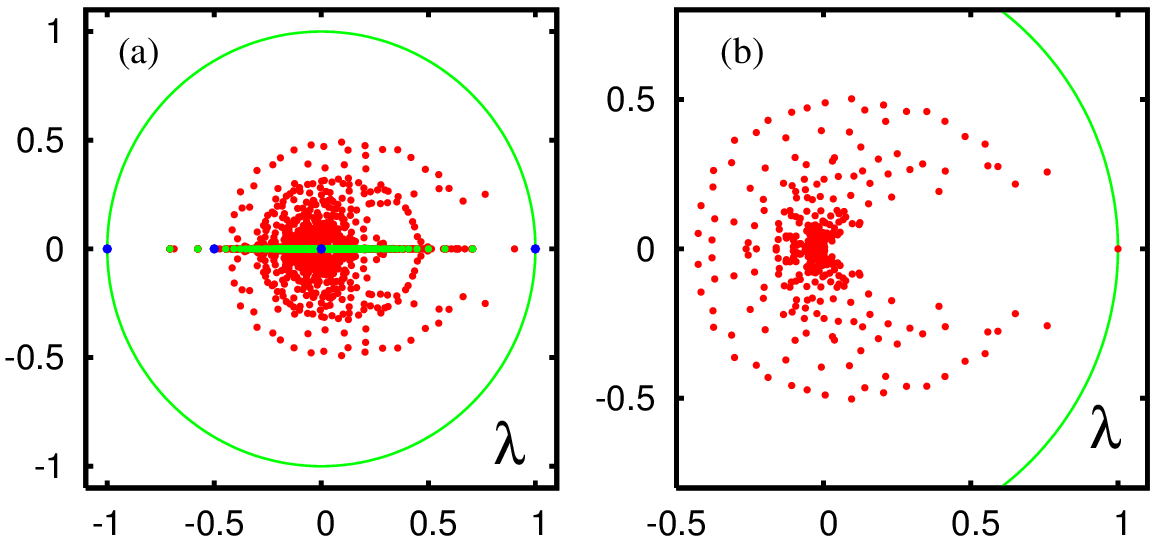}
\end{center}
\caption{(Color online)
(a) Most accurate spectrum of eigenvalues for the full 
Physical Review network;
red/gray dots represent the core space eigenvalues obtained by the rational 
interpolation method with the numerical precision of $p=16384$ binary 
digits, $n_R=2500$ eigenvalues; green (light gray) dots 
show the degenerate subspace 
eigenvalues of the matrix $S_0$ which are also eigenvalues of $S$ with a 
degeneracy reduced by one (eigenvalues of the first group);
blue/black  dots show the direct subspace eigenvalues of $S$. 
(b) Spectrum of numerically accurate 352 non-vanishing 
eigenvalues of the Google matrix for the triangular reduced 
Physical Review network determined 
by the Newton-Maehly method applied to the reduced polynomial 
(\ref{eq_polyred}) with a high-precision calculation of 
256 binary digits; note the absence of subspace eigenvalues for this case. 
In both panels the green/gray curve represents the unit circle. 
After \cite{frahm:2014b}.
\label{fig12_4}}
\end{figure}  

The numerical high precision spectra obtained by the semi-analytic methods 
for both cases, triangular reduced and 
full citation network, are shown in Fig.~\ref{fig12_4}. One may mention 
that it is also possible to implement the Arnoldi method using 
the high precision library GMP for both cases and the resulting eigenvalues 
coincide very accurately with the semi-analytic spectra for 
both cases \cite{frahm:2014b}. 

\begin{figure}[H]
\begin{center}
\includegraphics[width=0.48\textwidth]{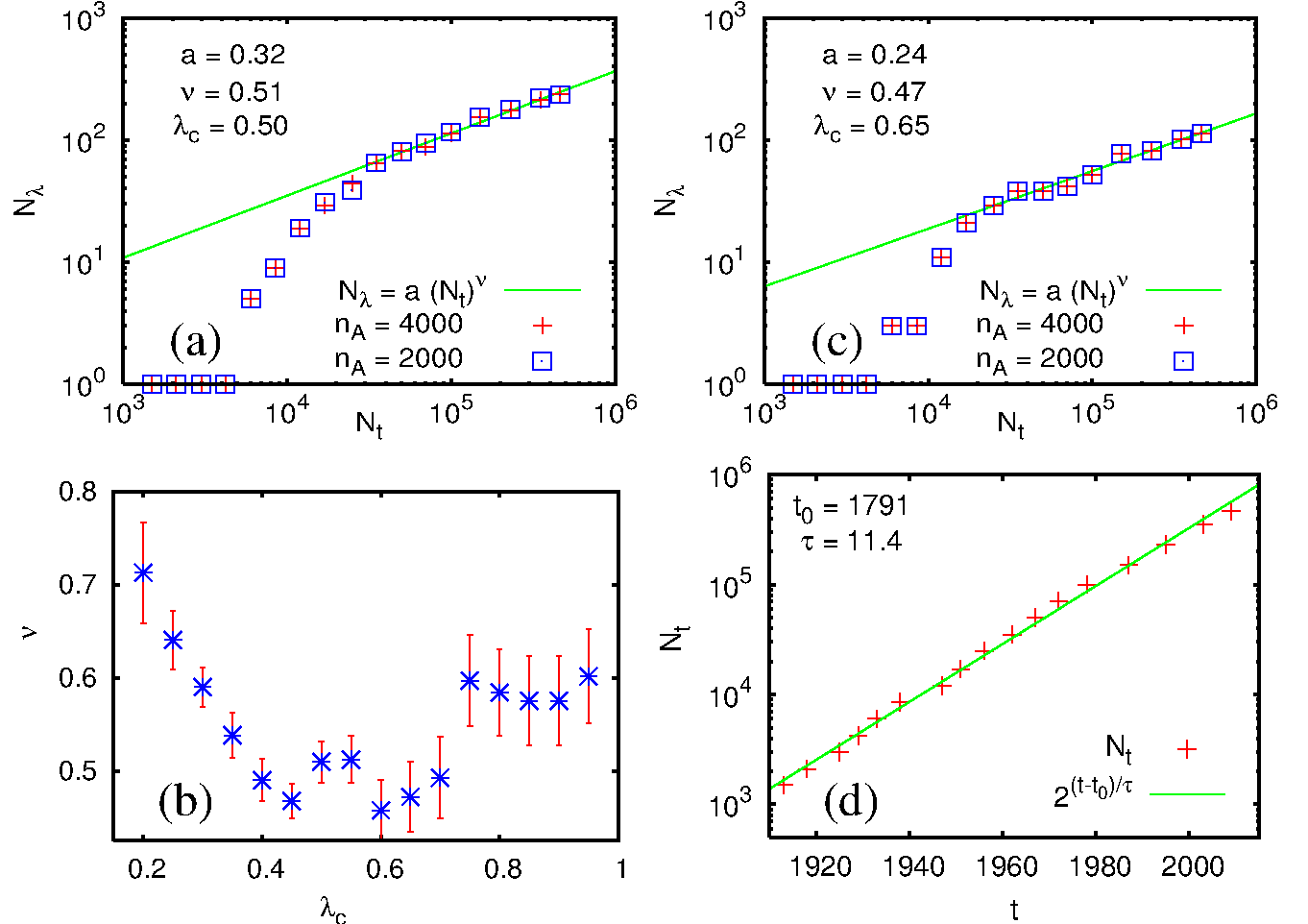}
\end{center}
\caption{(Color online)
Data for 
the whole CNPR at different moments of time.
 Panel (a) (or (c)): 
shows the number $N_\lambda$ of eigenvalues 
with $\lambda_c\leq \lambda\leq 1$
for $\lambda_c=0.50$ (or $\lambda_c=0.65$) versus the effective 
network size $N_t$ where the nodes with publication times after a 
cut time $t$ are removed from the network. 
The green/gray line shows the fractal Weyl law $N_\lambda=a\,(N_t)^\nu$ 
with parameters $a=0.32\pm 0.08$ ($a=0.24\pm 0.11$)
and $\nu=0.51\pm 0.02$ ($b=0.47\pm 0.04$) obtained from a fit 
in the range $3\times 10^4\le N_t < 5 \times 10^5$. 
The number $N_\lambda$ includes both 
exactly determined invariant subspace eigenvalues 
and core space eigenvalues obtained from the 
Arnoldi method with double-precision (52 binary 
digits) for $n_A=4000$ (red/gray crosses) and $n_A=2000$ (blue/black squares). 
Panel (b):  exponent $b$ with error bars 
obtained from the fit $N_\lambda=a\,(N_t)^\nu$ in the 
range $3 \times 10^4 \le N_t < 5 \times 10^5$ versus cut value $\lambda_c$. 
Panel (d): effective network size $N_t$ versus cut time $t$ (in years). 
The green/gray line shows the exponential fit $2^{(t-t_0)/\tau}$ with 
$t_0=1791\pm 3$ and $\tau=11.4\pm 0.2$ representing the number of years 
after which the size of the network (number of papers published 
in all Physical Review journals) is effectively doubled. 
After \cite{frahm:2014b}.
\label{fig12_5}}
\end{figure}

When the spectrum of $G$ is determined with a good accuracy
we can test the validity of the fractal Weyl law (\ref{eq5_1}) 
changing the matrix size 
$N_t$ by considering articles published from the beginning
to a certain time moment $t$ measured in years. 
The data presented in Fig.~\ref{fig12_5}
show that the network size grows 
approximately exponentially as $N_t = 2^{(t-t_0)/\tau}$
with the fit parameters $t_0=1791$, $\tau=11.4$. 
The time interval considered in Fig.~\ref{fig12_5} is 
$1913 \leq t \leq 2009$ since the first data point corresponds 
to $t=1913$ with $N_t=1500$ papers published between 1893 and 1913. 
The results, for the number $N_\lambda$ of eigenvalues
with $|\lambda_i|>\lambda$, show that its growth
is well described by the relation $N_\lambda=a\,(N_t)^\nu$ 
for the range when the number of articles becomes
sufficiently large $3\times 10^4\le N_t < 5 \times 10^5$.
This range is not very large and probably due to that
there is a certain dependence of the exponent $\nu$ on the range 
parameter $\lambda_c$. At the same time
we note that the maximal matrix size $N$
studied here is probably the largest one used in numerical
studies of the fractal Weyl law.
We have $0.47 < \nu <0.6$ 
for all $\lambda_c\ge 0.4$ that is definitely smaller than unity
and thus the fractal Weyl law
is well applicable to the Phys. Rev. network. 
The value of $\nu$ increases up to $0.7$ 
for the data points with $\lambda_c<0.4$ but this is due to 
the fact here $N_\lambda$ also includes some numerically incorrect 
eigenvalues related to the numerical instability of the Arnoldi method 
at standard double-precision (52 binary digits) as discussed 
above. 

We conclude that the most appropriate choice for the 
description of the data is obtained at $\lambda_c=0.4$
which from one side excludes small, partly numerically incorrect, values of 
$\lambda$ and on the other side gives sufficiently large
values of $N_\lambda$. Here we have $\nu=0.49 \pm 02$ corresponding to the
fractal dimension $d=0.98\pm 0.04$. Furthermore, for 
$0.4\le \lambda_c\le 0.7$ we have a rather constant value $\nu \approx 0.5$ 
with $d_f\approx 1.0$. Of course, it would be
interesting to extend this analysis to a larger
size $N$ of citation networks of various type and not only for Phys. Rev.
We expect that the fractal Weyl law is a generic feature of citation networks.

Further studies of the citation network of Physical Review concern 
the properties of eigenvectors (different from the PageRank) associated 
to relatively large complex eigenvalues, the fractal Weyl law, 
the correlations between PageRank and CheiRank (see also 
subsection \ref{s4.3}) and the notion of ``ImpactRank'' \cite{frahm:2014b}. 
To define the ImpactRank one may ask the question how a paper influences 
or has been influenced by other papers. For this one considers an 
initial vector $v_0$, localized on a one node/paper. Then 
the modified Google matrix $\tilde G=\gamma\,G+(1-\gamma)\,v_0\,e^T$ 
(with a damping factor $\gamma \sim 0.5 - 0.9$) produces a ``PageRank'' 
$v_f$ by the propagator $v_f=(1-\gamma)/(1-\gamma G)\,v_0$. In the 
vector $v_f$ the leading nodes/papers have strongly influenced the initial 
paper represented in $v_0$. Doing the same for $G^*$ one obtains 
a vector $v_f^*$ where the leading papers have been influenced by the initial 
paper represented in $v_0$. This procedure has been applied 
to certain historically important papers \cite{frahm:2014b}. 

In summary, the results of this section
show that the phenomenon of the Jordan error enhancement 
(\ref{eqjordan}), induced by finite accuracy of computations
with a finite number of digits, can be resolved 
by advanced numerical methods described above.
Thus the accurate eigenvalues $\lambda$ can be obtained
even for the most difficult case of
quasi-triangular matrices. We note that for other networks
like WWW of UK universities, Wikipedia and Twitter
the triangular structure of $S$ is much less pronounced
(see e.g. Fig.~\ref{fig1_1}) that gives a reduction of Jordan
blocks so that the Arnoldi method with double precision
computes accurate values of $\lambda$.

\section{Random matrix models of Markov chains}
\label{s13}

\subsection{Albert-Barab\'asi model of directed networks}
\label{s13.1}

There are various preferential attachment models generating
complex scale-free networks 
(see e.g. \cite{albert:2002,dorogovtsev:2010}). Such undirected networks
are generated by the Albert-Barab\'asi (AB) procedure \cite{albert:2000}
which builds networks by  an iterative process. Such a procedure
has been generalized to generate directed networks
in \cite{giraud:2009} with the aim to study properties of the Google 
matrix of such networks. The procedure is working as follows:
starting from $m$ nodes, at each step $m$ links are added to the
existing network with probability $p$, or $m$ links are rewired 
with probability $q$,
or a new node with $m$ links is added with probability $1-p-q$.  
In each case the end
node of new links is chosen with preferential attachment, i.e. with probability 
$(k_i+1)/\sum_j(k_j+1)$ where $k_i$ is the total number 
of ingoing and outgoing links
of node $i$.  This mechanism generates directed networks having 
the small-world and scale-free
properties, depending on the values of $p$ and $q$.  The results  are averaged
over $N_r$ random realizations of the network to improve the statistics.

The studies \cite{giraud:2009} are done mainly for 
$m=5$, $p=0.2$ and two values of $q$ corresponding to scale-free
 ($q=0.1$) and exponential ($q=0.7$) regimes of link distributions 
(see Fig.~1 in \cite{albert:2000} for undirected networks).  
For the generated directed networks at $q=0.1$, one finds 
properties close to the behavior for the WWW
with the cumulative distribution of ingoing
links showing algebraic decay $P_c^{\rm \,in}(k) \sim 1/k$
and average connectivity $\langle k \rangle \approx 6.4$. 
For $q=0.7$ one finds $P_c^{\rm \,in}(k) \sim \exp(-0.03k)$ and 
$\langle k \rangle \approx 15$.
For outgoing links, the numerical data are compatible with 
an exponential decay in both cases
with $P_c^{\rm \,out}(k) \sim \exp(-0.6 k)$ for $q=0.1$ and 
$P_c^{\rm \,out}(k) \sim \exp(-0.1 k)$
for $q=0.7$. It is found that small variations of parameters $m, p,q$ near 
the chosen values
do not qualitatively affect the properties of $G$ matrix.  

It is found that the eigenvalues of $G$ for the AB model have one $\lambda=1$
with all other $|\lambda_i| <0.3$ at 
$\alpha=0.85$ (see Fig.~1 in \cite{giraud:2009}).
This distribution shows no significant modification
with the growth of matrix size $2^{10} \leq N \leq 2^{14}$. 
However, the values of
IPR $\xi$ are growing with $N$ for typical values $|\lambda| \sim 0.2$.
This indicates a delocalization of corresponding eigenstates
at large $N$. At the same time the PageRank probability 
is well described by the algebraic 
dependence $P \sim 1/K$ with $\xi$ being practically
independent of $N$.

These results for directed AB model network shows that
it captures certain features of real directed networks,
as e.g. a typical PageRank decay with the exponent $\beta \approx 1$.
However, the spectrum of $G$ in this model is characterized by 
a large gap between $\lambda=1$ and other eigenvalues
which have $\lambda \leq 0.35$ at $\alpha=1$.
This feature is in a drastic difference with spectra of 
such typical networks at WWW of universities,
Wikipedia and Twitter (see Figs.~\ref{fig8_1},\ref{fig9_1},\ref{fig10_2}).
In fact the AB model has no subspaces and no isolated or weakly coupled 
communities. In this network all sites can be reached from a given site
in a logarithmic number of steps that generates a large
gap in the spectrum of Google matrix and a rapid relaxation
to PageRank eigenstate. In real networks there are plenty
of isolated or weakly coupled communities and the introduction
of damping factor $\alpha <1$ is necessary 
to have a single PageRank eigenvalue at $\lambda=1$.
Thus the results obtained in \cite{giraud:2009}
show that the AB model is not able to capture the important
spectral features of real networks.

Additional studies in \cite{giraud:2009} analyzed the model
of a real WWW university network with rewiring procedure
of links,  which consists in 
randomizing the links of the network keeping fixed the 
number of links at any given node.  Starting from a single network, 
this creates an ensemble of randomized networks of same size,
where each node has
the same number of ingoing and outgoing links as for the original network.
The spectrum of such randomly rewired networks is also characterized by a 
large gap in the spectrum of $G$ showing that rewiring 
destroys the communities existing in original networks.
The spectrum and eigenstate properties are studied in the related work 
on various real networks of moderate size $N < 2 \times 10^4$
which have no spectral gap \cite{georgeot:2010}.

\subsection{Random matrix models of directed networks}
\label{s13.2}

Above we saw that the standard models of scale-free networks
are not able to reproduce the typical properties of spectrum of
Google matrices of real large scale networks.
At the same time we believe that it is important to 
find realistic matrix models of WWW and other networks.
Here we discuss certain results for certain random  
matrix models of $G$.

Analytical and numerical studies of random unistochastic 
or orthostochastic matrices 
of size $N=3$ and $4$ lead to triplet and cross structures in the 
complex eigenvalue spectra \cite{zyczkowski:2003} 
(see also Fig.~\ref{fig8bis}). However, the size of such matrices is too small.

\begin{figure}[H]
\begin{center}
\includegraphics[width=0.48\textwidth]{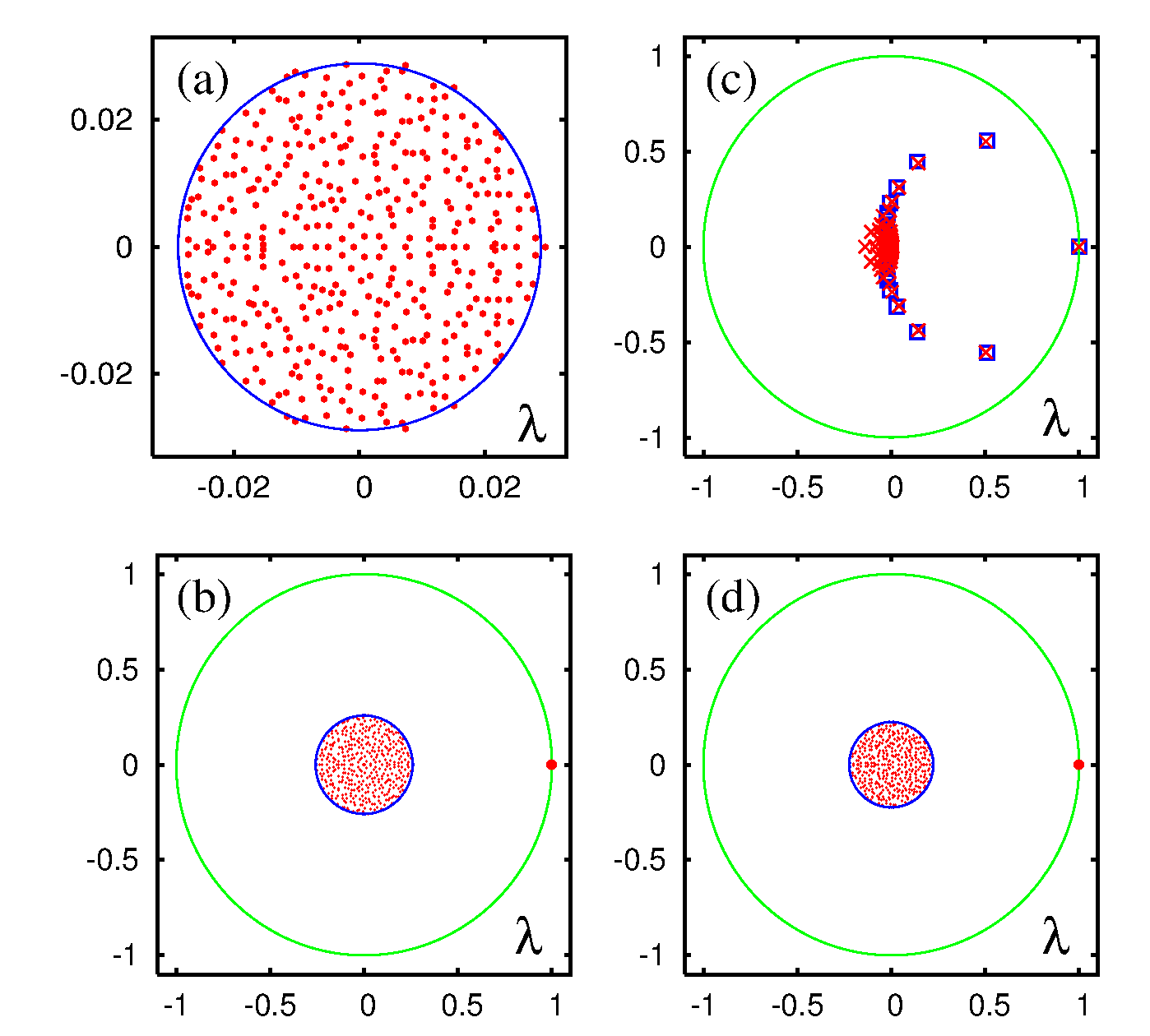}
\end{center}
\vglue -0.4cm
\caption{ (Color online)
Panel $(a)$ shows the  spectrum (red/gray dots) of 
one realization of a full uniform RPFM with 
dimension $N=400$ and matrix elements uniformly distributed in 
the interval $[0,2/N]$; the blue/black circle represents the theoretical 
spectral border with radius $R=1/\sqrt{3N}\approx 0.02887$. 
The unit eigenvalue $\lambda=1$ is not shown due to the zoomed 
presentation range. 
Panel $(c)$ shows the spectrum of 
one realization of triangular RPFM (red/gray crosses) with 
non-vanishing matrix elements 
uniformly distributed in the interval $[0,2/(j-1)]$ 
and a triangular matrix with 
non-vanishing  elements  $1/(j-1)$ (blue/black squares); 
here $j=2,3,\ldots,N$ is the index-number of non-empty columns 
and the first column with $j=1$ corresponds to a dangling node with 
elements $1/N$ for both triangular cases. 
Panels $(b), (d)$ show the complex eigenvalue spectrum (red/gray dots) 
of a sparse RPFM with dimension $N=400$ and $Q=20$ non-vanishing elements 
per column at random positions. Panel $(b)$ (or $(d)$)  corresponds 
to the case of uniformly distributed non-vanishing elements in 
the interval $[0,2/Q]$ (constant non-vanishing elements being $1/Q$);
the blue/black circle represents the theoretical 
spectral border with radius $R=2/\sqrt{3Q}\approx 0.2582$ 
($R=1/\sqrt{Q}\approx 0.2236$). In  panels $(b), (d)$ 
$\lambda=1$ is shown by a larger red dot for better visibility. 
The unit circle is shown by green/gray curve (panels $(b), (c), (d)$).
After \cite{frahm:2014b}.
\label{fig13_1}}
\end{figure}

Here we consider other
examples of random matrix models of Perron-Frobenius operators 
characterized by non-negative matrix elements and column sums normalized 
to unity. We call these models Random Perron-Frobenius Matrices (RPFM).
A number of RPFM, with arbitrary 
size $N$, can be constructed by drawing $N^2$ independent matrix elements 
$0\le G_{ij}\le 1$ from a given distribution $p(G_{ij})$ with finite variance 
$\sigma^2=\langle G_{ij}^2\rangle-\langle G_{ij}\rangle^2$ and normalizing 
the column sums to unity \cite{frahm:2014b}. 
The average matrix $\langle G_{ij}\rangle=1/N$ is just a projector on the 
vector $e$ (with unity entries on each node, see also 
Sec.~\ref{s12.0}) and has the two eigenvalues $\lambda_1=1$ 
(of multiplicity $1$) and $\lambda_2=0$ (of multiplicity $N-1$). Using 
an argument of degenerate perturbation theory on 
$\delta G=G-\langle G\rangle$ 
and known results on the eigenvalue density of non-symmetric 
random matrices \cite{akemann:2011,guhr:1998,mehta:2004} one 
finds that an arbitrary 
realization of $G$ has the leading eigenvalue $\lambda_1=1$ and 
the other eigenvalues are uniformly distributed on the complex unit circle 
of radius $R=\sqrt{N}\sigma$ (see Fig.~\ref{fig13_1}). 

Choosing different distributions $p(G_{ij})$ one obtains different variants 
of the model \cite{frahm:2014b}, for example $R=1/\sqrt{3N}$ 
using a full matrix with uniform $G_{ij}\in [0,\,2/N]$. 
Sparse models with $Q\ll N$ 
non-vanishing elements per column can be modeled by a distribution where 
the probability of $G_{ij}=0$ is $1-Q/N$ and for non-zero $G_{ij}$ 
(either uniform in $[0,\,2/Q]$ or constant $1/Q$) is $Q/N$ leading to 
$R=2/\sqrt{3Q}$ (for uniform non-zero elements) or $R=1/\sqrt{Q}$ 
(for constant non-zero elements). The circular eigenvalue density with these 
values of $R$ is also very well confirmed by numerical simulations 
in Fig.~\ref{fig13_1}. Another case is a power law $p(G)=D/(1+aG)^{-b}$ 
(for $0\le G\le 1$) with $D$ and $a$ to be determined by normalization and 
the average $\langle G_{ij}\rangle =1/N$. For $b>3$ 
this case is similar to a full 
matrix with $R\sim 1/\sqrt{N}$. However for $2<b<3$ one finds 
that $R\sim N^{1-b/2}$. 

The situation changes when one imposes a triangular structure on 
$G$ in which case the complex spectrum of $\langle G\rangle$ is already quite 
complicated and, due to non-degenerate perturbation theory, 
close to the spectrum of $G$ with modest fluctuations, mostly 
for the smallest eigenvalues \cite{frahm:2014b}. 
Following the above discussion about triangular networks (with 
$G_{ij}=0$ for $i\ge j$) we also study numerically a triangular 
RPFM where for $j\ge 2$ and $i<j$ the matrix elements $G_{ij}$ are 
uniformly distributed in 
the interval $[0,2/(j-1)]$ and for $i\ge j$ we have $G_{ij}=0$. 
Then the first column is empty, that means it 
corresponds to a dangling node and it needs to be replaced by $1/N$ entries. 
For the triangular RPFM the situation changes completely since here the 
average matrix $\langle G_{ij}\rangle=1/(j-1)$ (for $i<j$ and $j\ge 2$)
has already a nontrivial structure and eigenvalue spectrum. Therefore the 
argument of degenerate perturbation theory which allowed to apply the results 
of standard full non-symmetric random matrices does not apply here. 
In Fig.~\ref{fig13_1} one clearly sees that for $N=400$ the spectra for 
one realization of a triangular RPFM and its average are very similar 
for the eigenvalues with large modulus but both do not have at all a uniform 
circular density in contrast to the RPRM models without the triangular 
constraint discussed above. 
For the triangular RPFM the PageRank behaves as $P(K)\sim 1/K$ 
with the ranking index $K$ being close to the natural order of nodes 
$\{1,2,3,\ldots\}$ that reflects the fact that the node 1 has the maximum 
of $N-1$ incoming links etc.

The above results show that it is not so simple to propose
a good random matrix model which captures 
the generic spectral features of real directed networks.
We think that investigations in this direction should be continued.

\subsection{Anderson delocalization of PageRank?}
\label{s13.3}

The phenomenon of Anderson localization of electron transport in
disordered materials \cite{anderson:1958} is now a well-known effect studied in detail
in physics (see e.g. \cite{evers:2008}). In one and two dimensions
even a small disorder leads to an exponential localization of
electron diffusion that corresponds to an insulating phase.
Thus, even if a classical electron dynamics is diffusive
and delocalized over the whole space, the effects of quantum interference
generates a localization of all eigenstates of the Sch\"odinger equation.
In higher dimensions a localization is preserved at a sufficiently
strong disorder, while a delocalized metallic phase appears
for a disorder strength being smaller a certain critical value
dependent on the Fermi energy of electrons. This phenomenon
is rather generic and we can expect that a somewhat similar
delocalization transition can appear in the small-world networks.

\begin{figure}[H]
\begin{center}
\includegraphics[width=0.48\textwidth]{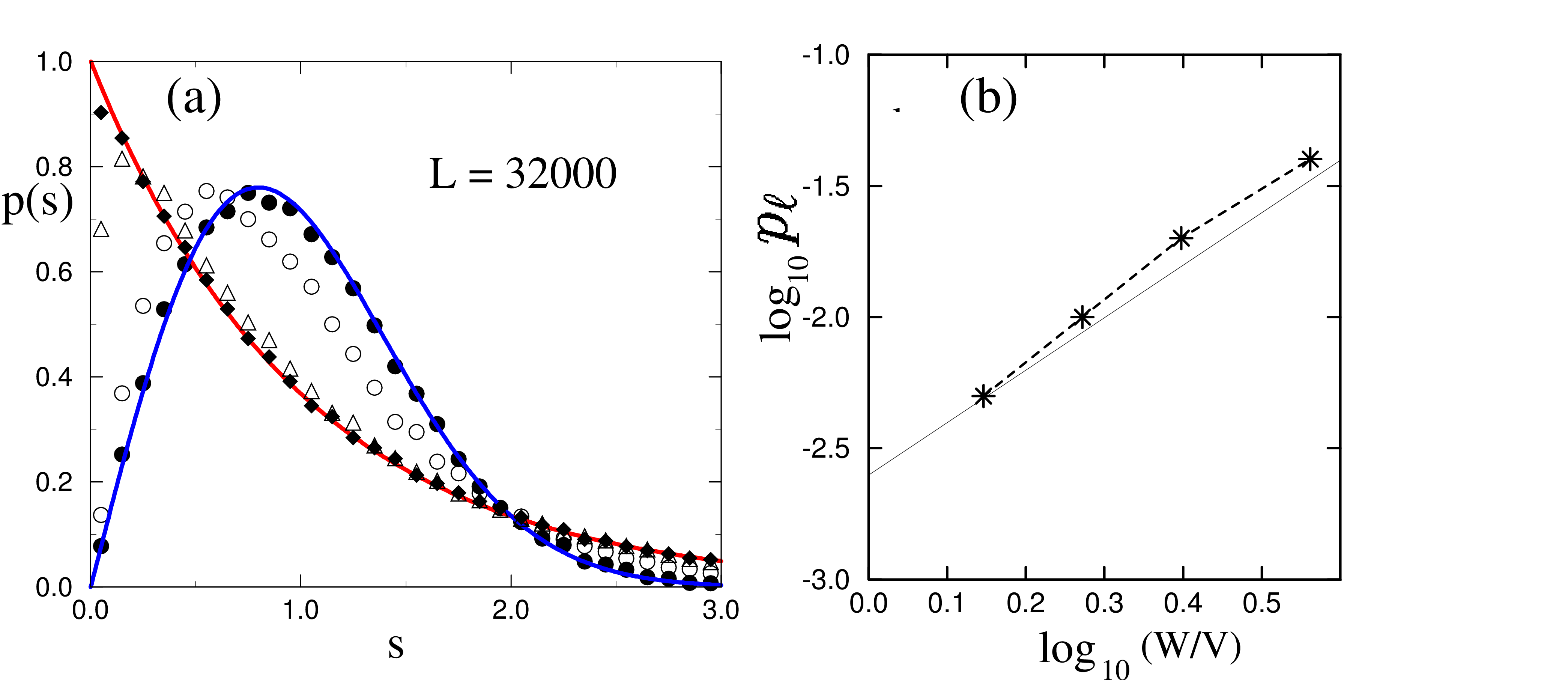}
\end{center}
\vglue -0.3cm
\caption{(Color online) 
(a) The red/gray and blue/black curves represent the Poisson and 
Wigner surmise distributions.
Diamonds, triangles, circles and black disks represent  respectively 
the level spacing statistics $p(s)$ at 
$W/V = 4, 3, 2, 1$; $p_\ell=0.02$, $L=32000$;
averaging is done over $60$ network realizations.
(b) Stars give dependence of $p_\ell$ on a disorder strength
$W/V$  at the critical point when $\eta_{\ell}(W, p_\ell) = 0.8$, and 
$p_\ell = 0.005, 0.01, 0.02, 0.04$ at fixed $L = 8000$;
the straight line corresponds to $p_\ell = p_c = 1 / 4 \ell_1 \approx 
(W / V)^2/400$; the dashed curve is drown to adapt an eye.
After \cite{chepelianskii:2001}.
\label{fig13_2}}
\end{figure}

Indeed, it is useful to consider the $1D$ Anderson model on a ring
with a certain  number of shortcut links, 
described by the Sch\"odinger equation
\begin{equation}
\label{eqquantsmallworld1}
\epsilon_n \psi_n +V(\psi_{n+1}+\psi_{n-1})+
V\sum_S(\psi_{n+S} + \psi_{n-S})= E \psi_n \; ,
\end{equation}
where $\epsilon_n$ are random on site energies homogeneously distributed
within the interval $-W/2 \leq \epsilon_n \leq W/2$, and $V$ is
the hopping matrix element. The sum over $S$ is taken over 
randomly established shortcuts from a site $n$ to any other random
site of the network. The number of such shortcuts is
$S_{\rm tot}=p_{\ell}L$, where $L$ is the total number of sites on a ring
and $p_\ell$ is the density of shortcut links. 
This model had been introduced in 
\cite{chepelianskii:2001}. 
The numerical study, reported there,
showed that the level-spacing statistics
$p(s)$ for this model has a transition 
from the Poisson distribution
$p_{\rm Pois}(s)=\exp(-s)$, typical for the Anderson localization phase,
to the Wigner surmise distribution $p_{\rm Wig}(s)=\pi s/2 \exp(-\pi s^2/4)$,
typical for the Anderson metallic phase \cite{guhr:1998,evers:2008}.
The numerical diagonalization was  done via the Lanczos algorithm
for the sizes up to $L=32000$ and the typical parameter range
$0.005 \leq p_{\ell} < 0.1$ and $1 \leq  W/V \leq 4$.  
An example, of the variation of $p_\ell(s)$ with a decrease of $W/V$
is shown in Fig.~\ref{fig13_2}(a). We see that the Wigner surmise
provides a good description of the numerical data
at $W/V=1$, when the maximal localization length 
$\ell_1 \approx 96(V/W)^2 \approx 96$ in the 1D Anderson
model (see e.g. \cite{evers:2008})
is much smaller than the system size $L$.

To identify a transition from one limiting case
$p_{\rm Pois}(s)$ to  another $p_{\rm Wig}(s)$
it is convenient to introduce the parameter 
$\eta_s = \int_0^{s_0}
(p(s)-p_{\rm Wig}(s)) ds / \int_0^{s_0} (p_{\rm Pois}(s)-p_{\rm Wig}(s)) ds$,
where $s_0=0.4729...$ is the intersection point
of $p_{\rm Pois}(s)$ and $p_{\rm Wig}(s)$. In this way $\eta_s$ varies from $1$
(for $p(s)=p_{\rm Pois}(s)$) to $0$ (for $p(s)=p_{\rm Wig}(s)\;$)
(see e.g. \cite{shepelyansky:2001}). From the variation of $\eta_s$
with system parameters and size $L$, the critical 
density  $p_\ell = p_c$ can be determined by the condition
$\eta_s(p_c,W/V)=\eta_c = 0.8 =const.$ being independent of $L$.
The obtained dependence of $p_c$ on $W/V$ obtained at a fixed 
critical point $\eta_c =0.8$
is shown in Fig.~\ref{fig13_2}(b).
The Anderson delocalization transition takes place 
when the  density of shortcuts becomes larger
than a critical density $p_\ell > p_c \approx 1/(4 \ell_1)$
where $\ell_1 \approx 96 (V/W)^2$ is 
the length of Anderson localization in $1D$.
A simple physical interpretation of this result is that
the delocalization takes place when the localization length
$\ell_1$ becomes larger than a typical distance $1/(4 p_{\ell} )$
 between shortcuts.
The further studies of time evolution
of wave function $\psi_n(t)$ 
and  IPR $\xi$ variation also confirmed the existence of quantum delocalization
transition on this quantum small-world network \cite{giraud:2005}.

Thus the results obtained for the quantum small-world networks 
\cite{chepelianskii:2001,giraud:2005}
show that the Anderson transition can take place in such systems.
However, the above model represents an undirected network
corresponding to a symmetric matrix with a real spectrum
while the typical directed networks are characterized by
asymmetric matrix $G$ and complex spectrum.
The possibility of existence of localized states of $G$
for WWW networks was also discussed by \cite{perra:2009}
but the fact that in a typical case the spectrum of $G$
is complex has not been analyzed in detail.

Above we saw certain indications on 
a possibility of Anderson type delocalization transition
for eigenstates of the $G$ matrix. 
Our results clearly show that certain eigenstates 
in the core space are exponentially localized 
(see e.g. Fig~\ref{fig8_2}(b)). Such states
are localized only on a few nodes touching other nodes
of network only by an exponentially small tail.
A similar situation would appear in the 1D Anderson model
if an absorption would be introduced on one end of the chain.
Then the eigenstates located far away from this place would
feel this absorption only by exponentially small tails
so that the imaginary part of the eigenenergy would have 
for such far away states only an exponentially small imaginary part.
It is natural to expect that such localization can be destroyed 
by some parameter variation.
Indeed, certain eigenstates  with $|\lambda|<1$ 
for  the directed network of the AB model
have IPR $\xi$ growing with the matrix size $N$
(see Sec.~\ref{s13.1} and \cite{giraud:2009})
even if for the PageRank the values of $\xi$
remain independent of $N$. The results for the Ulam network
from Figs.~\ref{fig6_6},~\ref{fig6_7}
provide an example of directed network where
the PageRank vector becomes delocalized when
the damping factor is decreased from $\alpha=0.95$ to $0.85$
\cite{zhirov:2010}.
This example demonstrates a possibility of PageRank delocalization
but a deeper understanding of the conditions required for
such a phenomenon to occur are still lacking.
The main difficulty is an absence of
well established random matrix models
which have properties similar to the available examples of real networks.

Indeed, for Hermitian and unitary matrices
the theories of random matrices, mesoscopic
systems  and quantum chaos allow to capture main universal properties
of spectra and eigenstates
\cite{guhr:1998,akemann:2011,mehta:2004,evers:2008,haake:2010}.
For asymmetric Google matrices  the spectrum is complex and
at the moment there are no good random matrix models
which would allow to perform analytical analysis of 
various parameter dependencies. It is possible that non-Hermitian
Anderson models in $1D$, which naturally generates a complex spectrum
and may have delocalized eigenstates,
will provide new insights in this direction \cite{goldsheid:1998}.
We note that the recent random
Google matrix models studied in \cite{zhirov:2015}
give indications on appearance of 
the Anderson transition for  Google matrix eigenstates
and a mobility edge contour in a plane of complex eigenvalues. 

\section{Other examples of directed networks}
\label{s14}

In this section we discuss additional examples of real directed networks.

\subsection{Brain neural networks}
\label{s14.1}

In 1958 John von Neumann traced 
first parallels between architecture of the 
computer  and the brain   \cite{neumann:1958}.
Since that time computers became an unavoidable element of the modern
society forming a computer network connected by the WWW
with about $4 \times 10^{9}$ indexed 
web pages spread all over the world (see e.g. 
 http://www.worldwidewebsize.com/). This number starts to become 
comparable with $10^{10}$ neurons in a human 
brain where each neuron can be viewed 
as an independent processing unit connected with 
about $10^4$ other neurons
by synaptic links (see e.g. \cite{sporns:2007}).
About 20\% of these links are unidirectional
\cite{felleman:1991} and hence the brain can be viewed
as a directed network of neuron links.
At present, more and more experimental information
about neurons and their links becomes available and
the investigations of properties of neuronal networks
attract an active interest 
(see e.g. \cite{bullmore:2009,zuo:2012}).
The fact that enormous sizes of WWW and brain networks
are comparable gives an idea that the Google matrix analysis 
should find useful application in brain science 
as it is the case of WWW.

First applications of methods of Google matrix methods
to brain neural networks was done in \cite{shepelyansky:2010b}
for a large-scale thalamocortical model 
\cite{izhikevich:2008} based on experimental measures
in several mammalian species. The model spans three anatomic
scales. (i) It is based on global (white-matter) thalamocortical
anatomy obtained by means of diffusion tensor imaging  of
a human brain. (ii) It includes multiple thalamic nuclei and 
six-layered cortical microcircuitry based on in vitro labeling and 
three-dimensional reconstruction of single neurons of cat visual cortex.
(iii) It has 22 basic types of neurons with appropriate laminar 
distribution of their branching dendritic trees. 
According to \cite{izhikevich:2008} the
model exhibits behavioral regimes of normal brain activity that
were not explicitly built-in but emerged spontaneously as the 
result of interactions among anatomical and dynamic processes.

\begin{figure}[H]
\begin{center}
\includegraphics[width=0.48\textwidth]{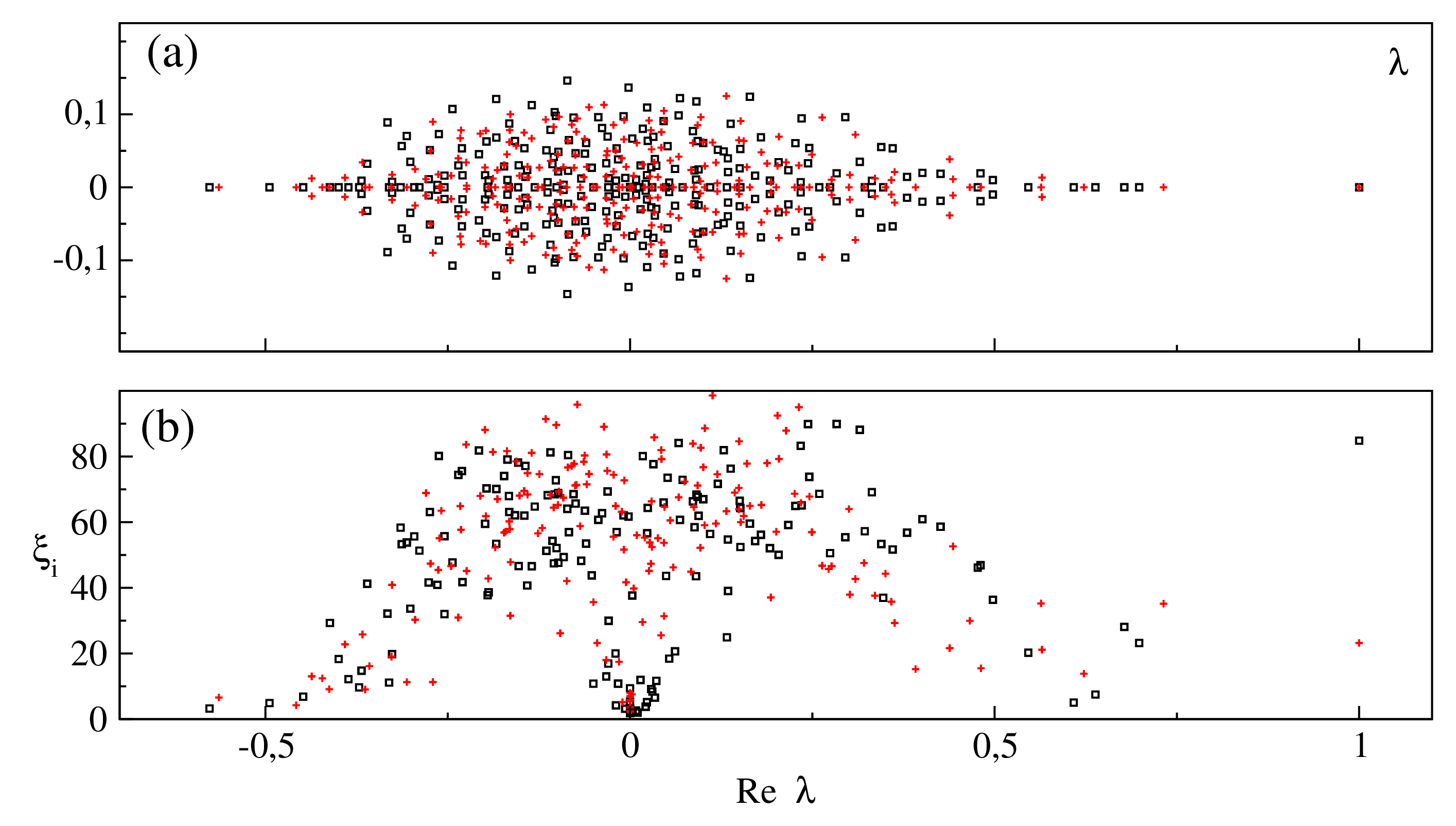}
\end{center}
\vglue -0.3cm
\caption{ (Color online)
(a) Spectrum of eigenvalues $\lambda$ for the Google matrices 
$G$ and $G^*$ at $\alpha=0.85$ 
for the neural network of {\it C.elegans}
(black and red/gray symbols). 
(b) Values of IPR $\xi_i$ of eigenvectors $\psi_i$ 
are shown
as a function of corresponding $Re \lambda$ (same colors).
After \cite{kandiah:2014a}.
\label{fig14_1}}
\end{figure}

The model studied in \cite{shepelyansky:2010b}
contains $N=10^4$ neuron with
$N_\ell=1960108$. The obtained results show that
PageRank and CheiRank vectors have rather large
$\xi$ being comparable with the whole network size
at $\alpha=0.85$.
The corresponding probabilities have very flat
dependence on their indexes showing that 
they are close to a delocalized regime.
We attribute these features to a rather large
number of links per node $\zeta \approx 196$
being even larger
than for the Twitter network.
At the same time the PageRank-CheiRank correlator
is rather small $\kappa = -0.065$. 
Thus this network is structured in such a way
that functions related to order signals 
(outgoing links of CheiRank) and
signals bringing orders 
(ingoing links of PageRank) 
are well separated and independent of each other
as it is the case for the Linux Kernel
software architecture. The spectrum of $G$ 
has a gapless structure showing that long
living excitations can exist in this neuronal network.

Of course, model systems of neural networks
can provide a number of interesting insights
but it is much more important to study examples of real 
neural networks. In \cite{kandiah:2014a}
such an analysis is performed for 
the neural network of {\it  C.elegans} (worm).
The full connectivity of this directed network
is known and well documented at WormAtlas \cite{wormatlas}.
The number of linked neurons (nodes) is $N = 279$ with the
number of synaptic connections and gap junctions (links) between
them being $N_\ell = 2990$.

\begin{figure}[H]
\begin{center}
\includegraphics[width=0.48\textwidth]{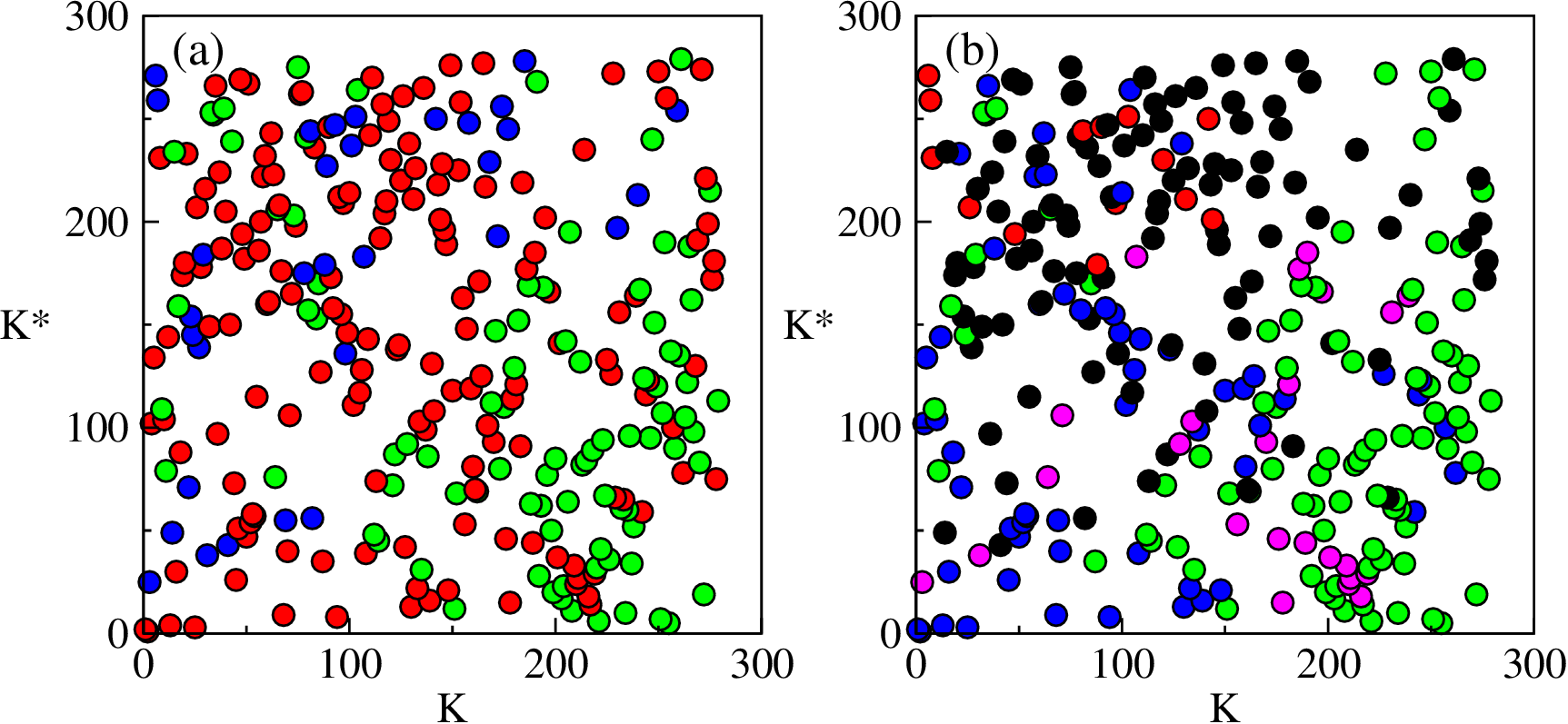}
\end{center}
\vglue -0.3cm
\caption{ (Color online)
PageRank - CheiRank plane  
$(K,K^*)$ showing distribution of neurons 
according to their ranking. 
(a): soma region coloration - head (red/gray), middle (green/light gray), 
tail (blue/dark gray). (b): neuron type coloration - 
sensory (red/gray), motor (green/light gray), interneuron (blue/dark gray), 
polymodal (purple/light-dark gray) and unknown (black). 
The classifications and colors 
are given according to WormAtlas \cite{wormatlas}.
After \cite{kandiah:2014a}.
\label{fig14_2}}
\end{figure}

The Google matrix $G$ of {\it C.elegans} is constructed using 
the connectivity matrix elements $S_{ij}=S_{{\rm syn},ij}+S_{{\rm gap},ij}$,
where $S_{\rm syn}$ is an asymmetric matrix of synaptic 
links whose elements are $1$ 
if neuron $j$ connects to neuron $i$ through 
a chemical synaptic connection and $0$ otherwise. 
The matrix part $S_{\rm gap}$ is a symmetric matrix describing gap junctions 
between pairs of cells, $S_{{\rm gap},ij}=S_{{\rm gap},ji}=1$ if neurons $i$ and $j$ 
are connected through a gap junction and $0$ otherwise.
Then the matrices $G$ and $G^*$ are constructed following
the standard rule (\ref{eq3_1}) at $\alpha=0.85$.
The connectivity properties of this network 
are similar to those of WWW of Cambridge and Oxford
with approximately the same number of links per node.

The spectra of $G$ and $G^*$ are shown in Fig.~\ref{fig14_1}
with corresponding IPR values of eigenstates. The imaginary part of
$\lambda$ is relatively small $|{\rm Im}(\lambda)| < 0.2$
due to a large fraction of symmetric links. 
The second by modulus
eigenvalues are $\lambda_2= 0.8214$ for $G$
and $\lambda_2= 0.8608$ for $G^*$.
Thus the network relaxation time 
$\tau = 1/|\ln \lambda_2|$ is approximately
$5, 6.7$ iterations of $G, G^*$.
Certain IPR values $\xi_i$ of eigenstates of $G, G^*$ have rather large
$\xi \approx N/3$ while others have $\xi$ located only on about ten nodes.

We have a large value $\xi \approx 85$ for PageRank
and a more moderate value $\xi \approx 23$ for CheiRank vectors.
Here we have the algebraic decay exponents being
$\beta \approx 0.33$ for $P(K)$ and 
$\beta \approx 0.50$ for $P^*(K^*)$.
Of course, the network size is not large
and these values are only approximate.
However, they indicate an interchange between 
PageRank and CheiRank showing importance 
of outgoing links. It is possible that such an inversion is 
related to a significant importance of
outgoing links in neural systems:
in a sense such links transfer orders,
while ingoing links bring instructions 
to a given neuron from other neurons.
The correlator $\kappa =0.125$
is small and thus, the network structure allows to perform 
a control of information flow in a more efficient way
without interference of errors between orders and
executions. We saw already in Sec.~\ref{s7.1}
that such a separation of concerns emerges in software architecture.
It seems that the neural networks also adopt such a structure.

We note that a somewhat similar 
situation appears for networks
of Business Process Management 
where {\it Principals} of a company
are located at the top CheiRank position
while the top PageRank positions
belong to company {\it Contacts} \cite{abel:2011}.
Indeed, a case study of  a real company structure
analyzed in \cite{abel:2011} also
stress the importance of company managers who
transfer orders to other structural units.
For this network the correlator is also small
being $\kappa = 0.164$.
We expect that brain neural networks
may have certain similarities with 
company organization. 

Each neuron $i$  belongs to two ranks $K_i$ and $K^*_i$
and it is convenient to represent the distribution of neurons
on PageRank-CheiRank plane $(K,K^*)$  shown in Fig.~\ref{fig14_2}.
The plot confirms that there are little correlations between both ranks
since the points are scattered over the whole plane. 
Neurons ranked at top $K$ positions of PageRank
have their soma located mainly 
in both extremities of the worm (head and tail) 
showing that neurons in those regions 
have important connections coming from many other neurons
which control head and tail movements. 
This tendency is even more visible 
for neurons at top $K^*$ positions of CheiRank 
but with a preference for head and middle regions. 
In general, neurons, that have their soma 
in the middle region of the worm,
are quite highly ranked in CheiRank but not in PageRank. 
The neurons located at the head region
have top positions in CheiRank and also 
PageRank, while the middle region
has some top CheiRank indexes but 
rather large indexes of PageRank
(Fig.~\ref{fig14_2} (a)).
The neuron type coloration 
(Fig.~\ref{fig14_2} (b)) also reveals 
that sensory neurons are at top PageRank positions
but at rather large CheiRank indexes, 
whereas in general motor neurons are in the opposite situation.

Top nodes of PageRank and CheiRank 
favor important signal 
relaying  neurons such as $AVA$ and $AVB$ 
that integrate signals from crucial nodes 
and in turn pilot other crucial nodes.
Neurons $AVAL,AVAR$, $AVBL,AVBR$ and $AVEL,AVER$ are considered 
to belong to the rich club analyzed in \cite{towlson:2013}.
The top neurons in 2DRank are AVAL, AVAR, AVBL, AVBR, PVCR
that corresponds to a dominance of interneurons. 
More details can be found in \cite{kandiah:2014a}.

The technological progress allows to obtain now more and more detailed
information about neural networks 
(see e.g. \cite{bullmore:2009,towlson:2013,zuo:2012}) 
even if it is not easy to get information about link directions.
In view of that we expect that the methods of directed network analysis
described here will find useful future applications for 
brain neural networks.

\subsection{Google matrix of DNA sequences}
\label{s14.2}

The approaches of Markov chains and Google matrix can be also 
efficiently used for analysis of  statistical properties of DNA sequences.
The data sets are publicly available at \cite{ensemblegenome}. 
The analysis of Poincar\'e recurrences
in these DNA sequences \cite{frahm:2012c} shows their
similarities with the statistical properties of
recurrences for dynamical trajectories in 
the Chirikov standard map and other symplectic maps \cite{frahm:2010}. 
Indeed, a DNA sequence can be 
viewed as a long symbolic trajectory
and hence, the Google matrix, constructed from it,
highlights the statistical features of DNA
from a new viewpoint.

An important step in the statistical analysis of DNA sequences
was done in \cite{mantegna:1995} applying methods of
statistical linguistics and determining the frequency of 
various words composed of up to 7 letters.
A first order Markovian models have been also
proposed and briefly discussed in this work.
The Google matrix analysis
provides a natural extension of this approach.
Thus the PageRank eigenvector gives most frequent words of given length.
The  spectrum and eigenstates of $G$ characterize the relaxation
processes of different modes in the Markov process
generated by a symbolic DNA sequence.
Thus the comparison of word ranks of different species
allows to identify their proximity. 

\begin{figure}[H]
\begin{center}
\includegraphics[width=0.48\textwidth]{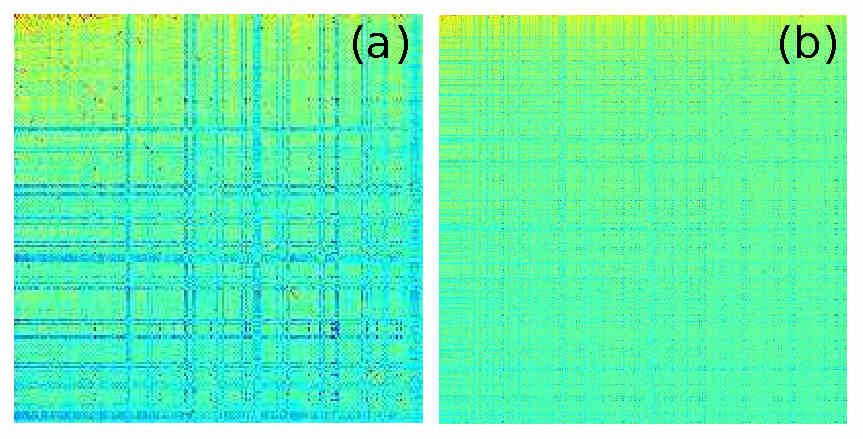}
\end{center}
\vglue -0.3cm
\caption{ (Color online)
DNA Google matrix of Homo sapiens (HS) constructed for words of 6-letters 
length. Matrix elements $G_{KK'}$ are shown 
in the basis of PageRank index $K$ (and $K'$). Here, $x$ and $y$ axes show 
$K$ and $K'$ within the range $1 \leq K,K' \leq 200$ (a) and 
$1 \leq K,K' \leq 1000$ (b). The element $G_{11}$ at
 $K=K'=1$ is placed at top left corner.
Color marks the amplitude of 
matrix elements changing from blue/black for minimum zero value 
to red/gray at maximum value.
After \cite{kandiah:2013}.
\label{fig14_3}}
\end{figure}

The statistical analysis is
done for DNA sequences
of  the species:  Homo sapiens (HS, human), 
Canis familiaris (CF, dog), Loxodonta africana (LA, elephant),
Bos Taurus (bull, BT),  Danio rerio (DR, zebrafish) \cite{kandiah:2013}.
For HS DNA sequences are represented as a single string of length
$L \approx 1.5 \cdot 10^{10}$ base pairs (bp)
corresponding to 5 individuals.
Similar data are obtained for  BT  ($2.9 \cdot 10^9$ bp),
CF ($2.5 \cdot 10^9$ bp), LA  ($3.1 \cdot 10^9$ bp), DR ($1.4 \cdot 10^9$ bp).
All strings are composed of 4 letters $A,G,G,T$
and undetermined letter ${\it N_l}$. 
The strings can be found from \cite{kandiah:2013}.

For a given sequence we fix the words $W_k$ of $m$ letters length
corresponding to the number of states $N=4^m$.
We consider that there is a transition 
from a state $j$ to state $i$ inside this basis $N$
when  we move along the string from left to right
going from a word $W_k$ to a next word $W_{k+1}$.
This transition adds one unit in the transition
matrix element $T_{ij} \rightarrow T_{ij}+1$.
The words with letter ${\it N_l}$
are omitted, the transitions are counted only between nearby words
not separated by words with $N_l$.
There are approximately $N_t \approx L/m$
such transitions for the whole length $L$
since the fraction of undetermined letters ${\it N_l}$ is small.
Thus we have $N_t=\sum_{i,j=1}^{N} T_{ij}$.
The Markov matrix of transitions $S_{ij}$ is
obtained by normalizing matrix elements in such a way that their sum
in each column is equal to unity: $S_{ij}=T_{ij}/\sum_i T_{ij}$.
If there are columns with all zero elements (dangling nodes)
then zeros of such columns are replaced by $1/N$.
Then the Google matrix $G$ is constructed from $S$
by the standard rule (\ref{eq3_1}).
It is found that the spectrum of $G$ has a significant gap
and a variation of $\alpha$ in a range $(0.5,1)$
does not affect significantly the PageRank probability.
Thus all DNA results are shown at $\alpha=1$.

\begin{figure}[H]
\begin{center}
\includegraphics[width=0.48\textwidth]{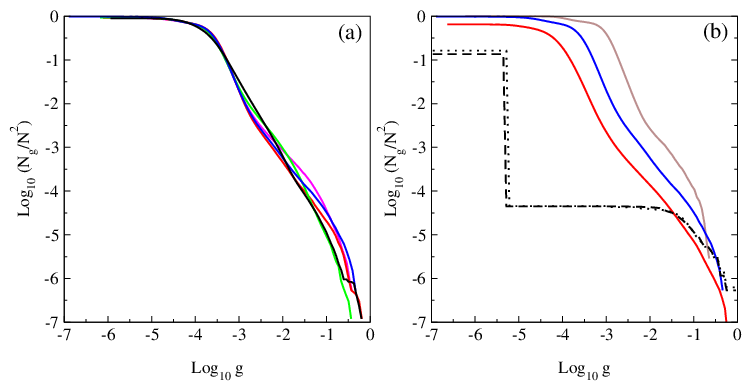}
\end{center}
\vglue -0.3cm
\caption{ (Color online)
Integrated fraction $N_g/N^2$ of  Google matrix elements with $G_{ij} > g$ 
as a function of $g$.
(a) Various species with 6-letters word length: 
elephant LA (green), zebrafish DR(black), dog CF (red),
bull BT (magenta), and Homo sapiens HS (blue)
(from left to right at $y=-5.5$).
(b) Data for HS sequence with 
words of length $m= 5$ (brown), $6$ (blue), $7$ (red)
(from right to left at $y=-2$);
for comparison black dashed and dotted curves
show the same distribution for the WWW networks
of Universities of Cambridge and Oxford in 2006
respectively.
After \cite{kandiah:2013}.
\label{fig14_4}}
\end{figure}

\begin{figure}[H]
\begin{center}
\includegraphics[width=0.48\textwidth]{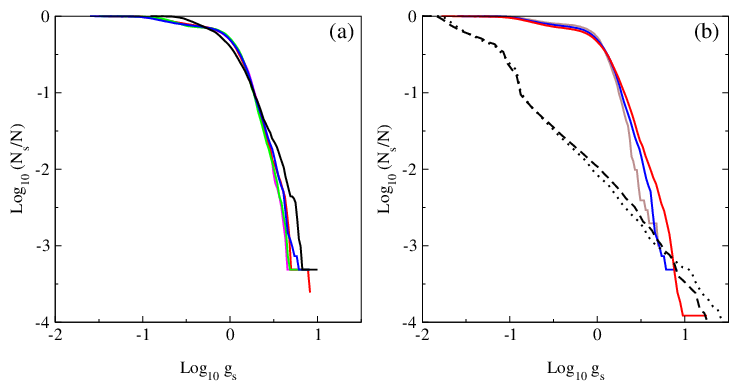}
\end{center}
\vglue -0.3cm
\caption{ (Color online)
Integrated fraction $N_s/N$ of sum of ingoing matrix elements 
with $\sum_{j=1}^NG_{i,j} \ge g_s$. Panels (a) and (b)
show the same cases as in Fig.~\ref{fig14_4} in same colors.
The dashed and dotted curves are shifted in $x$-axis 
by one unit left to fit the figure scale.
After \cite{kandiah:2013}.
\label{fig14_5}}
\end{figure}

\begin{figure}[H]
\begin{center}
\includegraphics[width=0.48\textwidth]{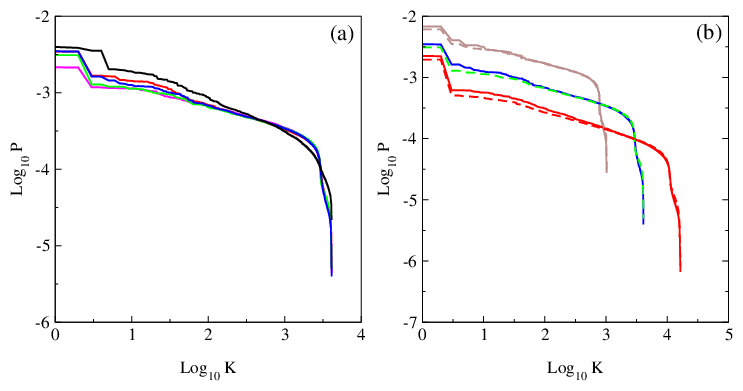}
\end{center}
\vglue -0.3cm
\caption{ (Color online)
Dependence of PageRank probability $P(K)$ on PageRank index $K$. 
(a) Data for different species for word length of  6-letters: 
zebrafish DR (black), dog CF (red), Homo sapiens HS (blue),
elephant LA (green) and bull BT (magenta)
(from top to bottom at $x=1$).
(b) Data for HS (full curve) and LA (dashed curve)
for word length $m=5$  (brown), $6$ (blue/green), $7$ (red)
(from top to bottom at $x=1$).
After \cite{kandiah:2013}.
\label{fig14_6}}
\end{figure}

The image of matrix elements $G_{KK'}$ is shown in Fig.~\ref{fig14_3}
for HS with $m=6$. We see that almost all matrix is
full that is drastically different from the WWW and other networks
considered above. The analysis of statistical properties of matrix elements
$G_{ij}$ shows that their integrated distribution follows a power law
as it is seen in Fig.~\ref{fig14_4}. Here  $N_g$ is the number of
matrix elements of the matrix $G$ with
values $G_{ij} > g$. The data show that the number 
of nonzero matrix elements $G_{ij}$ 
is  very close to $N^2$.
The main fraction of elements has values $G_{ij} \leq 1/N$
(some elements $G_{ij} < 1/N$ since 
for certain $j$ there are many  transitions
to some node $i'$ with $T_{i'j} \gg N$
and e.g. only one transition to other $i''$ with $T_{i''j}=1$).
At the same time there are also transition elements $G_{ij}$
with large values whose fraction decays
in an algebraic law $N_g \approx  A N/g^{\nu-1}$
with some constant $A$ and an exponent $\nu$. 
The fit of numerical data  in the range 
 $-5.5 < \log_{10}g < -0.5$ of algebraic decay
gives for $m=6$: $\nu=2.46 \pm 0.025$ (BT),
$2.57 \pm 0.025$ (CF),  $2.67 \pm 0.022$ (LA),
$2.48 \pm 0.024$ (HS), $2.22 \pm 0.04$ (DR).
For HS case we find
$\nu = 2.68 \pm 0.038$ at $m=5$ and
$\nu = 2.43 \pm 0.02$ at $m=7$ with the average  
$A \approx 0.003$ for $m=5,6,7$.
There are visible oscillations in the algebraic
decay of $N_g$ with $g$ but in global
we see that on average all species are 
well described by a universal decay law
with the exponent $\nu \approx 2.5$.
For comparison we also show the distribution
$N_g$ for the WWW networks of University of Cambridge
and Oxford in year 2006. We see that in these
cases the distribution $N_g$ has a very short 
range in which the decay is at least 
approximately algebraic ($-5.5 < \log_{10}(N_g/N^2) < -6$).
In contrast to that for the DNA sequences
we have a large range of algebraic decay.

Since in each column we have the sum of
all elements equal to unity we can say that
the differential fraction
$d N_g/dg \propto 1/g^{\nu}$ gives the distribution of
outgoing matrix elements which is similar to the distribution
of outgoing links extensively studied for the WWW networks.
Indeed, for the WWW networks all links in a column
are considered to have the same weight
so that these matrix elements are given by 
an inverse number of outgoing links
with the decay exponent $\nu \approx 2.7$.
Thus, the obtained data show that
the distribution of DNA matrix elements
is similar to the distribution of outgoing links
in the WWW networks. Indeed, for outgoing links
of Cambridge and Oxford networks
the fit of numerical data gives the exponents
$\nu = 2.80 \pm 0.06$ (Cambridge) and
$2.51 \pm 0.04$ (Oxford).

As discussed above, on average the probability of
PageRank vector is proportional to the number of
ingoing links that works satisfactory
for sparse $G$ matrices. For DNA we have a situation where 
the Google matrix is almost full
and zero matrix elements are practically absent.
In such a case an analogue of number of ingoing
links is the sum of ingoing matrix elements
$g_s=\sum_{j=1}^N G_{ij}$. 
The integrated distribution of
ingoing matrix elements with 
the dependence of $N_s$ on $g_{s}$ is shown in Fig.~\ref{fig14_5}.
Here $N_s$ is defined as the number of nodes with the sum of ingoing 
matrix elements being larger than $g_{s}$.
A significant part of this dependence, corresponding to large
values of $g_s$ and determining
the PageRank probability decay,
is well described by a power law
$N_s \approx B N /g_{s}^{\mu-1}$.
The fit of data at $m=6$ gives
$\mu=5.59 \pm 0.15$ (BT), $4.90 \pm 0.08$ (CF),
$5.37 \pm0.07$ (LA), $5.11 \pm 0.12$ (HS),
$4.04 \pm 0.06$ (DR).
For HS case at $m=5,7$ we find respectively
$\mu=5.86 \pm 0.14$ and $4.48 \pm 0.08$.
For $HS$ and other species we have an average 
$B \approx 1$.

For WWW one usually have $\mu \approx 2.1$.
Indeed, for the ingoing matrix elements of
Cambridge and Oxford networks 
we find respectively the exponents
$\mu = 2.12 \pm 0.03$ and $2.06 \pm 0.02$ 
(see curves in Fig.~\ref{fig14_5}).
For ingoing links distribution of
Cambridge and Oxford networks we obtain respectively
$\mu =2.29 \pm 0.02$ and  $\mu =2.27 \pm 0.02$
which are close to the usual WWW value 
$\mu \approx 2.1$.
In contrast  the exponent
$\mu$ for DNA Google matrix elements
gets significantly larger value
$\mu \approx 5$. This feature marks a significant difference
between DNA and WWW networks.

The PageRank vector can be obtained by a direct diagonalization.
The dependence of probability $P$ on index $K$
is shown in Fig.~\ref{fig14_6} for various species and
different word length $m$. The probability $P(K)$
describes  the steady state of random walks on the Markov chain
and thus it gives the frequency of appearance of various
words of length $m$ in the whole sequence $L$.
The frequencies or probabilities 
of words appearance in the sequences
have been obtained in \cite{mantegna:1995}
by a direct counting of words along the sequence 
(the available sequences $L$ were shorted at that times).
Both methods are mathematically equivalent 
and indeed our distributions $P(K)$ are in good agreement 
with those found in \cite{mantegna:1995}
even if now we have a significantly better statistics.

\begin{figure}[H]
\begin{center}
\includegraphics[trim=0 0 0 0cm,clip=true,width=0.48\textwidth]{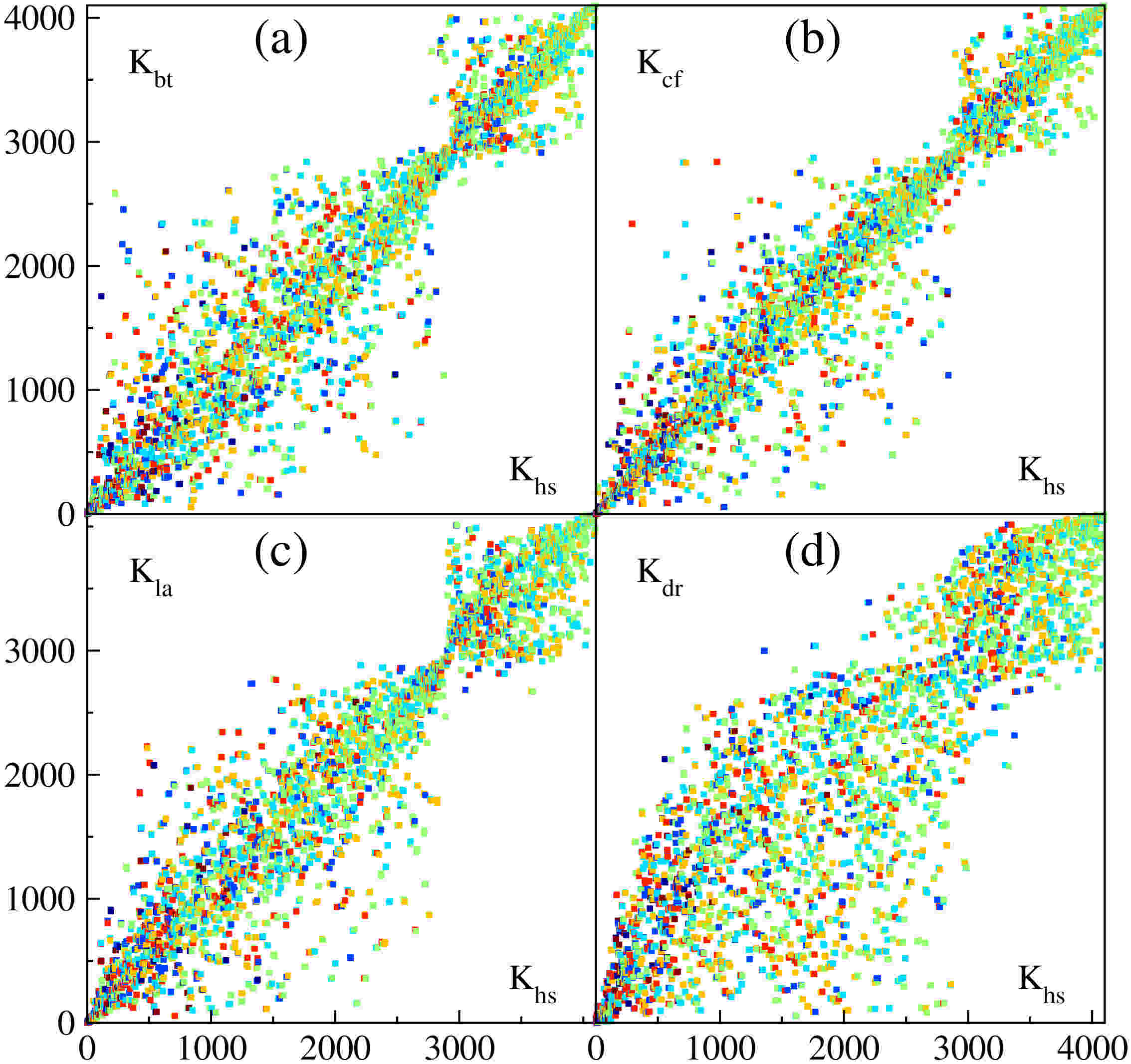}
\end{center}
\vglue -0.3cm
\caption{ (Color online)
PageRank proximity $K-K$ plane diagrams 
for different species in comparison with Homo sapiens: 
(a) $x$-axis shows PageRank index $K_{hs}(i)$ of a word $i$ and
$y$-axis shows PageRank index of the same word $i$ 
with $K_{bt}(i)$ of bull, (b) $K_{cf}(i)$ of dog, 
(c) $K_{la}(i)$ of elephant  and (d) $K_{dr}(i)$ of zebrafish;
here the word length is $m=6$. 
The colors of symbols marks the purine content in a word $i$ 
(fractions of letters $A$ or $G$ in any order);
the color varies from red/gray at maximal content, via brown, 
yellow, green, light blue,
to blue/black at minimal zero content.
After \cite{kandiah:2013}.
\label{fig14_7}}
\end{figure}

The decay of $P$ with $K$
can be approximately described by a power law
$P \sim 1/K^{\beta}$. Thus for example for HS sequence at $m=7$
we find  $\beta=0.357 \pm 0.003$
for the fit range $1.5 \leq \log_{10} K \leq 3.7$
that is rather close to the exponent found in \cite{mantegna:1995}.
Since on average the PageRank probability is proportional
to the number of ingoing links, or the sum
of ingoing matrix elements of $G$, one has the relation between
the exponent of PageRank $\beta$ and exponent of ingoing
links (or matrix elements): $\beta = 1/(\mu -1)$.
Indeed, for the HS DNA case at $m=7$ we have
$\mu=4.48$ that gives $\beta = 0.29$ being close to the
above value of $\beta =0.357$ obtained from the direct fit
of $P(K)$ dependence. The agreement is
not so perfect since there is a visible curvature in the
log-log plot of $N_s$ vs $g_{s}$  and also
since  a small value of $\beta$ gives a moderate variation
of $P$ that produces a reduction of accuracy of numerical fit procedure.
In spite of this only approximate agreement
we conclude that in global the relation between $\beta$ and $\mu$
works correctly.

It is interesting to plot a PageRank index 
 $K_{s}(i)$ of a given species $s$
versus the index $K_{hs}(i)$ of HS for the same word $i$.
For identical sequences one should have all points
on diagonal, while the deviations from diagonal
characterize the differences between species.
The examples of such PageRank proximity $K-K$ diagrams
are shown in Fig.~\ref{fig14_7}
for words at $m=6$. A visual impression is that
CF case has less deviations from HS rank
compared to BT and LA. The non-mammalian DR case
has most strong deviations from HS rank.

The fraction of purine letters $A$ or $G$
in a word of $m=6$ letters 
is shown by color in Fig.~\ref{fig14_7} for all words
ranked by PageRank index $K$.
We see that these letters are approximately 
homogeneously distributed over the whole range
of $K$ values. To determine the proximity between different species
or different HS individuals we compute 
the average dispersion 
\begin{equation}
\label{eq_average_dispersion}
\sigma(s_1,s_2) = \sqrt{\frac1N\sum_{i=1}^N 
\Bigl(K_{s_1}(i)- K_{s_2}(i)\Bigr)^2}
\end{equation}
between two species (individuals) $s_1$ and $s_2$.
Comparing the words with length $m=5,6,7$ we find
that the scaling  $\sigma \propto N$
works with a good accuracy (about 10\% when $N$
is increased by a factor 16). To represent the result
in a form independent of $m$ we compare the values of
$\sigma$ with the corresponding random model
value $\sigma_{rnd}$. This value is computed assuming
a random distribution of $N$ points in 
a square $N \times N$ when  only one point appears 
in each column and each line
(e.g. at $m=6$ we have $\sigma_{rnd} \approx 1673$ 
and $\sigma_{rnd} \propto N$). The dimensionless dispersion
is then given by $\zeta(s_1,s_2)=\sigma(s_1,s_2)/\sigma_{rnd}$.
From the ranking of different species we obtain the following values
at $m=6$:
$\zeta(CF,BT)=0.308$; $\zeta (LA,BT)=0.324$,  $\zeta(LA,CF)=0.303$;
$\zeta(HS,BT)=0.246$, $\zeta(HS,CF)=0.206$, $\zeta(HS,LA)=0.238$;
$\zeta(DR,BT)=0.425$, $\zeta(DR,CF)=0.414$, $\zeta(DR,LA)=0.422$,
$\zeta(DR,HS)=0.375$ (other $m$ have similar values).
According to this statistical analysis of PageRank proximity
between species we find that $\zeta$ value is minimal between
CF and HS showing that these are two most similar 
species among those considered here.
The comparison of two  HS individuals 
gives the value $\zeta(HS1,HS2) = 0.031$ being significantly smaller
then the proximity correlator 
between different species \cite{kandiah:2012}.

The spectrum of $G$ is analyzed in detail in \cite{kandiah:2012}.
It is shown that it has a relatively large gap due to which
there is a relatively rapid relaxation of probability of a random surfer
to the PageRank values.

\begin{figure}[H]
\begin{center}
\includegraphics[width=0.48\textwidth]{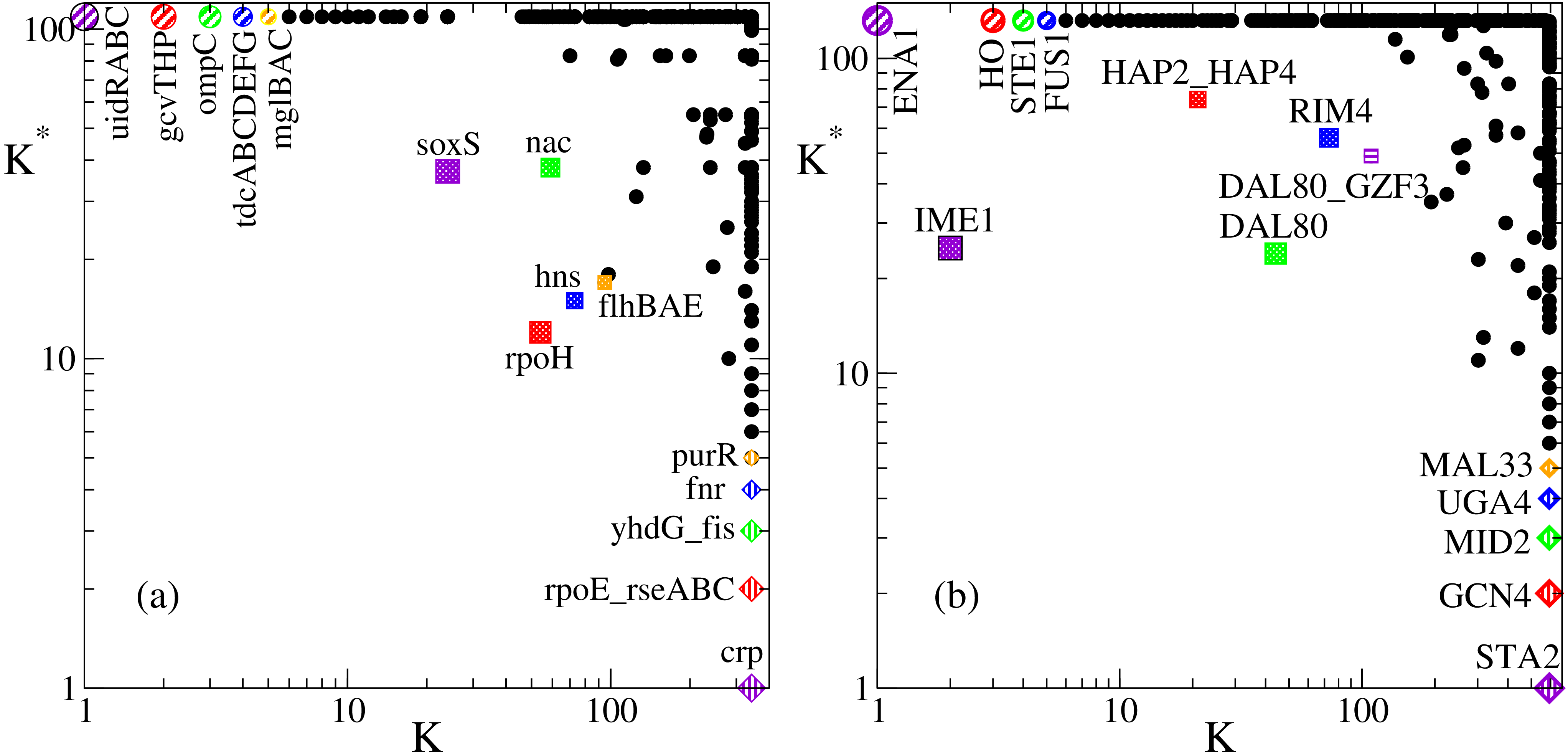}
\end{center}
\vglue -0.3cm
\caption{ (Color online)
Distribution of nodes in the PageRank-CheiRank plane $(K,K^*)$ 
for Escherichia Coli v1.1 (a), and Yeast (b) gene
transcription networks on
(network data are taken from \cite{shen-orr:2002,milo:2002}
and \cite{alon:2014}).
The nodes with five top probability values of PageRank, 
CheiRank and 2DRank 
are labeled by their corresponding operon (node) names; 
they correspond to 5 lowest values of indexes $K, K_2, K^*$.
After \cite{ermann:2012a}.
\label{fig14_8}}
\end{figure}

\subsection{Gene  regulation networks}
\label{s14.3}

At present the analysis of gene transcription regulation networks
and recovery of their control biological functions
becomes an active research field of bioinformatics
(see e.g. \cite{milo:2002}). Here, following 
\cite{ermann:2012a}, we provide two simple examples of 
2DRanking analysis 
for gene transcriptional regulation networks of Escherichia Coli 
($N=423$, $N_\ell=519$  \cite{shen-orr:2002})
and Yeast ($N=690$, $N_\ell=1079$  \cite{milo:2002}).
In the construction of $G$ matrix the outgoing links to all 
nodes in each column are  taken with the same weight, 
$\alpha=0.85$. 

The distribution of nodes in PageRank-CheiRank plane is shown 
in Fig.~\ref{fig14_8}. The  top 5 nodes, with their operon names,
are given there for indexes of PageRank $K$, CheiRank $K^*$ 
and 2DRank $K_2$. This ranking selects operons
with most high functionality in communication ($K^*$), popularity ($K$)
and those that combines these both features ($K_2$).
For these networks the correlator $\kappa$
is close to zero ($\kappa=-0.0645$ for  Escherichia Coli
and $\kappa=-0.0497$ for Yeast, see Fig.~\ref{fig4_2}))
that indicates the statistical independence between outgoing and ingoing
links being quite similarly to the case of the PCN for the Linux Kernel.
This may indicate that a slightly negative correlator $\kappa$ 
is a generic property for the data flow network of control and regulation 
systems. A similar situation appears
for networks of business process management and brain neural networks. 
Thus it is possible that the networks performing control functions
are  characterized in general by  small correlator $\kappa$  values.
We expect that 2DRanking will find further useful applications
for  large scale gene regulation networks.

\subsection{Networks of game go}
\label{s14.4}

The complexity of the well-known game go is such that no computer 
program has been able to beat a good player, in contrast
with chess where world champions have been bested by
game simulators. It is partly due to the fact that the
total number of possible allowed positions in go is about
$10^{171}$, compared to e.g. only $10^{50}$ for chess \cite{go1}.

It has been argued that the complex network analysis can give useful insights
for a better understanding of this game.
With this aim a network,  modeling the game of go, has been defined  
by a statistical analysis of the data bases of several important historical 
professional and amateur Japanese go tournaments \cite{georgeot:2012}. 
In this approach moves/nodes are defined as all possible patterns in $3\times 3$ 
plaquettes on a go board of $19\times 19$ intersections. Taking into account 
all possible obvious symmetry operations the number of non-equivalent moves is 
reduced to $N=1107$. 
Moves which are close in space (typically a maximal distance 
of 4 intersections) are assumed to belong to the same tactical fight
generating transitions on the network.

Using the historical data of many games, the transition probabilities 
between the nodes may be determined leading to a directed network 
with a finite size Perron-Frobenius operator 
which can be analyzed by tools of PageRank, 
CheiRank, complex eigenvalue spectrum, properties of certain 
selected eigenvectors and also certain other quantities 
\cite{georgeot:2012,kandiah:2014b}. 
The studies are done for plaquettes of different sizes
with the corresponding network size changing from $N= 1107$
for plaquettes squares with $3 \times 3$ intersections
up to maximal $N=193995$ for diamond-shape plaquettes
with  $3 \times 3$ intersections plus the four at distance
two from the center in the four directions left, right, top,
down. It is shown that
the PageRank leads to a 
frequency distribution of moves which obeys a Zipf law with exponents 
close to unity but this exponent may slightly vary if the network is 
constructed with shorter or longer sequences of successive moves. 
The important nodes in certain eigenvectors may correspond to certain 
strategies, such as protecting a stone and eigenvectors are also different 
between amateur and professional games.
It is also found that the different phases of the game go are
characterized by a different spectrum of the $G$ matrix.
The obtained results show that with the help of the Google matrix analysis 
it is possible  to extract communities 
of moves which share some common properties. 

The authors of these studies \cite{georgeot:2012,kandiah:2014b}
argue that the Google matrix analysis can find a number
of interesting applications in the theory of games
and the human decision-making processes.

\subsection{Opinion formation on directed networks}
\label{s14.5}

Understanding the nature and origins of mass opinion formation is 
an outstanding challenge of democratic societies \cite{zaller:1999}.
In the last few years the enormous development 
of such social networks as LiveJournal, 
Facebook, Twitter, and VKONTAKTE, with up to hundreds of millions of users, 
has demonstrated the growing influence of these networks on
social and political life. The small-world scale-free structure of 
the social networks, combined with
their rapid communication facilities, leads to a very fast information 
propagation over networks of electors, consumers, and
citizens, making them very active on instantaneous social events. 
This invokes the need for new theoretical models which
would allow one to understand the opinion formation process in modern society 
in the 21st century.

The important steps in the analysis of opinion formation have been done 
with the development of various voter models, described in great detail in 
\cite{castellano:2009,krapivsky:2010}.
This research field became known as sociophysics \cite{galam:1986,galam:2008}.
Here, following \cite{kandiah:2012}, we analyze the opinion formation
process introducing several new aspects which take into account 
the generic features of social networks. 
First, we analyze the opinion
formation on real directed networks
such as WWW of Universities of Cambridge and Oxford (2006), Twitter (2009)
and LiveJournal. This allows us to incorporate the correct scale-free network
structure instead of unrealistic regular lattice networks, 
often considered in voter models. Second, we assume that
the opinion at a given node is formed by the opinions of
 its linked neighbors weighted with the PageRank probability of
these network nodes. 
The introduction of such a weight represents 
the reality of social networks where 
network nodes are characterized by the PageRank vector 
which provides a
natural ranking of node importance, or elector or society member importance.
In a certain sense, the top nodes of PageRank correspond to a political 
elite of the social network whose opinion influences
the opinions of other members of the society \cite{zaller:1999}. 
Thus the proposed PageRank opinion formation (PROF) model takes into
account the situation in which an opinion of 
an influential friend from high ranks 
of the society counts more than an opinion
of a friend from a lower society level. We argue that the PageRank 
probability is the most natural form of ranking of society
members. Indeed, the efficiency of PageRank rating 
had been well demonstrated for 
various types of scale-free networks.

The PROF model is defined in the following way. 
In agreement with the standard PageRank algorithm
we determine the probability $P(K_i)$ for each node
ordered by PageRank index $K_i$ (using $\alpha=0.85$).
In addition, a network node $i$ is characterized 
by an Ising spin variable $\sigma_i$ which
can take values $ +1$ or $-1$, coded also by red or blue color, respectively. 
The sign of a node $i$ is determined by its direct
neighbors $j$, which have PageRank probabilities $P_j$ . 
For that we compute the sum $\Sigma_i$ over 
all directly linked neighbors $j$ of node $i$:
\begin{equation}
\label{eqopinion1}
\begin{array}{cc}
\Sigma_i = a \sum_j ({P^+}_{j,\rm in} - {P^-}_{j,\rm in}) + \\
           b \sum_j ({P^+}_{j,\rm out} - {P^-}_{j,\rm out}) \; , \; \; a+b=1 \; ,
\end{array}
\end{equation}
where $P_{j,\rm in}$ and $P_{j,\rm out}$ denote the PageRank probability $P_j$ of 
a node $j$ pointing to node $i$ (ingoing link) and a node $j$ 
to which node $i$ points to (outgoing link), respectively. 
Here, the two parameters $a$ and $b$ 
are used to tune the importance of ingoing
and outgoing links with the imposed relation
$a+b=1$ ($0 \leq a,b \leq 1$). 
The values $P^+$ and $P^-$ correspond to red and blue
nodes, and the spin $\sigma_i$ takes the value 
$1$ or $-1$, respectively, for $\Sigma_i >0$  or $\Sigma_i<0$. 
In a certain sense we can say that
a large value of parameter $b$ corresponds to a conformist society 
in which an elector $i$ takes an opinion of other electors to
which he/she points. In contrast, a large
value of $a$ corresponds to a tenacious society 
in which an elector $i$ takes mainly the opinion of those electors 
who point to him/her. A standard random number generator 
is used to create an initial
random distribution of spins $\sigma_i$ on a given network. 
The time evolution then is
determined by the relation (\ref{eqopinion1})
applied to each spin one by one. 
When all $N$ spins are turned following (\ref{eqopinion1})
a time unit $t$ is changed to $t \rightarrow t+1$.
Up to $N_r=10^4$ random initial generations of
spins are used to obtain statistically stable results.
We present results for the number  of red nodes since other nodes are blue.

\begin{figure}[H]
\begin{center}
\includegraphics[width=0.48\textwidth]{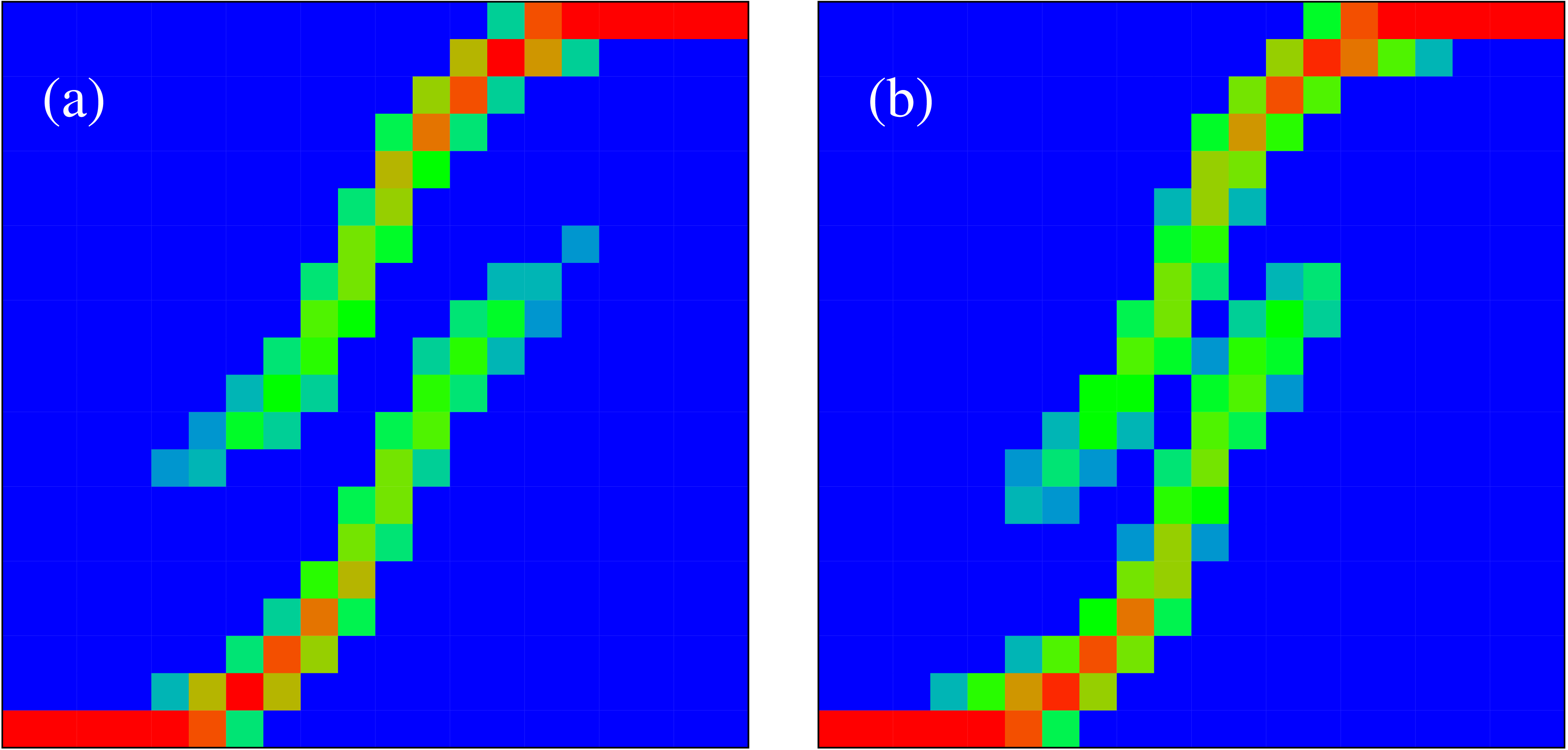}
\end{center}
\vglue -0.3cm
\caption{ (Color online)
Density plot of probability $W_f$ 
to find a final red fraction $f_f$, shown in $y-$axis,
in dependence on an initial red fraction $f_i$, shown in $x-$ axis;
data are shown inside the unit square $0\leq f_i,f_f \leq 1$.
The values of $W_f$ are defined as a relative number of
realisations found inside each of $20\times20$ cells
which cover  the whole unit square. Here $N_r=10^4$ realizations of
randomly distributed colors are used to obtained $W_f$ values;
for each realization the
time evolution is followed up the convergence time with up to 
$t=20$ iterations;
(a) Cambridge network;
(b) Oxford network at $a=0.1$.
The probability $W_f$ is proportional to
color changing from zero (blue/black) to unity (red/gray).
After \cite{kandiah:2012}.
\label{fig14_9}}
\end{figure}

The main part of studies is done for 
the WWW of Cambridge and Oxford discussed above.
We start with a random realization of a given fraction
of red nodes $f_i=f(t=0)$ which evolution in time
converges to a steady state with a final fraction of red nodes $f_f$
approximated after time $t_c \approx 10$. However, different 
initial realisations with the same $f_i$ value 
evolve to different final fractions $f_f$
clearly showing a bistability phenomenon.
To analyze how the final fraction of red nodes 
$f_f$ depends on its initial fraction $f_i$, 
we study the time evolution $f(t)$ for a large number 
$N_r$  of initial random realizations of colors 
following it up to the convergence time for each realization. 
We find that the final red nodes are homogeneously distributed in 
PageRank index $K$. Thus there is no specific preference 
for top society levels for an initial random distribution. 
The probability distribution $W_f$ of final fractions $f_f$ 
is shown in Fig.~\ref{fig14_9} as a function of initial
fraction $f_i$ at $a=0.1$. 
The results show two main features of the model: 
a small fraction of red opinion
is completely suppressed if $f_i<f_c$ and 
its larger fraction dominates completely for $f_i>1-f_c$; 
there is a bistability phase for the initial opinion range 
$f_b \leq f_i \leq 1-f_b$. 
Of course, there is a symmetry in respect to exchange of red and blue colors.
For the small value $a=0.1$ we have $f_b \approx f_c$ with 
$f_c \approx 0.25$. For the larger value $a=0.9$ we have 
$f_c \approx 0.35$, $f_b \approx 0.45$ \cite{kandiah:2012}.

\begin{figure}[H]
\begin{center}
\includegraphics[width=0.48\textwidth]{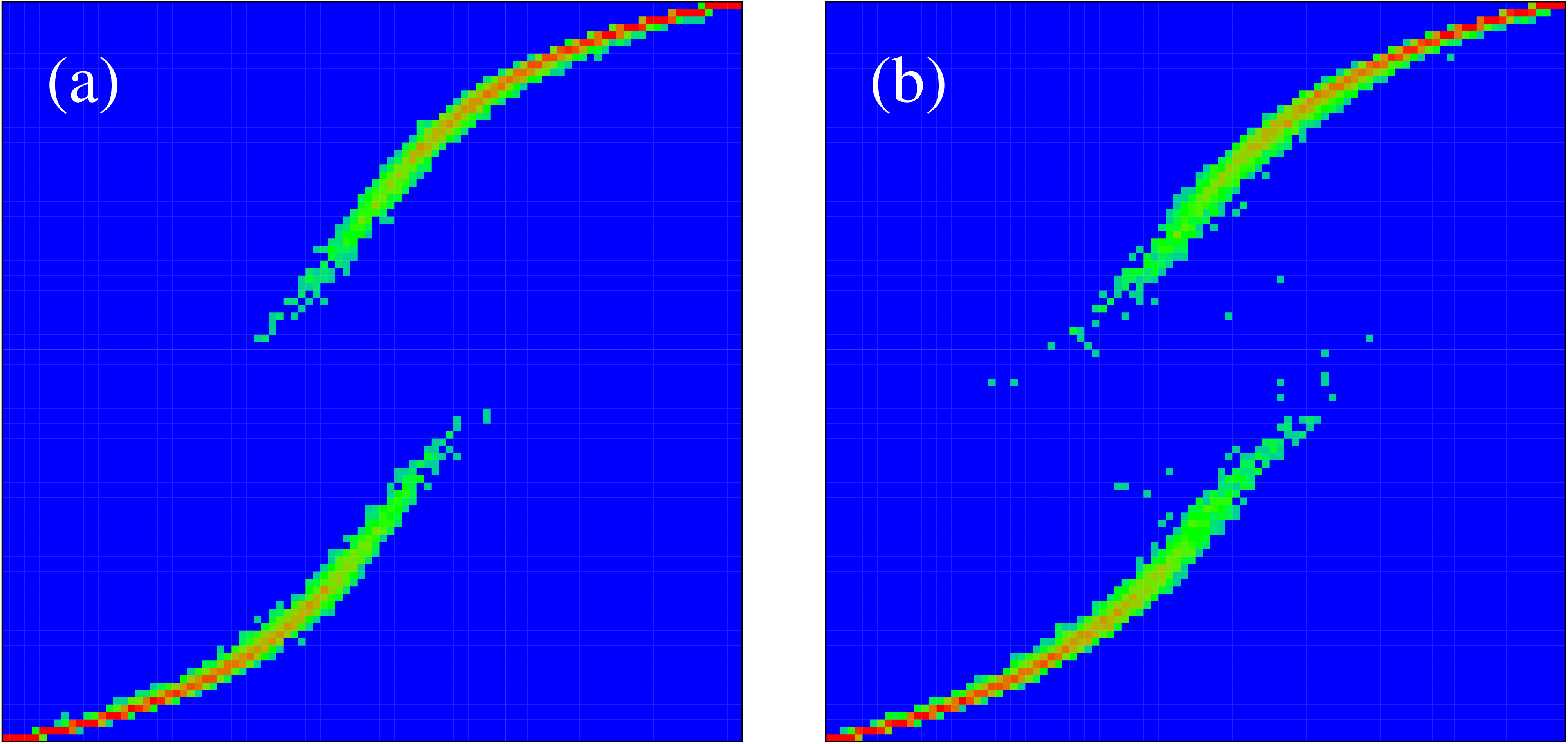}
\end{center}
\vglue -0.3cm
\caption{ (Color online)
PROF-Sznajd model, option 1:
density plot of probability $W_f$ 
to find a final red fraction $f_f$, shown in $y-$axis,
in dependence on an initial red fraction $f_i$, shown in $x-$ axis;
data are shown inside the unit square $0\leq f_i,f_f \leq 1$.
The values of $W_f$ are defined as a relative number of
realizations found inside each of $100\times100$ cells
which cover  the whole unit square. Here $N_r=10^4$ realizations of
randomly distributed colors are used to obtained $W_f$ values;
for each realization the
time evolution is followed up the convergence time with up to 
$\tau=10^7$ steps.
(a) Cambridge network;
(b) Oxford network;
here $N_g=8$.
The probability $W_f$ is proportional to
color changing from zero (blue/black) to unity
(red/gray).
After \cite{kandiah:2012}.
\label{fig14_10}}
\end{figure}

Our interpretation of these results is the following. 
For small values of $a \ll 1$ the opinion of a given society member
is determined mainly by the PageRank of 
neighbors to whom he/she points (outgoing links).
The PageRank probability $P$ of nodes to which 
many nodes point is usually high, 
since $P$ is proportional to the number of ingoing links.
Thus at $a \ll 1$ the society is composed of members 
who form their opinion by listening to an elite opinion. 
In such a society its elite
with one color opinion can impose this opinion on 
a large fraction of the society. 
Indeed, the direct analysis of the case, where the top $N_{top}=2000$
nodes of PageRank index have the same red color,
shows that this 1\% of the society elite can  
impose its opinion to about 50\% of the whole society at small $a$ values
(conformist society)
while at large $a$ values (tenacious society) this fraction drops significantly
(see Fig.4 in \cite{kandiah:2012}).
We attribute this to the fact that in Fig.~\ref{fig14_9}
we start with a randomly distributed opinion, 
since the opinion of the elite has two fractions of 
two colors this creates a bistable situation when
the two fractions of society follow the opinions 
of this divided elite, which makes the situation bistable on a larger interval
of $f_i$ compared to the case of a tenacious society at $a \rightarrow 1$.
When we replace in (\ref{eqopinion1}) $P$ by $1$ then 
the bistability disappears. 

However, the detailed understanding of the opinion formation
on directed networks still waits it development. Indeed, the 
results of PROF model for the LiveJournal and Twitted networks
show that the bistability in these networks practically disappears.
Also e.g. for the Twitter network studied in Sec.~\ref{s10.1},
the elite of $N_{top} =35000$ (about 0.1\% of the whole society)
can impose its opinion to 80\% of the society at small $a < 0.15$
and to about 30\% for $a>0.15$ \cite{kandiah:2012}.
It is possible that a large number of links 
between top PageRank nodes in Twitter
creates a stronger tendency to a totalitarian opinion 
formation comparing to the case of University networks.
At the same time the studies of opinion formation with the PROF model
on the Ulam networks \cite{levon}, which have not very  
large number of links, show practically no bistability in opinion
formation. It is expected that a small number of loops 
is at the origin of such a difference in respect to university networks.

Finally we discuss a more generic version of opinion formation
called the PROF-Sznajd model \cite{kandiah:2012}. 
Indeed, we see that in the PROF model
on university network opinions of small groups of red nodes
with $f_i < f_c$ are completely suppressed that
seems to be not very realistic.
In fact, the Sznajd model \cite{sznajd}
features the idea of resistant groups of a society 
and thus incorporates a well-known 
trade union principle ``United we stand, divided we fall''.
Usually the Sznajd model is studied on regular lattices.
Its generalization for directed networks
is done on the basis of the notion
of group of nodes $N_g$ at each discrete time step $\tau$. 

The evolution of group is defined by the following rules:

$(a)$ we pick in the network by random a node $i$ and consider 
the polarization of
$N_g-1$ highest PageRank nodes pointing to it;

$(b)$ if node $i$ and all other $N_g-1$ nodes have 
the same color (same polarization),
then these $N_g$ nodes form a group whose effective 
PageRank value is the sum of all 
the member values $P_g=\sum_{j=1}^{N_g}P_j$; 

$(c)$ consider all the nodes 
pointing to any member of the group
and check all these nodes $ n$ directly linked to the group: if 
an individual node PageRank value $P_n$ is less than the defined
above $P_g$ , the node joins the group by taking the same color
(polarization) as the group nodes and increase $P_g$ by the value
of $P_n$; if it is not the case, a node is left unchanged.

The above time step is repeated many times during time $\tau$, 
counting the number of steps and choosing a random node $i$ on
each next step. 

The time evolution of this PROF-Sznajd model converges
to a steady state approximately after $\tau \approx 10 N$ steps.
This is compatible with the results obtained for the PROF model.
However, the statistical fluctuations in the steady-state regime 
are present keeping the color distribution only on average.
The dependence of the final fraction of red nodes 
$f_f$ on its initial value $f_i$ is shown by the density 
plot of probability $W_f$ in Fig.~\ref{fig14_10}
for the university networks. 
The probability $W_f$ is obtained from many initial random realizations in
a similar way to the case of Fig.~\ref{fig14_9}.
We see that there is a significant difference compared to the PROF model: 
now even at small values of $f_i$ we find small but finite values of 
$f_f$, while in the PROF model the red color disappears at $f_i<f_c$.
This feature is related to the essence of the Sznajd model: 
here, even small groups can resist against the totalitarian opinion. Other
features of Fig.~\ref{fig14_10} are similar to those found for the PROF model: 
we again observe bistability of opinion formation. The
number of nodes $N_g$, 
which form the group, does not significantly affect the distribution 
$W_f$ (for studied $3 \leq N_g \leq 13$). 

The above studies of opinion formation models on scale-free networks
show that the society elite, corresponding to the top PageRank nodes, can
impose its opinion on a significant fraction of the society.
 However, for a homogeneous
distribution of two opinions, there exists a bistability 
range of opinions which depends on
a conformist parameter characterizing the opinion formation. 
The proposed PROF-Sznajd model shows
that totalitarian opinions can be escaped from by small subcommunities. 
The enormous development of social networks in the last few years 
definitely shows that the analysis of opinion
formation on such networks requires further investigations.

\section{Discussion}
\label{s15}

Above we considered many examples of real directed networks 
where the Google matrix analysis finds useful applications.
The examples belong to various sciences varying from 
WWW, social and Wikipedia networks,
software architecture to world trade, games, DNA sequences
and Ulam networks. It is clear that the concept of Markov chains 
and Google matrix represents now the mathematical foundation
of directed network analysis. 

For Hermitian and unitary
matrices there are now many universal concepts, developed
in theoretical physics, so that the main properties of
such matrices are well understood. Indeed, such characteristics
as level spacing statistics, localization and delocalization
 properties of eigenstates, Anderson transition \cite{anderson:1958},
quantum chaos features can 
be now well handled by various theoretical methods
(see e.g. \cite{akemann:2011,evers:2008,guhr:1998,haake:2010,mehta:2004}).
A number of generic models has been developed in this area
allowing to understand the main effects via numerical simulations
and analytical tools.

In contrast to the above case of Hermitian or unitary matrices, the
studies of matrices of Markov chains of directed networks are now only at 
their initial stage. In this review,  
on examples of real networks we illustrated  
certain typical properties of such matrices.
Among them there is the fractal Weyl law, which has certain
 traces in the field of quantum chaotic scattering,
but the main part of features are new ones. In fact, the spectral 
properties of Markov chains had not been investigated 
on a large scale. We try here to provide an introduction to 
the properties of such matrices which contain
all information about large scale directed networks.
The Google matrix is like {\it The Library of Babel} 
\cite{borges}, which contains everything.
Unfortunately, we are still not able to find generic Markov matrix  models
which reproduce the main features of the real networks.
Among them there is the possible spectral degeneracy at damping $\alpha=1$,
absence of spectral gap, algebraic decay of eigenvectors.
Due to absence of such  generic models it is still difficult to capture the
main properties of real directed networks and to understand or predict their
variations with a change of network parameters.
At the moment the main part of real networks
have an algebraic decay of PageRank vector with an exponent
$\beta \approx 0.5 - 1$. However, certain examples of Ulam networks
(see Figs.~\ref{fig6_6},~\ref{fig6_7})
show that a delocalization of PageRank probability over the whole network
can take place. Such a phenomenon looks to be similar 
to the Anderson transition for electrons in disordered solids.
It is clear that if an Anderson delocalization of PageRank
would took place, as a result of further developments
of the WWW, the search engines based on the PageRank 
would loose their efficiency since the ranking would become 
very sensitive to various fluctuations. 
In a sense the whole world would go blind 
the day such a delocalization takes place.
Due to that a better understanding of the fundamental properties of Google
matrices and their dependencies on various system parameters
have a high practical significance.
We believe that the theoretical research in this direction
should be actively continued. In many respects, as {\it the Library of Babel},
the Google matrix still keeps its secrets
to be discovered by researchers from various fields of science.
We hope that a further research will allow 
``{\it to formulate a general theory of the Library 
and solve satisfactorily the problem which no conjecture had deciphered: 
the formless and chaotic nature of almost all the books.}'' \cite{borges}

\section{Acknowledgments}
\label{s16}

We are grateful to 
our colleagues M.~Abel, A.~D.~Chepeliankii, Y.-H.~Eom, B.~Georgeot,
O.~Giraud,  V.~Kandiah, O.~V.~Zhirov for fruitful collaborations
on the topics included in this review.
We also thank our partners of
the EC FET Open project NADINE
A.~Bencz\'ur, N.~Litvak, S.~Vigna and  colleague A.Kaltenbrunner
for illuminating discussions.
Our special thanks go to Debora Donato
for her insights at our initial
stage of this research.

Our research presented here is supported 
in part by the EC FET Open project 
``New tools and algorithms for directed network analysis''
(NADINE $No$ 288956). This work was granted access 
to the HPC resources of 
CALMIP (Toulouse) under the allocation 2012-P0110.
We also thank the
United Nations Statistics Division for provided
help and friendly access to the UN COMTRADE database.


\end{document}